\documentclass[twocolumn]{aastex62}
\usepackage{color}
\usepackage{framed}
\usepackage{natbib}
\usepackage{textcomp}
\usepackage{rotating}
\usepackage{longtable}
\usepackage{graphicx}
\usepackage{float}

\newcommand{\APF}{\emph{APF}} 
\newcommand{\C}{\ensuremath{^{\circ}C\;}}

\newcommand{\Mearth}{\ensuremath{M_\earth}}
\newcommand{\Msun}{\ensuremath{M_\sun}}
\newcommand{\PFS}{\emph{PFS}}
\newcommand{\Pcyc}{$P_{\rm cyc}$}
\newcommand{\Porb}{$P_{\rm orb}$}
\newcommand{\Prot}{$P_{\rm rot}$}
\newcommand{\RVSearch}{{\tt RVSearch}}
\newcommand{\RVsearch}{{\tt RVSearch}}
\newcommand{\Radvel}{{\tt RadVel}}

\newcommand{\radvel}{{\tt RadVel}}
\newcommand{\Rearth}{\ensuremath{R_\earth}}
\newcommand{\TESS}{\emph{TESS}}
\newcommand{\Teff}{$T_{\rm eff}$}

\newcommand{\feh}{\ensuremath{\left[{\rm Fe}/{\rm H}\right]}}

\newcommand{\gaia}{\emph{Gaia}}

\newcommand{\habex}{\emph{HabEx}}
\newcommand{\HARPS}{\emph{HARPS}}
\newcommand{\hipparcos}{\emph{Hipparcos}}
\newcommand{\HIRES}{\emph{HIRES}}

\newcommand{\kms}{km\,s$^{-1}$}

\newcommand{\logg}{\ensuremath{\log g}}

\newcommand{\logl}{log($L/L_{\odot}$)}
\newcommand{\logrphk}{log\,$R^{\prime}_{HK}$}
\newcommand{\loglxlbol}{log($L_X$/$L_{\rm bol}$)}

\newcommand{\lsun}{\ensuremath{L_\sun}}
\newcommand{\luvoir}{\emph{LUVOIR}}

\newcommand{\luvoirA}{\emph{LUVOIR-A}}
\newcommand{\luvoirB}{\emph{LUVOIR-B}}
\newcommand{\masyr}{\ensuremath{\rm mas\,yr^{-1}}}

\newcommand{\mearth}{\ensuremath{M_\earth}}

\newcommand{\mj}{\ensuremath{\,M_{\rm J}}}
\newcommand{\Mjup}{\ensuremath{\,M_{\rm J}}}

\newcommand{\msini}{$m$\,sin\,$i$}

\newcommand{\msun}{\ensuremath{M_\sun}}
\newcommand{\ms}{\ensuremath{\rm m\,s^{-1}}}

\newcommand{\prot}{$P_{\rm rot}$}

\newcommand{\rsun}{\ensuremath{R_\sun}}
\newcommand{\rvsearch}{{\tt RVSearch}}
\newcommand{\serval}{{\tt SERVAL}}

\newcommand{\starshade}{\emph{Starshade Rendezvous}}
\newcommand{\teff}{$T_{\rm eff}$}
\newcommand{\UCLES}{\emph{UCLES}}

\newcommand{\vsini}{$v$sin$i$}

\shorttitle{EPRV Archival RVs}
\shortauthors{Laliotis et al.}

\begin{document}
\title{Doppler Constraints on Planetary Companions to Nearby Sun-like Stars:\\ 
An Archival Radial Velocity Survey of {Southern} Targets for Proposed NASA Direct Imaging Missions
\footnote{This paper includes data gathered with the 6.5 meter Magellan Telescopes located at Las Campanas Observatory, Chile.}}

\correspondingauthor{Katherine Laliotis}
\email{laliotis.2@osu.edu}

\author[0000-0002-6111-6061]{Katherine Laliotis}
\affiliation{Department of Physics, The Ohio State University, 191 West Woodruff Ave., Columbus, OH 43210, USA}

\author[0000-0002-0040-6815]{Jennifer A. Burt}
\affil{Jet Propulsion Laboratory, California Institute of Technology, 4800 Oak Grove Drive, Pasadena, CA 91109, USA}

\author[0000-0003-2008-1488]{Eric~E.~Mamajek}
\affiliation{Jet Propulsion Laboratory, California Institute of Technology, 4800 Oak Grove Drive, Pasadena, CA 91109, USA}
\affiliation{Department of Physics and Astronomy, University of Rochester, Rochester, NY 14627-0171, USA}

\author[0000-0002-4860-7667]{Zhexing Li}
\affiliation{Department of Earth and Planetary Sciences, University of California, Riverside, CA 92521, USA}

\author{Volker Perdelwitz}
\affiliation{Kimmel fellow, Department of Earth and Planetary Sciences, Weizmann Institute of Science, Rehovot, 76100, Israel}
\affiliation{Hamburger Sternwarte, Universit{\"a}t Hamburg, Gojenbergsweg 112, D-21029 Hamburg, Germany}

\author[0000-0001-5290-2952]{Jinglin~Zhao}
\affil{Department of Astronomy \& Astrophysics, The Pennsylvania State University, 525 Davey Lab, University Park, PA 16802, USA}

\author[0000-0003-1305-3761]{R.~Paul~Butler}
\affiliation{Earth \& Planets Laboratory, Carnegie Institution for Science, 5241 Broad Branch Road, NW, Washington, DC 20015, USA}

\author[0000-0002-6153-3076]{Bradford Holden}
\affil{UCO/Lick Observatory, Department of Astronomy and Astrophysics, University of California at Santa Cruz, Santa Cruz, CA, 95064, USA}

\author[0000-0001-8391-5182]{Lee Rosenthal}
\affiliation{Cahill Center for Astronomy \& Astrophysics, California Institute of Technology, Pasadena, CA 91125, USA}

\author[0000-0003-3504-5316]{B. J. Fulton}
\affiliation{Cahill Center for Astronomy \& Astrophysics, California Institute of Technology, Pasadena, CA 91125, USA}
\affiliation{IPAC-NASA Exoplanet Science Institute, Pasadena, CA 91125, USA}

\author[0000-0001-6039-0555]{Fabo Feng}
\affiliation{Tsung-Dao Lee Institute, Shanghai Jiao Tong University, 800 Dongchuan Road, Shanghai 200240, People's Republic of China}
\affiliation{Department of Astronomy, School of Physics and Astronomy, Shanghai Jiao Tong University, 800 Dongchuan Road, Shanghai 200240, People's Republic of China}

\author[0000-0002-7084-0529]{Stephen R. Kane}
\affiliation{Department of Earth and Planetary Sciences, University of California, Riverside, CA 92521, USA}

\author[0000-0002-5726-7000]{Jeremy Bailey}
\affiliation{School of Physics, University of New South Wales, Sydney, NSW 2052, Australia}

\author[0000-0003-0035-8769]{Brad Carter}
\affiliation{University of Southern Queensland, Centre for Astrophysics, USQ Toowoomba, QLD 4350, Australia}

\author[0000-0002-5226-787X]{Jeffrey~D.~Crane}
\affiliation{The Observatories of the Carnegie Institution for Science, 813 Santa Barbara Street, Pasadena, CA 91101, USA}

\author[0000-0001-9800-6248]{Elise~Furlan}
\affiliation{NASA Exoplanet Science Institute, Caltech/IPAC, Mail Code 100-22, 1200 E. California Blvd., Pasadena, CA 91125, USA}

\author[0000-0003-2519-6161]{Crystal~L.~Gnilka}
\affil{NASA Ames Research Center, Moffett Field, CA 94035, USA}

\author[0000-0002-2532-2853]{Steve~B.~Howell}
\affil{NASA Ames Research Center, Moffett Field, CA 94035, USA}

\author[0000-0002-3253-2621]{Gregory Laughlin}
\affil{Department of Astronomy, Yale University, New Haven, CT, 06511, USA}

\author[0000-0002-8681-6136]{Stephen~A.~Shectman}
\affiliation{The Observatories of the Carnegie Institution for Science, 813 Santa Barbara Street, Pasadena, CA 91101, USA}

\author{Johanna~K.~Teske} \affiliation{Earth \& Planets Laboratory, Carnegie Institution for Science, 5241 Broad Branch Road, NW, Washington, DC 20015, USA}

\author[0000-0002-7595-0970]{C. G. Tinney}
\affiliation{School of Physics and Australian Centre for Astrobiology, University of New South Wales, Sydney 2052, Australia}

\author[0000-0001-7177-7456]{Steven S. Vogt}
\affil{UCO/Lick Observatory, Department of Astronomy and Astrophysics, University of California at Santa Cruz, Santa Cruz, CA, 95064, USA}

\author[0000-0002-6937-9034]{Sharon Xuesong Wang}
\affiliation{Department of Astronomy, Tsinghua University, Beijing 100084, People's Republic of China}

\author[0000-0001-9957-9304]{Robert A. Wittenmyer}
\affiliation{University of Southern Queensland, Centre for Astrophysics, USQ Toowoomba, QLD 4350, Australia}

\begin{abstract}
Directly imaging temperate rocky planets orbiting nearby, Sun-like stars with a {6-m-class IR/O/UV} space telescope, {recently dubbed the {\it Habitable Worlds Observatory}}, is a high priority goal of the Astro2020 Decadal Survey. To prepare for future direct imaging surveys, the list of potential targets should be thoroughly vetted to maximize efficiency and scientific yield. We present an analysis of archival radial velocity data for southern stars from the NASA/NSF Extreme Precision Radial Velocity Working Group's list of high priority target stars for future direct imaging missions (drawn from the \habex, \luvoir, and \starshade\, studies). For each star, we constrain the region of companion mass and period parameter space we are already sensitive to based on the observational baseline, sampling, and precision of the archival RV data. Additionally, for some of the targets we report new estimates of magnetic activity cycle periods, rotation periods, improved orbital parameters for previously known exoplanets, and new candidate planet signals that require further vetting or observations to confirm. Our results show that for many of these stars we are not yet sensitive to even Saturn-mass planets in the habitable zone, let alone smaller planets, highlighting the need for future EPRV vetting efforts before the launch of a direct imaging mission. We present evidence that the candidate temperate super-Earth exoplanet HD 85512 b is most likely due to the star's rotation, and report an RV acceleration for $\delta$ Pav which supports the existence of a distant giant planet previously inferred from astrometry. 
\end{abstract}

\keywords{
Exoplanet astronomy (486), 
Exoplanet systems (484), 
Radial velocity (1332)}

\section{Introduction \label{sec:intro}}

In order to further push the boundaries of the search for life elsewhere in the universe, astronomers must advance our capabilities to detect temperate, terrestrial planets orbiting Sun-like stars and to characterize their atmospheres.
To detect and spectrally characterize many such planets in reflected light, it is expected that future space-based direct imaging (DI) missions will employ starlight suppression technologies such as coronagraphs or starshades, and survey 
{of order $\sim$100}
of the nearest Sun-like stars \citep{ESS,Astro2020}. 
Habitable Exoplanet Observatory (\habex) and the Large Ultraviolet Optical Infrared Surveyor (\luvoir), two proposed space-based DI mission concepts considered by the {\it 2020 Decadal Survey on Astronomy and Astrophysics} (Astro2020\footnote{https://www.nationalacademies.org/our-work/decadal-survey-on-astronomy-and-astrophysics-2020-astro2020}), were designed to obtain direct atmospheric spectra and enable atmospheric characterization of small, temperate planets \citep{Gaudi2020,LUVOIRFinalReport}. 
{\it Pathways to Habitable Worlds} was a priority science theme in the Astro2020 Decadal Survey\footnote{Released November 2021, near the completion of this study}, which included a recommendation\footnote{{\it ``Recommendation: After a successful mission and technology maturation program, NASA should embark on a program to realize a mission to search for biosignatures from a robust number of about $\sim$25 habitable zone planets and to be a transformative facility for general astrophysics. If mission and technology maturation are successful, as determined by an independent review, implementation should start in the latter part of the decade, with a target launch in the first half of the 2040s}."} for NASA to work {\footnote{NASA Administator Bill Nelson recently announced plans in December 2022 to the National Academies on the 50th anniversary of the Apollo 17 mission to proceed with the Decadal mission concept for a large UV/Vis/IR space telescope with the name {\it Habitable Worlds Observatory}.}} towards launching a 6-m class UV/Visible/IR space observatory in the early 2040s to spectrally search for biosignatures in the atmospheres of directly-imaged temperate rocky planets orbiting nearby stars\footnote{We note that an alternative approach to discovering and characterizing rocky temperate exoplanets via space-based mid-infrared nulling interferometry is being pursued for the Large Interferometer for Exoplanets (LIFE) concept for ESA's Voyage 2050 program \citep{Quanz2022,Dannert2022,Konrad2022,Hansen2022}.}.
The scale of the proposed observatory is intermediate between the {\it HabEx} and {\it LUVOIR-B} concepts, and will require further technology and science maturation and trade studies to converge on an architecture before project implementation later in the 2020s. 
To inform trade studies and simulate mission yields, and ultimately to fulfill the Astro2020 Decadal goal of searching for biosignatures in the atmospheres of $\sim$25 imaged exo-Earths, one requires a prioritized and carefully vetted target list.\\

The stars that are chosen for a future DI mission must meet criteria related to their \teff, brightness (e.g. $V_{\rm{mag}}$), luminosity, multiplicity, and distance from Earth.
Of primary importance is the maximum separation that a potentially habitable planet can achieve in its orbit around its star, as seen from Earth.
The habitable zone annuli scale as the square root of the luminosity, such that one can define the Earth Equivalent {Insolation} Distance (EEID) as $\sqrt{L/L_{\odot}}$ au, or in angular separation $\theta_{EEID}$ = $\sqrt{L/L_{\odot}}$ $D_{pc}^{-1}$ = $\sqrt{L/L_{\odot}}$ $\varpi$, for a star at distance $D_{pc}$ ({parsecs}) (=1/$\varpi$) and parallax $\varpi$ ({arcseconds}). 
For a planet to be visible, at least part of its orbit must be outside the inner working angles (IWA) for the observatory's means of starlight suppression -- usually either a coronagraph or starshade\footnote{For LUVOIR's ECLIPS instrument (a coronagraph with imaging and imaging spectroscopy), IWA = 3.5$\lambda$/$D$ for wavelength $\lambda$ (between 0.2-2.0\,$\mu$m) and aperture size $D$ (15\,m for LUVOIR-A, 8\,m for LUVOIR-B) \citep{LUVOIRFinalReport}. For \habex\, ($D$ = 4\,m), IWA = 2.4$\lambda$/$D$ for the coronagraph ($\lambda$ range 0.48-1.8\,$\mu$m, corresponding to 62 mas for 0.5\,$\mu$m) and 58\,mas for the starshade over $\lambda$ = 0.3-1.0\,$\mu$m. Astro2020 recommended an observatory with $D$ $\simeq$ 6\,m, however since multiple architectures will likely be considered, the IWA is not yet defined -- but should ultimately be a low multiple of $\lambda$/$D$.} -- so that the incoming starlight does not overwhelm the planet's signal.
Instrumentation limits constrain the primary candidate stars' locations to be within a range where a $\sim$1 AU orbit would have a minimum on-sky separation of $\sim$40 mas for \luvoir\,  \citep{LUVOIRFinalReport} and {larger} separations are required for \habex\, and \starshade. 
Given the limited number of Sun-like stars in the solar neighborhood, and the sizes of their habitable zones, this places tight limits on the distance ranges and sample sizes of targets well suited to Earth-analog searches with a DI mission. An ideal observation candidate must both exhibit Sun-like characteristics and be close enough that its habitable zone would be accessible to a direct imaging mission's instruments. 

The accuracy to which the properties of the exoplanets' atmosphere can be determined will rely on precise measurements of the planets' surface gravities, which in turn require precise planet mass measurements \citep{Batalha2019}. 
{There are two practical methods for the foreseeable future for measuring the masses of temperate rocky exoplanets orbiting Sun-like stars -- radial velocity and astrometry -- both of which would require considerable advancement to achieve this goal \citep{Lovis2010,Quirrenbach2010,ESS}.} It is generally acknowledged, however, that Extreme Precision Radial Velocity (EPRV) measurements obtained via spectrographs with single measurement precisions \textless\ 10\,c\ms\ are likely the most direct route to detecting Earth analogs around Sun-like stars \citep{ESS}. The current generation of RV instruments are beginning to demonstrate single measurement precisions at the 30-50\,c\ms\ level \citep[see, e.g., ][]{Pepe2021, Brewer2020, Trifonov2021, Seifahrt2018}. Further improvements to reach the 10 c\ms\ level will require advances in sustained instrument stability, wavelength calibration, and data extraction and analysis techniques with a focus on methods for mitigating stellar variability.

Exploring solutions to these challenges that prevent mass measurements for Earth analogs was the goal of the Extreme Precision Radial Velocity (EPRV) Working Group chartered by NASA and the NSF\footnote{https://exoplanets.nasa.gov/exep/NNExplore/EPRV/.}. The EPRV Working Group was charged with devising a path towards developing methods and facilities that will be capable of accurately measuring the masses of temperate terrestrial exoplanets orbiting Sun-like stars\footnote{The final NASA concept study reports for the \habex, \luvoir, and \starshade\, reports for the Astro 2020 Decadal Survey are posted at:  https://science.nasa.gov/astrophysics/2020-decadal-survey-planning}. {The Working Group's final report \citep{Crass2021}} includes recommendations for advancements in stellar activity and telluric mitigation, instrument efficiency and accuracy, and research and analysis techniques.

In parallel to these advancements in RV instrumentation and analysis, there must also be a concerted effort to better understand the potential DI target stars. One way to do this is by studying archival RV data sets taken using previous, less precise ($\sigma_{\rm{RV}}$ \textless\ 5\,\ms), generations of RV spectrographs. The EPRV Working Group curated a list of $\sim$100 nearby, Sun-like (F9-K7) stars that are promising candidates for a future DI mission, many of which have already been included in multiple RV surveys over the past three decades as the community's interests have often been tied to bright G and K dwarfs.

In this study, we analyze archival radial velocity and stellar activity data from the \HARPS, \HIRES, \UCLES, \APF, and \PFS\ instruments for {49} southern hemisphere stars identified as promising future DI mission targets. We perform planet injection and recovery tests to assess the completeness of the existing RV data as a function of planet mass and orbital period, so as to identify regions of mass/period parameter space in which planet signals might still be hiding. Any mass/period gaps identified in this work can then be filled by directed future observations, contributing to the completeness of the target star list data. In preparing the RV data sets for the injection/recovery analysis we first identify and remove significant signals from the RV time series. This results in the detection of numerous, previously confirmed exoplanets; a number of new planet candidates; and rotation and magnetic activity cycles within the data.

The structure of this paper is as follows. In \S2, we discuss the stars chosen for this project and the types and sources of the data analyzed. {In \S3 we detail the sources and treatments of the data sets used in this work. In \S4,} we present the different methods of analysis used to characterize the stars' existing RV sensitivity. In \S5-7 we explain the results of our analysis. Each star on the target list has a subsection including updates to parameters of any known planets, evidence of strong stellar activity cycles, and any new signals recovered. We address  those targets which lacked any significant signals in \S7. {\S8 contains a general discussion of our results, including highlights of the analysis carried out in this work and exploration of major gaps we have identified in the archival RV data.} Finally, in \S9 we cover the conclusions drawn from this work, and {identify} future work necessary before any target list is finalized for a direct imaging mission concept. The full set of figures for each target, including radial velocity, {S-index}, completeness contour, and, if relevant, {H$\alpha$} activity and speckle imaging plots can be found in the online journal.

\section{Stellar Target List}

\begin{figure}[tbp]
\includegraphics[width=.46 \textwidth]{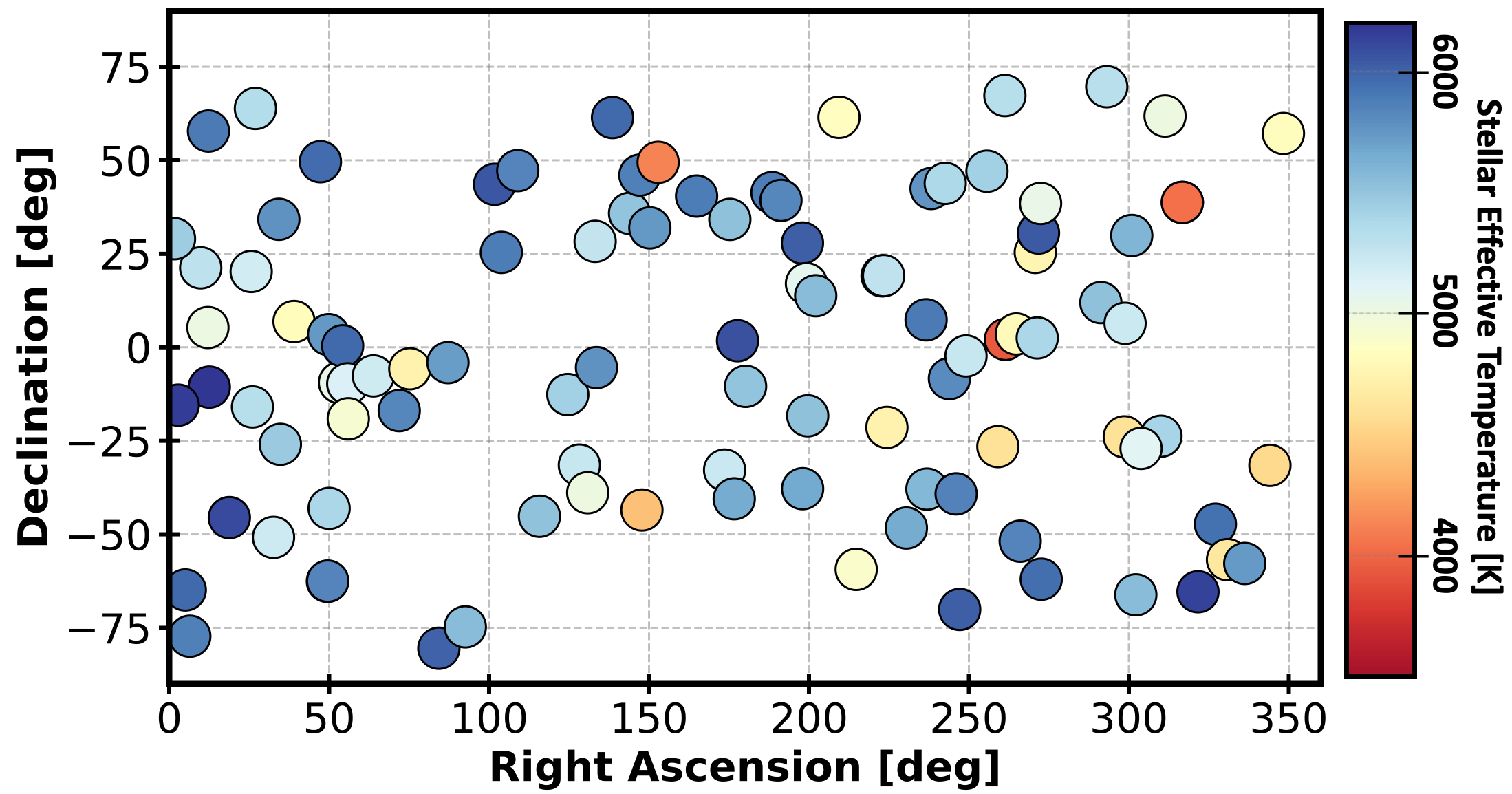}
\caption{Stars identified by the EPRV Working Group as potential targets for future direct imaging missions such as HabEx and LUVOIR that would aim to detect and characterize Earth analog exoplanets. This work focuses primarily on the stars located in the southern hemisphere.
\label{fig:targetlist}}
\end{figure}

Our list of target stars is drawn from the EPRV Working Group, and a full description of the selection process and criteria considered can be found in its final presentation\footnote{https://exoplanets.nasa.gov/internal$\_$resources/1556/} and report \citep{Crass2021}. 
In brief, the Working Group cross-matched target lists provided by the \habex, \luvoirA, \luvoirB, and \starshade\, teams \citep{Gaudi2020, LUVOIRFinalReport, Seager2018} to assemble a combined list of potential target stars. 
It then compiled information on the stars' effective temperatures, apparent magnitudes, rotational velocities, metallicities, and surface gravities, among other traits.
From this catalog, they culled those stars with spectral types from F7-K9, projected rotational velocities \vsini\,$<$\,5\,\kms, and that appear on at least two of the mission concept target lists. 
Stars were not eliminated based on knowledge of their stellar activity levels as the characterization and mitigation of stellar variability is an active field and we may yet overcome the obstacles it presents. 
The resulting list includes 101 stars (Figure \ref{fig:targetlist}).

\begin{figure}[tbp]
\includegraphics[width=.48 \textwidth]{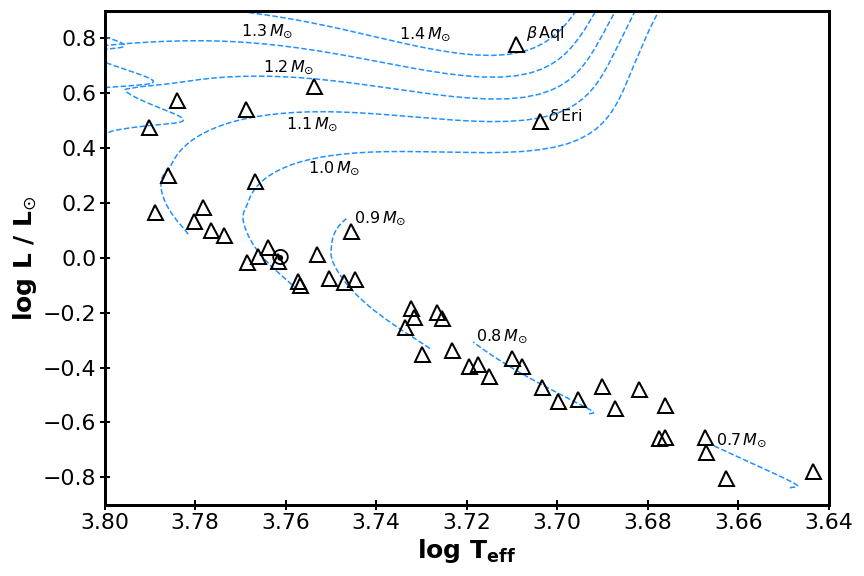}
\caption{HR diagram containing the target stars in this study. The dashed blue lines are MIST V1.2
evolutionary tracks
from \citet{Choi2016} covering masses 0.7 - 1.4 M$_{\odot}$ over age range 100Myr - 14 Gyr. The tracks adopts protosolar mix (initial $Y = 0.2703$, $Z = 0.0142857$, [$\alpha$/Fe]=0, $v/v_{crit}=0${)}.
The Sun is depicted using the $\odot$ symbol.
\label{fig:HRDiagram}}
\end{figure}

For this work, we have chosen to focus primarily on the {53} stars from this list located in the southern hemisphere{\footnote{It should be noted that this list was assembled before the Astro2020 Decadal Survey was released, which recommended a 6-m-class space telescope. From subsequent analysis and literature survey (Mamajek \& Stapelfeldt, in prep.), ten of the sample stars in Table 1 may be undesirable targets for a survey for potentially habitable exoplanets with a 6-m-class space telescope: 4 systems have habitable zones prohibitively close to their stars (HD 85512 [EEID\,=\,36\,mas], HD 23356 [EEID\,=\,40\,mas], HD 125072 [EEID\,=\,49\,mas], HD 196761 [EEID\,=\,36\,mas]), 5 stars have close
($<$3\arcsec) companions (HD 16160, HD 147584, HD 13445A, HD 160346), one has high luminosity and its HZ rocky planets may have planet-star brightness ratios that are prohibitive low (HD 188512). These ten stars will be omitted from a NASA ExEP mission target list meant to inform precursor science for the {\it Habitable Worlds Observatory}.}}
(Figure \ref{fig:HRDiagram}) due to the recent publication of archival HARPS RV data {that is now available on the RVBank website} \citep{Trifonov2020}.

\startlongtable
\begin{deluxetable*}{llccrrllrl}
\tablecaption{Sample and Stellar Parameters \label{tab:targets}}
\tablehead{\colhead{HD    } & \colhead{GJ    } & \colhead{\teff\,} & \colhead{\logg\,} & \colhead{\feh\,} & \colhead{Ref.} & \colhead{Spec. Type} & \colhead{Ref.} & \colhead{$\log{L/\lsun}$} & \colhead{Ref.(L)}}
\startdata
693    & 10	  & 6169 & 4.07	& -0.34	& 1	 & F8V Fe-0.8 CH-0.5 & 2	&  0.477 & 3\\
1581   & 17	  & 5977 & 4.51	& -0.18	& 4	 & F9.5V 	         & 2	&  0.101 & 3\\
2151   & 19	  & 5873 & 3.98	& -0.04	& 5	 & G0V 	             & 2	&  0.541 & 27\\
4628   & 33	  & 5009 & 4.62	& -0.24	& 1	 & K2V 	             & 6	& -0.523 & 3\\
7570   & 55   & 6111 & 4.36	&  0.17	& 7	 & F9V Fe+0.4 	     & 2	&  0.302 & 3\\
13445  & 86A  & 5217 & 4.56	& -0.23	& 9	 & K1V 	             & 2	& -0.389 & 3\\
14412  & 95	  & 5368 & 4.55	& -0.47	& 10 & G8V 	             & 2	& -0.351 &	3\\
16160  & 105A &	4866 & 4.66	& -0.12	& 11 & K3V 	             & 8	& -0.549 &	3\\
20766  & 136  &	5713 & 4.48	& -0.20 & 12 & G2V 	             & 2	& -0.100 &	3\\
20794  & 139  &	5398 & 4.41	& -0.41	& 13 & G8V 	             & 2	& -0.184 &	3\\
20807  & 138  &	5837 & 4.47	& -0.22	& 12 & G1V 	             & 8	&  0.007 &	3\\
22049  & 144  &	5050 & 4.60 & -0.09	&  5 & K2V 	             & 8	& -0.471 &	3\\
23249  & 150  &	5057 & 3.80	&  0.08	& 15 & K1IV 	         & 6	&  0.500 &	16\\
23356  & ...  &	4960 & 4.60 & -0.09	&  9 & K2.5V 	         & 2	& -0.515 &	3\\
26965  & 166A &	5128 & 4.37	& -0.37	& 17 & K0.5V 	         & 2	& -0.364 &	3\\
30495  & 177  &	5870 & 4.54	&  0.04	& 18 & G1.5V CH-0.5 	 & 2	& -0.015 &	3\\
32147  & 183  &	4745 & 4.57	&  0.19	&  9 & K3+V 	         & 6	& -0.537 &	3\\
38858  & 1085 &	5719 & 4.49	& -0.23	& 13 & G2V 	             & 6	& -0.083 &	3\\
39091  & 9189 &	6003 & 4.42	&  0.09	&  4 & G0V 	             & 2	&  0.186 &	3\\
50281  & 250A &	4758 & 4.92	&  0.14	& 11 & K3.5V 	         & 6	& -0.658 &	3\\
69830  & 302  &	5390 & 4.39	& -0.05	& 12 & G8+V 	         & 2	& -0.216 &	3\\
72673  & 309  &	5243 & 4.46	& -0.41	&  4 & G9V 	             & 2	& -0.394 &	3\\
75732  & 324A &	5328 & 4.58	&  0.46	&  1 & K0IV-V 	         & 6	& -0.197 &	3\\
76151  & 327  &	5776 & 4.54	&  0.11	& 19 & G2V 	             & 8	& -0.013 &	3\\
85512  & 370  &	4400 & 4.36	& -0.26	& 13 & K6V(k) 	         & 2	& -0.778 &	3\\
100623 & 432A &	5189 & 4.68	& -0.37	& 11 & K0-V 	         & 2	& -0.432 &	3\\
102365 & 442A &	5629 & 4.44	& -0.29	&  4 & G2V 	             & 2	& -0.074 &	3\\
102870 & 449  &	6083 & 4.08	&  0.24	&  5 & F8.5IV-V          & 20	&  0.576 &	3\\
114613 & 9432 &	5672 & 3.95	&  0.12	& 15 & G4IV 	         & 2	&  0.626 &	3\\
115617 & 506  &	5556 & 4.35	& -0.02	& 21 & G6.5V 	         & 8	& -0.078 &	3\\
125072 & 542  &	4899 & 4.55	&  0.28	&  9 & K3IV 	         & 2	& -0.466 &	3\\
131977 & 570A &	4744 & 4.76	&  0.12	& 11 & K4V 	             & 8	& -0.653 &	3\\
136352 & 582  &	5664 & 4.39	& -0.34	&  4 & G2-V 	         & 2	&  0.012 &	3\\
140901 & 599A &	5586 & 4.45	&  0.09	&  9 & G7IV-V 	         & 2	& -0.088 &	3\\
146233 & 616  &	5808 & 4.44	&  0.04	& 18 & G2Va 	         & 8	&  0.039 &	3\\
147584 & 624  &	6030 & 4.43	& -0.08	&  9 & F9V 	             & 2	&  0.134 &	3\\
149661 & 631  &	5289 & 4.61	&  0.05	&  1 & K0V(k) 	         & 2	& -0.335 &	3\\
156026 & 664  &	4600 & 4.70 & -0.34	&  9 & K5V(k) 	         & 2	& -0.803 &	3\\
160346 & 688  &	4808 & 4.56	& -0.08	&  9 & K2.5V 	         & 6	& -0.480 &	22\\
160691 & 691  &	5845 & 4.27	&  0.35	&  5 & G3IV-V 	         & 2	&  0.278 &	3\\
165341 & 702A &	5314 & 4.51	&  0.05	&  1 & K0-V 	         & 8	& -0.221 &	23\\
188512 & 771A &	5117 & 3.64	& -0.19	& 15 & G8IV 	         & 8	&  0.780 &	16\\
190248 & 780  &	5566 & 4.24	&  0.32	& 13 & G8IV 	         & 2	&  0.097 &	3\\
192310 & 785  &	5104 & 4.54	&  0.06	&  9 & K2+V 	         & 2	& -0.394 &	3\\
196761 & 796  &	5415 & 4.43	& -0.31	&  4 & G7.5IV-V          & 8	& -0.252 &	3\\
203608 & 827  &	6150 & 4.35	& -0.66	&  9 & F9V Fe-1.4 CH-0.7 & 2	&  0.166 &	3\\
207129 & 838  &	5937 & 4.49	&  0.00 &  4 & G0V Fe+0.4        & 2	&  0.082 &	3\\
209100 & 845  &	4649 & 4.63	& -0.19	&  9 & K4V(k) 	         & 2	& -0.654 &	3\\
216803 & 879  &	4647 & 4.88	&  0.07	& 24 & K4+Vk 	         & 2	& -0.707 &	3\\
\enddata
\tablecomments{
References:
(1) \citet{Takeda2005},
(2) \citet{Gray2006}, 
(3) \citet{Stassun2019},
(4) \citet{Sousa2008},
(5) \citet{Jofre2014},
(6) \citet{Gray2003},
(7) \citet{Ramirez2014}, 
(8) \citet{Keenan1989}, 
(9) \citet{Ramirez2013}, 
(10) \citet{Santos2004}, 
(11) \citet{Valenti2005}, 
(12) \citet{Adibekyan2016}, 
(13) \citet{Tsantaki2013},
(14) \citet{Gonzalez2010}, 
(15) \citet{Maldonado2016}, 
(16) \citet{Brewer2016}
(17) \citet{Montes2018},
(18) \citet{Spina2016},
(19) \citet{Mahdi2016},
(20) \citet{Gray2001},
(21) \citet{Sousa2018},
(22) \citet{Luck2017},
(23) \citet{Schofield2019},
(24) \citet{Santos2001},
(25) Luminosity for this star was calculated using the Virtual Observatory SED Analyzer version 7.0 \citep[VOSA;][]{Bayo2008} assuming zero extinction, $log(g)=4.0$, and BT-Settl-AGSS2009 model spectra. This produced a best fit log(L/\lsun) value of $+0.5407\pm0.0065$ for HD 2151.}
\end{deluxetable*}

\section{Data}\label{sec:Data}

\subsection{Radial Velocities}\label{subsec:RVs}

In this work, we include data sets consisting of unbinned radial velocity measurements from five different instruments: the High Accuracy Radial Velocity Planet Searcher \citep[HARPS, on the ESO 3.6m telescope,][]{Mayor2003}, the HIgh REsolution Spectrometer \citep[HIRES, on the 10m Keck I telescope,][]{Vogt1994},  the Levy spectrometer \citep[on the 2.4m Automated Planet Finder (APF) telescope,][]{Vogt2014}, the Planet Finder Spectrometer \citep[PFS, on the 6.5m Magellan Clay telescope,][]{Crane2006, Crane2008, Crane2010}, and the University College London Échelle Spectrograph \citep[UCLES, on the 3.9m Anglo-Australian Telescope,][]{Diego1990}. Table \ref{tab:RV_Obs} lists the number of {archival RV epochs (binned at a 12 hour cadence)} acquired by each facility. 
All radial velocity and activity indicator measurements for each star will be provided in a machine readable table alongside this paper {in their original, unbinned, form} (Table \ref{tab:Full_Timeseries}).

The \HARPS\, instrument uses multiple observing fibers; one directed at the stellar target, and one directed instead at a Th-Ar calibration lamp. The calibration lamp serves as a wavelength reference for the stellar spectra. \HARPS\, has a resolving power of $\sim$115,000 and a spectral grasp of {3800-6900\,\AA\,} \citep{Pepe2002, Cosentino2012}. 
All \HARPS\, RV data used in this work were downloaded from the \HARPS RVBank archive\footnote{https://www2.mpia-hd.mpg.de/homes/trifonov/HARPS$\_$RVBank.html} \citep{Trifonov2020}. These RVBank velocities were generated using the \serval\, pipeline \citep{Zechmeister2020} which uses a template matching approach \citep{Zechmeister2018}. 
For each star, \serval\, creates a high S/N template spectrum by shifting and co-adding all individual spectra of that star. 
The template is then used to derive RVs from the same observed spectra by using a $\chi^{2}$-minimization approach. 
The final velocities were checked for any nightly systematic errors that can be corrected in order to increase the precision of the RV data set. 

\HIRES, \APF, \UCLES, and \PFS\, are Iodine-based instruments, meaning that they each include a cell of gaseous I$_2$ within the converging beam of their respective telescopes. Incoming stellar spectra are imprinted in a high-density forest of I$_2$ lines in the 5000-6200\AA\ band pass. These lines act both as a calibrator for the wavelengths of the stellar spectra and as a representative for the point-spread function (PSF) of each instrument. After extraction of the iodine region of the spectrum, the stellar spectrum must be deconvolved from the I$_2$ absorption lines such that the wavelengths, instrument PSFs, and Doppler shifts may be extracted. This is accomplished by splitting each iodine region into 2\AA\ chunks, and then analyzing them via the spectral synthesis technique outlined in \citet{Butler1996}. A weighted mean of all the Doppler velocities of the individual chunks is taken, and serves as the final Doppler velocity for each individual observation. The standard deviation of all the 2\AA\ chunks ($\sim$800 for \PFS\, and $\sim$700 for the \APF\, and \HIRES\,) constitutes the total internal uncertainty for each velocity measurement. The timestamps for each iodine-based RV are converted from their pipeline produced MJD values to BJD$_{\rm{TDB}}$ timestamps using the Pexo modeling package \citep{Feng2019a}.

We note that three of the above spectrographs have undergone instrumental upgrades since their deployment. The \HIRES\, detector was replaced with a new mosaic CCD in August 2004, \HARPS\, moved to the use of an octagonal science fiber in 2015 and in 2018, the \PFS\, detector was replaced with a smaller pixel 10k$\times$10k detector and the slit used for I$_2$ observations was changed from 0.5'' to 0.3''. {In all three cases, we treat the data taken before and after the upgrade as coming from two separate instruments, identified in our RV data sets and figures as -Pre and -Post velocities.}

The instruments cover spectral ranges of 3700-8000\AA\ for \HIRES, 3700-9000\AA\ for \APF, 3900-6700\AA\ for \PFS, and 4800-8400\AA, for \UCLES, however the radial velocity measurements are made using only the 5000-6200\AA\ wavelength region. The typical spectral resolutions for each instrument are: $R$ $\simeq$ 90,000 for the \APF, 60,000 for \HIRES, 45,000 for \UCLES, and 80,000/130,000 for \PFS\, pre-/post-upgrade, respectively.

The \HIRES\, data was obtained from the public Earth Bound Planet Search archive\footnote{https://ebps.carnegiescience.edu/data/hireskeck-data}, which provides updates to the \citet{Butler2017} \HIRES\, data catalog. {The final HIRES data point included in this analysis was taken on 26 December 2017.} The \APF, \PFS, and \UCLES\, data were provided by the corresponding instrument teams.

For each instrument's data set for a given star, we apply a robust sigma clipping where any points further than 5$\sigma$ from the mean are discarded as outliers. We visually inspect the points identified as outliers in each case, and find that in practice this analysis flags 1-3 data points per instrument per star, which is generally a small percentage of the overall data. Once each instrument's outliers are removed, we combine the cleaned data sets from each instrument into a single list. We include a column of data tracking which instrument was used to generate each measurement, so that later analysis can determine offsets between instruments.

\begin{table*}[!htbp]\footnotesize
\centering
\caption{Number of Archival RV Epochs Analyzed for Each Target Star \label{tab:RV_Obs}}
\begin{tabular}{llcccc||llcccc}
\hline\hline
HD & $^{a}$HARPS & HIRES & UCLES & $^{a}$PFS & APF & HD & $^{a}$HARPS & HIRES & UCLES & $^{a}$PFS & APF\\
\hline \vspace{2pt}
HD693 & 16 [16/0] & 0 [0/0] & 0 & 0 [0/0] & 0 &	HD85512 & 580 [517/63] & 7 [0/7] & 31 & 44 [38/6] & 0 \\
HD1581 & 329 [262/67] & 0 [0/0] & 119 & 0 [0/0] & 0 & HD100623 & 4 [4/0] & 64 [16/48] & 104 & 40 [34/6] & 0 \\
HD2151 & 34 [34/0] & 0 [0/0] & 163 & 0 [0/0] & 0 & HD102365 & 82 [78/4] & 13 [0/13] & 187 & 33 [22/11] & 0 \\
HD4628 & 42 [37/5] & 117 [0/117] & 0 & 0 [0/0] & 71 & HD102870 & 8 [8/0] & 0 [0/0] & 0 & 0 [0/0] & 59 \\
HD7570 & 19 [19/0] & 0 [0/0] & 60 & 0 [0/0] & 0 & HD104304 & 26 [0/26] & 42 [0/42] & 0 & 0 [0/0] & 14 \\
HD13445 & 11 [0/11] & 0 [0/0] & 74 & 0 [0/0] & 0 & HD114613 & 20 [13/7] & 45 [0/45] & 244 & 39 [27/12] & 0 \\
HD14412 & 26 [0/26] & 139 [24/115] & 28 & 12 [11/1] & 0 & HD115617 & 229 [224/5] & 157 [0/157] & 169 & 31 [28/3] & 0 \\
HD16160 & 45 [45/0] & 76 [0/76] & 0 & 0 [0/0] & 83 & HD125072 & 74 [55/19] & 0 [0/0] & 86 & 0 [0/0] & 0 \\
HD20766 & 26 [26/0] & 0 [0/0] & 58 & 0 [0/0] & 0 & HD131977 & 22 [22/0] & 0 [0/0] & 0 & 0 [0/0] & 0 \\
HD20794 & 260 [187/73] & 0 [0/0] & 147 & 21 [18/3] & 0 & HD136352 & 249 [242/7] & 28 [0/28] & 169 & 24 [21/3] & 0 \\
HD20807 & 99 [76/23] & 0 [0/0] & 99 & 16 [13/3] & 0 & HD140901 & 27 [27/0] & 0 [0/0] & 117 & 27 [23/4] & 0 \\
HD22049 & 28 [24/4] & 89 [0/89] & 0 & 0 [0/0] & 0 &	HD146233 & 177 [119/58] & 112 [28/84] & 81 & 15 [15/0] & 0 \\
HD22484 & 26 [0/26] & 8 [0/8] & 0 & 0 [0/0] & 71 &	HD147584$^{b}$ & 1 [1/0] & 0 & 0 & 0 [0/0] & 0\\
HD23249 & 116 [76/40] & 55 [20/35] & 95 & 0 [0/0] & 29 & HD149661 & 12 [12/0] & 43 [31/12] & 14 & 0 [0/0] & 0 \\
HD23356 & 14 [14/0] & 71 [12/59] & 0 & 0 [0/0] & 0 & HD156026 & 0 [0/0] & 27 [18/9] & 11 & 54 [35/19] & 59 \\
HD26965 & 103 [82/21] & 163 [7/156] & 112 & 24 [20/4] & 13 & HD160346 & 34 [34/0] & 0 [0/0] & 0 & 0 [0/0] & 0 \\
HD30495 & 44 [35/9] & 6 [0/6] & 0 & 0 [0/0] & 0 & HD160691 & 163 [161/2] & 0 [0/0] & 178 & 14 [12/2] & 0 \\
HD32147 & 41 [37/4] & 157 [0/157] & 0 & 27 [22/5] & 65 & HD165341$^{b}$ & 7 [7/0] & 0 & 0 & 0 [0/0] & 0\\
HD38858 & 103 [91/12] & 69 [16/53] & 0 & 0 [0/0] & 0 & HD188512 & 17 [17/0] & 57 [7/50] & 0 & 0 [0/0] & 0 \\
HD39091 & 49 [42/7] & 0 [0/0] & 77 & 37 [0/37] & 0 & HD190248 & 391 [279/112] & 0 [0/0] & 236 & 0 [0/0] & 0 \\
HD43834 & 26 [0/26] & 0 [0/0] & 140 & 24 [21/3] & 0 & HD192310 & 432 [348/84] & 137 [0/137] & 171 & 19 [19/0] & 0 \\
HD50281 & 12 [12/0] & 52 [29/23] & 0 & 0 [0/0] & 33 & HD196761 & 37 [27/10] & 63 [30/33] & 49 & 29 [21/8] & 0 \\
HD69830 & 273 [265/8] & 154 [0/154] & 24 & 29 [29/0] & 87 &	HD203608$^{b}$ & 7 [7/0] & 0 & 0 & 0 [0/0] & 0\\
HD72673 & 158 [115/43] & 77 [21/56] & 63 & 15 [15/0] & 0 & HD207129 & 111 [98/13] & 0 [0/0] & 123 & 22 [15/7] & 0 \\
HD75732 & 2 [2/0] & 220 [23/197] & 0 & 0 [0/0] & 25 & HD209100 & 137 [100/37] & 0 [0/0] & 0 & 0 [0/0] & 0 \\
HD76151 & 7 [7/0] & 0 [0/0] & 0 & 0 [0/0] & 0 &	HD216803 & 11 [11/0] & 16 [6/10] & 15 & 0 [0/0] & 0 \\
\hline\hline
\end{tabular}
\tablenotetext{a}{The total number of \HARPS\, \HIRES\, and \PFS\, {RV epochs} are followed by a break down of how many data points were taken before and after the instruments' upgrades (see Section \ref{sec:Data}) as the pre- and post- upgrade time series are treated as coming from two different instruments. $^{b}$These stars have thousands of individual observations all taken over observational baselines covering less than 1 week of time, making them incompatible with exoplanet search and injection/recovery analyses. We therefore remove them from further consideration in this paper.}
\end{table*}
\begin{table*}[!htbp]\footnotesize
\centering
\caption{Full Timeseries Data for Each Target Star \label{tab:Full_Timeseries}}
\begin{tabular}{lccccccccc}
\hline\hline
Starname & BJD$_{TBD}$ & RV [\ms\,] & RV$_{err}$ [\ms\,] & Instrument & S-index & S-index$_{err}$ & H$\alpha$ & H$\alpha_{err}$ & File Name\\
\hline \vspace{2pt}
HD115617 & 2453026.86393 & -3.12 & 0.95 & HARPS-Pre & 0.1404 & 0.0026 & -1.0 & -1.0 & 2004-01-22T08:41:20 \\
HD115617 & 2453026.86533 & -3.98 & 0.98 & HARPS-Pre & 0.1426 & 0.0032 & -1.0 & -1.0 & 2004-01-22T08:43:22 \\
HD115617 & 2453026.86674 & -2.59 & 0.96 & HARPS-Pre & 0.1490 & 0.0028 & -1.0 & -1.0 & 2004-01-22T08:45:26 \\
HD115617 & 2453026.86822 & -3.02 & 0.94 & HARPS-Pre & 0.1420 & 0.0027 & -1.0 & -1.0 & 2004-01-22T08:47:29 \\ 
HD115617 & 2453026.86967 & -5.38 & 0.95 & HARPS-Pre & 0.1471 & 0.0025 & -1.0 & -1.0 & 2004-01-22T08:49:33 \\
HD115617 & 2453026.87111 & -2.8 & 0.95 & HARPS-Pre & 0.1414 & 0.0028 & -1.0 & -1.0 & 2004-01-22T08:51:38 \\
HD115617 & 2453026.87254 & -4.59 & 0.95 & HARPS-Pre & 0.1506 & 0.0026 & -1.0 & -1.0 & 2004-01-22T08:53:42 \\
HD115617 & 2453026.87393 & -3.23 & 0.94 & HARPS-Pre & 0.1423 & 0.0025 & -1.0 & -1.0 & 2004-01-22T08:55:46 \\
HD115617 & 2453026.87541 & -1.22 & 0.94 & HARPS-Pre & 0.1468 & 0.0025 & -1.0 & -1.0 & 2004-01-22T08:57:49 \\
HD115617 & 2453026.87686 & -4.58 & 0.94 & HARPS-Pre & 0.1481 & 0.0024 & -1.0 & -1.0 & 2004-01-22T08:59:54 \\
\vdots & \vdots & \vdots & \vdots & \vdots & \vdots & \vdots & \vdots & \vdots & \vdots \\
\hline\hline
\end{tabular}
\tablenotetext{}{This table contains, for each target, the full set of observations of RVs, S-indices, and H-alpha from each instrument included in this study. Complete time series data for each target in the format above can be found in the online journal.}
\end{table*}

There are three stars that, despite having a significant number of \HARPS\, observations, cover time baselines incompatible with our science case. HD 203608 and HD 165341 were both observed over a single week, and HD 147584 was observed over a single night. None of these stars have been targeted by the other instruments in our study, and so we remove them from further consideration.

\subsection{{S-index Measurements}}\label{subsec:ActivityIndicators_Sindex}

A major challenge when classifying periodic signals seen in Doppler velocity data is determining whether those signals are due to planetary companions or the star itself. Stellar variability is produced by a variety of surface phenomena that occur and evolve across a range of time scales, but they can be grouped into four broad categories. Acoustic waves within the star cause patches of the surface to rise and fall periodically, creating RV oscillations at the few \ms\ level over timescales of minutes \citep{BouchyCarrier2001, Nordlund2009}. Granulation is due to motion within stellar convective cells as hot plasma wells up to the surface before radiatively cooling and sinking back down via intergranular lanes. This process takes anywhere from 20 minutes to 1 day depending on the granule size and again results in RV shifts at the few \ms\ level \citep{Cegla2019, Meunier2015, Meunier2019}. Active regions are areas of increased magnetic flux on the stellar surface such as star spots, plages, and faculae that transit across the visible hemisphere of the star as it rotates. They generally persist for multiple rotation periods and induce cyclic RV variations at the 1-10\, \ms\ level each time they pass over the star’s face \citep{SaarDonahue1997, Lockwood2007, Haywood2016}. Finally, stellar magnetic cycles are driven by stellar dynamos, which are maintained through differential rotation at the tachocline – the interface between a star’s radiative and convective layers. These magnetic cycles vary slowly, generally exhibiting periods of 5-15 years for Sun-like stars, and can induce RV variations up to 20 \ms\ over that time span \citep{Meunier2010, Makarov2010, Dumusque2011}. In our case of working primarily with radial velocity data sets taken at relatively low cadence (e.g., once per week or month) the latter two varieties, active regions and magnetic cycles, produce the highest rate of false positive signals.

A well-established method for tracing a star's variability level is the use of stellar activity indicators, which compare the amount of flux inside activity sensitive lines to the flux in nearby continuum regions. The most common stellar activity indicators for Sun-like stars are derived from measurement of the emission reversal at the cores of the Fraunhofer H and K lines of Ca II located at at 3968 {\AA} and 3934 {\AA}, respectively, which trace chromospheric activity. As the Ca II line core emission is generated in regions of concentrated magnetic fields, these lines serve as a proxy for the number of spots on the star, and often show variations with the star's rotational period. Because stars in the active phases of their magnetic cycles tend to produce more sunspots \citep{Schwabe1843}, activity indicators based on the Ca II lines can also act as a tracer of long-term magnetic cycles.

For Sun-like stars, the best known Ca II activity indicator is the S-index which compares the flux in the cores of the H \& K lines to two nearby continuum regions denoted as the R and V filters \citep{Wilson1968, Duncan1991}. The S-index generally takes the form:
\begin{equation}
    S_{\rm{index}} = \frac{H + K}{R + V}
\end{equation}
and is often calibrated to the original Mt. Wilson S-index survey which ran from 1966-1983 \citep{Duncan1991} to allow for comparisons between facilities. Over the Mt. Wilson survey's two-decade span, 111 F2-M2 stars were monitored continuously from Mt. Wilson and ~60\% were seen to exhibit magnetic cycles on a 5-15 year time scale \citep{Baliunas1995}. These time series make clear that the range of variability exhibited by the continuously monitored Mt. Wilson stars is much more diverse than what we observe in the Sun. 

To search for evidence of these long term magnetic cycles, in addition to shorter term rotational periods, in our RV data sets we first derive an S-index value from each RV spectrum. For the \HIRES\, \APF\, and \PFS\, data sets, these S-index values are generated automatically as part of the data reduction pipelines and further details can be found in \citet{Butler2017} and \citet{Burt2021}. We determine errors for each instrument's S-index values taking into account photon noise. The resulting uncertainty, by error propagation, is:
\begin{equation}
\sigma_{\rm{S}} = \rm{S}\cdot\sqrt{\frac{\sigma^{2}_{H}+\sigma^{2}_{K}}{(H+K)^{2}} + \frac{\sigma^{2}_{R}+\sigma^{2}_{V}}{(R+V)^{2}}}
\end{equation}

In each case we have used a set of overlapping target stars to calibrate the instrument's S-index measurements to the Mt. Wilson survey so that they can be considered together without concerns for large scaling offsets. {In some cases, however, the Mt. Wilson calibration is based on a small number of stars and may introduce non-astrophysical offsets between the instruments. To account for this, our analysis allows us to fit for offsets between the S-index data sets, as described more thoroughly in Section \ref{subsec:RVsearchOverview}.}

{The \HARPS\, RVBank data does not yet provide S-index measurements, and so we instead make use of the methodology described in \citet{Perdelwitz2021}. Specifically, we use a set of narrow bands close to the Ca II line cores, along with PHOENIX synthetic spectra \citep{Husser2013}, to derive $R_{\mathrm{HK}}^\prime$.  We then convert these into S-Indices using the prescription given by \citet{GomesdaSilva2021} and calibrate the results to the Mount Wilson scale by cross matching with the mean S-indices derived by \cite{Duncan1991}. {The S-index errors are calculated via a Monte Carlo approach where, in each trial, the flux in each bin of the measured spectrum is randomly displaced within a Gaussian distribution with width of the flux error. The S-index is then evaluated for each trial, and the error is taken to be the standard deviation of the resulting set.} This approach yields S-indices for $\sim93.5\%$ of the \HARPS\ spectra present in the RVBank archives.}

{Attempts to use these Monte Carlo S-index errors for the HARPS data alongside the photon-noise limited errors derived from the HIRES, PFS, and APF data results in an uneven weighting in favor of the iodine-based instruments. While we report all of these errors in the data tables that accompany this publication for reference, in practice we adopt a third, alternative method for determining S-index errors so that all four instruments' data sets are treated in the same way. We begin by selecting five stars (HD 69830, HD 196761, HD 114613, HD 4628, and HD 39091) all of which have at least two dozen observations from at least two of the instruments. We assign all of the S-index data for each star the same error bar of $\sigma_{S}$ = 0.01 and carry out an initial uninformed fit with \RVSearch. We then combine all of the residual values from each instrument across all stars, measure the standard deviation, and assign that as the global error for that instrument. Those values are: HARPS-Pre : 0.010; HARPS-Post : 0.006; HIRES-Pre : 0.010; HIRES-Post : 0.014; PFS-Pre : 0.005; PFS-Post : 0.009, APF : 0.007. As we expect the S-index measurement to be systematics dominated (e.g. from the deblazing and continuum normalization) rather than photon-noise dominated, this empirical approach to measuring the uncertainty provides a more homogeneous error estimate.}

When summarizing the properties of our target stars below, we reference \logrphk\ values for stars where it has been reported in the literature. This metric is also derived using the Ca II H\&K absorption lines, but \logrphk\ removes the basal (rotation independent) photospheric flux \citep{Noyes1984, Schrijver1987,Mittag2013}. This photospheric flux, which can contaminate the S-index filters, introduces a dependency on stellar effective temperature. By removing it, the \logrphk\ metric produces a measure of activity that can be compared across spectral types.

\subsection{{H$\alpha$ Equivalent Width Measurements}\label{subsec:ActivityIndicators_Halpha}}

{The \UCLES\, spectrograph cannot simultaneously cover the Iodine region necessary for precise wavelength calibration of the stellar spectra and the Ca II H \& K region necessary for extracting the S-index measurements.} To provide a stellar activity check on the RVs derived from the AAT dataset, {we instead make use of the H$\alpha$ absorption line, using measurements of the line's equivalent width (EW) to detect variations related to the long-term stellar magnetic activity cycle.} This EW$_{H \alpha}$ analysis follows the methodology of \citet{Wittenmyer2017} which is similar to that presented by \citet{Robertson2014}, except for the addition of an automated algorithm for continuum normalization and telluric contamination identification near the H$\alpha$ line. 

\begin{figure}[H]
\includegraphics[width=.45 \textwidth]{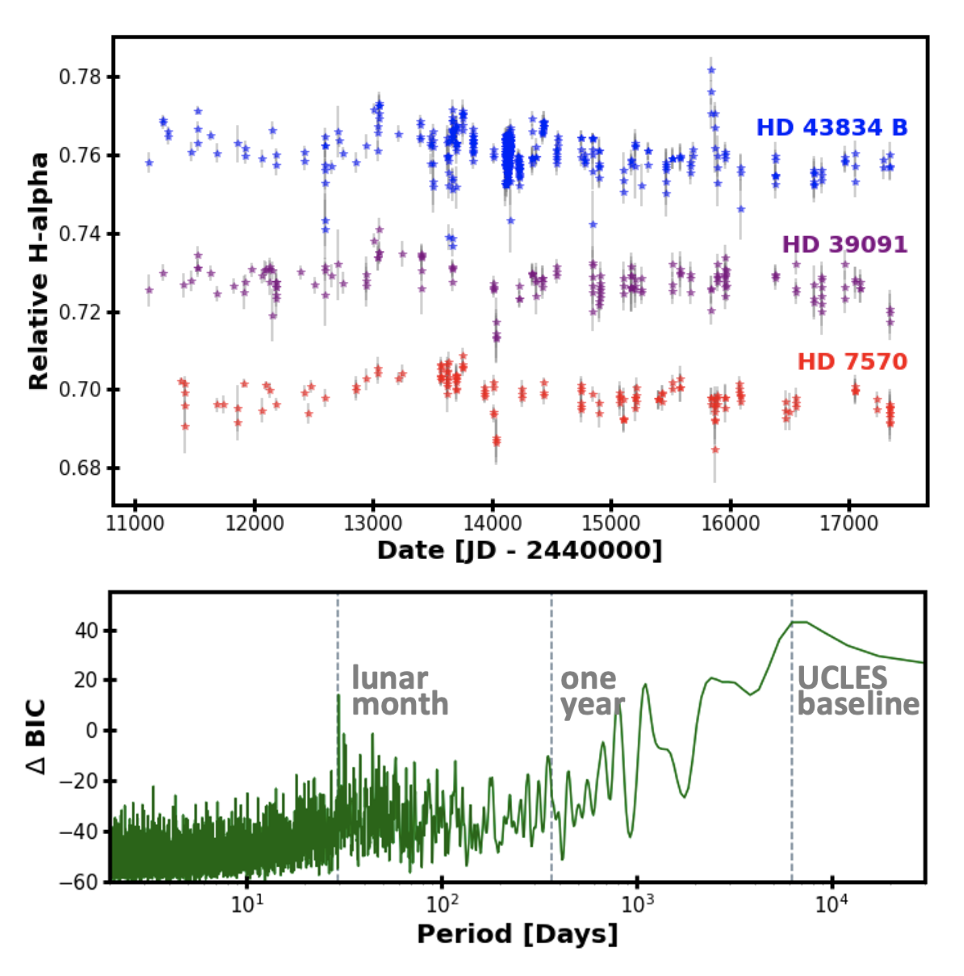}
\caption{Top: EW$_{H \alpha}$ time series for three stars observed with the \UCLES\, spectrograph, offset vertically from one another for ease of viewing. Similar long term behavior is present in each of these three data sets, suggesting that the cause is not stellar but rather instrumental or environmental. Bottom: Stacked periodograms from 6 stars with at least 100 \UCLES\, H$\alpha$ measurements spread over the majority of the instrument's $\sim$6000 day observational baseline, where no significant signals were detected by \rvsearch. The long term structure seen in the top panel emerges as two humps in this composite periodogram, one at \UCLES' observing baseline (6000 days) and another at half that (3000 days). The impact of the monthly lunar cycle is also evident via a narrow peak at 29.5 days. We therefore treat H$\alpha$ periods detected around any of these three periods with some degree of caution.}
\label{fig:Halpha-Comp}
\end{figure}

{A visual comparison of the resulting EW$_{H \alpha}$ time series reveals the presence of very similar structured variations among each of the resulting data sets (Figure \ref{fig:Halpha-Comp}). Given the shared trends between the stars, the source is likely either instrumental or environmental in nature. One potential cause is variations in the water content of the atmosphere. As our EW$_{H \alpha}$ calculation algorithm does not actively correct for the telluric lines, it is reasonable to assume the EW$_{H \alpha}$ measurements are subject to effects from atmospheric water content. However, when compared to the historic precipitable water vapor measurements from Siding Spring (e.g. those in \citet{Haslebacher2022}), no strong correlation is evident.}

{While the exact cause of the variations is not clear, we note that the stacked periodograms of a dozen stars (each with dozens of \UCLES\, H$\alpha$ data points but no significant detections) all linearly interpolated onto the same period grid shows prominent peaks at $\sim$1 year, $\sim$3000 day (approximately half the \UCLES\, observation time span), and $\sim$6000 day (approximately equal to the \UCLES\, observation time span) periods. We therefore advise caution when interpreting the results of the H$\alpha$ \rvsearch analysis, especially in the case of long period signals. Shorter period detections, those in the 10 - 100 day range where we generally look for evidence of stellar rotation for these F - K dwarfs stars, seem to be unaffected by these long period variations.}

\subsection{Speckle Imaging}\label{subsec:speckleimaging}

{For a handful of the stars studied in this work, high resolution speckle imaging observations to search for and/or rule out nearby stellar companions were obtained.} The speckle imaging observations were carried out using the `Alopeke and Zorro instruments at Gemini-North and Gemini-South, respectively \citep{Scott2021}. These instruments observe simultaneously in two bands, (832$\pm$40\,nm and 562$\pm$54\,nm) obtaining diffraction limited images with inner working angles of 0\arcsec.026 and 0\arcsec.017, respectively. All targets were observed using a sequence of short 60~ms exposures. These images were combined using Fourier analysis techniques, examined for stellar companions, and used to produce reconstructed speckle images (see \citet{Howell2011} and \citet{Horch2021}). {We summarize the observations and the resulting sensitivity to companions in Table~\ref{tab:AATwork}. and Figure Set 8 (available in the online journal).}

\section{Analysis}

\subsection{{Overview of RVSearch}\label{subsec:RVsearchOverview}}

\RVSearch\ \citep{Rosenthal2021} is a recently released Python package based on \Radvel\, \citep{Fulton2018}, built specifically to perform uninformed searches for Keplerian signals in RV data and to perform injection and recovery analysis of RV time series data. 

{\rvsearch\,'s uninformed search function is used to identify candidate signals in our compiled radial velocity, S-index, and H$\alpha$ data sets. For each data set we bin the input velocities / activity measurements to nightly data points to {decrease the computational requirements} and set a minimum search period of 2 days and a maximum search period of three times the total observational baseline days. In addition to any Keplerian signals, \RVSearch\ also fits for a constant offset between each instrument's data set and for the `jitter' of each instrument. This jitter term is used to address the unmodeled instrumental effects or stellar variability that induce additional scatter in the RV time series and encompasses uncorrelated signals that occur on timescales shorter than the observational baseline.}

\RVSearch\ implements an iterative fitting approach when searching for periodic signals in a time series. It first tests for the presence of a linear or quadratic slope in the data, before beginning the Keplerian fitting process by generating a single planet with undefined orbital parameters to become the initial likelihood model. With the initial model in hand, \RVSearch\ defines a set of periods to test\footnote{The periods are spaced such that the frequency offset between adjacent grid points is 1/2$\pi\tau$, where $\tau$ is the observational baseline of the full RV time series.} and computes a $\Delta$BIC goodness-of-fit periodogram by fitting a sinusoid to the data at each fixed period. The Bayesian Information Criterion, or BIC, is used for model selection when considering a finite set of models and is calculated as:

\begin{equation}
\rm{BIC} = -2\ \rm{ln}\ \mathcal{L}_{\rm{max}}\ + \it{n}\ \rm{ln}\ \it{N}
\end{equation}

where $\mathcal{L}_{\rm{max}}$ is the maximum likelihood, \emph{n} is the number of model free parameters, and \emph{N} is the number of data points \citep[see, e.g.,][for details]{KassRaftery1995}. Models with lower BIC values are generally preferred. The $\Delta$BIC value at each period is the difference between the best-fit, $n$+1-planet model with the given fixed period, and the $n$-planet fit to the data. 

{Once the $\Delta$BIC periodogram has been calculated, a linear fit is applied the data, and a histogram of periodogram power values is plotted on a log scale. A detection threshold is then constructed such that only 0.1\% of periodogram peaks are expected to be high enough powered to exceed it. This threshold is the empirical False Alarm Probability (FAP) of 0.1\% \citep{Rosenthal2021}.} Any signal above a 0.1\%\, empirical False Alarm Probability (FAP) is considered significant. {For our S-index search, we enforce an additional requirement that the $\Delta$BIC value be at least 10 for a signal to be added to the system's model, even if that corresponds to a FAP value $<$1\%. This prevents the inclusion of a nonphysical number of short period signals in sparse data sets, while still being a generous inclusion criteria as the field standard for considering a signal worthy of consideration for publication is more often $\Delta$BIC $>$ 25.}

If a significant detection is made, \RVSearch\ refines the fit of the signal's Keplerian orbit by performing a maximum a posteriori (MAP) fit with all model parameters free, including eccentricity, and records the BIC of that best-fit model. The search algorithm then adds an additional planet to the model and repeats the fitting and evaluation process. In the n+1 planet fit, the signals are treated simultaneously, so that the change in the BIC can again be evaluated to compare the $n$-planet fit to the $n$+1-planet fit. {We note here that when analyzing our S-index and H$\alpha$ data sets, the `planet' detections instead refer to activity-driven periodicities in the data sets.} If the new planet is supported by the data, the search continues. The uninformed search continues to iterate on the time series {until no additional significant signals are present in the periodogram.}

{At this point, \rvsearch\ returns the max-likelihood estimates of the orbital model parameters the the dataset and the model posteriors are sampled via an affine-invariant sampling that is implemented in \radvel\ using the \texttt{emcee} package \citep{ForemanMackey2013}. The resulting parameter estimates and uncertainties, reported as the median and $\pm$1$\sigma$ intervals are visible on the summary figures produced by \rvsearch\ and in our summary tables.}

{One complication encountered in the fitting process, across both the RV and S-index applications, is the treatment of signals with periods on order or greater than the total observational baseline. While the $\Delta$BIC periodogram approach used in the first phase of \rvsearch\,'s process can only fully resolve periods shorter than the observational baseline, the posterior sampling is not subject to similar constraints. Thus in cases where there is a prominent peak in the periodogram that has a peak close to or beyond the observational baseline, the MCMC will sometimes suggest that the true period is 2x the periodogram peak, or in some cases many times larger. In these instances the traditional MCMC method fails to return a well-sampled model posterior and the resulting period uncertainty is as large as, if not many times larger than, the period itself. We note these types of detections as long period signals (`LPS') in our summary tables and report just the initial $\Delta$BIC periodogram peak instead of the final MCMC fit and its corresponding uncertainties, as they are non-physical. For these signals, note that additional data is required to fully reveal and constrain the underlying signal.}

\subsection{{Identification of Candidate Signals in the Radial Velocity data}}

\begin{figure*}
\makebox[\textwidth]{\includegraphics[width=.4\paperwidth]{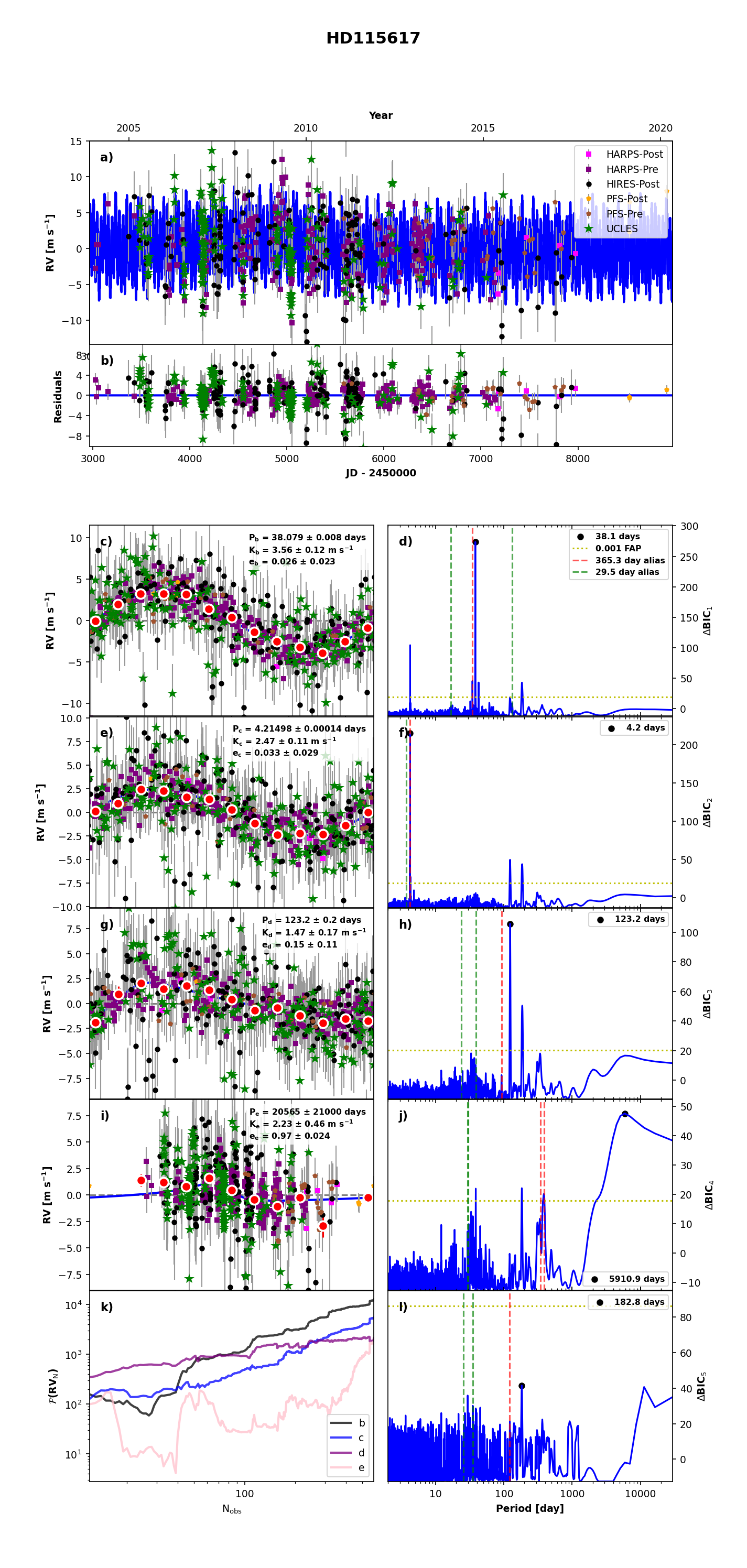}}
\caption{\RVSearch\ results for HD 115617. Panel (a) shows the
    initial radial velocity time series with the best-fit model
    plotted in blue and panel (b) shows the RV residuals. Panels (c),
    (e), (g) show phased RV curves for the three known planets in the
    system, and report the best-fit parameters for each orbit. Panels
    (d), (f), and (h) show the periodograms associated with each
    planet detection. The yellow horizontal dotted line marks the
    minimum $\Delta$BIC for a 1\%\, FAP, while the vertical dotted
    lines show monthly and yearly aliases. Panels (i) and (j) show
    the periodogram and best fit curve to a fourth, much longer and
    highly eccentric signal that is likely driven by stellar
    variability. Panel (k) shows the RV significance of each signal
    relative to the number of observations considered {and is
    calculated using the best-fit orbits shown in the left side
    panels above it}, and panel (l) shows the residual periodogram,
    indicating that no further planets are found in the data set.
    The complete set of Radial Velocity summary plots (46 figures) 
    can be found in the online journal.
\label{fig:HD115617_Summary}}
\end{figure*}

An example \rvsearch\ {fit to the radial velocity data for HD 115617} is shown in Figure \ref{fig:HD115617_Summary}. The HD 115617 planetary system was first published in \citet{Vogt2010} using data from the \HIRES\, and \UCLES\, spectrographs. Three planets were discovered with periods and RV semi-amplitudes of 4.215$\pm$0.0006\,d and $2.12\pm0.23$\,\ms\, $38.021\pm0.034$\,d and $3.62\pm0.23$\,\ms, and $123.01\pm0.55$\,d and $3.25\pm0.39$\,\ms\, for planets b, c, and d, respectively. Revisiting the system with the available archival data, we supplement the published data with an additional 275 \HIRES\, points, 159 \UCLES\, points, 1248 \HARPS\, points, and 11 \PFS\, points taken between 2004 and 2020. 

\figsetstart
\figsetnum{4}
\figsettitle{Radial Velocity Analysis Summary Plots}

\figsetgrpstart
\figsetgrpnum{4.30}
\figsetgrptitle{HD 115617}
\figsetplot{HD115617_Resubmission.png}
\figsetgrpnote{Radial velocity analysis results for HD 115617 from \rvsearch, including radial velocity timeseries, residuals, and periodogram. See individual discussion sections in the paper for further details on interpreting these plots.}
\figsetgrpend

\figsetend

\begin{figure*}
\makebox[\textwidth]{\includegraphics[width=.6\paperwidth]{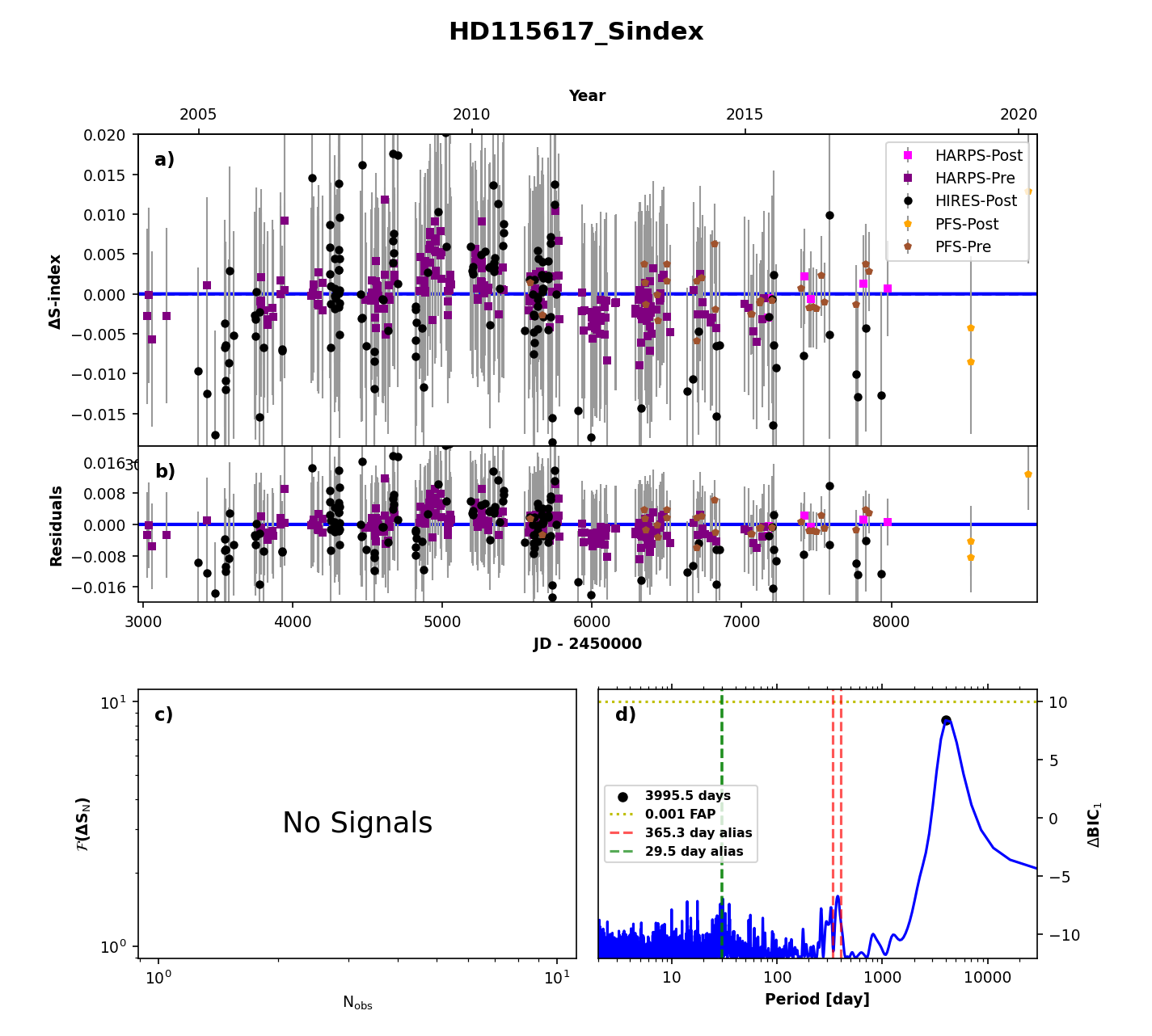}}
\caption{\RVSearch\, results for the relative S-index measurements of
    HD 115617, {following the same plot image structure as in
    Figure \ref{fig:HD115617_Summary}. Panel (a) shows the S-index
    time series with the best-fit model plotted behind them while
    panel (b) shows the S-index residuals. Panel (c) would present
    the phase folded curves for any signals identified by \rvsearch,
    but as seen in panel (d) no signals in the periodogram rise above
    the $\Delta$BIC $>$ 10 requirement imposed on the S-index search
    (yellow horizontal dotted line). The red and green vertical
    dotted lines show the one month and one year aliases of the
    tallest peak in the periodogram.}
    The complete set of S-Index summary plots (46 figures) can be 
    found in the online journal.
\label{fig:HD115617_Activity_Summary}}
\end{figure*}

\figsetstart
\figsetnum{5}
\figsettitle{S-Index Activity Summary Plots}

\figsetend

Incorporating this additional RV data produces a fit consistent with the \cite{Vogt2010} results{. A}ll three previously published planets are again detected at statistically significant levels and at very similar period and semi-amplitude values. The uncertainties on those values, however, are notably improved in the updated fit; the RV semi-amplitude uncertainty decreases by a factor of two, thereby doubling the detection significance, and the period uncertainty decreases by {factors of two to four} across the three planets. 

\rvsearch\ also identifies a fourth, much longer period signal that rises above the detection threshold{, with {$P=20565\pm21000$} days (Figure 3 panel i)}. The uncertainty on the best-fit orbital period is of order the period itself and it overlaps broadly with a long period signal in the star's activity data (see: Section \ref{sec:activity_analysis_sindex}). Additionally, the strength of this fourth RV signal varies quite noticeably as the number of data points increases. This suggests that only specific clumps of data are providing additional power in the periodogram, as compared to the roughly monotonic increase that is expected for a Keplerian signal (similar to {the} \citet{Mortier2017} stacked periodogram technique). {Finally, we note that the period of the peak which is actually being fit by this signal is of approximately the same length as the observation baseline for this target. As discussed in Section \ref{subsec:RVsearchOverview}, this results in unphysical MCMC fitting to the signal.} These concerns, combined with the fact that the best-fit model's high eccentricity would produce a semi-minor axis of $\sim$0.5~AU that would disturb the three shorter period planets that have been robustly vetted, leads us to conclude the final signal detected by \RVSearch\, is not planetary in nature. {This signal is classified as LPS in Table \ref{tab:AllSignals}.}

For all previously discovered planetary systems, HD 115617 included, we report our best-fit results as updates to the published orbital parameters in Section \ref{sec:updated_params}.

\startlongtable
\begin{deluxetable*}{llllllr}
\tablecaption{Keplerian RV Signals Identified by \rvsearch\ (Updated for Resubmission) \label{tab:AllSignals}}
\tablehead{\colhead{ID} & \colhead{Period [days]} & \colhead{K [\ms]} & \colhead{Ecc.} & \colhead{Msini [M$_{\oplus}$]} & \colhead{FAP} & \colhead{Interp}}
\startdata							
HD1581 I	&	635.0$\pm$4.4	&	0.89$\pm$0.14	&	0.55$\pm$0.13	&	10.08$^{+1.22}_{-1.17}$	&	7.24e-09	&	SRC\\
HD1581 II	&	15.653$\pm$0.005	&	0.662$\pm$0.096	&	0.106$\pm$0.097	&	2.56$^{+0.37}_{-0.38}$	&	1.85e-05	&	ACT-R\\
HD1581 III	&	29.4661$\pm$0.0041	&	1.6$\pm$1.1	&	0.89$\pm$0.12	&	3.53$^{+1.15}_{-0.85}$	&	8.26e-04	&	ACT\\		
HD2151 I	&	5365$\pm$1400	&	3.21$\pm$0.58	&	0.54$\pm$0.15	&	81.41$^{+12.95}_{-13.22}$	&	8.90e-07	&	ACT\\
HD13445 I	&	88080$\pm$46000	&	3117$\pm$750	&	0.68$\pm$0.12	&	201858.32$^{+79293.94}_{-71306.73}$	&	2.28e-16	&	Binary\\
HD13445 b	&	15.764862$\pm$4.3e-05	&	377.58$\pm$0.77	&	0.0485$\pm$0.0018	&	1271.19$^{+25.51}_{-25.72}$	&	3.56e-83	&	KP\\
HD16160 I	&	22999$\pm$1200	&	702.5$\pm$2.9	&	0.6075$\pm$0.0092	&	20304.22$^{+434.82}_{-436.28}$	&	1.23e-26	&	Binary\\
HD20766 I	&	5643.5	& -- &   --	&    --	&    --	&	LPS\\				
HD20794 b	&	18.305$\pm$0.0052	&	0.807$\pm$0.089	&	0.17$\pm$0.11	&	2.83$\pm$0.31	&	2.20e-11	&	KP\\
HD20794 d	&	89.766$\pm$0.085	&	0.86$\pm$0.12	&	0.27$\pm$0.11	&	5.02$^{+0.66}_{-0.64}$	&	7.38e-11	&	KP\\
HD20807 I	&	3180$\pm$130	&	2.9$\pm$0.4	&	0.23$\pm$0.11	&	62.48$^{+8.81}_{-8.74}$	&	2.73e-07 &	SRC\\		
HD22049 b	&	2832$\pm$120	&	11.1$\pm$1.2	&	0.09$\pm$0.08	&	211.16$^{+23.57}_{-24.34}$	&	8.55e-11	&	KP\\	
HD23249 I	&	596.6$\pm$2.6	&	3.0$\pm$1.1	&	0.65$\pm$0.14	&	33.33$^{+7.89}_{-5.6}$	&	8.18e-08 &	SRC\\		
HD26965 I	&	42.303$\pm$0.025	&	1.4$\pm$0.22	&	0.37$\pm$0.17	&	5.94$\pm$0.79	&	1.48e-08	&	ACT*\\
HD26965 II	&	37.33$\pm$0.02	&	1.17$\pm$0.19	&	0.14$\pm$0.12	&	5.14$^{+0.84}_{-0.86}$	&	7.45e-05	&	Alias\\	
HD26965 III	&	367.9$\pm$3.1	&	1.63$\pm$0.88	&	0.46$\pm$0.27	&	13.9$^{+5.13}_{-2.95}$	&	1.37e-05	&	FP \\		
HD32147 I	&	2866$\pm$140	&	1.8$\pm$0.21	&	0.34$\pm$0.13	&	32.02$^{+3.54}_{-3.49}$	&	3.94e-12	&	SRC\\
HD38858 I	&	2893$\pm$150	&	2.8$\pm$0.3	&	0.19$\pm$0.12	&	58.15$^{+6.19}_{-6.01}$	&	1.41e-13	&	ACT-M\\	
HD39091 b	&	2089.05$\pm$0.46	&	196.5$\pm$0.6	&	0.6428$\pm$0.0017	&	3225.56$^{+58.95}_{-59.18}$	&	1.10e-20	&	KP\\
HD39091 d	&	125.58$\pm$0.27	&	2.16$\pm$0.42	&	0.16$\pm$0.15	&	17.56$^{+3.49}_{-3.31}$	&	4.26e-04	&	KP\\	
HD69830 b	&	8.66897$\pm$0.00028	&	3.4$\pm$0.1	&	0.128$\pm$0.028	&	10.1$^{+0.38}_{-0.37}$	&	2.15e-62	&	KP\\		
HD69830 c	&	31.6158$\pm$0.0051	&	2.6$\pm$0.1	&	0.03$\pm$0.027	&	12.09$^{+0.55}_{-0.54}$	&	1.47e-84	&	KP\\
HD69830 d	&	201.4$\pm$0.4	&	1.5$\pm$0.1	&	0.08$\pm$0.071	&	12.26$^{+0.89}_{-0.88}$	&	1.89e-36	&	KP\\
HD75732 b	&	14.65157$\pm$0.00015	&	70.39$\pm$0.37	&	0.0069$\pm$0.0047	&	254.81$^{+4.79}_{-4.81}$	&	4.13e-97 &	KP\\
HD75732 d	&	14951$\pm$5100	&	54$\pm$5	&	0.515$\pm$0.086	&	1686.0$^{+229.11}_{-264.51}$	&	9.03e-20	&	KP\\
HD75732 c	&	44.39$\pm$0.01	&	9.95$\pm$0.37	&	0.22$\pm$0.041	&	50.78$^{+2.05}_{-2.0}$	&	2.37e-34	&	KP\\	
HD75732 e	&	0.736546$\pm$5e-06	&	6.26$\pm$0.34	&	0.039$\pm$0.035	&	8.35$^{+0.48}_{-0.47}$	&	6.05e-28	&	KP\\	
HD75732 f	&	260.88$\pm$0.36	&	5.68$\pm$0.48	&	0.585$\pm$0.057	&	43.4$^{+3.61}_{-3.5}$	&	3..51e-14	&	KP\\	
HD85512 I	&	3891	&	-- &	--	&	--	&	1.90e-16	&	LPS \\	
HD85512 II	&	51.195$\pm$0.073	&	0.438$\pm$0.079	&	0.3$\pm$0.19	&	1.86$^{+0.31}_{-0.3}$	&	6.98e-06	&	ACT-R*\\
HD102365 b	&	121.3$\pm$0.25	&	1.38$\pm$0.23	&	0.28$\pm$0.15	&	9.34$^{+1.52}_{-1.5}$	&	1.58e-04	&	KP\\
HD114613 I	&	6622$\pm$270	&	7.29$\pm$0.45	&	0.291$\pm$0.061	&	239.94$^{+13.54}_{-13.5}$	&	2.87e-34	&	ACT-M*\\
HD114613 II	&	73.141$\pm$0.056	&	2.54$\pm$0.45	&	0.51$\pm$0.14	&	16.72$^{+2.38}_{-2.42}$	&	4.31e-04	&	SRC\\
HD114613 III	&	1954$\pm$39	&	2.98$\pm$0.52	&	0.6$\pm$0.11	&	54.01$^{+7.59}_{-7.5}$	&	2.09e-04	&	SRC\\
HD115617 b	&	4.21498$\pm$0.00014	&	2.47$\pm$0.11	&	0.033$\pm$0.029	&	5.98$^{+0.3}_{-0.29}$	&	5.01e-61	&	KP\\
HD115617 c	&	38.079$\pm$0.008	&	3.56$\pm$0.12	&	0.026$\pm$0.023	&	17.94$^{+0.73}_{-0.7}$	&	2.93e-46	&	KP\\
HD115617 d	&	123.2$\pm$0.2	&	1.47$\pm$0.17	&	0.15$\pm$0.11	&	10.82$^{+1.23}_{-1.03}$	&	5.63e-22	&	KP\\
HD115617 I	&	5910.9	&	--	&	--	&	--	&	1.22e-10 &	LPS\\	
HD136352 b	&	11.5767$\pm$0.0015	&	1.65$\pm$0.11	&	0.05$\pm$0.045	&	5.5$\pm$0.38	&	3.67e-38	&	KP\\	
HD136352 c	&	27.5845$\pm$0.0064	&	2.49$\pm$0.12	&	0.041$\pm$0.036	&	11.12$\pm$0.57	&	3.04e-24	&	KP\\	
HD136352 d	&	107.5$\pm$0.14	&	1.44$\pm$0.12	&	0.072$\pm$0.061	&	10.08$^{+0.87}_{-0.85}$	&	2.06e-23	&	KP\\
HD136352 I	&	121.66$\pm$0.26	&	0.68$\pm$0.13	&	0.22$\pm$0.19	&	4.69$^{+0.87}_{-0.86}$	&	9.76e-04	&	ACT\\
HD140901 I	&	5084$\pm$1200	&	11.6$\pm$2.4	&	0.44$\pm$0.25	&	269.81$^{+43.83}_{-42.02}$	&	7.32e-04 &	SRC\\
HD146233 I	&	2374$\pm$47	&	5.47$\pm$0.33	&	0.21$\pm$0.07	&	111.72$^{+6.67}_{-6.59}$	&	1.31e-25	&	ACT-M*\\
HD146233 II	&	6256$\pm$370	&	4.96$\pm$0.57	&	0.59$\pm$0.06	&	114.86$^{+12.0}_{-11.43}$	&	7.39e-14	&	ACT-M\\	
HD146233 III	&	19.8777$\pm$0.0062	&	1.73$\pm$0.26	&	0.38$\pm$0.16	&	6.77$\pm$0.86	&	7.23e-09 &	Candidate\\		
HD160346 I	&	83.7286$\pm$0.0005	&	5690.3$\pm$2.3	&	0.2048$\pm$0.0003	&	35280.0$^{+706.83}_{-716.46}$	&	1.42e-15	&	Binary\\
HD160691 b	&	644.93$\pm$0.28	&	35.7$\pm$0.2	&	0.0499$\pm$0.0082	&	528.58$^{+11.05}_{-11.13}$	&	2.16e-46 &	KP\\
HD160691 c	&	9.6394$\pm$0.0008	&	2.8$\pm$0.2	&	0.132$\pm$0.069	&	10.22$\pm$0.73	&	5.38e-98	&	KP\\
HD160691 d	&	308.4$\pm$0.23	&	12.7$\pm$0.3	&	0.074$\pm$0.016	&	147.23$^{+4.63}_{-4.56}$	&	8.24e-131	&	KP\\
HD160691 e	&	4035$\pm$21	&	22.25$\pm$0.24	&	0.026$\pm$0.013	&	607.79$^{+14.0}_{-13.99}$	&	2.84e-32	&	KP\\	
HD190248 I	&	360.8$\pm$1.9	&	1.21$\pm$0.43	&	0.29$\pm$0.15	&	12.96$^{+5.08}_{-3.76}$	&	5.14e-04	&	FP \\
HD192310 b	&	74.278$\pm$0.035	&	2.484$\pm$0.098	&	0.032$\pm$0.027	&	14.28$^{+0.64}_{-0.63}$	&	8.11e-50	&	KP\\
HD192310 c	&	549.1$\pm$4.5	&	1.3$\pm$0.1	&	0.078$\pm$0.073	&	14.96$^{+1.21}_{-1.18}$	&	3.64e-27	&	KP\\
HD192310 I	&	3836$\pm$240	&	1.48$\pm$0.11	&	0.34$\pm$0.15	&	29.3$^{+3.33}_{-3.07}$	&	1.64e-49	&	ACT-M\\
HD192310 II	&	43.614$\pm$0.023	&	0.93$\pm$0.13	&	0.5$\pm$0.1	&	3.83$\pm$0.44	&	2.41e-13	&	ACT-R\\	
HD192310 III	&	39.509$\pm$0.059	&	1.0$\pm$0.1	&	0.22$\pm$0.11	&	4.48$\pm$0.46	&	1.43e-09	&	ACT-R\\	
HD192310 IV	&	24.559$\pm$0.016	&	0.6$\pm$0.1	&	0.16$\pm$0.12	&	2.46$^{+0.39}_{-0.4}$	&	7.74e-06	&	Candidate\\	
HD207129 I	&	1964$\pm$49	&	4.02$\pm$0.61	&	0.44$\pm$0.16	&	72.95$^{+8.37}_{-8.07}$	&	3.88e-11	&	ACT-M\\
HD209100 I	&	13138.7	&	--	&	--	&	--	&	1.38e-37	&	LPS\\		
\enddata
\tablenotetext{}{This table contains, for each target, any significant signals identified by \rvsearch. We report the period, semi-amplitude, eccentricity, minimum mass, {False Alarm Probability (FAP)}, and a classification of each signal (KP: Known Planet, Candidate: promising planet candidate signal, Binary: Binary Star, SRC: Source Requiring Confirmation, ACT-R: Stellar Rotation, ACT-M: Magnetic Activity Cycle, {LPS: Long Period Signal that does not return a well sampled posterior and therefore lacks a full orbital solution prompting us to report only the initial $\Delta$BIC periodogram peak,} FP: False positive from aliasing or window function of another detection). *: This signal was reported as a planet in other previous works but we believe it to be stellar in nature, see corresponding discussion sections for details.}
\end{deluxetable*}

\begin{table}[!htbp]
\centering
\caption{Linear{/Quadratic} RV Trends \label{tab:linearrvsignals}}
\begin{tabular}{lll}
\hline\hline
HD     & Alias & RV Trend \\
\hline \vspace{2pt}
100623 & 20 Crt  & 0.00514923 \ms\,day$^{-1}$\\ 
131977 & GJ 570A & -0.0116872 \ms\,day$^{-1}$\\ 
188512 & $\beta$ Aql & 0.00262165 \ms\,day$^{-1}$\\
190248 & $\delta$ Pav & -0.00055 \ms\,day$^{-1}$\\
\hline\hline
\end{tabular}
\tablenotetext{}{Stars from our sample for which the preferred \rvsearch\ model included a linear and/or quadratic trend. All appear to be due to known stellar companions except for $\delta$ Pav.}
\end{table}

\subsection{Identification of Candidate Activity Signals in the S-index Data}\label{sec:activity_analysis_sindex}

We also use \rvsearch\ to carry out an uninformed search on each star's combined S-index measurements, similar to the RV fitting described above. By providing \rvsearch\ with data sets composed of the observation time stamp, S-index, and {the empirical instrument-by-instrument S-index errors} described in Section \ref{subsec:ActivityIndicators_Sindex} from each observation, we are able to determine whether the S-index data contain significant periodic signals.  Further, if the empirical errors are underestimated, \rvsearch\,'s use of a jitter term will adjust them to more accurately capture the scatter on a star-by-star basis. A list of all of the detected S-index signals is provided in Table \ref{tab:ActivitySignals_Sindex}. 

We then carry out a side-by-side comparison of the signals found in the activity search to the signals found in the radial velocity search. In instances of overlapping periods between the two search results, we assert that the signal in the radial velocities is likely caused by stellar activity, rather than by the gravitational effects of an orbiting exoplanet. New RV signals that show evidence of this period overlap are reported in Table \ref{tab:AllSignals} as `Activity'. Instances where there is no overlap between the significant periods detected in the RV and S-index data sets are treated on a more individual basis. If the signal detected in the radial velocity periodogram peak is well defined, the strength of the RV signal increases roughly monotonically with the number of observations, and we do not find a correlated activity signal, then we mark the signal as a `Candidate' in Table \ref{tab:AllSignals}. 

For less obvious cases, where the RV signal is one of a set of numerous peaks clustered in a narrow period range but we find no corresponding signal in the activity data, we then consider the star's spectral type and known activity history to decide whether the signal is likely to be due to activity. These cases are discussed in detail in each star's subsection. We adopt the classification Source Requiring Confirmation ('SRC') for signals that do not yet have enough evidence to be classified as either Candidate or Activity. Signals marked as SRC in Table \ref{tab:AllSignals} require further followup analysis to determine their nature.

Results for HD 115617's \RVSearch\ analysis of the S-index data are shown in Figure \ref{fig:HD115617_Activity_Summary}. {No significant detections are made, although we note that a strong signal is present at $P=3995.5$\,days in the periodogram. The strength of this signal is limited by the time span of observations for this target; we expect that given several more years' worth of data, we would be able to state conclusively whether this signal is physical or not.}

When compared with the radial velocity analysis in Figure \ref{fig:HD115617_Summary}, we see no overlapping periods between the three previously published planet signals the S-index signal of growing strength in the periodogram. We thus affirm that the planets HD 115617 b, c, and d are not false signals caused by activity, but rather true exoplanets.

We include summary figures like Figure \ref{fig:HD115617_Activity_Summary} from each star's S-index data in a figure set in the online journal. As referenced in \ref{subsec:ActivityIndicators_Halpha}, the \UCLES\, spectrograph does not cover a wide enough wavelength range for the Ca II H \& K lines to be observed simultaneously with the iodine region. For targets whose RV signals are largely driven by \UCLES\ data, this causes the corresponding {S-index} activity analysis to be less definitive as a planet vetting step. {Instead, we use EW$_{H\alpha}$ measurements as an activity indicator for those stars.}

\startlongtable
\begin{table*}[!htbp]
\centering
\caption{S-index Signals Identified by \rvsearch\
\label{tab:ActivitySignals_Sindex}}
\begin{tabular}{lll||lll}
\hline\hline
ID & Period [days] & S-index Amp. & ID & Period [days] & S-index Amp.\\
\hline 
HD4628 I & 3699$\pm$310 & 0.0161$\pm$0.0016 &	HD85512 VIII & 51.74$\pm$0.06 & 0.0152$\pm$0.0023 \\
HD14412 I & 2312$\pm$73 & 0.013$\pm$0.0034 &	HD100623 I & 3741$\pm$93 & 0.0228$\pm$0.0025 \\
HD14412 II & 5686$\pm$1600 & 0.0191$\pm$0.0079 & HD114613 I * & 6722.80 & -- \\
HD16160 I & 4232$\pm$310 & 0.0417$\pm$0.0073 &	HD125072 I & 2989$\pm$100 & 0.098$\pm$0.011 \\
HD16160 II & 3204$\pm$110 & 0.0253$\pm$0.0061 &	HD125072 II & 40.49$\pm$0.04 & 0.0336$\pm$0.0072 \\
HD20766 I * & 1553.62 & -- &	HD131977 I & 22.77$\pm$0.0 & 0.29$\pm$0.17 \\
HD22049 I & 1086.7$\pm$7.1 & 0.0496$\pm$0.0048 &	HD131977 II & 3.88$\pm$0.0 & 0.192$\pm$0.065 \\
HD26965 I & 3177$\pm$84 & 0.0206$\pm$0.0018 &	HD131977 III & 2.09$\pm$0.0 & 0.067$\pm$0.014 \\
HD30495 I & 71.46$\pm$0.11 & 0.0303$\pm$0.0046 &	HD146233 I & 2812$\pm$290 & 0.0094$\pm$0.0032 \\
HD32147 I & 3774$\pm$250 & 0.063$\pm$0.016 &	HD146233 II & 5272$\pm$1500 & 0.0116$\pm$0.0043 \\
HD32147 II & 3204$\pm$310 & 0.043$\pm$0.016 &	HD149661 I & 1649$\pm$55 & 0.0423$\pm$0.0065 \\
HD32147 III & 381.7$\pm$2.4 & 0.0093$\pm$0.0019 &	HD149661 II & 3874$\pm$1200 & 0.068$\pm$0.095 \\
HD32147 IV & 343.2$\pm$2.7 & 0.0088$\pm$0.0018 &	HD156026 I & 378.9$\pm$2.2 & 0.05$\pm$0.01 \\
HD32147 V & 95.6$\pm$0.24 & 0.005$\pm$0.0016 &	HD160346 I & 2975$\pm$600 & 0.0883$\pm$0.0094 \\
HD50281 I & 2264$\pm$11 & 0.0748$\pm$0.0042 &	HD160346 II & 392.6$\pm$3.2 & 0.05$\pm$0.013 \\
HD50281 II & 2102$\pm$12 & 0.065$\pm$0.005 &	HD160346 III & 7.96$\pm$0.01 & 0.0313$\pm$0.0093 \\
HD50281 III & 139.42$\pm$0.05 & 0.0345$\pm$0.0039 &	HD160346 IV & 2.54$\pm$0.0 & 0.0177$\pm$0.0081 \\
HD50281 IV & 12.48$\pm$0.0 & 0.0266$\pm$0.0039 &	HD190248 I * & 6810.18 & -- \\
HD50281 V & 16.5$\pm$0.0 & 0.022$\pm$0.0036 &	HD192310 I & 3817$\pm$60 & 0.0409$\pm$0.0013 \\
HD50281 VI & 5.39$\pm$0.0 & 0.083$\pm$0.084 &	HD192310 II & 345.34$\pm$0.48 & 0.0093$\pm$0.0058 \\
HD50281 VII & 2.7$\pm$0.0 & 0.0169$\pm$0.0047 &	HD192310 III & 44.01$\pm$0.11 & 0.0044$\pm$0.0011 \\
HD69830 I & 3989$\pm$190 & 0.0146$\pm$0.0017 &	HD192310 IV & 432.6$\pm$3.4 & 0.015$\pm$0.0031 \\
HD69830 II & 731$\pm$31 & 0.0038$\pm$0.0018 &	HD192310 V & 40.8$\pm$0.1 & 0.00383$\pm$0.00088 \\
HD69830 III & 2530$\pm$180 & 0.008$\pm$0.002 &	HD192310 VI & 34.6$\pm$0.03 & 0.00521$\pm$0.00082 \\
HD72673 I & 3217$\pm$200 & 0.0097$\pm$0.0016 &	HD192310 VII & 133.38$\pm$0.43 & 0.0069$\pm$0.0015 \\
HD75732 I & 3801$\pm$130 & 0.0263$\pm$0.0015 &	HD192310 VIII & 33.73$\pm$0.05 & 0.00563$\pm$0.00097 \\
HD85512 I & 4245$\pm$52 & 0.2106$\pm$0.0029 &	HD207129 I * & 1897.99 & -- \\
HD85512 II & 1294$\pm$14 & 0.0443$\pm$0.0035 &	HD209100 I & 2063$\pm$160 & 0.0588$\pm$0.0046 \\
HD85512 III & 478.1$\pm$2.2 & 0.0324$\pm$0.0027 &	HD209100 II & 32.87$\pm$0.07 & 0.045$\pm$0.036 \\
HD85512 IV & 322.05$\pm$0.85 & 0.0351$\pm$0.0032 &	HD216803 I & 3.89$\pm$0.0 & 0.066$\pm$0.008 \\
HD85512 V & 45.52$\pm$0.04 & 0.0187$\pm$0.0021 &	HD216803 II & 4.08$\pm$0.0 & 0.051$\pm$0.016 \\
HD85512 VI & 44.18$\pm$0.03 & 0.0188$\pm$0.0016 &	HD216803 III & 2.8$\pm$0.3 & 0.019$\pm$0.013 \\
HD85512 VII & 104.3$\pm$0.15 & 0.022$\pm$0.0034 &	
\end{tabular}
\tablenotetext{}{This table contains all significant S-index signals identified by \rvsearch. We report the period and S-Index semi-amplitude (in units of the Mt. Wilson S-index) of each signal. For signal interpretations, see each star's individual discussion section. Signals with a * designation failed to return  well constrained MCMC results, and so we instead report the MAP fits for their orbital periods.}
\end{table*}

\subsection{Identification of Candidate Activity Signals in \UCLES\ EW$_{H \alpha}$ Data} \label{sec:activity_analysis_halpha}

{Stellar variability in the \UCLES\, time series, which does not provide coverage of the Ca II H \& K lines, was instead assessed using measurements of the equivalent width of the H$\alpha$ absorption line (EW$_{H \alpha}$). Although the H$\alpha$ line is usually more informative for cooler, M dwarf stars \citep[see, e.g.,][]{Robertson2013} it is still sensitive to some activity variations in hotter, Sun-like stars. We subjected the EW$_{H \alpha}$ time series to the same uninformed search process with \rvsearch\, described for the S-index data above, again beginning with the removal of any 5+$\sigma$ outliers within each star's data set. We run the cleaned time series through \rvsearch\, and record any significant periodicities so that they can be compared with the periods (if any) detected in that star's RV data. We find that the long term structure present in EW$_{H \alpha}$ is not sufficiently periodic to show up in the \rvsearch\ $\Delta$BIC periodograms, but caution that it may still obscure some lower amplitude activity signals within the data. A list of all of the detected EW$_{H \alpha}$ signals is provided in Table \ref{tab:ActivitySignals_Halpha}.} 

\begin{figure*}
\makebox[\textwidth]{\includegraphics[width=.6\paperwidth]{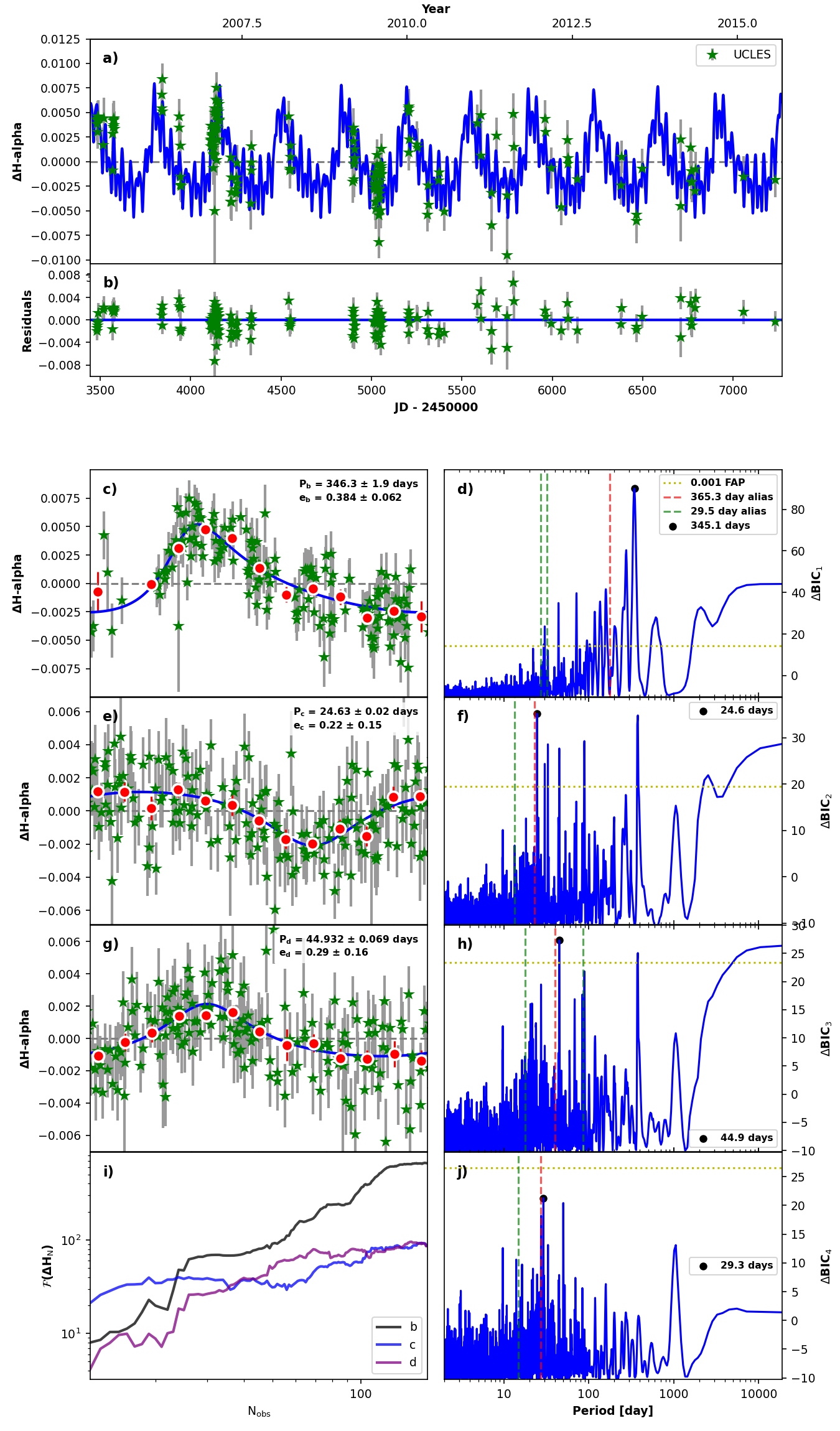}}
\caption{\RVSearch\, results for the relative EW$_{H \alpha}$
    measurements of HD 115617. Panel (a) shows the EW$_{H \alpha}$
    time series with the best-fit model plotted behind them while
    panel (b) shows the EW$_{H \alpha}$ residuals. Panels (c), (e),
    and (g) show phase folded curves for the signals identified by
    \rvsearch, and panels (d), (f), and (h) show the periodograms
    associated with each signal, with the yellow horizontal dotted
    line marking the minimum $\Delta$BIC for a 1\%\, FAP significance
    and the red and green vertical dotted lines showing the one month
    and one year aliases of the tallest peak. Panel (i) shows the
    effective strength of each signal as a function of the number of
    observations, and panel (j) shows the residuals periodogram,
    indicating that no further signals are found in the data set.
    The complete set of ${H \alpha}$ activity summary plots (31
        figures) can be found in the online journal.
\label{fig:HD115617_Halpha_Summary}}
\end{figure*}%
\figsetstart
\figsetnum{6}
\figsettitle{H$\alpha$ Activity Analysis Summary Plots}

\figsetend

Figure \ref{fig:HD115617_Halpha_Summary} shows an example summary
    figure from H$\alpha$ analysis of HD 115617. Our analysis returns
    three detections, only one of which we believe to be of
    astrophysical causes. The first signal, with $P=346.3\pm1.9$\,d,
    is extremely close to one year. Just as with \HARPS\, data, we
    expect to see yearly systematics within the H$\alpha$ data caused
    by the observing cadence. This signal is therefore attributed to
    systematics. H$\alpha$ signal II is close to the rotation period
    predicted for this star. Because we have a longer observation
    baseline and more precise measurements than were used for
    previous rotation period estimates in the literature, we report
    H$\alpha$ signal II as an update to the stellar rotation period.
    H$\alpha$ signal III is most likely too long-period to be caused
    by differential rotation, though we discuss the possibility in
    detail in \S5.14. It also does not correspond to any peaks in RV
    or S-index data. We leave it to future, more in-depth studies of
    stellar activity to characterize the cause of this detection.
    The resulting summary figures for all 31 stars in this study observed
    by \UCLES\, are available in a figure set in the online journal.

\startlongtable
\begin{table*}[!htbp]
\centering
\caption{H $\alpha$ Signals Identified by \rvsearch\
\label{tab:ActivitySignals_Halpha}}
\begin{tabular}{lll||lll}
\hline\hline
ID & Period [days] & EW$_{H \alpha}$ Amp. & ID & Period [days] & EW$_{H \alpha}$ Amp.\\
\hline 
HD20794 I & 2204$\pm$16 & 0.0057$\pm$0.0041 &	HD115617 III & 44.93$\pm$0.07 & 0.00151$\pm$0.00026 \\
HD20794 II & 1753$\pm$46 & 0.00283$\pm$0.00095 &	HD125072 I * & 7137.76 & 0.00630685 \\
HD20807 I * & 2859.91 & 0.00367565 &	HD136352 I * & 376.779 & 0.00913149 \\
HD23249 I & 49.57$\pm$0.1 & 0.00241$\pm$0.00049 &	HD136352 II * &  7207.58 & 0.00537 \\
HD26965 I & 43.5$\pm$0.07 & 0.00316$\pm$0.00054 &	HD140901 I & 7161$\pm$2100 & 0.01205$\pm$0.00082 \\
HD72673 I & 341.2$\pm$3.6 & 0.00508$\pm$0.00076 &	HD140901 II & 19.99$\pm$0.02 & 0.0036$\pm$0.0007 \\
HD100623 I & 3205$\pm$130 & 0.01136$\pm$0.00043 &	HD160691 I * & 30888.2 [peak=5293.7] & 0.00379764 \\
HD102365 I * & 369.144 & 0.00752353 &	HD160691 II * & 362.611 & 0.00247693 \\
HD102365 II * & 18549.80 [peak=7273.6] & 0.00322183 &	HD190248 I & 352.9$\pm$1.5 & 0.00261$\pm$0.00032 \\
HD102365 III * & 49.68 & 0.00130454 &	HD190248 II & 1171$\pm$36 & 0.0021$\pm$0.00033 \\
HD114613 I * & 27460.5 [peak=7652.9] & 0.00360532 &	HD192310 I * & 13621.7 & 0.00676791 \\
HD114613 II * & 365.429 & 0.00211492 &	HD192310 II * & 363.678 & 0.00499708 \\
HD115617 I & 346.3$\pm$1.9 & 0.00376$\pm$0.00027 &	HD207129 I & 5455$\pm$1900 & 0.0036$\pm$0.00036 \\
HD115617 II & 24.63$\pm$0.02 & 0.00152$\pm$0.00025 &	HD207129 II & 1726$\pm$71 & 0.00309$\pm$0.00061 \\
\end{tabular}
\tablenotetext{}{This table contains all significant EW$_{H \alpha}$ signals identified by \rvsearch. We report the period and EW$_{H \alpha}$ semi-amplitude of each signal. For signal interpretations, see each star's individual discussion section. Systems with a * designation failed to return well constrained MCMC results, generally producing period uncertainties larger than the median period value, and so we instead report their MAP orbital solutions. In three cases, even the MAP period for the long period signal is more than 3x larger than the significant peak in the $\Delta$BIC periodogram and so we note the period of the original peak in brackets next to the MAP result.}
\end{table*}

\subsection{Injection and Recovery Analysis}

After conducting and analyzing the uninformed radial velocity and stellar activity indicator searches, we use \rvsearch\, to execute an injection/recovery (I/R) analysis for each star's residual RV data set. I/R analyses characterize the completeness of each star's RV time series, quantifying what combinations of companion orbital periods and minimum masses we are currently sensitive to. The results of these I/R efforts make clear what types of planets we would expect to be able to detect in each star's Habitable Zone given the current data sets, and can help to prioritize future RV surveys that aim to push sensitivity limits to lower mass temperate planets.
While we are primarily interested in the current RV sensitivity within each star's Habitable Zone, we use this exercise as an opportunity to quantify our planet sensitivity across the entirety of the orbital period space covered by the combined RV data sets. 

To accomplish this, 5000 synthetic planet signals are injected into the RV residuals of each star's uninformed search results. These ``planets" are assigned orbits and \msini\, values drawn from log-uniform distributions. The corresponding periods and RV semi-amplitudes span 2 to 10,000 days and 0.1 to 1,000 \ms, respectively. Key properties of the data set such as observation baseline, measurement values, and uncertainties are preserved. Following the results of \citet{Kipping2013}, which examined the eccentricities  of the population of RV detected exoplanets, the synthetic planets have eccentricities drawn from a $\beta$ distribution. The same planet search algorithm used in the uninformed search is then run on these modified data sets to determine whether the injected signals can be recovered. This quantifies the planet sensitivity of the existing data, calculating the probability that a planet of a given \msini\, and orbital period would be detected within the data. 

\begin{figure}[H]
\includegraphics[width=.49 \textwidth]{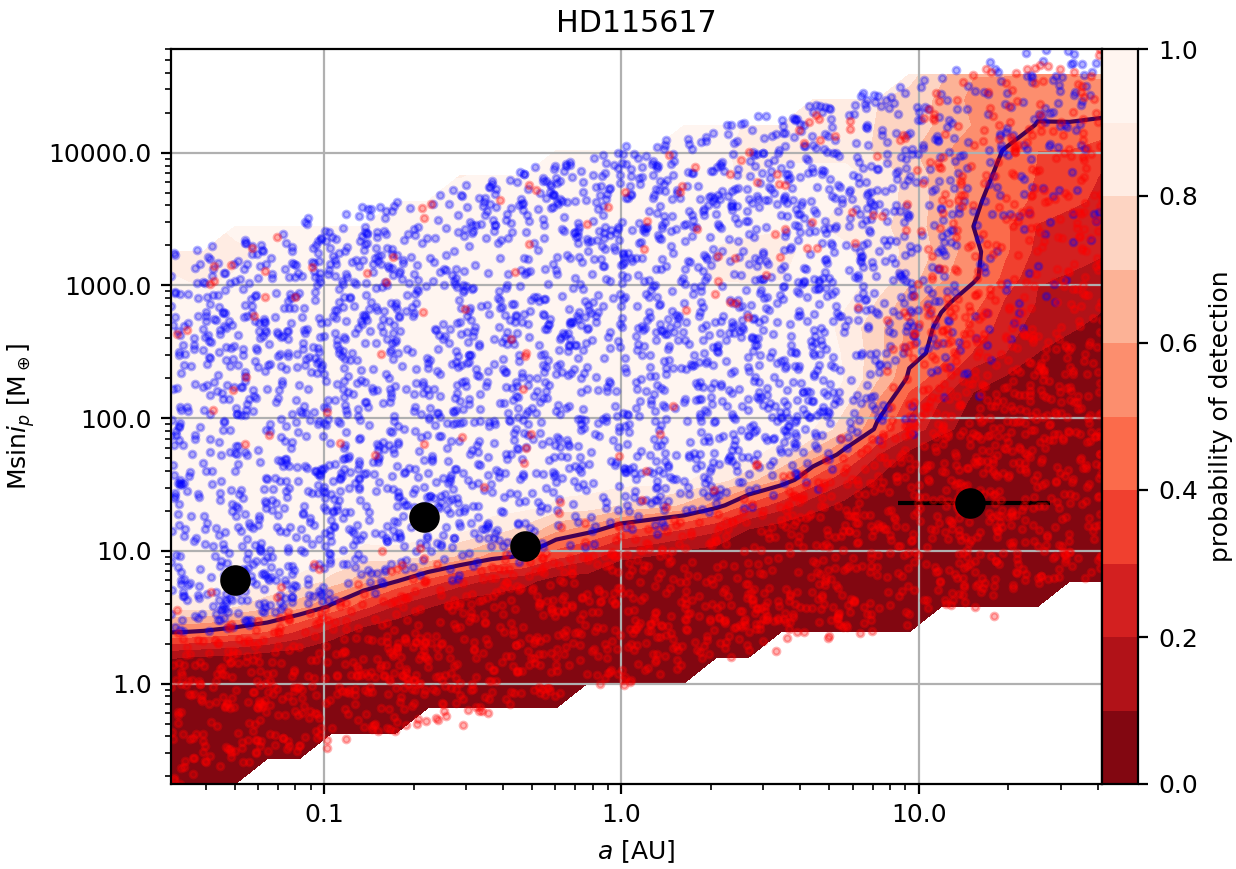}
\caption{The \RVSearch\, completeness contour plot for HD 115617. The
    large black dots indicate the periodic signals identified by
    \rvsearch\, in the archival RV data (see figure
    \ref{fig:HD115617_Summary}). The colored points depict the
    synthetic planets that were injected into the RV residuals --
    blue points represent planets that were successfully recovered,
    while red points were not recovered. The red contours display the
    probability of detection averaged over small regions of
    semi-major axis and \msini\, space. The black line is the 50\%
    detection probability contour. 
    {The complete set of Injection and Recovery Analysis plots
        (46 figures) can be found in the online journal.}
\label{fig:recoveries}}
\end{figure}

A completeness contour plot is generated, demonstrating what regions of \msini\, and orbital period space we are already sensitive to with the existing data. Figure \ref{fig:recoveries} shows an example of a completeness contour plot for HD 115617. The three planets published in \citet{Vogt2010} and detected in the uninformed search stage of our analysis are depicted as black circles, and lie within the region of \msini\, and semi-major axis space where we expect to be sensitive to Keplerian signals. The three regions of red points above the detected signals are remnants of the way the injection and recovery analysis works. Once an injected planet is recovered, it is compared with the already-fit model to see whether it would be in a reasonably stable orbit location compared with what has already detected. If the injected planet has the same orbit as an already-fit signal, \rvsearch\ would not include the planet in the model due to orbital stability constraints, and so the injected signal is ``not recovered." This results in columns of non-recovered injected planets that align with previously detected/removed Keplerian signals. The black circle corresponding to the final candidate detected in the uninformed search, the long period signal that overlaps with a period in the activity search, is located in a regime with a much lower probability of detection. {While the probability of detecting a synthetic at this period and semi-amplitude is $<$10\% according to the figure, the False Alarm Probability (FAP) of this long period signal detected in the RV time series is 1.22e-10, making it very unlikely to be a result of random fluctuations in the data.}

\figsetstart
\figsetnum{7}
\figsettitle{Injection and Recovery Analysis Contour Plots}

\figsetgrpstart
\figsetgrpnum{7.30}
\figsetgrptitle{HD 115617}
\figsetplot{HD115617_recoveries.png}
\figsetgrpnote{Injection and recovery analysis contour plot for HD 115617 from \rvsearch. See individual discussion sections in the paper for further details on interpreting these plots.}
\figsetgrpend

\figsetend

Based on Figure \ref{fig:recoveries}, we would expect that for planets on a 0.1 AU orbit the existing RV data would be sensitive down to planet masses of \msini\, = 2.6\,\mearth, a mass that could include terrestrial planets. But when considering planets on more temperate orbits near 0.914 AU (this star's EEID) the existing RV data is only sensitive to planets with masses of $\geq$15\,\mearth. Thus efforts to detect an Earth analog around HD 115617 would not succeed with the existing data set. {However,} we can say with some confidence that if a 15\,\mearth\, or larger planet were orbiting the star within its habitable zone, which would preclude the existence of an Earth-analog on a similar orbit, that we would already be able to detect it. Thus, at the moment, there is nothing to eliminate the possibility that HD 115617 could host an Earth-analog.

Table \ref{tab:IR_Sensitivity} describes the results of these injection and recovery tests. For each star in the set, we report what regions of \msini\, and period space we are already sensitive to using the compiled archival data. We include minimum mass values for several semi-major axes, marking the limits of sensitivity for each star's archival data set.

\subsection{Speckle Imaging Analysis}
\begin{figure}[H]
\includegraphics[width=.49 \textwidth]{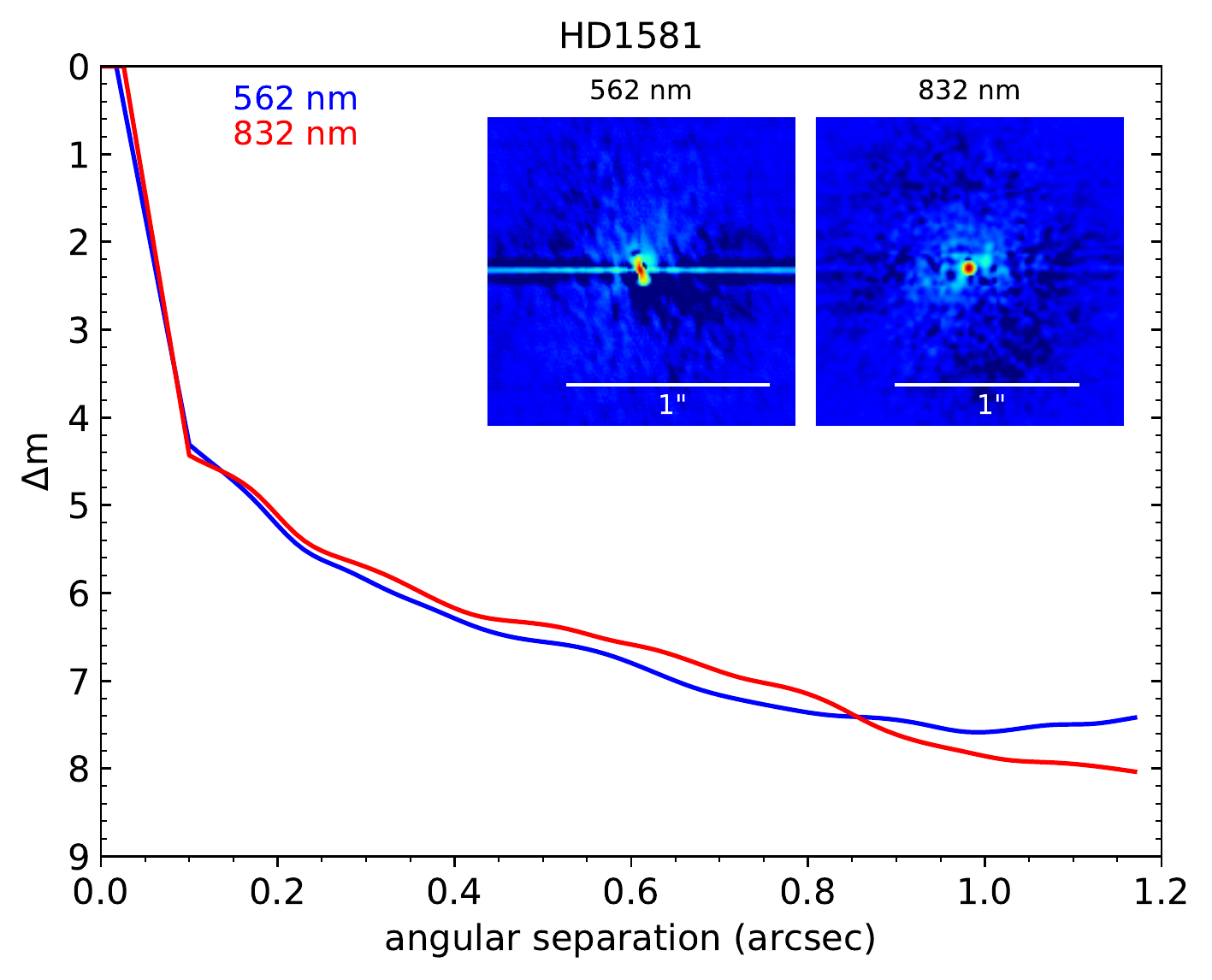}
\caption{Speckle Imaging Analysis plot for HD 1581. 
    {Top right: the reconstructed Speckle images for HD 1581 for each wavelength. Larger plot: 5$\sigma$ flux detection limit relative to image backgrounds, measured in concentric circular annuli from the center of the image. The complete set of Speckle Imaging Analysis plots
        (6 figures) can be found in the online journal.}
\label{fig:speckle}}
\end{figure}

In an attempt to identify any stellar companions from the speckle imaging observations obtained for some of the target stars, reconstructed images derived from the image reduction process were used \citep{Howell2011,Horch2011}. Distribution of all local maxima and minima in the background of the images as a function of separation were examined by drawing five concentric annuli each with width of 0.2\arcsec\, centered at radii of 0\arcsec.2, 0\arcsec.4, 0\arcsec.6, 0\arcsec.8, and 1\arcsec.0 from the primary star. Standard deviations of these extrema from the mean background in each annulus were computed by averaging the values obtained from both maxima and minima. A 5$\sigma$ detection limit, which is five times brighter than the mean background within each annulus, was then estimated. Any peak in the image that was above the 5$\sigma$ limit at a specific angular separation was considered a companion candidate for further study. For the 6 targets that had speckle imaging observations, no such peaks were found and therefore no stellar companions were identified. For these non-detections, 5$\sigma$ limits derived for each annulus in terms of instrumental magnitude difference ($\Delta$m$_{i}$, where $i$ is filter type) were used as a conservative upper limit above which stars should be detected, thus providing a constraint of the possible undetected low mass companions nearby. Since $\Delta$m$_{i}$ varies as a function of separation from the primary where at smaller separations $\Delta$m$_{i}$ is slightly smaller, we reported the estimated $\Delta$m$_{i}$ at both 0\arcsec.1 and 1\arcsec.1 from the primary in Table \ref{tab:AATwork}. Figure \ref{fig:speckle} summarizes these results for HD 1581.

\figsetstart
\figsetnum{8}
\figsettitle{Speckle Imaging Analysis Plots}
\figsetgrpstart
\figsetgrpnum{8.1}
\figsetgrptitle{HD 1581}
\figsetplot{HD1581_20201029_562_832_final.pdf}
\figsetgrpnote{Speckle Imaging Analysis plot for HD 1581. Top right: reconstructed images in two wavebands. Full plot: the 5$\sigma$flux detection limit relative to the image background as a function of angular distance from the image center.}
\figsetgrpend

\figsetend

\begin{table*}
\begin{center}
\caption{Speckle Imaging Results \label{tab:AATwork}}
\begin{tabular}{c|c|c|c|c|c|c|c}
HD & Instrument & Date (UT) & $\Delta$m$_{562}$ (0.1\arcsec) & $\Delta$m$_{562}$ (1.1\arcsec) & $\Delta$m$_{832}$ (0.1\arcsec) & $\Delta$m$_{832}$ (1.1\arcsec) & EW$_{H \alpha}$ Correlation \\ \hline\hline
1581 & Zorro & 2020 Oct 29 & 4.31 & 7.50 & 4.43	& 7.95 & N/A \\ \hline
20766 & Zorro & 2020 Oct 23 & 5.02 & 7.36 & 4.82 & 8.00 & -0.48\\ \hline
20807 & Zorro & 2020 Oct 23 & 4.80 & 6.29 & 4.70 & 8.22 & -0.17\\ \hline
140901 & Zorro & 2021 Jul 22 & N/A & N/A & 4.44	& 7.79 & -0.68\\ \hline
146233 & `Alopeke & 2021 Jun 27 & 4.66 & 6.56 & 4.74 & 9.33 & -0.06\\ \hline
196761 & `Alopeke & 2021 Jun 27 & 4.97 & 6.02 & 5.05 & 8.22 & N/A\\ \hline\hline
\end{tabular}
\tablecomments{All columns without N/A values presents speckle imaging details, except for the last column that includes Pearson coefficient values of correlation between \UCLES\, RVs and H$\alpha$ EWs. Some \UCLES\, RVs returned no significant signals when analyzed alone and so no correlation is calculated, thus N/A. For HD 140901, no data was acquired for the blue channel due to alignment issue.}
\end{center}
\end{table*}

\section{Systems with updated parameters}\label{sec:updated_params}

{In this section, we present results from targets for which we recover previously published planetary systems, stellar companions, and activity cycles. For the cases where we have additional data or increased precision, we cite the current accepted values and report updates to these systems' parameters. Table \ref{tab:updates} contains former and new (this work) parameters for previously reported exoplanet systems.}

\subsection{HD 13445 (GJ 86 A) \label{sec:HD13445}}

HD 13445 (GJ 86 A, HR 637, HIP 10138) is a nearby K1V star \citep{Gray2006} at $d$\,=\,10.76 pc \citep[$\varpi$\,=\,92.9251\,$\pm$\,0.0461 mas; ][]{GaiaEDR3}.
GJ 86A has both a known stellar companion (GJ 86B, WD 0208-510) and an exoplanet (GJ 86Ab, HD 13445b). 
\citet{Farihi2013} characterize the white dwarf companion GJ 86B and constrain its orbit -- estimating a spectral type of DQ6, mass = 0.59\,$\pm$\,0.01 \msun, orbital period of $P=120-481$\,yr, and adopted system age of 2.5 Gyr. 

{We detect two signals in the RVs}. One is the known exoplanet, and the other {may be caused by the binary companion, but is too poorly constrained to say for certain}. \citet{Butler2001} published the planet with 15.76 day period. We derive orbital parameters for GJ 86Ab of {$P_{b}=15.764862\pm0.000043$\,d, $K_{b}=377.58\pm0.77$\,\ms, $e_{b}=0.0485\pm0.0018$. }
{The second significant detection has a peak in the periodogram with $P=20504$\,days. Because this is much longer than the observation baseline for the target, the MCMC fit for the signal is not physical. We categorize this signal as LPS and note that more data would be required to constrain this signal further. Further discussion of the LPS category of detection can be found in Section \ref{subsec:RVsearchOverview}}.

{Finally, we note that H$\alpha$ analysis shows strength in the periodogram at $P=2001.7$\,days, though the detection does not cross the False Alarm Probability threshold.}

\subsection{HD 16160 (GJ 105 A) \label{sec:HD16160}}

HD 16160 (GJ 105 A, HR 753, HIP 12114) is a nearby K3V spectral standard star \citep{Keenan1989} in a triple system, located at $d=7.23$ pc \citep[$\varpi=138.2084\pm0.1436$ mas;][]{GaiaDR2}. 
The star is a $5.1\pm1.1$ Gyr-old thin disk \citep{Ramirez2012}, low-activity \citep[\logrphk = -4.87;][]{GomesdaSilva2021} star with magnetic activity cycle of \Pcyc\, $\simeq$ 12.18\,yr \citep{Willamo2020} or $12.7\pm0.11$\,yr \citep{BoroSaikia2018}. 
From analysis of the Mt. Wilson survey data, \citet{Donahue1996} reports an average rotation period of \Prot\, = 48.0\,d over 5 seasons, with individual seasonal rotation periods ranging from 42.2\,d to 51.5\,d (i.e, pronounced differential rotation). 

GJ 105 A has a faint M7V companion GJ 105 C (HD 16160B, WDS J02361+0653B) observed at separations between 1\arcsec.7 and 3\arcsec.3 \citep{Golimowski1995,Golimowski2000,Mason2001}\footnote{References to the Washington Double Star (WDS) catalog \citep{Mason2001} are actually referring to the regularly updated WDS table at Vizier (https://cdsarc.unistra.fr/viz-bin/cat/B/wds).}, and the M4.0V star GJ 105 B star on a very wide orbit at 164\arcsec\, separation \citep{Mann2015,vanMaanen1938}.  
Astrometric perturbations attributed to a low-mass stellar companion to GJ 105 A (BD+6$^{\circ}$ 398) were first reported by \citet{Lippincott1973}, who measured photographic plate positions from the Sproul astrometric program taken between 1937 and 1968 to estimated the orbit to have $P = 50$\,yr and $e = 0.6$, and predicted the companion to be a 0.10\,\Msun\, star of type M6. 

\citet{Golimowski2000} concluded that the companion C first observed in 1993 with the Palomar 60" Adaptive Optics Coronagraph \citep{Golimowski1995} was consistent with (1) the astrometric perturbations with $\sim$50-60\,yr periodicities reported by \citet{Lippincott1973} (and later refined by \citet{Ianna1992} and \citet{Heintz1994}), and (2) the 11\,m\,s$^{-1}$\,yr$^{-1}$ radial velocity trend observed during the 1990s by \citet{Cumming1999}.
\citet{Ianna1992} analyzed photographic plate positions for GJ 105 A between 1915 and 1992 and estimated $P = 59.5$\,yr, astrometric amplitude $\alpha$\,=\,0\arcsec.293, $e=0.35$, and companion mass $M_C =0.13$\,\msun.
Our analysis of the archival RV data with \rvsearch\, yields a long-period signal with {$P=22999\pm1200$\,d\, ($63.0\pm3.3$\,yr), $K=702.5\pm2.9$\,\ms, and $e=0.6075\pm0.0092$,} which is reasonable for the orbit of GJ 105 C.

These values are not far from the most recent astrometric-only orbital analysis from \citet{Heintz1994}, who estimated $P=61$\,yr, $e=0.67$, $i=49^{\circ}$ (consistent with companion mass $M_C = 0.10$\,\Msun). They also align well with the values reported in the recent \citet{Rosenthal2021}, who found $a=16.37\pm0.28$\,AU and $e=0.6427^{+0.0038}_{-0.0039}$. Using the stellar mass adopted in this paper, we calculate the period to be approximately $P=77$\,years. This {also aligns with the} signal we recover, though our detection is significantly less well constrained than in other works. A joint analysis of the radial velocity, astrometric, and imaging data over the past century could yield stronger orbital constraints and a more accurate dynamical mass estimate, but is beyond the scope of this study.

{Analysis of S-indices using \RVSearch\, returns two significant activity detections, with parameters: $P_{I}=4232\pm310$\,d, $e_{I}=0.17\pm0.055$ and $P_{II}=3204\pm110$\,d, $e_{II}=0.413\pm0.092$. The first of these aligns fairly well with the magnetic activity cycle reported by \citet{Willamo2020}. We recommend a more in-depth study of this star's activity to fully characterize the sources of these periodic signals in the S-indices.}

\subsection{HD 20794 (82 Eri) \label{sec:HD20794}}

82 Eri (GJ 139, HD 20794, HR 1008, HIP 15510) is a G8V star \citep{Gray2006} at $d=6.00$ pc \citep[$\varpi=166.5242\pm0.0784$ mas;][]{GaiaEDR3}. 
The star is somewhat cooler than the Sun (\teff$\,=5398$ K), metal poor ([Fe/H]$\,=-0.41$) \citep{Tsantaki2013}, and very inactive (\logrphk\, = -5.025) \citep{Lovis2011}.
\citet{Lovis2011} reported a magnetic activity cycle of \Pcyc\, = 751$_{-25}^{+290}$\,d\, {($2.06^{+0.79}_{-0.07}$\,yr)} based on 197 \logrphk\, measurements over a span of 2694\,d. 
The star was reported to host three planets by \citet{Pepe2011}, with orbital periods of $P_{b}=18.1$  d, $P_{c}=40.1$\,d, and $P_{d}=90.3$\,d, based upon their analysis of 173 \HARPS\ RV data points taken between 2003 and 2011. A reanalysis of the system was published in 2017, which made use of an updated \HARPS\ data set containing 713 RV epochs obtained between 2003 and 2013 \citep{Feng2017}. The \citet{Feng2017} results confirm the Keplerian nature of the 18 and 90 day signals put forth in \citet{Pepe2011} and identify two additional planet candidates with orbital periods of 147 and 330 days. {They find} only weak evidence of the $\sim$40 day signal reported by \citet{Pepe2011}, however, and assert that more data are necessary to determine the nature of this signal.

Our data set for HD 20794 contains 763 \HARPS\ epochs, spanning 2003 - 2016, along with 549 \UCLES\ points, and 77 \PFS\ points. Running this combined RV data set through \rvsearch, we confirm HD 20794 b {($P_{b}=18.305\pm0.0052$\,d, $K_{b}=0.807\pm0.089$ \ms\, $e_{b}=0.17\pm0.11$)} and HD 20794 d ($P_{d}=89.766\pm0.085$\,d, $K_{d}=0.86\pm0.12$ \ms\, $e_{d}=0.27\pm0.11$) in Tables \ref{tab:updates} and \ref{tab:AllSignals}. 

Similarly to \citet{Feng2017}, we do not register a detection of the $\sim$40 day signal attributed to HD 20794 c, though we note that the residuals periodogram shows significant power for a signal at 40.2 days, which likely corresponds to the reported 40.1-day period of HD 20794 c in \citet{Pepe2011}. If the signal were Keplerian in nature, however, we would expect its statistical significance to increase as more RV data points are added to the analysis. This is especially true for a star that exhibits the low levels of RV scatter we see in the HD 20794 RVs, where RMS = 1.99 \ms\, and 1.00 \ms\, for \HARPS\, and \PFS, respectively.

Another possibility is that the 40 day signal is tied to stellar variability. The star's low chromospheric activity (\logrphk = -5.03) is consistent with a rotation period of \prot\,$>$\,34 days for a star of its color \citep[using activity-rotation relations of ][]{Mamajek2008}, and so rotational modulation at a period of roughly 40 days would not be surprising. Yet applying \rvsearch\, to the star's assembled S-index data does not reveal significant power at or near a 40 day period. 
We also do not significantly detect planet candidates e, f, or g, as reported by \citet{Feng2017}. There is another peak in the {RV} residuals periodogram close to 330 days, the orbital period of candidate f, but similarly to the 40 day signal it does not cross the threshold of being detected by \rvsearch. Given the more sophisticated treatment of stellar variability and correlated noise in the that paper, however, we do not view our non-detections as a refutation of these candidates. 

{S-index analysis returns no significant signals, but analysis of H$\alpha$ in the \UCLES\, data yields two significant detections: $P_{I}=2204\pm16$\,d, $e_{I}=0.886\pm0.049$ and $P_{II}=1753\pm46$\,d, $e_{II}=0.68\pm0.15$. The high eccentricities fit to these signals are cause for skepticism regarding their exactness, but we regard them as good evidence for the existence of an approximately 7-8 year activity cycle for this star.}

\subsection{HD 22049 ($\epsilon$ Eri) \label{sec:HD22049}}

$\epsilon$ Eri (GJ 144, HD 22049, HR 1084, HIP 16537, Ran) is a young, active K2V spectral standard star \citep{Keenan1989} at $d=3.22$\,pc \citep[$\varpi=310.5773\pm0.1355$ mas;][]{GaiaEDR3} with a candidate planet and debris disks \citep{Mawet2019}.
Analysis of the Mt. Wilson Ca II H \& K data of $\epsilon$ Eri by \citet{Donahue1996} found strong evidence for differential rotation, with season-averaged rotation periods ranging from \Prot\,=\,11.04 to 12.18\,d over 9 seasons, with average \Prot\,=\,11.68\,d.
An archival analysis of 45 years of chromospheric activity data by \citet{Metcalfe2013} identified two prominent activity cycles for $\epsilon$ Eri at
$P_{cyc1} = 2.95\pm0.03$\,yr and $P_{cyc2} = 12.7\pm0.3$\,yr, at approximately 0.68$\times$ and 2.94$\times$ planet orbital period reported by \citet{Mawet2019}. 

We recover the one confirmed planet \citep[$P=2690\pm30$\,d = $7.365\pm0.082$\,yr;][]{Mawet2019}, but with a less
certain period of {$P_{b}=2832\pm120$\,d ($7.76\pm0.33$\,yr), and semiamplitude and eccentricity $K_{b}=11.1\pm1.2$ \ms, $e_{b}=0.09\pm0.08$}.
The periodogram residuals show a signal at 12.4 days, which agrees well with rotation periods reported by \citet{Donahue1996}. 

We detect one {S-index} activity signal with parameters {$P_{I}=1086.7\pm7.1$\,d (2.98$\pm$0.02 years) and $e_{I}=0.268\pm0.081$.} This agrees with the 2.95 year activity cycle reported by \citet{Mawet2019}. 

We note that we do not detect the false positive $P=773.4^{+4.7}_{-4.8}$d signal reported by \citet{Rosenthal2021}.

\subsection{HD 26965 (40 Eri A) \label{sec:HD26965}}

40 Eri A ($o^2$ Eri A, GJ 166 A, HD 26965, HR 1325, Keid) 
is a famous nearby ($d=4.98$ pc) \citep[$\varpi=200.62\pm0.23$ mas;][]{vanLeeuwen2007} K0.5V standard star \citep{Keenan1989} in a triple system with a white dwarf (B) and M dwarf (C) component.  
From time series analysis of chromospheric activity data from the Mt. Wilson survey, 
the rotation period of the star has been previously
measured to be  
43\,d \citep[][]{Baliunas1996} and
42\,d \citep[][]{Frick2004}, 
and {\it predicted} rotation periods 
(based on \logrphk\, values and correlations
with rotation for other cool dwarfs) 
have been reported to be 37.1\,d  \citep{Saar1997},
$42.2\pm4.4$\,d \citep{Lovis2011}, 
and 43\,d \citep{Isaacson2010}. 
Long term monitoring of Ca H \&\, K emission from 
40 Eri A has revealed a magnetic activity period with
measured period $P_{cyc} = 10.1\pm0.1$\,yr \citep{Baliunas1995}, $10.4$\,yr \citep[3800 d;][]{Frick2004},
$9.18^{+2.20}_{-1.48}$\,yr \citep{Lovis2011}, 
and $10 (9.57-10.5)$\,yr \citep{Olah2016}, or $10.23\pm0.07$\,yr \citep{BoroSaikia2018}.\\

\citet{Diaz2018} presented an extensive analysis of $\sim$1100 spectra taken using \HIRES, \PFS, CHIRON, and \HARPS, and reported a strong signal at $P = 42.364\pm0.015$\,d, 
$K = 1.59\pm0.15$\,\ms, $e = 0.017\pm0.046$,
but found it challenging to distinguish this signal from the
star's rotation. 
Shortly after, \citet{Ma2018} conducted a reanalysis of the
\citet{Diaz2018} data combined with 133 new spectroscopic observations taken with the TOU instrument.
\citet{Ma2018} found that
while there were signals in the star's activity indices
at $41.2\pm0.9$\,d\, and $39.2\pm0.7$\,d\, likely corresponding 
to (differential) stellar rotation, the well-defined 
$P = 42.38$\,d\, signal persisted over the seasons and between activity states -- concluding that the signal was most likely
due to a planet.
\citet{Rosenthal2021} reported a signal at 
$P = 42.305^{+0.015}_{-0.019}$\,d 
($K = 1.82^{+0.43}_{-0.31}$\,\ms) and considered
it a false positive attributed to the star's rotation, 
and another longer signal at $P = 3560^{+200}_{-580}$\,d
($K = 1.89^{+0.37}_{-0.32}$\,\ms) attributed to long-period magnetic activity cycle.\\

We detect a strong significant RV signal with 
{$P_I=42.303\pm0.025$\,days, 
$K_I=1.40\pm0.22$\,\ms, 
$e_I=0.37\pm0.17$, very similar to that reported previously
by \citet{Diaz2018} and \citet{Ma2018}. Analysis of H$\alpha$ data for this target returns a well-correlated detection with $P=43.504\pm0.066$\,d, $e=0.37\pm0.18$. The extreme proximity of these two detections leads us to classify this RV detection conclusively as activity.}
Additionally, we detect {RV signals with $P_{II}=37.33\pm0.02$\,d, $K_{II}=1.17\pm0.19$\,\ms, 
$e_{II}=0.14\pm0.12$, and $P_{II}=367.9\pm3.1$\,d, 
$K_{III}=1.63\pm0.88$\,\ms, 
$e_{III}=0.46\pm0.27$.} Looking closely at the periodogram, we note that the 37-day period signal is extremely close to the yearly alias of the 42-day signal, and report it as such. The 365-day signal is likely driven by the window function of this star as the phase folded fit makes clear that a significant (\textgreater 25\%) portion of the orbital phase space is unpopulated due to seasonal observing constraints. We therefore classify this as a false positive signal.

In the {S-index} activity analysis, we find a signal with {$P_{I}=3177\pm84$\,d ($8.70\pm0.23$\,yr) and $e_{I}=0.059\pm0.051$,}  which {agrees well with \citet{Rosenthal2021} and which} we report as an update to the 10-year magnetic cycles previously published.

\subsection{HD 39091 ($\pi$ Men) \label{sec:HD39091}}

$\pi$ Men (GJ 9189, HD 39091, HR 2022, HIP 26394) 
is a G0V star \citep{Gray2006} at $d=18.28$ pc \cite[$\varpi = 54.6825\pm0.0354$ mas;][]{GaiaEDR3}. 
$\pi$ Men has three published planets. 
$\pi$ Men b was first published in \citet{Jones2002}, and was discovered using radial velocity data from the \UCLES\, instrument. 
We recover $\pi$ Men b 
{in our radial velocity data with $P_{b}=2089.05\pm0.46$\,d, 
$K_{b}=196.5\pm0.6$\,\ms\, and 
$e_{b}=0.6428\pm0.0017$. }
These parameters are comparable to recent estimates
by \citet{Huang2018}, \citet{Gandolfi2018}, and \citet{Xuan2020}. Our analysis includes newly released \PFS\, data, building upon the \HARPS\, + \UCLES\, orbital fits performed in the previous works, and so we report our detection as an update to the orbital parameters of $\pi$ Men b.

$\pi$ Men c was the first new transiting planet discovered by NASA's Transiting Exoplanet Survey Satellite \citep[TESS;][]{Huang2018}. The planet was not robustly detected by \rvsearch\,'s uninformed search of the RVs, although there is a well defined peak in the residuals periodogram at the expected period of $P_{c}=6.2$\,d. 

$\pi$ Men d is a recently detected, sub-Neptune mass planet candidate reported to have $P_{d}=124.64^{+0.48}_{-0.52}$\,days, $K_{d}=1.68\pm0.17$\,\ms, and $e_{d}=0.22\pm0.079$ \citep{Hatzes2022}. These parameters are driven largely by observations taken as part of intensive \HARPS\, and \emph{ESPRESSO} observing campaigns, the data for which is not included in this analysis.

We detect a similar signal, consistent to within 1.5$\sigma$ on all parameters, albeit with larger uncertainties on the planet's RV semi-amplitude. Our best fit results for this third signal are {$P_{d} = 125.58\pm0.27$\,d, $K_{d} = 2.16\pm0.42$\,\ms, and $e_{d} = 0.16\pm0.15$.}

Activity analysis {of both S-index and H$\alpha$ data} for this target recovers no significant signals.

\subsection{HD 69830 (GJ 302) \label{sec:HD69830}}
HD 69830 (GJ 302, HR 3259, HIP 40693) is a well-studied star of type G8+V \citep{Gray2006} {at distance} $d$\,=\,12.58 pc \citep[$\varpi$\,=\,79.4953\,$\pm$\,0.0400 mas;][]{GaiaEDR3}, famous for hosting a planetary system of three Neptunes \citep{Lovis2006} and a dusty debris disk \citep{Beichman2005}.  
The stellar rotation period has been estimated by \citet{Isaacson2010} to be 42 days, while \citet{Simpson2010} report $35.1\pm0.8$\,d.

With 1515 additional RV measurements since \citet{Lovis2006}, we recover all three of the same planets with slightly different periods and amplitudes (see Table \ref{tab:updates} for a full comparison).  
For HD 69830 b, we report {$P_b=8.66897\pm0.00028$\,d, $K_b=3.4\pm0.1$\,\ms, and $e=0.128\pm0.028$. For HD 69830 c: $P_c=31.6158\pm0.0051$\,d, $K_c=2.6\pm0.1$\,\ms, and $e_c=0.030\pm0.027$, 
and for HD 69830 d: $P_d=201.4\pm0.4$\,d, $K_d=1.5\pm0.1$\,\ms, and $e_d=0.080\pm0.071$.}
The uncertainties on our derived orbital periods are slightly smaller than those recently reported by \citet{Rosenthal2021}, which used data from \HIRES\ and the \APF\ but not the other instruments included here, and appear to be the most precise yet reported. 

\citet{Rosenthal2021} find two false positives in their analysis with periods of 201 and 382 days, which they attribute to systematic errors. {However, the $P=201$\,d signal is in fact a detection of \citet{Lovis2006}'s planet d, and its inclusion in \citet{Rosenthal2021}'s false positive table (Table 7) is a typo.} We do not recover the 382 day false positive reported by \citet{Rosenthal2021}, but our inclusion of multiple instruments' data which are not included in \citet{Rosenthal2021} may dilute individual facilities' systematics.

Additionally, we recover three significant signals in the S-index activity analysis. S-index signal I has {$P_I=3989\pm190$\,d ($10.93\pm0.52$\,yr)}, which is similar to the Sun's own 11-year activity cycle \citep[e.g.][]{Hathaway2015}. 
\citet{Lovis2011} reported a poorly constrained activity cycle period of \Pcyc\, = 5865$_{-1235}^{+\infty}$\,d, which is 1.46$\sigma$ longer than our measured activity signal I, but they are likely detections of the same long-term magnetic activity cycle. 
We report S-index activity signal I as a magnetic activity cycle. 
S-index activity signal II has {$P_{II}=731\pm31$\,d ($2.00\pm0.08$\,yr) which we note is almost twice the expected \HARPS\, yearly systematic. Attribution of this signal to a \HARPS\, systematic is further supported by the complete lack of corresponding signal in the \UCLES\, H$\alpha$ analysis for this target, which not only returns no significant detections but shows almost no strength in the periodogram at this period.}

Finally, S-index activity signal III has {$P_{III}=2530\pm180$\,d ($6.93\pm0.49$\,yr)}. This may be another magnetic activity cycle, {though we note that there appears to be a minimum in the H$\alpha$ periodogram at this period. We recommend further investigation by future work to understand this signal.}

Comparison of the star's rotation period and cycle periods with other nearby Sun-like stars in Fig. 9 of \citet{BoroSaikia2018} indicate that HD\,69830 may be a rare case of a slow-rotating star with two detected activity cycles {($10.93\pm0.52$ yr and $6.93\pm0.49$ yr}, both of which are near the ``inactive branch" locus in $P_{rot}$ vs. $P_{cyc}$ space).{\footnote{A similar example from \citet{BoroSaikia2018} is the K2V star HD\,149661, with activity cycles of $15.3\pm0.4$ and $7.7\pm0.12$\,yr \citep[see also][]{Saar1999}.}}\\

\subsection{HD 75732 (55 Cnc) \label{sec:HD75732}}

55 Cnc ($\rho^1$ Cnc, GJ 324 A, HD 75732, HR 3522, HIP
43587, Copernicus) is a famous, K0IV-V \citep{Gray2003} exoplanet host star at $d=12.58$ pc \citep[$\varpi=79.4482\pm0.0429$ mas;][]{GaiaEDR3}. 55 Cnc also has a wide separation (85\arcsec, $\sim$1060\,AU ) low-mass stellar companion 55 Cnc B. 
\citet{Bourrier2018} presents an extensive review of the 55 Cancri system and its five exoplanets \citep[see also e.g.,][]{Fischer2008,Endl2012,Fischer2018}. \citet{Bourrier2018}'s analysis contains 1552 RV measurements from a combination of both first generation and more modern {precise} RV spectrographs spanning 25 years. In comparison, this work includes only modern {precise} RV data sets and contains 837 RV measurements taken over 18 years. Our analysis does, however, have longer \HIRES\, and \APF\, baselines than present in \citet{Bourrier2018}.
We recover signals corresponding to all five reported planets around 55 Cnc and report updates to the parameters of planets b, c, e, and f in Table \ref{tab:AllSignals}.

{Our detection of the long period planet 55 Cnc d suffers from our more limited observational baseline and the months of time between the \HIRES\,-Pre and -Post data sets. The $\Delta$BIC periodogram peak suggests a 4421 day period, which is notably shorter than the P$_{d}$ = 5574.2$^{+93.8}_{-88.6}$ result from \citet{Bourrier2018}. After fitting the full system, \rvsearch\, arrives at a best-fit model of $P_d=14951\pm5100$\,d, $K_d=54\pm5$\,\ms, and $e_d=0.515\pm0.086$ for this long period signal. This overly long, poorly constrained result exhibits similar behavior to the other `LPS' signals in this work due to the lack of full orbital phase coverage. Due to the period of the initial periodogram peak we attribute this signal to 55 Cnc d, but we do not report this as an updated orbital parameter result.}

\citet{Baluev2015} report an activity cycle for 55 Cnc of period $P_{cyc} =12.6^{+2.5}_{-1.0}$\,yr, with a prediction that an activity minimum would occur around 2014-2015. 
{This correlates with our S-index detection at $P=3801\pm130$\,d ($10.4\pm0.36$\,yr), so we report this signal as an update to the previously published activity cycle.}
The activity cycle prediction from \citet{Baluev2015} also does not align with the signal reported in \citet{Rosenthal2021}.

\subsection{HD 85512 (GJ 370) \label{sec:HD85512}}

HD 85512 (GJ 370, HIP 48331) is a nearby, somewhat
metal poor ([Fe/H] \,=\, -0.26) \citep{Tsantaki2013}, 
inactive \citep[\logrphk = -4.976;][]{Costes2021} K6V(k) star \citep{Gray2006} at $d=11.28$ pc \citep[$\varpi$\,=\,461.446 mas;][]{GaiaEDR3}. The star has one previously reported planet at $P_b =58.43\pm0.13$\,days \citep{Pepe2011}, recovered through radial velocity analysis of 185 \HARPS\, data points. 

Our \RVSearch\, analysis, run on an additional 1,127 data points, does {\it not} detect a 58-day signal, but rather a shorter period signal with parameters: {$P_{b}=51.195\pm0.073$\,d, $K_{b}=0.438\pm0.079$\,\ms, and $e_{b}=0.3\pm0.19$.} This change, from $P_b =58.43\pm0.13$\,days in \citet{Pepe2011} to {$P_{b}=51.195\pm0.073$} days in this study amounts to a 48$\sigma$ difference, well beyond any expected planetary orbit refinement. The reported amplitudes are also somewhat inconsistent, with \citet{Pepe2011} reporting $K_{b}=0.769\pm0.090$\,\ms\, a 2.8$\sigma$ difference from our \rvsearch\, result. 

\begin{figure}[htb!]
\includegraphics[width=.49 \textwidth]{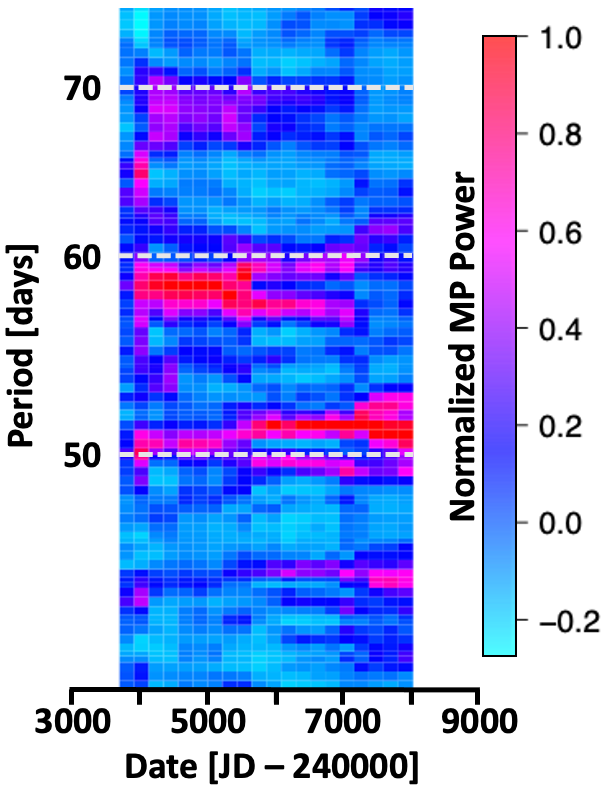}
\caption{Moving periodogram (MP) for the combined HD 85512 radial velocity data sets. The colors encode the scaled MP power, which is truncated to optimize the visualization of signals. The previously reported 58.43 day planet is visible as a dark red horizontal band on the left side of the MP, denoting its significance in the first half of the RV time series, but bifurcates and then disappears in later years. A second signal at $P\simeq51$\,d takes over in the latter half of the time series, but does not exhibit a tight enough period range to be seriously considered as a planet candidate.}
\label{fig:MP_HD85512}
\end{figure}

We note that there is a suggestive similarity between the reported Doppler periods from both \citet{Pepe2011} and our work and the predicted rotation period for the star. Based on HD 85512's chromospheric activity, \citet{Pepe2011} predicted \prot\, = $47.13\pm6.98$\,d, and \citet{Lovis2011} predicted \prot\, = $50.9\pm7.0$\,d\, -- i.e. within 0.58$\sigma$ and 0.04$\sigma$ of the RV signal we measure ({$P_{b}=51.195\pm0.073$\,d}). \citet{Pepe2011} searched for power in the \logrphk\, activity indicator and the CCF line bisector (BIS) but did not detect any excess power consistent with stellar rotation between 50 and 100 days. {Our analysis of the S-index measurements detects significant periods at 44, 45, and 51 days, causing us to suspect rotation as the cause of this signal.}

To investigate the consistency of these two periods over time, we generated a moving Bayes Factor Periodogram (BFP) using the AGATHA software suite \citep{Feng2017b}. Since RV data are typically not measured in a uniform way, especially when combining results from different surveys, the consistency of a true Keplerian signal may depend on the sampling cadence even if the power is normalized. Moving periodograms can help to identify false positives if a signal is found to be inconsistent even during spans where data was taken at a high cadence and over a number of nights comparable with or longer than the signal period. 

The moving periodogram results for HD 85512 (Figure \ref{fig:MP_HD85512}) show a prominent peak in the 58 day region when looking at the first half of the RV time series. This signal, however, bifurcates in roughly 2017 (JD $\simeq$ 2458000) and splits into two weaker periodicities of 59 and 57 days. At approximately the same time a more prominent peak appears at the $P\simeq51$\,day period identified by \RVSearch, and becomes the most significant period for the duration of our observational baseline.

As low-mass stars can manifest differential rotation \citep[e.g.][]{Donahue1996}, the reported 58\,d and 51\,d periods could be due to active regions rotating at different latitudes. 
{The trend in differential rotation among G/K dwarfs from the Mt. Wilson survey shown by \citet{Donahue1996} (their Fig. 3) shows that K dwarfs with \Prot\,$\simeq$\,50\,d could exhibit differential rotation at the $\Delta P\,\simeq\,9$\,d level. Indeed the previously discussed K3V star HD 16160 (see Sec. \ref{sec:HD16160}) was the Mt. Wilson survey poster child for such extreme differential rotation, exhibiting seasonal mean rotation periods ranging from 42.2 to 51.5\,d over five seasons. So it is certainly reasonable for a slow-rotating mid-K dwarf like HD 85512 to manifest differential rotation over \Prot\,$\simeq$\,44-58\,d.}

Given the lack of periodic consistency for both signals across the RV time series, we assert that the reported companion HD 85512 b from \citet{Pepe2011} is not caused by a physical planet, but rather that the signal is due to the star's rotation. {We adopt the notation of calling this HD 85512 RV Signal II (rather than HD 85512 b) in Table \ref{tab:AllSignals}, since it is the second RV signal fit by the \rvsearch\, algorithm.}

\rvsearch\, finds one additional significant signal in the RV data, with an initial $\Delta$BIC periodogram peak of 3891 days. The best fit period {for RV Signal I }after running the RV data through \rvsearch\,'s MCMC analysis is $P_I = 9646\pm5500$ days, which we categorize as a `LPS' due to the large error bars and period that stretches beyond the baseline of the combined RV data.

While \citet{Pepe2011} presented time series \logrphk\, data for the star over a span of 2745\,d, and sinusoidal-looking activity variability was observed, they did not estimate a magnetic cycle period. By eye, interpreting the \citet{Pepe2011} \logrphk\, data near JD 2453000 as a minimum and JD 2454500 as a maximum, one can infer \Pcyc\,$\simeq$\,3000\,d. And indeed, when analyzing the majority of the \citet{Pepe2011} data set (175 of the 185 observations) \citet{Lovis2011} estimated an activity cycle of \Pcyc\, = $3793^{+806}_{-566}$\,d\, ($10.38^{+2.21}_{-1.55}$\,yr). 

Here we compile in total 1312 S-index measurements taken over a baseline of $\sim$7000 days. With this longer baseline, we see a subsequent activity minimum around JD 2456600 and a maximum around JD 2458800 and \RVsearch\, identifies a significant S-index signal with $P = 4245\pm52$\,d ($11.62\pm0.14$\,yr). This falls well within the 1$\sigma$ uncertainties of the \citet{Lovis2011} activity cycle, and better constrains the period by a factor of 10. It is also consistent with the $\Delta$BIC periodogram peak in the RV data, and so we note that it is likely that the LPS in the RVs is caused by a magnetic activity cycle, however more data is needed to definitively characterize the nature of this signal.

Our S-Index analysis returns a multitude of additional signals falling between the rotation period and magnetic activity cycle of HD 85512.  To fully characterize the significance of all these signals, a much more in depth study of the system is required than is covered within the scope of this work. The full list of signals can be found in the activity summary plot contained in the figure set in the online journal. We leave further analysis of the activity results from this target to future work.

\subsection{HD 102365 (GJ 442A) \label{sec:HD102365}}

HD 102365 (GJ 442 A, HR 4523, HIP 57443) is a G2V star \citep{Gray2006} at $d=9.32$ pc \citep[$\varpi = 107.3024 \pm 0.0873$ mas;][]{GaiaEDR3}. 
{The star is just slightly cooler (\teff$\,=5618\pm14$\,K) and of similar chromospheric activity to the Sun (\logrphk\,=\,-4.94) \citep{Meunier2022}, however the star is substantially more metal poor ([Fe/H]$=-0.31\pm0.02$) compared to the Sun \citep[][]{Soubiran2022}.}
{Recent age estimates for the star make it ancient: $11.3\pm0.9$\,Gyr \citep{Nissen2020}, $12.46^{+1.04}_{-1.42}$\,Gyr \citep{GaiaDR3}, $13.1\pm1.5$\,Gyr \citep{Casali2020}.}
The star has a low-mass stellar companion of type M4V \citep{Henry2002} at projected separation 22\arcsec.72 or 211\,AU \citep{Tian2020}. 
The masses of the stars A and B are $0.88^{+0.02}_{-0.03}$\,\msun\, \citep{AguileraGomez2018} and 0.192 \Msun\, \citep{Mugrauer2019}, respectively. 
\citet{Tinney2011} reported an exoplanet with orbital period \Porb$=122.1\pm0.3$\,d, $K=2.40\pm0.35$\,d, and eccentricity $e=0.34\pm0.14$, corresponding to a Neptune-like predicted mass of $16.0\pm2.6$\,\Mearth. 
No subsequent orbital solution has been reported over the past decade. 

We recover this same planet {as \citep{Tinney2011}}, with updated orbital parameters thanks to our additional RV observations. We report parameters for HD 102365 b of {$P=121.3\pm0.25$}\,d, $K=1.38\pm0.23$\,\ms, and $e=0.28\pm0.15$.
 Note that our RV amplitude is only 58\%\, of that reported by \citet{Tinney2011}, resulting in a significantly lower \msini\, of $9.34^{+1.52}_{-1.50}$\,\Mearth\, (Table \ref{tab:AllSignals}).

{We find no significant signals in analysis of the S-index data, but the H$\alpha$ data analysis yields three periodic signals. The first has a period of approximately one year, so we assert that it is most likely caused by the seasonal availability of the star. The second H$\alpha$ signal is fit to a periodogram peak of 7273.6 days, longer than the baseline of the \UCLES\, data. This peak is most likely the same long-period trend present in most of the \UCLES\, H$\alpha$ data as described in Section \ref{subsec:ActivityIndicators_Halpha}, and so we disregard from consideration as astrophysical. The third H$\alpha$ peak has period $P_{III}=49.68$ d and $e_{III}=0.08$. While the peak is sharp and well-defined, and could reasonably correspond to the stellar rotation period for this ancient star, it does not show any corroborating signal in the S-indices. We note that the MCMC fit to the EW$_{H \alpha}$ data settles on a solution for the longest signal that is nonsensical ($P=36122\pm64000$ days) and so we report the MAP best fit solution in the summary table. Even the MAP result suffers from the similarity between the long period signal and the total observational baseline of the \UCLES\, data, however, and the resulting fit lands on an orbital period of P=18549 days; clearly well beyond the scope of our data set.}

\subsection{HD 114613 (GJ 9432) \label{sec:HD114613}}
HD 114613 (GJ 9432, GJ 501.2, HR 4979, HIP 64408) is a G4IV star \citep{Gray2006} at $d = 20.46$ pc \citep[$\varpi=48.8691\pm0.1058$ mas;][]{GaiaEDR3}.
\citet{Brewer2016} finds that the star is more massive ($1.24\pm0.17$\,\msun) and metal-rich ([Fe/H] = 0.17) than the Sun, and also cooler (\teff\, = 5641\,K) and larger ($2.14\pm0.06$ \rsun) with lower surface gravity (\logg\,=3.87). 
\citet{Lovis2011} report the star to have very low chromospheric activity (\logrphk\, = -5.509) with a magnetic activity cycle of \Pcyc\,$=897_{-53}^{+61}$\,d {(2.46$^{+0.15}_{-0.17}$\,yr)}.  

\citet{Wittenmyer2014} reported a planet with $P_{\rm{b}} = 3827\pm105$\,days and a semi-amplitude of $K_{\rm{b}} = 5.4\pm0.4$\,\ms\,. Our search of the updated RV data set, which contains 980 additional velocities from \HARPS, \HIRES, and \PFS, does not recover this planet but rather reveals a much longer period signal at $P_I = 6622\pm270$\,d ($18.13\pm0.74$\,yr) with a semi-amplitude of $K_I = 7.29\pm0.44$\,\ms, {suggestive of a planet candidate with \msini\,$\simeq$\,0.74\,\Mjup.} 

{However, {\it both} the S-index data and the EW$_{H \alpha}$ periodograms identify significant signals at similar periods to the RV periodogram, with the S-index peak appearing at 7563 days and the EW$_{H \alpha}$ peak at 7653 days. Both of these signals struggle with similarity in duration to the baselines of their respective data sets, however, as the lack of a single activity indicator that covers the full 8500 day span of radial velocities prevents a clean resolution of the activity signals. Instead, both the S-index signal and the EW$_{H \alpha}$ signal are pushed to much larger and nonsensical values: 81942$\pm$190000\,days for the S-index data, and 29213$\pm$41000\,days for the EW$_{H \alpha}$ data. Because of this, we opt to disregard the MCMC analyses and proceed with the signals evident in the initial $\Delta$BIC periodograms. We take the agreement of these two activity indicator periodogram peaks and their overlap with the RV signal (which is well resolved) to be sufficient evidence that all three signals are manifestations of the same underlying magnetic cycle. }

{Given this overlap, and the lack of a re-detection of the original 3827 day signal in the radial velocities, we assert that the previously claimed HD 114613 b from \citet{Wittenmyer2014} is not in fact a planet, but 
may be attributed to a long period magnetic cycle.}

\RVSearch\ detects two additional RV signals with period and semi-amplitude pairings of $P_{II}=73.14\pm0.06$\,d, $K_{II}=2.54\pm0.48$\,\ms\, and $P_{III}=1954\pm39$\,d, $K_{III}=2.98\pm0.52$\,\ms. There are no corresponding signals in the S-index or EW$_{H \alpha}$ periodograms, and so we label each of these signals as SRC. As a subgiant star, we would expect HD 114613 to exhibit higher levels of RV jitter \citep{Luhn2020}, which may drive some of the scatter seen in the RV summary figure.

\subsection{HD 115617 (61 Vir) \label{sec:HD115617}}

61 Vir (GJ 506, HD 115617, HR 5019, HIP 64924) is a G7V star \citep{Gray2006} at $d=8.53$ pc \citep[$\varpi=117.1726\pm0.1456$ mas;][]{GaiaEDR3}. The star is slightly less active than the Sun, with reported \logrphk\, values between -4.93 and -5.03 \citep[e.g.][]{Baliunas1996,Hall2007,Isaacson2010,Vogt2010,Wittenmyer2006,Lovis2011,Brewer2016,Meunier2017}.
{\citet{Baliunas1996} report an average rotation period of \Prot\,=\,29\,d based on analysis of the Mt. Wilson survey data, while \citet{Lovis2011} reported a chromospheric activity cycle of \Pcyc\, = $1548^{+266}_{-811}$\,d\, and a predicted rotation period of \Prot\, = $33.9\pm3.6$\,d.} 
According to \citet{Vogt2010}, 61 Vir is a three-planet system with $P_b = 4.21$\,days, $P_c = 38.021\pm0.034$\,days, and $P_d = 123.01\pm0.55$\,days.

We recover the same three planets {as \citet{Vogt2010}} with slightly updated parameters: 
{$P_b=4.21498\pm0.00014$\,d, 
$K_b=2.47\pm0.11$\,\ms,  
$e_b=0.033\pm0.029$; 
$P_c=38.079\pm0.008$\,d, 
$K_c=3.56\pm0.12$\,\ms, 
$e_c=0.026\pm0.023$; 
$P_d=123.2\pm0.2$\,d, 
$K_d=1.47\pm0.17$\,\ms, 
$e_d=0.15\pm0.11$.}
We report these parameters as an improvement to the previously reported ones, due to an additional 2473 RV observations since \citet{Vogt2010}.

After analysis of residual signals, \citet{Rosenthal2021} report the 123-day period signal as a yearly alias. We do not see similar evidence in our data as they report, after examining an additional ten years of data. We therefore report this signal as an update to the currently confirmed planet d.

We recover one additional RV signal, RV signal I, with parameters {$P = 20565\pm21000$\,d, $K = 2.23\pm0.46$\,\ms, and $e = 0.97\pm0.024$. The original periodogram peak being fit by this Keplerian is at 5910.9 days, which is close to the observation baseline, so the fit is poorly constrained. The presence of the peak is evidence of a long-period trend which is not yet well defined, so we classify this signal as LPS.}

{Analysis of the S-index activity does not yield any detections, but we note that there is significant strength in the S-index periodogram at 3995 days. This may be indicative of an approximately 11 year magnetic activity cycle, but there is insufficient data to push this signal past the false alarm probability threshold. Continued study of this target would allow for further understanding of potential origins of this signal.}

{Analysis of H$\alpha$ data returns three detections. The first of these has $P_{I}=346.3\pm1.9$\,d, which we attribute to observation cadence effects. The second and third signals have $P_{II}=24.63\pm0.02$\,d and $P_{III}=44.932\pm0.069$\,d.
The observed H$\alpha$ periodicities II and III are bit shorter and longer, respectively, than the predicted rotation period (\Prot\,=\,$33.9\pm3.6$\,d) from \citet{Lovis2011}.}

{H$\alpha$ signal II is close to the rotation period reported by \citet{Baliunas1996}, though it is shorter by several days. Because the \citet{Lovis2011} rotation period is predicted, and the \citet{Baliunas1996} \Prot\, is 26 years old, we assert that it is possible that H$\alpha$ signal II is due to stellar rotation.}

{H$\alpha$ signal III does not correlate with any periodicity in Ca H \& K data nor RV data, and is unlikely to be caused by differential rotation, as its period is $\sim$20\,d greater than signal II. However, we note that it is not impossible that H$\alpha$ signal III is caused by differential rotation. Quantifying the differential rotation in terms of $\alpha = \vert P_2 - P_1 \vert/P_{max} = 0.45$, this would suggest surface shear approximately twice that of the Sun \citep[$\alpha_\odot$ = 0.2;][]{Reinhold2013}.
The observed differential rotation trend for nearby solar-type stars from \citet{Donahue1996} predicts that for a mean rotation period of $\sim$35 days, one would predict observing $\Delta P$ $\simeq$ 7.6\,d. However, the data from \citet{Donahue1996} also show that there are cases for rotators with $\sim$month-long periods of having $\Delta P$ as high as $\sim$18\,d! Because of this, we believe it is somewhat plausible that H$\alpha$ signals II and III could be hinting at strong differential rotation, but further observations would be needed to test this idea further.}

\subsection{HD 136352 ($\nu^2$ Lup) \label{sec:HD136352}}

$\nu^2$ Lup (HD 136352, GJ 582, HR 5699, HIP 75181) is a nearby G2-V \citep{Gray2006} star at $d=14.74$\,pc \citep[$\varpi=67.8467\pm0.0601$ mas;][]{GaiaEDR3}. 
The star is similar to the Sun in temperature and gravity, but considerably more metal poor: 
\teff$\,=5664\pm14$\,K, 
\logg$\,=4.39\pm0.02$, 
$[Fe/H]=-0.34\pm0.01$
\citep{Sousa2008}.
Given its high velocity, low metallicity, and $\alpha$-element enhancement ([$\alpha$/Fe] $\simeq$ 0.17) \citep[][]{Soubiran2005}, the star is widely classified as a thick disk star \citep[e.g.][]{Ibukiyama2002,Soubiran2005,Adibekyan2012,Hinkel2017,Kane2020}\footnote{The only known star brighter in $V$ and $G$ band than HD 136352 with transiting exoplanets in the NASA Exoplanet Archive is HD 219134, which is an $\alpha$-poor thin disk star of approximately solar metallicity \citep{Mishenina2004,Ramirez2012}. The only known star brighter in $K_s$ band than HD 136352 is 55 Cnc, which is a metal-rich thin disk star \citep{Mishenina2004,Ramirez2012}. 
Hence, HD 136352 ($\nu^2$ Lup) appears to be the brightest -- either in visible or near-IR bands -- thick disk star known to have transiting exoplanets.}.
\citet{Lovis2011} report a magnetic activity cycle of \Pcyc\, = 1041$_{-97}^{+581}$\,d ($2.85^{+1.59}_{-0.27}$ yr), with a predicted rotation period of \Prot\, = $25.0\pm3.1$\,d\, based on the mean activity level (\logrphk\, = -4.986)\footnote{Independently, \citet{Isaacson2010} predicts the rotation period of HD 136352 to be 23\,d based on \logrphk.}. 
\citet{Udry2019} reported three planet signals in \HARPS\, radial velocity observations for HD 136352.
\citet{Kane2020} present a detailed study of $\nu^2$ Lupi, reporting \TESS\, observations that planets $b$ and $c$ were observed to be transiting, with their derived radii and densities consistent with being on either side of the planet radius gap. 

$\nu^2$ Lup b is a $4.62^{+0.45}_{-0.44}$\,\Mearth\, $1.482^{+0.058}_{-0.056}$\,\Rearth\, planet with period $P=11.57779^{+0.00091}_{-0.0011}$\,d -- likely the stripped core of a sub-Neptune (now a "super-Earth"), and 
c is $11.29^{+0.73}_{-0.69}$\,\Mearth, $2.608^{+0.078}_{-0.077}$\,\Rearth\, exoplanet with period $P=27.5909^{+0.0028}_{-0.0031}$\,d -- a "sub-Neptune" \citep{Kane2020}. 
$\nu^2$ Lup d is a planet with radius $2.56\pm0.09$\,\Rearth\, and mass $8.82\pm0.94$\,\Mearth\, with orbital period $P=107.245$\,d \citep{Delrez2021}.

We recover all three of these planets, but defer to \citet{Kane2020} and \citet{Delrez2021} for the most accurate parameters. 
We recover one additional RV signal with {$P=121.66\pm0.26$\,d, $K=0.68\pm0.13$, $e=0.22\pm0.19$}. 
\citet{Udry2019} recovered a similar signal in their RV analysis, with a period of 123\,d, and discarded it as a three-planet fit was favored over four in their analysis. 
Our search results find that the signal just crosses the threshold for being considered a valid additional signal, however the significance of the signal in the running periodogram wanes notably as more and more RV data points are added, which suggests a non-Keplerian origin. 
Additionally, a 121-day planet would be dynamically inconsistent with the confirmed 107-day planet, ruling this out as a planetary signal, and leaving only the possibility of an activity signal. 
{Because of this we classify the fourth signal as being due to stellar activity. We note, also, that \citet{Rosenthal2021} detect a false positive signal with a period of 244 days, which is quite nearly double that of the fourth signal detected here and in \citet{Udry2019}.}

{The \RVsearch analysis of the S-index data returns no significant detections. Despite our large increase in observation timeline since \citet{Lovis2011}, from 2543 days to 5771 days, we find no evidence for the \Pcyc\,2.85\,yr activity cycle reported in that work.}

{Analysis of the EW$_{H \alpha}$ data returns two significant periodicities. Signal I has period $P=364.7$ days in the periodogram, which is clearly close to one year and we suspect is due to sampling effects. The signal also has high eccentricity, but there is a significant gap in the orbital phase coverage, which \rvsearch\, addresses by using a high eccentricity solution to try and fit a Keplerian curve to this jump. Signal II is detected based on a periodogram peak close to 6000 days, which is approximately the observation baseline for the \UCLES\, data. The MCMC Keplerian fit is poorly constrained because of this fact, so we attribute this signal to the long-period \UCLES\, trend and disregard it as significant. When run through \RVsearch/'s MCMC analysis, the long period signal produces nonsensical error bars of $P=19924\pm33000$. We therefore choose to report the best fit MAP orbital solution in the EW$_{H \alpha}$ summary table.}

\subsection{HD 160346 (GJ 688) \label{sec:HD160346}}

HD 160346 (GJ 688, HIP 86400) is a nearby K2.5V \citep{Gray2003} at $d$\,=\,11.00 pc \citep[$\varpi$\,=\,90.91\,$\pm$\,0.67 mas;][]{vanLeeuwen2007}.
The star has published chromospheric activity estimates ranging from \logrphk\,=\,-4.766 \citep{Meunier2017} to -4.85 \citep{Gondoin2020} -- comparable to the active Sun.
Analysis of the Mt. Wilson Ca II H \& K survey data by \citet{Donahue1996} detected seasonal rotation periods ranging from \Prot\, = 35.4 to 37.8\,d, with an average over 5 seasons of \Prot\,=\,36.4\,d.
\citet{BoroSaikia2018} report a Ca II H \& K activity cycle of \Pcyc\,=\,$7.19\pm0.04$\,yr. 
GJ 688 is a SB1 with three published orbits listed in the SB9 catalog \citep{Pourbaix2004}, with orbits by \citet{Tokovinin1991}, \citet{Katoh2013}, and \citet{Halbwachs2018}. The latter provides: $P=83.7140\pm0.0120$\,d, $e=0.2100\pm0.0120$, $K =5644\pm57$\,\ms.  

Our updated RV analysis produces best-fit orbital parameters of {$P=83.7286\pm0.0005$\,d, $e=0.2048\pm0.00033$, and $K =5690.3\pm2.3$\,\ms}, shrinking the error bars on all three parameters by over an order of magnitude. 

{The S-index analysis detects a number of significant signals, the first two of which have periods of $P=2975\pm600$\,d and $392.6\pm3.2$\,d. Comparison of their \RVSearch\ periodograms, however, suggests that one of these signals is likely an alias of the other. The longer period signal seems more likely to be due to a decade-long magnetic cycle much like the Sun's, however we caution that our analysis does not include the detailed phase analysis necessary to identify which of these signals is the true manifestation of the star's activity.}

{The additional two S-index signals detected by \rvsearch\ have much shorter periods of $P=7.9567\pm0.0055$\,d and $P=2.54223\pm0.00068$\,d. While these are both too short to be due to HD 160346's rotation, there is a possibility that one of the signals could be due to flux contributions from a fast rotating, low mass companion. But given the relative sparsity of the data, we do not have sufficient evidence to say anything truly definitive about their origins.}

\subsection{HD 160691 ($\mu$ Ara) \label{sec:HD160691}}
$\mu$ Ara (HD 160691, GJ 691, HR 6585, HIP 86796, Cervantes) is a G3IV-V \citep{Gray2006} star at $d=15.6$ pc \citep[$\varpi=64.0853\pm0.0904$ mas;][]{GaiaEDR3}, with 4 previously reported planets \citep{Pepe2007}. 
The star is metal-rich ([Fe/H] = $0.27\pm0.05$) and magnetically inactive \citep[\logrphk\, = -5.11;][]{GomesdaSilva2021}, with slightly lower surface gravity than typical G dwarfs \citep[\logg\, = $4.20\pm0.02$;][]{Ramirez2013}. 
Combining asteroseismic observations with evolutionary models, \citet{Soriano2010} find that $\mu$ Ara is most likely near the beginning of its subgiant branch phase, with a mass of $1.10\pm0.02$\,\msun\, and age of $6.34\pm0.80$ Gyr. 
The most recent parameters for this system are from {\citet{Benedict2022}}. We recover the same four signals with minor revisions but generally good agreement on the best fit values, and significant improvements to all but one of the parameters' uncertainties:
{
$P_b =644.93\pm0.28$\,d, 
$K_b =35.7\pm0.2$\,\ms, 
$e_b =0.0499\pm0.0082$; 
$P_c =9.6394\pm0.0008$\,d, 
$K_c =2.8\pm0.2$\,\ms, 
$e_c =0.132\pm0.069$; 
$P_d =308.4\pm0.23$\,d, 
$K_d =12.7\pm0.3$\,\ms, 
$e_d =0.074\pm0.016$; 
$P_e =4035\pm21$\,d, 
$K_e =22.25\pm0.24$\,\ms, 
$e_e =0.026\pm0.013$.}

{We detect no significant S-index activity signals, but do find two signals in the H$\alpha$ data. The first of the H$\alpha$ signals is fit from a $\Delta$BIC periodogram peak at 5293.7\,d, which is close to the \UCLES\, observation baseline, so we attribute this to the long-period \UCLES\, systematic present in all the H$\alpha$ data (see Section \ref{subsec:ActivityIndicators_Halpha} for further discussion). H$\alpha$ Signal II has a period of $P=362.4\pm1.6$\,d, close to one year. This signal is likely caused by the star's seasonal availability and the observing cadence, so we disregard it as a significant detection for this system.}

\subsection{HD 192310 (GJ 785) \label{sec:HD192310}}
HD 192310 (HR 7722, GJ 785, HIP 99825) is a
K2+V star \citep{Gray2006} of roughly solar metallicity
([Fe/H] = $-0.03\pm0.04$) \citep{Tsantaki2013}
at $d=8.81$ pc \citep[$\varpi=113.4872\pm0.0516$ mas; ][]{GaiaEDR3}, and with two previously reported planets \citep{Pepe2011}.
\citet{Lovis2011} detect a magnetic activity cycle of \Pcyc\, = 3792$_{-566}^{+806}$\,d and predict the rotation period to be \Prot\, = 43.7$\pm$4.9\,d based on the their estimate of the
star's mean chromospheric activity (\logrphk\, = -4.996). 
Combining the star's mean chromospheric activity levels (\logrphk\, = -4.993) recently reported by \citet{Meunier2017}, with its color \citep[$B-V$ = 0.884;][]{Mermilliod1991} and the rotation-activity relations from \citet{Mamajek2008}, one predicts the star's rotation to be approximately \Prot\,$\simeq$\,48\,d.

We detect two RV signals that appear to correspond to the previously reported planets b and c, with
$P_b =74.278\pm0.035$\,d, 
$K_b =2.484\pm0.098$\,\ms, 
$e_b =0.032\pm0.027$, and 
$P_c =549.1\pm4.5$\,d, 
$K_c =1.3\pm0.1$\,\ms, 
$e_c =0.077\pm0.073$. 
{These appear to be the most accurate periods yet derived, with \citet{Rosenthal2021} recently reporting $P_{b}$ = $74.062\pm0.085$\,d, and \citet{Pepe2011} reporting $P_{c}$ = $525.8\pm9.2$\,d. And our derived amplitude for $b$ is three times more precise than that derived by \citet{Rosenthal2021} ($2.49^{+0.35}_{-0.33}$\,\ms), thanks to the addition of the \HARPS\ and \UCLES\ data. We note that our amplitude for planet $c$ is less than half that reported by \citet{Pepe2011} ($2.27\pm0.28$\,\ms). Given the increase in observational baseline and the number of instruments contributing data, and the additional signals resolved in the RV data, some shifting in the semi-amplitude is expected. However, as this is an almost 3.5$\sigma$ offset from the \citet{Pepe2011} value, a more thorough analysis that treats the RV and activity indicator data simultaneously is well warranted.}

\rvsearch\ also identifies four additional RV signals, which we designate as Signals I, II, III, and IV. 
Signal I has $P=3836\pm240$\,d, $K=1.48\pm0.11$\,\ms, and $e=0.34\pm0.15$. We suspect that Signal I is caused by activity due to a corresponding peak in the S-index data at $P=3817\pm60$\,d ($10.45\pm0.16$\,yr), which matches the magnetic activity cycle period (\Pcyc\, = $3792_{-566}^{+806}$\,d = $10.38^{+2.21}_{-1.55}$\,yr) reported by \citet{Lovis2011}. We therefore attribute it to a magnetic activity cycle. 

RV signals II and III have similar periods: $P=43.614\pm0.023$\,d\, for signal I and $P =39.509\pm0.059$\,d\, for signal II.
Analysis of S-index data using \rvsearch\, returns various signals with periods between 35 and 50 days. Recalling that \citet{Lovis2011} predicted the star's rotation period to be \Prot\,=\,$43.7\pm4.9$\,d\, we attribute these signals II and III to differential rotation of the star, as the appearance of active regions at various latitudes over the course of the star's magnetic activity cycle could lead to a wide range of measured periods.

Finally, RV signal IV has parameters $P=24.559\pm0.016$\,d, $K=0.6\pm0.1$\,\ms, and $e=0.16\pm0.12$. The periodogram peak for this signal is sharp and well defined, and the RV fit is very well constrained. The period is sufficiently distinct from the rotation period that it is unlikely to be a signature of differential rotation. We therefore report RV Signal IV as a Candidate in Table \ref{tab:AllSignals}, and recommend further investigation of this signal to determine whether it is planetary in origin.

{In addition to the differential rotation signals, the \rvsearch\ S-index analysis returns three well-defined signals with $P=345.34\pm0.048$\,d, $P=432.6\pm3.4$\,d, and $P=133.38\pm0.043$\,d. As these are logarithmically almost half way between the star's rotation period and its magnetic activity cycle, no obvious activity-based explanation exists for these signals. We note that the eccentricities of these signals ($e$ = 0.918, 0.7, and 0.78, respectively) are significantly higher than those of the other S-index detections ($e$ = 0.26 for the magnetic activity cycle, and 0.14 on average for the four rotation-associated periods). This could suggest that these intermediate periods are being driven by small amounts of outlier points that were not far enough removed from the mean to be rejected by our 5$\sigma$ outlier clipping.}

\citet{Rosenthal2021} detect a false positive signal at $P = 1630^{+51}_{-53}$\,d\, and $K = 1.95^{+0.49}_{-0.36}$\,\ms, which they attribute to a long-period magnetic activity cycle. We do not recover this same signal.

{The EW$_{H \alpha}$ data for this star produce two significant signals when analyzed with \rvsearch. The first is a very long period signal, with an initial periodogram peak of 30,840 days (well beyond the \UCLES\ baseline) and a best-fit period of $P=62641\pm110000$ days. Given the extreme nature of both the signal duration and the corresponding uncertainty on the period, we report the best fit MAP solution in place of the MCMC solution. That gives $P=13627.7$\,d for the longer period signal, and $P=363.678$\,d for the second signal which seems to be driven by seasonal observing impacts that leave $\sim$1/3 of the orbital phase with little to no data, and therefore make it easier for high eccentricity signals to be fit to the data.}

\subsection{HD 209100 ($\epsilon$ Ind A) \label{sec:HD209100}} 
$\epsilon$ Indi A (HD 209100, GJ 845, HR 8387, HIP 108870) is a well-studied K4V(k) star \citep{Gray2006} at $d$\,=\,3.64 pc \citep[$\varpi$\,=\,274.8431\,$\pm$\,0.0956 mas;][]{GaiaEDR3}.
{\citet{Lovis2011} report a magnetic activity cycle of \Pcyc\, = 1719$_{-315}^{+217}$\,d ($4.71^{+0.59}_{-0.86}$), and predicted the star's rotation period to be \Prot\,=\,$37.6\pm6.2$\,d\, based on the star's chromospheric activity (\logrphk\,=\,-4.806).}
It has one reported planet, a cold Jupiter-mass planet with a period of $P_b$\,=\,45.2\,yr \citep{Feng2019b}. 

We detect only a very poorly constrained, long-period signal in our RV data, with a $\Delta$BIC periodogram peak of 13138.7 days (35.97\,yr), and a best fit MCMC solution of $P=13626\pm110000$\,d ($37.3\pm30.1$\,yr). Due to the signal clearly stretching beyond the bounds of our RV baseline, we note this as an `LPS' in our detections table.
\citet{Feng2019b}, however, analyses a longer RV baseline than used here, as it includes data from previous generation RV instruments{ such as the Coude Echelle Spectrograph Long Camera and Very Long Camera \citep[see ][]{Zechmeister2013}, and the Ultraviolet and Visible Spectrometer \citep[UVES, ][]{Dekker2000}} and utilizes a combination of RVs and Hipparcos and Gaia astrometry to constrain the planet's orbit. It is therefore not surprising that we do not resolve the \citet{Feng2019b} planet in our RV time series, and we defer to their publication for orbital parameters on $\epsilon$ Indi A b.

In our S-index activity analysis for HD 209100, we detect a signal with period $P=2063\pm160$\,d ($5.65\pm0.44$\,yr) which matches the \citet{Lovis2011} magnetic activity cycle to within 1.5$\sigma$. We report this as an updated fit to the previously published activity cycle. We detect an additional, much shorter, S-index signal with $P=32.87\pm0.067$\,d which matches the \citet{Lovis2011} predicted rotation period to within 1$\sigma$. We therefore take this to be an updated, and better constrained, measurement of the star's rotation period.

\section{Targets with New Signals}\label{sec:newsignal}

{Here we present results from targets whose analyses returned signals that have not previously been published. Rather than the lettering system which is used to identify planets and companions, we refer to our new detections with Roman numerals for discussion purposes. Signals are interpreted and reported in Table \ref{tab:AllSignals}.}

\subsection{HD 1581 ($\zeta$ Tuc) \label{sec:HD1581}}

$\zeta$ Tuc (HD 1581, GJ 17, HR 77, HIP 1599) is an F9.5V type star \citep{Gray2006} at distance $d=8.59\pm0.05$ pc \citep[$\varpi=116.46\pm0.16$ mas;][]{vanLeeuwen2007}. 
The star is slightly hotter {(\teff$\,=5932\pm12$\,K), and more metal poor ([Fe/H]$=-0.21\pm0.01$) than the Sun, but with similar gravity (\logg\,$=4.43\pm0.03$) \citep{Soubiran2022}.}
\citet{Lovis2011} reports {a} magnetic activity cycle of \Pcyc\,=\,$1018_{-47}^{+51}$\,d ($2.79^{+0.14}_{-0.13}$) based on 127 \logrphk\, measurements over a 2625\,d\, span, and using the mean activity level (\logrphk\,=\,-4.954) they predict the rotation period to be \Prot\,=\,$16.7\pm2.6$\,d. 
There are no confirmed exoplanets for this system.

We recover three RV signals {with} \rvsearch. RV signal I has parameters {$P=635.0\pm4.4$\,d, $K=0.89\pm0.14$, and $e=0.55\pm0.13$.} This signal may correspond to magnetic activity, as the period is long, and the peak in the periodogram is somewhat broad and accompanied closely by several other peaks. {The next strongest peak, located at $P\simeq860$ days sits directly on top of the yearly alias for the detected 635 day signal, denoted by the red dashed line in the \rvsearch\ summary plot. Examination of the window function for this data set reveals a dramatic yearly period in the periodogram, further supporting the concept that one of these signals is indeed the yearly alias of the other. Given the similarity of their periodogram peaks, however, identifying which signal is the true Keplerian would require a full analysis of the phases of the peaks in the window function as seen in \citet{DawsonFabrycky2010}. This is beyond the scope of our current effort, but we encourage future investigation into the true nature of these two signals. For the time being, as there are no correlated peaks in the activity periodogram for the \RVSearch-identified $P=635.0$ day peak, report only this signal in our summary table and classify it as an SRC. The two remaining signals (II and III) are more well-defined periodogram peaks.}

Signal II has {$P=15.653\pm0.005$\,d, $K=0.662\pm0.096$, and $e=0.106\pm0.097$.} This aligns with the rotation period {predicted by} \citet{Lovis2011}. With {our increased} observation baseline {of 3600 RV measurements, we have the ability to measure the rotation period for this star much more accurately than previous works.} We report {RV signal II} as {a measurement of the} rotation period for HD 1581.

Signal III has parameters {$P=29.4661\pm0.0041$\,d, $K=1.6\pm1.1$\,\ms, and $e=0.89\pm0.12$.} {The period of this detection is} almost exactly twice the rotation period and we therefore suspect it has stellar and not planetary origins.

{A periodogram analysis of 456 measurements (spanning 6157 days) of the H$\alpha$ equivalent width measurements taken with \UCLES\, yielded a signal just below the detection threshold of False Alarm Probability 0.001 at $P=29.7$\,days, which further points to the 29-day RV signal being caused by activity. We therefore classify RV Signal III as Activity in Table \ref{tab:AllSignals}.
}

Analysis of the S-index activity data returns no significant detections.

\subsection{HD 2151 ($\beta$ Hyi) \label{sec:HD2151}}

$\beta$ Hyi (GJ 19, HD 2151, HR 98, HIP 2021) is a bright \citep[$V=2.82$;][]{ESA1997} G0V star \citep{Gray2006} 
at distance $d=7.46$\,pc \citep[$\varpi\,=\,134.07\pm0.11$][]{vanLeeuwen2007}. 
Asteroseismic analysis of $\beta$ Hyi by \citet{Brandao2011} yields a mass of $1.08\pm0.03$\,\Msun\, and age of $6.40\pm0.56$ Gyr.

We recover one {RV} signal with {$P=5365\pm1400$\,d, $K=3.21\pm0.58$\,\ms, and $e=0.54\pm0.16$.} We suspect that the signal is activity-induced due to the lack of consistent growth in the running periodogram which quantifies the signal's power as a function of number of RV data points included. {Analysis of H$\alpha$ measurements from \UCLES\, shows a long-period trend just under the detection threshold at $P=4957.8$\,d. We regard this signal with some skepticism as it is close to the observation baseline for the spectrograph (see discussion in Section \ref{subsec:ActivityIndicators_Halpha}) but note that it supports a conclusion that this signal may be caused by activity. S-index data is too sparse to make any significant detections, so we cannot completely corroborate that suspicion and report the signal as SRC rather than activity. A future, more in-depth study of stellar activity is recommended to completely characterize this signal.} 

{Finally, we note  that the} RV residuals periodogram contains a well-defined peak at 73.3 days, which may correspond to rotation, as this is a slightly evolved star.

\subsection{HD 20766 ($\zeta^{1}$ Ret) \label{sec:HD20766}}
$\zeta^{1}$ Ret (HD 20766, GJ 136, HR 1006, HIP 15330) was classified as G2.5V H$\delta$1 by \citet{Keenan1989} and lies at $d=12.04$ pc \citep[$\varpi=83.0240\pm0.0438$ mas;][]{GaiaEDR3}. 
{$\zeta^1$ Ret is the secondary of a wide binary\footnote{From the Gaia DR3 astrometry, \citet{Kervella2022} report that $\zeta^2$ and $\zeta^1$ Ret have projected separation 309\arcsec.11 (3720\,au) and $V_{tan}$ that agree within $0.40\pm0.01$\,\kms, with predicted escape velocity $v_{esc}\,=\,0.91\,$\kms. Using Gaia DR3, we estimate that the stars are co-distant to $\Delta d \,=\,1095\pm2240$\,au. The difference in the mean radial velocities reported by \citet{Soubiran2018} 
($11.953\pm0.0031$\,\kms for $\zeta^2$ Ret and $12.488\pm0.0019$\,\kms for $\zeta^1$ Ret) is $\Delta v_R$\,=\,$0.535\pm0.004$\,\kms. Ignoring possible differences due to gravitational redshift and convective blueshift as negligible, since the stars are nearly twins, we interpret the velocity offset as true orbital motion. The total relative orbital motion between $\zeta^2$ and $\zeta^1$ is then only $v_{orb}\,=\,662\pm9$\,\ms, and with current 3D separation of $s = 4100^{+1000}_{-350}$ (68\% CL). The system is consistent with being a bound binary with $a\, \simeq\, 4500$\,au and $P\, \simeq\, 220$\,kyr, although further analysis would be needed to constrain the orbit further.}
(309\arcsec) with $\zeta^{2}$ Ret (HD 20807; see Sec. \ref{sec:HD20807}).).}
A few conflicting estimates of the rotation period have been reported for this star: \prot\,$<$\,12.1\,d \citep{Cincunegui2007}, \prot$\,=14.81$\,d\, \citep{Oelkers2018}, and \prot/sin$i=15.9$\,d\, \citep{AmmlervonEiff2012}.
Recently, \citet{Flores2021} found evidence for an activity cycle of \Pcyc\,=\,$1527\pm43$\,d ($4.18\pm0.12$\,yr). 

{We report one significant RV detection. The periodogram peak occurs at $P=5643.5$\,d, which is fairly close to the observation baseline of approximately 6000 days for this target. As discussed in Section \ref{subsec:RVsearchOverview}, the \rvsearch\, MCMC fitting for this signal yields nonphysical results ($P=10218\pm10000$\,days, $K=12\pm2$\,\ms, $e=0.82\pm0.11$), so we record the periodogram peak as the best estimate of this signal and classify it as LPS.} The turnaround we see in the center of the RV time series was also reported by \citet{Zechmeister2013} based on the \HARPS\, data. 

We do not recover this signal in our S-index analysis, but the non-detection is unsurprising {given that the RV signal is driven by \UCLES\, data, which lack S-indices}. We report one S-index activity detection, {which encounters a similar period-to-baseline fitting issue as the detection in the RVs; our observation baseline is just over 1200 days, while the detection peaks at $P=1406$\,d (3.85\,yr) in the $\Delta$BIC periodogram. This appears to correspond to the $4.18\pm0.12$\,yr activity cycle reported by \citet{Flores2021}. In this case the MCMC fit cannot even reach a final fit solution, and so instead we report just the MAP period fit in the S-index table while noting the signal's LPS-esque behavior. This signal overlaps with one just barely below the detection threshold in the EW$_{H \alpha}$ data, with $P=1059$\,d (2.90\,yr).}

\movetabledown=1.0in
\begin{rotatetable*}
\begin{deluxetable*}{llllllll}
\tablecaption{Updated Parameters for Previously Reported Exoplanets \label{tab:updates}}
\tablewidth{700pt}
\tabletypesize{\footnotesize}
\tablehead{
\colhead{ID} & 
\colhead{$P$(new)} & 
\colhead{$K$(new)} & 
\colhead{$e$(new)} & 
\colhead{$P$(old)} & 
\colhead{$K$(old)} & 
\colhead{$e$(old)} & 
\colhead{Reference}\\
\colhead{\ldots} & 
\colhead{(days)} & 
\colhead{(\ms)} & 
\colhead{\ldots} & 
\colhead{(days)} & 
\colhead{(\ms)} & 
\colhead{\ldots} & 
\colhead{\ldots}
}
\startdata
HD 13445 b$^*$ & 15.764862$\pm$04.3e-05 & 377.58$\pm$0.77 & 0.0485$\pm$0.0018 & 15.76491$\pm$0.00039 & 376.7$\pm$2.9 & 0.04$\pm$0.01 & \citet{Butler2006}\\
HD 20794 b & 18.305$\pm$0.0052 & 0.807$\pm$0.089 & 0.17$\pm$0.11 & $18.33^{+0.01}_{-0.02}$ & $0.81^{+0.0}_{-0.24}$ & $0.27^{+0.04}_{-0.22}$ & \citet{Feng2017}\\
HD 20794 d & 89.766$\pm$0.085 & 0.86$\pm$0.12 & 0.27$\pm$0.11 & $88.90^{+0.37}_{-0.41}$ & $0.60^{+0.10}_{-0.18}$ & $0.25^{+0.16}_{-0.21}$ & \citet{Feng2017}\\
HD 22049 b & 2832$\pm$120 & 11.1$\pm$1.2 & 0.09$\pm$0.08 & 2690$\pm$30 & -- & 0.07$^{+0.06}_{-0.05}$ & \citet{Mawet2019}\\
HD 39091 b & 2089.05$\pm$0.46 & 196.5$\pm$0.6 & 0.6428$\pm$0.0017 & 2088.33$\pm$0.34 & 192.99$\pm$0.38 & 0.6396$\pm$0.0009 & \citet{Hatzes2022}\\
HD 39091 d & 125.58$\pm$0.27 & 2.16$\pm$0.42 & 0.16$\pm$0.15 & 124.64$^{+0.48}_{-0.52}$ & 1.68$\pm$0.17 & 0.22$\pm$0.079 & \citet{Hatzes2022}\\
HD 69830 b & 8.66897$\pm$0.00028 & 3.4$\pm$0.1 & 0.128$\pm$0.028 & 8.667$\pm$0.003 & 3.51$\pm$0.15 & 0.10$\pm$0.04 & \citet{Lovis2006}\\
HD 69830 c & 31.6158$\pm$0.0051 & 2.6$\pm$0.1 & 0.03$\pm$0.027 & 31.56$\pm$0.04 & 2.66$\pm$0.16 & 0.13$\pm$0.06 & \citet{Lovis2006}\\
HD 69830\,d & 201.4$\pm$0.4 & 1.5$\pm$0.1 & 0.08$\pm$0.071 & 197$\pm$3 & 2.20$\pm$0.19 & 0.07$\pm$0.07 & \citet{Lovis2006}\\
HD 75732 b$^*$ & 14.65157$\pm$0.00015 & 70.39$\pm$0.37 & 0.0069$\pm$0.0047 & 14.6516$\pm$0.0001 & 71.37$\pm$0.21 & 0.0$\pm$0.0 & \citet{Bourrier2018}\\ 
HD 75732 c$^*$ & 44.39$\pm$0.01 & 9.95$\pm$0.37 & 0.22$\pm$0.041 & $44.3989^{+0.0042}_{-0.0043}$ & 9.890$\pm$0.220 & 0.03$\pm$0.02 & \citet{Bourrier2018}\\
HD 75732 e$^*$ & 0.736546$\pm$5e-06 & 6.26$\pm$0.34 & 0.039$\pm$0.035 & $0.7365474^{+0.0000013}_{-0.0000014}$ & $6.02^{+0.24}_{-0.23}$ & 0.05$\pm$0.03 & \citet{Bourrier2018}\\
HD 75732 f$^*$ & 260.88$\pm$0.36 & 5.68$\pm$0.48 & 0.585$\pm$0.057 & 259.88$\pm$0.29 & $5.14^{+0.26}_{-0.25}$ & $0.080^{+0.05}_{-0.04}$ & \citet{Bourrier2018}\\
HD 102365 b & 121.3$\pm$0.25 & 1.38$\pm$0.23 & 0.28$\pm$0.15 & 122.1$\pm$0.3 & 2.30$\pm$0.35 & 0.34$\pm$0.14 & \citet{Tinney2011}\\
HD 115617 b$^*$ & 4.21498$\pm$0.00014 & 2.47$\pm$0.11 & 0.033$\pm$0.029 & 4.215$\pm$0.0006 & 2.12$\pm$0.23 & 0.12$\pm$0.11 & \citet{Vogt2010}\\
HD 115617 c$^*$ & 38.079$\pm$0.008 & 3.56$\pm$0.12 & 0.026$\pm$0.023 & 38.021$\pm$0.034 & 3.62$\pm$0.23 & 0.14$\pm$0.06 & \citet{Vogt2010}\\
HD 115617 d$^*$ & 123.2$\pm$0.2 & 1.47$\pm$0.17 & 0.15$\pm$0.11 & 123.01$\pm$0.55 & 3.25$\pm$0.39 & 0.35$\pm$0.09 & \citet{Vogt2010}\\
HD 136352 b & 11.5767$\pm$0.0015 & 1.65$\pm$0.11 & 0.05$\pm$0.045 & 11.5824$^{+0.0024}_{-0.0025}$ & 1.59$\pm$0.13 & 0.14$\pm$0.08 & \citet{Udry2019}\\
HD 136352 c & 27.5845$\pm$0.0064 & 2.49$\pm$0.12 & 0.041$\pm$0.036 & 27.5821$^{+0.0089}_{-0.0086}$ & 2.65$\pm$0.14 & 0.04$^{+0.05}_{-0.03}$ & \citet{Udry2019}\\
HD 136352 d & 107.5$\pm$0.14 & 1.44$\pm$0.12 & 0.072$\pm$0.061 & 107.5983$^{+0.2796}_{-0.2669}$ & 1.35$\pm$0.15 & 0.90$^{+0.1}_{-0.07}$ & \citet{Udry2019}\\
HD 160346 & 83.7286$\pm$0.0005 & 5690.3$\pm$2.3 & 0.2048$\pm$0.0003 & 83.7140 & -- & -- & \citet{Halbwachs2018}\\
HD160691 b & 644.93$\pm$0.28 & 35.7$\pm$0.2 & 0.0499$\pm$0.0082 & 645.3$\pm$0.3 & 336.1$\pm$0.2 & 0.036$\pm$0.007 & \citet{Benedict2022}\\
HD160691 c & 9.6394$\pm$0.0008 & 2.8$\pm$0.2 & 0.132$\pm$0.069 & 9.6392$\pm$0.0006 & 2.94$\pm$0.17 & 0.16$\pm$0.06 & \citet{Benedict2022}\\
HD160691 d & 308.4$\pm$0.23 & 12.7$\pm$0.3 & 0.074$\pm$0.016 & 307.9$\pm$0.3 & 12.23$\pm$0.27 & 0.091$\pm$0.014 & \citet{Benedict2022}\\
HD160691 e & 4035$\pm$21 & 22.25$\pm$0.24 & 0.026$\pm$0.013 & 3947$\pm$23 & 22.18$\pm$0.25 & 0.022$\pm$0.012 & \citet{Benedict2022}\\
HD 192310 b$^*$ & 74.278$\pm$0.035 & 2.484$\pm$0.098 & 0.032$\pm$0.027 & 74.72$\pm$0.1 & 3.00$\pm$0.12 & 0.13$\pm$0.04 & \citet{Pepe2011}\\
HD 192310 c$^*$ & 549.1$\pm$4.5 & 1.3$\pm$0.1 & 0.078$\pm$0.073 & 525.8$\pm$9.2 & 2.27$\pm$0.28 & 0.32$\pm$0.11 & \citet{Pepe2011}\\
\enddata
\tablenotetext{*}{Additional signals detected in our best fit solution, see Table \ref{tab:AllSignals} for additional information.}
\end{deluxetable*}
\end{rotatetable*}

\subsection{HD 20807 ($\zeta^2$ Ret) \label{sec:HD20807}}

$\zeta^2$ Ret (HD 20807, GJ 138, HR 1010, HIP 15371) is a slightly metal poor ([Fe/H]$=-0.215\pm0.010$) \citep{Adibekyan2016} G1V standard star \citep{Keenan1989}, and fairly nearby at distance 12.04 pc \citep[$\varpi=83.0606\pm0.0608$ mas;][]{GaiaEDR3}. 
{$\zeta^2$ Ret is the primary of a wide binary (309\arcsec) with $\zeta^1$ Ret (HD 20766; see Sec. \ref{sec:HD20766}).} 
\citet{Lovis2011} reported a magnetic activity cycle of \Pcyc\, = 1133$_{-65}^{+1090}$\,d\, (3.10$^{+2.98}_{-0.18}$\,yr) based on only 38 \logrphk\, measurements over a span of 2309 d. 
\citet{Flores2021} present an analysis of the time series chromospheric activity data for $\zeta^2$ Ret, finding an activity cycle of \Pcyc = 7.9$\pm$0.38\,yr {($\sim$$2885\,\pm\,139$,\d)}, and predicting a rotation period of \Prot$=16.5\pm1.8$\,d\, based on \logrphk.
\citet{Zechmeister2013} also reported correlations between the RVs and \logrphk\, FWHM, and BIS based on their limited HARPS data that spanned $\sim$1500 days. 
The star has been claimed to have far-IR excess (70, 100 $\mu$m) from a dusty debris disk \citep{Trilling2008, Eiroa2013, Gaspar2013, Sierchio2014}, however recent ALMA observations have shown that the mm emission in the vicinity of $\zeta^2$ Ret is likely to be attributable to background sources \citep{Faramaz2018}. 

{We detect one significant RV signal with 
$P=3180\pm130$\,d, 
$K\,=\,2.9\pm0.4$\,\ms, and
$e=0.23\pm0.11$.}
{This signal, corresponding to a period of 8.7 years, is just beyond the 1$\sigma$ error overlap with the activity cycle reported by \citet{Flores2021}, prompting suspicion about its nature. Our own S-index analysis does not yield any significant detections, and indeed the star appears to be very inactive. Our EW$_{H \alpha}$ analysis detects one significant signal with a periodogram peak at $P=2897$\,d (7.9 yr) which aligns well with the activity cycles reported in the literature. We note, however, that this is also approximately half the \UCLES\, observation baseline and a period where the stacked periodogram of EW$_{H \alpha}$ non-detections exhibits significant power (Figure \ref{fig:Halpha-Comp}). We therefore regard this activity detection with some uncertainty as discussed in Section \ref{subsec:ActivityIndicators_Halpha}. Because of this uncertainty and the lack of our own S-index detection, we report our RV Signal I as SRC rather than activity, and recommend a more in depth study of activity indicators for this target to confirm the nature of this signal.}

\subsection{HD 23249 ($\delta$ Eri) \label{sec:HD23249}}

$\delta$ Eri (HD 23249, GJ 150, HR 1136, Rana) is a K0+IV spectral standard star \citep{Keenan1989} at $d=9.09$ pc \citep[$\varpi=110.0254\pm0.1944$ mas;][]{GaiaEDR3}. 
The star is a slightly metal-rich, evolved star \citep[\teff\,=\,5045\,K, \logg\,=\,3.77\,$\pm$\,0.02, \feh\,=\,0.06\,$\pm$\,0.01;][]{Jofre2014}, slow-rotating \citep[\prot\,=\,71\,d, \vsini\, = $1.54\pm0.23$\,\kms;][]{Baliunas1996,Jofre2015}, and magnetically very inactive - both chromospherically \citep[\logrphk\,=\,-5.184;][]{Baliunas1996} and coronally \citep[\loglxlbol\, $=-7.14\pm0.18$;][]{Morel2004}.
Despite the very low activity, the star is oddly classified in the General Catalog of Variable Stars \citep{Samus2017} as an RS CVn variable (chromospherically active binary) - which typically implies a very magnetically active detached stellar binary with orbital period between $\sim$1 and $\sim$14 days \citep{Hall1976}. 
This RS CVn classification appears to be erroneous and can be traced to time series photometric observations which used a fast-rotating spotted star as a photometric standard. \citet{Fisher1983} reported $\delta$ Eri to be a suspected RS CVn variable based on detection of $\sim$0.02\,mag amplitude variability with period $\sim$10 days. Unfortunately the observations used $\epsilon$ Eri as a photometric standard, itself a spotted variable star with \prot\, $\simeq$ 10-12\,d\, and variability at the $\sim$0.01-0.03 mag level \citep{Frey1991,Croll2006}. Subsequent VLTI/VINCI interferometry measurements by \citet{Thevenin2005} ruled out the existence of any stellar companion down to about $\sim$2\%\, the luminosity of $\delta$ Eri. We concur with findings of \citet{Eaton1985}, \citet{Frey1991}, and \citet{Thevenin2005} that $\delta$ Eri is unlikely to be a RS CVn variable, and suggest that this four-decade-old misclassification for this bright nearby star be dropped from the GCVS and SIMBAD.

{\rvsearch\ recovers one significant RV signal, with parameters $P=596.6\pm2.6$\,d, $K=3.0\pm1.1$\,\ms\, and $e=0.65\pm0.14$.} Though the peak is well-defined, as expected for a planet, the eccentricity is a bit high and is being pulled quite strongly by a few \UCLES\, points. We thus classify this signal as SRC, and suggest future work investigate this signal more thoroughly. 

{No signals are detected in the S-index activity data. H$\alpha$ activity analysis returns one significant signal just over the detection threshold, with $P=49.568\pm0.097$\,d, $e=0.21\pm0.18$. This is substantially shorter than the reported rotation period from \citet{Baliunas1996} (71\,d). It seems possible that we could be seeing differential rotation ($\alpha=|P_2 - P_1|/P_{max}=0.43$) with surface shear approximately twice that of the Sun \citep[$\alpha_\odot$ = 0.2;][]{Reinhold2013}. 
There is a general trend that slower rotating stars exhibit enhanced differential rotation \citep[e.g.][]{Donahue1996}, however the behavior is not well-constrained observationally for periods longer than a $\sim$month, or for subgiants \citep[e.g.,][]{Reinhold2013}.}

\subsection{HD 32147 (GJ 183) \label{sec:HD32147}}

HD 32147 (GJ 183, HR 1614, HIP 23311) is a metal-rich \citep[\feh\, $=+0.29\pm0.02$;][]{Maldonado2012} K3+V star \citep{Gray2003} at $d=8.84$ pc \citep[$\varpi=113.0715\pm0.0222$ mas;][]{GaiaEDR3}. 
Although \citet{Baliunas1996} report the star to have low chromospheric activity (\logrphk\,=\,-4.948) and slow rotation (\prot\,=\,47\,d), it is  classified as a BY Dra variable with amplitude 0.03\,mag in the General Catalog of Variable Stars \citep{Samus2017}. 
More recently, \citet{Willamo2020} report \logrphk\,=\,-4.939 and rotation period \prot\,=\,33.7\,d, and activity cycle period \Pcyc\,=\,10.40\,yr ($\sim$3800\,d). 
\citet{BoroSaikia2018} estimate the activity period to be \Pcyc\,=\,$10.84\pm0.15$\,yr, whereas analysis of Mt. Wilson survey between 1967 and 2002 by \citet{Garg2019} reported two activity cycles of 9.33\,yr ($\sim$3408\,d) and 12.42\,yr ($\sim$4536\,d). 
We report one radial velocity signal with 
{$P=2866\pm140$\,d, 
$K=1.8\pm0.21$\,\ms, and 
$e=0.34\pm0.13$.}
\citet{Rosenthal2021} find a similar signal with $P=3444.0^{+91.0}_{-81.0}$\,d, which they classify as a false positive due to activity as well. Our signal is not quite close enough for us to consider it to be from the same source, and so we classify our RV Signal I as SRC rather than activity.

{Analysis of S-index activity data returns a multitude of signals. The first two of these signals have periods of $P=3774\pm250$\,d and $3204\pm310$\,d, which appear to be the same periodogram peak being fit multiple times. This signal correlates well with the 9.33 year (3405.5 days) signal from \citet{Garg2019} and the false positive from \citet{Rosenthal2021}, so we report it as that same cycle but make no update to the period as our detection is clearly not well constrained. We recover a second set of similar signals, with periods $P=381.7\pm2.4$\,d and $P=343.2\pm2.7$\,d. The periodograms clearly show that these are aliases of the respective first two signals, so we disregard these detections as having any significance.}

{We recover one further activity signal with parameters $P=95.6\pm0.24$\,d, $e=0.39\pm0.22$. This signal does not appear in the radial velocity data, though we note that it is approximately twice the rotation period of 47 days reported by \citet{Baliunas1996}. Additionally, the RV residual periodogram shows a strong peak at 44.4 days, which may correspond to the reported rotation period.}

The residual peak in the periodogram at 44.3 days likely corresponds to the rotation period, which \citet{Baliunas1996} measured to be 47 days. 

Additionally, \citet{Rosenthal2021} report a false positive at $P=51.997^{+0.078}_{-0.039}$\,d, which they attribute to an annual or instrumental systematic. Our data includes instruments not included in their study such as \HARPS\, and \PFS, so we expect not to detect this systematic from \HIRES\, as strongly.

\subsection{HD 38858 (GJ 1085) \label{sec:HD38858}}

HD 38858 (GJ 1085, HR 2007, HIP 27435) is a nearby star at 
distance 15.21 pc \citep[$\varpi=65.7446\pm0.0307$ mas;][]{GaiaEDR3} 
classified as G2V \citep{Gray2003}, with just slightly higher 
gravity (\logg$=4.51\pm0.01$) but more metal poor 
([Fe/H]$=-0.22\pm0.01$) \citep{Sousa2008}. 
\citet{Isaacson2010} and \citet{Lovis2011} predict rotation periods of 24\,d\, and $23.6\pm3.1$\,d\, based on mean chromospheric activity levels. 
The star's mean chromospheric activity level \citep[\logrphk\, =\, -4.948;][]{Lovis2011}) is similar to that of the Sun. 
{We detect one RV signal, with $P_{I}=2893\pm150$\,d, $K_{I}=2.8\pm0.3$\,\ms, and $e_{I}=0.19\pm0.12$. The long period and broad shape of this peak in the periodogram lead us to suspect that this signal is caused by magnetic activity. Though S-index analysis returns no significant detections, we note the presence of a growing signal in the periodogram at $P=2615.0$\,d. This signal does not meet our detection threshold of $\Delta BIC=10$, but is strong evidence supporting our classification of RV Signal I as a magnetic activity cycle.
}

\citet{Rosenthal2021} report a similar signal with $P=3113^{+82.0}_{-79.0}$\,d, $K=4.43^{+0.73}_{-0.64}$\,\ms, {which they attribute to an activity cycle as well.}

\subsection{HD 100623 (20 Crt) \label{sec:HD100623}} 

20 Crt (HD 100623, GJ 432 A, HR 4458, HIP 56452) is a K0-V star \citep{Gray2006} at distance $d=9.55$ pc \citep[$\varpi=104.6570\pm0.0267$ mas;][]{GaiaEDR3}. 
20 Crt is cooler (\teff$\,=5189$ K) and metal poor ([Fe/H]$=-0.37$) \citep{Valenti2005}.
It has a wide separation \citep[15\arcsec.3, projected separation 146 AU ;][]{Tian2020} white dwarf companion 20 Crt B (GJ 432B, HD 100623B, VB 4) of type DC10 \citep{Holberg2016}.
\citet{Kervella2019} analysis of \hipparcos, and \gaia\, astrometry finds 20 Crt A to have a tangential velocity anomaly of $41.26\pm5.38$ m\,s$^{-1}$ with a position angle of velocity anomaly vector of PA$=131^{\circ}.24\pm 5^{\circ}.18$, which is remarkably close to the observed PA to component B (PA\,=\,129$^{\circ}$) \citep{Mason2001}. 
Adopting fiducial masses of $M_A = 0.78$\,\Msun\, \citep{AguileraGomez2018} and $M_B = 0.66$\,\Msun\, \citep{GentileFusillo2019}, and assuming the projected separation is representative of the semi-major axis, one would estimate a system mass of $\sim$1.44\,\Msun, orbital period of $\sim$1470\,yr, and approximate orbital velocities of $\sim$1.4 and $\sim$1.6 \kms\, for A and B, respectively.

Analysis using \rvsearch\, fits the RV data using a linear trend rather than a Keplerian orbit. The signal is very evident in the radial velocity time series and we recover a best-fit trend of $0.00482\pm0.00022$ m\,s$^{-1}$\,d$^{-1}$ for HD 100623.

We assert that this signal is due to the companion, but our observation baseline is obviously not long enough to constrain its orbit well. \citet{Rosenthal2021} {also }report a long term linear trend of $\dot{\gamma}=0.00475\pm0.00028$ m\,s$^{-1}$\,day$^{-1}$ ($1.73\pm0.10$ m\,s$^{-1}$\,yr$^{-1}$), which {is consistent with our result, suggesting that our two signal detections are likely being caused by the same source}.

Additionally, we report one significant signal in the S-index activity analysis, with parameters $P=3729\pm89$\,d and $e=0.288\pm0.073$. The peak is fairly well-defined, and the long period makes this detection a plausible new magnetic activity cycle. 

\subsection{HD 131977 (GJ 570A) \label{sec:HD131977}}

HD 131977 (GJ 570 A, HR 5568, HIP 73184, KX Lib, Lalande 27173) is the primary in a complicated multiple star system with at least two other stellar companions situated 24\arcsec\, away \citep[HD 131976, resolved into the M dwarf pair GJ 570B and C;][]{Forveille1999}, and a distant substellar companion, GJ 570D, 274\arcsec\, away \citep{Burgasser2000}. 
HD 131977 is 5.89 pc away \citep[$\varpi$ = $169.8843\pm0.0653$ mas; ][]{GaiaEDR3} and classified K4V \citep{Keenan1989}.
There are two published rotation periods, \Prot\, = 44.6\,d \citet{Cincunegui2007} and 39.993\,d \citep{Fuhrmeister2022}. 
There is a surprisingly wide range of quoted metallicities for HD 131977, ranging from [Fe/H] $= -0.24\pm0.05$ \citep{Mishenina2012} to $0.12\pm0.03$ \citep{Valenti2005}. 

Through analysis with \rvsearch\,  we recover only a linear trend for this system, likely attributable to one of the (sub-)stellar companions. There are only 55 data points for this target, all from \HARPS, spanning $\sim$6 years. Because of these constraints, it is unsurprising that we do not recover full stellar companion orbits for this system and we recommend further observations to better constrain the parameters of the system.

{Our S-index analysis returns three significant detections. The first detection has $P=22.7657\pm0.0049$d, which we note is half the rotation period published by \citet{Cincunegui2007}. Detection of a P$_{\rm{rot}}$/2 signal can be caused by stellar spots on different hemispheres of the star being observed over multiple observing seasons, and so we attribute this signal to stellar rotation. The other two signals are extremely short-period ($P=3.87799\pm0.00054$d and $P=2.08913\pm0.00044$d) and we suspect that the relatively small amount of data for this target stretched over 6 years allows for Keplerian signals to fit multiple short-period cycles to the sparse sampling. We disregard these signals from being astrophysically significant at this point in time, and recommend more observations of this target to better characterize the star's activity.}

\subsection{HD 140901 (GJ 599 A) \label{sec:HD140901}}

HD 140901 (GJ 599 A, HR 5864, HIP 77358) is a G7IV-V type star \citep{Gray2006} with a high proper motion.
It is located at $d=15.25$ pc \citep[$\varpi$\,=\,65.5889\,$\pm$\,0.0342 mas;][]{GaiaEDR3}, {and has a 14\arcsec.6\, separation white dwarf companion HD 140901B (GJ 599 B).}
It is slightly cooler than the Sun (\teff$\,=\,5602\,\pm\,14$ K), and slightly more metal-rich at [Fe/H]$\,=\,0.10\pm0.02$ \citep{Soubiran2022}. 
There are no confirmed planets or published rotation periods for this star.

{Using the average \logrphk\, value from \citet{GomesdaSilva2014},  color from {\it Hipparcos} ($B-V$ = 0.715), and the activity-rotation calibration from \citet{Mamajek2008},
we predict that the rotation period of the star would be \Prot\,$\simeq21.5$\,d. }

Our radial velocity analysis in \rvsearch\, recovers one signal, with {$P=5084\pm1200$\,d, $K=11.6\pm2.4$\,\ms, and $e=0.44\pm0.25$. S-index analysis does not return any significant detections.} The majority of our radial velocity data comes from \UCLES, and because we do not have S-index activity data from this instrument, it makes sense that we do not see this same RV signal within the S-index data. 

{H$\alpha$ data analysis recovers two significant detections. The first of these signals is too long period to be well constrained by the Keplerian fit because its duration is on par with the \UCLES\, data observation baseline, so we defer to the original periodogram peak as the best estimator of this signal: $P_{I}=5431.8$\,d. This period agrees well with our RV detection, but because it also aligns with the long-period trend present in all the \UCLES\, data, we refrain from concluding definitively that RV Signal I is caused by magnetic activity. We classify it instead as SRC and recommend further study of this target to confirm the source of this signal.}

{The second H$\alpha$ signal has parameters $P_{II}=19.986\pm0.019$\,d and $e_{II}=0.27\pm0.19$. This is in good agreement with our prediction of a 21.5 day rotation period. We report H$\alpha$ activity signal II as a measurement of this star's rotation period.}

\subsection{HD 146233 (18 Sco) \label{sec:HD146233}}

18 Sco (HD 146233, GJ 616, HR 6060, HIP 79672) is a well-characterized solar twin and G2Va spectral standard star
\citep{Keenan1989} at $d=14.13$ pc
\citep[$\varpi=70.7371\pm0.0631$ mas;][]{GaiaEDR3}. 
\citet{Spina2018} reports stellar parameters extremely similar to those of the Sun:
\teff$\,=5808\pm3$ K,
\logg$\,=4.440\pm0.009$, 
\feh$\,=+0.041\pm0.003$,
$\tau=4.0\pm0.4$ Gyr, 
$M=1.022\pm0.004$\,\msun.
The star also has both rotation
\citep[\prot\,=\,22.9\,d;][]{Vidotto2014}
and chromospheric activity
levels \citep[\logrphk\,=\,-4.919;][]{Meunier2017},
very similar to the Sun as well. 
\citet{Lovis2011} report a magnetic activity cycle
of \Pcyc\, = $2803_{-392}^{+2663}$\,d, with predicted
rotation period \Prot\, $= 23.8\pm3.2$\,d based on the
mean activity level (\logrphk\, = -4.923). 
\citet{BoroSaikia2018} report a period of \Prot\,=\,22.7\,d and 
activity cycle of \Pcyc\,=\,$11.36\pm1.23$\,yr. 
It has no {reported} exoplanets.

We find {three} radial velocity signals within this system: 
{ 
$P_I=2374\pm47$\,d, 
$K_I=5.47\pm0.33$\,\ms, 
$e_I=0.21\pm0.07$; 
$P_{II}=6256\pm370$\,d ($17.1\pm1.01$\,yr), 
$K_{II}=4.96\pm0.57$\,\ms, 
$e_{II}=0.59\pm0.06$; 
$P_{III}=19.8777\pm0.0062$\,d, 
$K_{III}=1.73\pm0.26$\,\ms, 
$e_{III}=0.38\pm0.16$.}
{Additionally, there is one signal in the residuals periodogram at $P=10.5$\,d that falls just below the detection threshold.}

\citet{Butler2017} reported a planet candidate at roughly the same period as our Signal I (P$_{Butler}$ = 2528.8$\pm$ 105.5 days) and an S-index periodicity of 4190 days in their \HIRES\, data. Our detection of {$P_I=2374\pm47$\,}d corresponds directly to a signal recovered in the {S-index} activity data ({$P=2812\pm290$\,d}), so we report this signal as an update to the magnetic activity cycle in Table \ref{tab:updates}. {We also note the existence of a broad peak in the H$\alpha$ periodogram at around 2000 days, which further supports our conclusion that this signal is caused by magnetic activity rather than a planet as proposed by \citet{Butler2017}.} The discrepancy between our analysis and that of \citet{Butler2017} comes from their work including only data from \HIRES, while ours incorporates \HARPS, \HIRES, \PFS, and \UCLES\,. Our signal is mainly driven by \HARPS, and comparatively, the error bars on the \HIRES\, measurements are significantly larger. It makes sense that we recover the activity detection while the \HIRES\,-only work did not. This is confirmed by \citet{Rosenthal2021}, who report a similar signal with $P=2426.0^{+60.0}_{-42.0}$\,d\, as an activity cycle as well.

The {6256}-day signal we believe to be activity as well, due to its long period and periodogram peak shape. {The S-index activity data also yields a significant detection at $P=5272\pm1500$, which corresponds well to this long-period signal. They are not exact matches, but the presence of a 5000-day signal in both data sets further supports the conclusion that this signal is caused by magnetic activity.}

{RV Signal III has parameters $P_{III}=19.877\pm0.0062$\,d, $K_{III}=1.73\pm0.26$\,\ms, and $e=0.38\pm0.16$. Rotation periods reported by \citet{Lovis2011} and \citet{Vidotto2014} are both $>20$d, and this signal is fit to extremely high precision to 19.877 days. A signal caused by rotation should also appear in the Ca H\&K data, but there is no strength in the S-index periodogram around 19 days. Additionally, the periodogram peak is very sharp and well-defined, which would be highly unusual if the signal was caused by rotation-- rotation signals in RV come from observing stellar spots as the star rotates, and spots migrate and change slightly over time, so we expect to see some level of imprecision or variation in these RV measurements. The definition in this periodogram peak suggests no variation in the period over our approximately 20 year observation baseline, which is highly unusual. We therefore classify this signal as a Candidate, and recommend further study of this signal to confirm whether it is planetary in origin.}

{Though it is not fit by \rvsearch, we note that the 10.5\,d peak in the residual periodogram is also very well-defined and extremely close to the false alarm probability line. A future, more in-depth study of this target could investigate this signal further to address the cause of this significant period.}

{Analysis of H$\alpha$ activity from the \UCLES\, instrument returns no significant detections.}

\subsection{HD 188512 ($\beta$ Aql) \label{sec:HD188512}}

$\beta$ Aql (HD 188512, GJ 771 A, HR 7602, HIP 98036, Alshain) is a high proper motion star at $d=13.69$ pc \citep[$\varpi$\,=\,$73.00\pm0.20$ mas;][]{vanLeeuwen2007}, and is the primary spectral standard for type G8IV \citep{Johnson1953,Keenan1989}. 
The star is the most luminous star in our sample, and is somewhat cooler than the Sun (\teff$\,=5117\pm10$ K), less metal-rich ([Fe/H]\,=\,$-0.19\pm0.01$), and somewhat evolved with a lower surface gravity (\logg\,=\,$3.64\pm0.03$) \citep{Maldonado2016}.
\citet{Butkovskaya2017} report a magnetic activity cycle of \Pcyc $=969\pm27$\,d ($2.653\pm0.074$\,yr) and a surprisingly short rotation period of \Prot $= 5.08697\pm0.00031$\,days. 
\citet{Corsaro2012} report asteroseismic analysis of radial velocity data for $\beta$ Aql showing intra-night oscillations at the $\sim$5-10 \ms\, amplitude level. 
The star appears to be an evolved star somewhat more massive than the Sun \citep[$1.36\pm0.17$\,\msun, $1.337\pm0.021$\,\msun;][]{Corsaro2012,GomesdaSilva2021}, hence the relatively fast rotation for a subgiant likely reflects that the star spent its main sequence life blueward of the Kraft break.
The star appears to be consistent with being an intermediate-mass star, with isochronal age estimates consistently slightly younger than the Sun \citep[$\sim$3-4 Gyr;][]{Maldonado2013, Jofre2015, daSilva2015, Brewer2016, GomesdaSilva2021}, and chromospheric age estimates which had assumed that the star was a typical solar-type dwarf  \citep[9.6, 11.4 Gyr;][]{Mamajek2008} are likely to be significantly overestimated. 

After finding and subtracting a linear trend (0.00225 m\,s$^{-1}$\,day$^{-1}$) in  the California Planet Search data set for $\beta$ Aql A, \citet{Luhn2020} reports a Doppler signal ``$b$" with $P=10524.603$\,d and velocity amplitude $K=5.43$\,\ms, which would correspond to a $m$sin$i$ = 0.167\,\mj\, companion at $a = 10.18$ au.
$\beta$ Aql is in the Washington Double Star Catalog \citep{Mason2001} with components A, B, and C (WDS J19553+0624 = STT 532), although component C at separation $214\arcsec$ (TYC 493-72-1) is reported to be an unrelated interloper \citep{Kiyaeva2008}. 
$\beta$ Aql B is a M2.5V star \citep{Montes2018} at projected separation 13\arcsec.27 (182 au), and clearly a physical companion sharing similar proper motion and parallax \citep[$\varpi = 73.3889\pm0.0215$ mas;][]{GaiaEDR3}. 
\citet{Kervella2022} reports that the inferred tangential velocity calculated from Gaia EDR3 astrometry differs from that of $\beta$ Aql A by only 1.60 \kms.
However, the astrometric perturbation on $\beta$ Aql A, in the form of the tangential velocity anomaly as estimated through comparing {\it Hipparcos} and {\it Gaia} astrometry, appears to be negligible \citep[$5.74\pm10.65$\,\ms;][]{Kervella2022}. 

Analysis of two decades' worth of RV data with \rvsearch\, returns a linear trend rather than a full Keplerian signal. We find a best-fit RV trend of {0.00262 m\,s$^{-1}$\,d$^{-1}$, in good agreement with the \citet{Luhn2020} result.} The long-period trend is undoubtedly associated with the perturbation induced by the M dwarf companion B at separation at $\sim$180 au.
As the position angle of the $\beta$ Aql binary has changed by only 23$^{\circ}$ between 1838 and 2016, this suggests the AB orbital period to be of order few thousand years.

{Analysis of the S-index data for this target returns no significant detections. The star does not have any \UCLES\ observations, so there are no EW$_{H \alpha}$ measurements to study.}

\subsection{HD 190248 ($\delta$ Pav) \label{sec:HD190248}}

$\delta$ Pav (HD 190248, GJ 780, HR 7665, HIP 99240)
is a G8IV \citep{Gray2006} star at $d=6.10$ pc \citep[$\varpi=163.9544\pm0.1222$ mas;][]{GaiaEDR3} and chromospherically quite inactive (\logrphk\,=\,-5.10) \citep{GomesdaSilva2014}. 
\citet{Ramirez2013} report the star to have \teff\, = $5517\pm60$ K, \logg\, = $4.28\pm0.03$, and to be fairly metal-rich ([Fe/H] $= 0.33\pm0.07$). 
The star's rotation period has been estimated to be \Prot\, = $21.4\pm9.3$\,d\, \citep{Hojjatpanah2020}. 

\RVSearch\, identifies one Keplerian signal in the combined RV data for this star, with $P_I=360.8\pm1.9$\,days and $K=1.21\pm0.43$\,\ms. The \HARPS\, and \UCLES\, data exhibit significant disagreements with one another in the phase folded plot, however, and the signal seems to be driven strongly by the seasonality of the \HARPS\, data as evidenced by the sudden increase in the strength of the signal as a function of observation (see HD 190248's \RVSearch\, final summary in the accompanying figure set). We therefore suspect that this signal is due to observational sampling effects and not a planet.

\RVSearch\, also detects a linear trend 
in the data, with $dv_{r}/dt$ = $-0.00055 \pm 0.00009$\,\ms\,day$^{-1}$ ($-0.201\pm0.033$\,\ms\,yr$^{-1}$). Such trends are often suggestive of long period sub-stellar or giant planet companions. We can compute initial estimates of the minimum mass and semi-major axis for this 
companion by considering the linear trend to fall in the non-quadrature portion of an RV phase curve. In this case, we assume that the period of such a companion must be at least twice our observational baseline, as otherwise we would have expected to see some level of quadratic or sinusoidal curvature by now, and that its RV semi-amplitude must be at least half of the total RV span covered by the linear trend in the data set. That sets P$_{\rm{min}}$ = 37 years and K$_{\rm{min}}$ = 1.85 \ms. Folding in our knowledge of the host star's stellar mass, M$_{\star}$ = 1.001 M$_{\odot}$, we find that the planet must be at least 69 M$_{\oplus}$
(0.22 $M_{Jup}$) and on an orbit with a minimum semi-major axis $a_{min} = 11.1$\,au. Comparing with the \RVSearch\, injection/recovery summary plot, this combination of planet mass and orbital distance falls into a region that is not reliably recovered and so it is not surprising that the potential companion inducing this signal is not yet detectable with our current RV data set. 

\citet{Makarov2021} recently reported the detection of an astrometric perturbation for $\delta$ Pav which they interpret as being likely due to a long-period giant planet. 
They compare the short-baseline Gaia EDR3 proper motions \citep{GaiaEDR3} for $\delta$ Pav with long baseline astrometric parameters ($\sim$22-26 yr) combining {\it Hipparcos} with ground-based astrometry USNO Robotic Astronomic Telescope \citep[URAT;][]{Zacharias2015}.
Combining the Gaia EDR3, {\it Hipparcos} and URAT data, \citet{Makarov2021} estimate the perturbation of the tangential velocity for $\delta$ Pav to be (17.4, -13.2) \ms\, in $\alpha$ and $\delta$, respectively (0.995 and 0.958 confidence levels).
Removing the ground-based data, and using only {\it Hipparcos} and Gaia EDR3, \citet{Makarov2021} find the signal to be small but still significant: (7.7, -6.2) \ms\, in $\alpha$ and $\delta$, respectively (at combined confidence level 0.999).
Simply subtracting the proper motions from {\it Hipparcos} (epoch 1991.5) from Gaia EDR3 (epoch 2016.0) yields $\Delta\mu_{\alpha}$, $\Delta\mu_{\delta}$ = $0.731\pm0.149$, $-0.187\pm0.167$ \masyr, which at the distance of $d$ = 6.099 pc (1/$\varpi$ from Gaia EDR3) yields differences in the tangential motions of $21.1\pm4.3$, $-5.4\pm4.8$\,\ms\, in $\alpha$ and $\delta$, respectively. 
{Over the 24.5-yr baseline between the mean epochs for {\it Hipparcos} and Gaia EDR3, the averaged tangential accelerations are then
{$a_{\alpha}, a_{\delta}$ = $0.861\pm0.176$, $-0.220\pm0.196$\, m\,s$^{-1}$\,yr$^{-1}$, or total tangential acceleration $a_{tan}$ = $0.889\pm0.263$ m\,s$^{-1}$\,yr$^{-1}$. }
{Combining the measured radial acceleration
($a_{rad}$ = $-0.201\pm0.033$\,\ms\,yr$^{-1}$) with the tangential acceleration ($a_{tan}$) yields a total inferred acceleration on $\delta$ Pav of $a_{tot}$ = $0.911\pm0.265$
m\,s$^{-1}$\,yr$^{-1}$ ($2.89\pm0.84$ $\times$10$^{-8}$ m\,s$^{-2}$).}}

{Analysis of S-index data returns one significant period, with an initial $\Delta$BIC periodogram peak at 6375 days and an initial MAP fit of 6810.18 days. This is suggestive of a $\sim$17 year magnetic cycle, but attempts to fully characterize the signal via \rvsearch\,'s MCMC analysis fail -- likely due to insufficient sampling of the full orbital phase space. We therefore note the signal as an `LPS' in the S-index detections table and report just the MAP period, but encourage further monitoring of this star in the coming years to help fully resolve the star's long term magnetic activity.}

{The star's EW$_{H \alpha}$ data contains two significant signals according to \rvsearch, one with a period P = 352.9$\pm$1.5 days, and the other with P = 1171$\pm$36 days. The first signal suffers from the star's seasonal availability, leaving $\sim$ 1/3 of its orbital phase curve much less populated than the rest, and we suspect it is due to observational cadence constraints. The longer period signal is well defined in the $\Delta$BIC periodogram but falls logarithmically between the periods expected for the star's rotation period and its potential magnetic cycle. As HD 190248 is a very inactive star, much like the sun at solar minimum, this $\sim$1200 day signal prompts a question of whether we are seeing less obvious activity phenomena \citep[e.g., meridional flows][]{MeunierLagrange2020} that operate on intermediate time scales.} 

\subsection{HD 207129 (GJ 838) \label{sec:HD207129}}

HD 207129 (GJ 838, HR 8323, HIP 107649) is a nearby star at distance $d=15.56$ pc \citep[$\varpi=64.2717\pm0.0430$ mas;][]{GaiaEDR3} classified as G0V Fe+0.4 \citep{Gray2006}, and famous for having a resolved dusty debris disk \citep{JourdaindeMuizon1999,Krist2010}.
The star is a dwarf (\logg$\,=4.49\pm0.02$) of solar metallicity ([Fe/H]$\,=0.00\pm0.01$), just slightly hotter than the Sun (\teff$\,=5937\pm13$ K) \citep{Sousa2008}. 
\citet{Marshall2011} estimate the rotation period of the star to be \Prot\, $\simeq$ 12.6\,d\, based on the star's \vsini. 
\citet{Watson2011} and \citet{Lovis2011} predict the rotation period to be \Prot\, = $17.13\pm1.61$\,d\, and 17.6$\pm$2.8 based on the star's chromospheric activity. 
\citet{Lovis2011} report a magnetic activity cycle with period \Pcyc\, = $1520_{-139}^{+171}$\,d\, using 79 observations of \logrphk\, measured over an 1876\,d\, span. 

We recover one significant RV signal, with parameters 
$P=1964\pm49$\,d 
($5.38\pm 0.134$\,yr), 
$K=4.02\pm0.61$\,\ms, 
$e=0.44\pm0.16$.

{We find a single significant signal in the S-index analysis, with an initial periodogram peak of $P_I$=1886 days, and a MAP fit of 1898 days. This signal does not converge when subjected to \rvsearch\,'s affine-invariant sampling, and so we interpret it as an LPS. Despite this, the MAP period of the S-index is within 2$\sigma$ of the signal detected in the RVs, and so we report RV Signal I as a magnetic activity cycle. Our estimate of the activity cycle period is marginally consistent with that reported by \citet{Lovis2011} (2.2$\sigma$ difference). Our signal has a longer period than the baseline of the \citet{Lovis2011} study, and so this difference between our best-fit models does not raise significant concerns.}

{The EW$_{H \alpha}$ data for this star produces two significant detections, the first at $P_{I}$=5455$\pm$1900 days and the second at $P_{II}$=1726$\pm$71 days. The longer signal is close to the \UCLES\ observational baseline extent and has a large uncertainty, so we interpret it as an LPS and do not assume that it is astrophysical in nature. The second signal, however, is well defined in period and similar in duration to both the \citet{Lovis2011} \logrphk\ detection and our own S-index detections. We therefore consider it to be additional evidence for a long period magnetic cycle in the star.}

{Given these S-index and H$\alpha$ detections, we report RV Signal I as an update to the previous, magnetic cycle driven, detection. }

\section{Targets Lacking RV Signals \label{sec:LackingRV}}

For the remaining {16} stars included in this study, \RVSearch\ did not recover any significant signals in the radial velocities. We further subdivide these targets into Section \ref{sec:ActivityOnly}, stars which returned only significant activity signals, and Section \ref{sec:NoDetections}, targets which failed to return any significant signals in either the RVs or the activity. Many of these had a very limited number of RV measurements. Future radial velocity surveys should focus primarily on these targets in order to build knowledge of their exoplanetary parameter space. The stars with no significant RV signals {but a nonzero number of activity detections are listed in Table \ref{tab:norvsignals}. Stars with no detections at all are listed in Table \ref{tab:nosignals}.} The number of measurements analyzed for each of these stars can be found in Table \ref{tab:RV_Obs}.

\begin{table}[!htbp]
\centering
\caption{Targets with Activity Detections Only \label{tab:norvsignals}}
\begin{tabular}{lll}
\hline\hline
Identifier &Identifier &Identifier\\
\hline \vspace{2pt}
 HD 4628 & HD 14412 & HD 30495\\
HD 50281 & HD 72673 & HD 125072\\
 {HD 149661} & {HD 156026} & {HD 216803} \\
\hline\hline
\end{tabular}
\tablenotetext{}{Stars from our sample for which \rvsearch\ did not detect any significant Radial Velocity signals, but did return significant detections in their S-index or EW$_{H \alpha}$ analyses. Activity detections can be found in Tables \ref{tab:ActivitySignals_Sindex} and \ref{tab:ActivitySignals_Halpha}.}
\end{table}

\subsection{Targets with Activity Detections Only \label{sec:ActivityOnly}}

\subsubsection{HD 4628 (GJ 33) \label{sec:HD4628}}

HD 4628 (GJ 33, HR 222, HIP 3765, Lalande 1299, Wolf 25) is a metal poor ([Fe/H] = $-0.24\pm0.03$) \citep{Takeda2005} K2V star \citep{Gray2003}
at only $d$ = 7.43 pc \citep[$\varpi = 134.4948\pm0.0578$ mas;][]{GaiaEDR3}. 
The star is fairly slow rotating, with differential rotation observed (seasonal periods ranging from 37.2 to 41.4\,d) and mean \Prot\, $\simeq$ 38.5\,d\, \citep{Donahue1996}. 
Analysis of the Mt. Wilson survey data by 
\citet{Donahue1996IAU176} yielded a mean
cycle period of of \Pcyc\, = 8.6\,yr ($\sim$1966-1995), and subsequent analysis of a longer baseline by \citet{Garg2019} yielded cycle periods of \Pcyc\, = 
8.67, 8.08, and 9.98\,yr (mean \Pcyc\,= 8.91\,yr). 
\citet{BoroSaikia2018} estimate the chromospheric activity cycle to
be \Pcyc\, = $8.47\pm0.05$ yr. 

We {recover one significant detection in the S-index data} and none in the radial velocities. 
{The fitted signal has $P=3699\pm310$\,d\, and
eccentricity $e=0.33\pm0.12$.}
This appears to correspond to the
activity cycle for the star (\Pcyc\, = $10.90\pm0.41$\,yr), although somewhat longer than the cycle periods
reported by the longer baseline 
Mt. Wilson survey data 
\citep[][]{Donahue1996IAU176, Garg2019}.

\subsubsection{HD 14412 (GJ 95) \label{sec:HD14412}}

HD 14412 (GJ 95, HR 683, HIP 10798) is a G8V type star \citep{Gray2006} at $d$ = 12.83 pc \citep[$\varpi = 77.9140\pm0.0295$ mas;][]{GaiaEDR3}. 
Rotation period \Prot\, estimates for HD 14412 range from$13.0\pm0.3$\,d\, \citep{Hojjatpanah2020} to 29\,d\, \citep{Isaacson2010} (from \logrphk), however the
13-day estimate seems surprisingly fast given the star's low chromospheric activity \citep[\logrphk = -4.839;][]{Isaacson2010}. 

We recover {two significant S-index} activity signals for this star: {$P_{I}=2312\pm73$d, $e_{I}=0.091\pm0.098$ and $P_{II}=5686\pm1600$d, $e_{II}=0.5\pm0.16$. The RV periodogram returns no significant detections but does contain one strong peak just under the detection threshold, with $P=2074.5$\,d.}
{ \citet{Howard2016} presented a S-value periodogram for HD 14412, showing a pronounced peak at 5.7\,yr (2082\,d). We report our Activity Signal I a magnetic activity cycle of \Pcyc\, = 
$2312\pm73$\,d\, ($6.33\pm0.2$\,yr), fairly consistent with that reported by \citet{Howard2016}. We suspect S-index activity signal II is caused by a magnetic activity cycle as well, due to the long period and broad shape of the peak.}

\subsubsection{HD 30495 (58 Eri) \label{sec:HD30495}}

58 Eri (HD 30495, GJ 177, HR 1532, HIP 22263, IX Eri) is a nearby star at distance 13.24 pc \citep[$\varpi=75.5289\pm0.0539$ mas;][]{GaiaEDR3} classified as G1.5V CH-0.5 \citep{Gray2006}. 
The star is a young ($\sim$1 Gyr) solar analog, with a rotation
period \Prot$=11.36\pm0.17$\,d, and manifesting both
short ($\sim$1.7\,yr) and long ($\sim$12.2\,yr) activity cycles \citep{Egeland2015}. 
\citet{Gaidos2000} reports time series photometry over 6 seasons, finding periods between 10.5 and 11.47 days, and reporting a mean rotation period of \Prot\,=\,11.3\,d.

\rvsearch\, finds no significant signals in the radial velocity data, but {one significant signal in the S-index activity data with $P=71.46\pm0.11$\,d, $e=0.31\pm0.12$}. There is a correlated peak in the radial velocity residual periodogram at 72 days, although it does not rise to the level of being a "significant detection". {This signal does not correspond to the published rotation period, nor to either of the published activity cycles referenced above. Because of this, we classify this signal as SRC and recommend further study of the activity data for this target in a future work}.

\subsubsection{HD 50281 (GJ 250A)\label{sec:HD50281}}

HD 50281 (GJ 250A, HIP 32984) is a K3.5V star \citep{Gray2003} at $d$ = 8.74 pc \citep[$\varpi = 114.3547\pm0.0418$ mas;][]{GaiaEDR3}. 
{The star is in a wide binary \citep[separation 58\arcsec.9;][]{Mason2001} with the M dwarf GJ 250B.} 
{HD 50281 is an active star (\logrphk\,=\,-4.554) \citep{Gondoin2020}, and \citet{Fuhrmeister2022} predict a rotation period of \Prot\,$=16.493$\,d based on the chromospheric activity.}

{Analysis of the RV periodogram yielded no significant signals.}
{The time series data in Ca H \& K shows a very complicated periodic pattern which resulted in 7 significant periodic signals detected.
As the last couple appear amid a forest of slightly lower power peaks, we believe that our statistical criterion may be inadequate for this very active star and picking out true signals from background noise. We focus on the interpretation of the first five prominent peaks, which had periods of $2264\pm11$\,d, $2102\pm12$\,d, $139.42\pm0.05$\,d, $12.47954\pm0.00046$\,d, and $16.49842\pm0.00089$\,d. The first three have similar semi-amplitudes in $\Delta$S at the $\sim$0.05-0.10 level, and appear to be attempts by our code to fit a single complicated activity cycle of \Pcyc\,$\simeq$\,2264\,d which is inadequately fit by a single Keplerian orbit model. The latter two are well-defined and similar to the predicted rotation period from \citet{Fuhrmeister2022}. Hence, we consider the $12.5$\,d and $16.5$\,d signals to be from differential rotation.}

\subsubsection{HD 72673 (GJ 309) \label{sec:HD72673}}

HD 72673 (GJ 309, HIP 41926) is a K1V star \citep{Keenan1989}, with no known companions or planets. 
The star is fairly inactive (\logrphk\, = -4.968) with
a slow predicted rotation period \citep[\Prot $=40.2\pm4.1$\,d;][]{Lovis2011}. 

We recover no significant RV signals {but one S-index and one H$\alpha$ activity detection.}
{The S-index activity detection has parameters $P=3217\pm200$d and $e=0.14\pm0.14$.} This signal matches the previously reported magnetic activity cycle period reported by \citet{Lovis2011} (\Pcyc $=3050_{-408}^{+558}$\,d), although the uncertainty in our cycle period is 7$\times$ smaller uncertainty. {We therefore report our detection as an update to this previously published magnetic activity cycle.}

{The H$\alpha$ activity detection is much shorter period, with parameters $P=341.2\pm3.6$d and $e=0.16\pm0.18$. This is obviously very close to one year, indicating a strong possibility that this signal is being driven by windowing effects similar to with the \HARPS instrument. The peak is extremely well-defined, however, and highly significant, so we refrain from decisively calling this detection a false positive.}

\subsubsection{HD 125072 (GJ 542) \label{sec:HD125072}}

HD 125072 (GJ 542, HIP 69972) 
is a K3V \citep{Houk1975} star at $d$ = 11.82 pc
\citep[$\varpi = 84.6029\pm0.0218$ mas][]{GaiaEDR3}.
\citet{Gray2003} classified the star as K3IV subgiant, however the star's spectroscopic parameters
\citep[\Teff $= 4899\pm48$ K, \logg\, $=4.55\pm0.03$, \feh\,= $0.28\pm0.08$ ;][]{Ramirez2013} and HR diagram position (\bv = 1.03, $M_V$ = 6.30, $\sim$0.44 mag above main sequence) clearly flag it as a very metal-rich dwarf. 

\citet{Lovis2011} report a magnetic activity cycle of \Pcyc\, =  1146$_{-70}^{+982}$\,d\, and a predicted rotation period of $42.0\pm5.9$\,d\, based on the low mean activity level (\logrphk\,=\,-4.941). 

We recover {no significant RV signals, two detections in the S-index activity data, and one in the EW$_{H \alpha}$ data}. S-index signal I, with {$P_I=2989\pm100$}\,d, loosely correlates with the magnetic activity cycle of \citet{Lovis2011}. S-index signal II has {$P_{II}=40.49\pm0.036$\,}d, which is most likely caused by stellar rotation,{ and agrees well with the predicted cycle from \citet{Lovis2011}.}

{The EW$_{H \alpha}$ data analysis yields one significant detection with an initial $\Delta$BIC period of 5468.5 days, but fails to produce a well constrained orbital fit during the MCMC analysis (instead giving $P=9483\pm9400$\,d). We therefore instead report the MAP best-fit solution, which has a period of 7137.76 days, which we attribute to the long-period \UCLES\, trend present in almost all the H$\alpha$ data for all targets.}

{We note additionally the presence of a signal in the RV residual periodogram that falls just below the detection threshold, at $P=13.5$\,d.}

\subsubsection{HD 149661 (12 Oph)}

{12 Oph (HD 149661, GJ 631, HR 6171, HIP 81300, V2133 Oph) is a K0V(k) \citep{Gray2006} star at $d$ = 9.89 pc
\citep[$\varpi = 101.0719\pm0.0501$ mas][]{GaiaEDR3}.}
{The star has dwarf surface gravity (\logg\,=\,$4.52\,\pm\,0.02$) 
and metallicity just slightly more than solar 
\citep[\feh\,= $0.03\pm0.01$ ;][]{Soubiran2022}.}
{Analysis of chromospheric activity levels (\logrphk\, index) show
that it varies widely over the past several decades.}
{During the Mt. Wilson survey period of 1967-1983, the star 
had an average \logrphk\, value of -4.583 \citep{Baliunas1996}, however the survey by \citet{Radick2018} during 1994-2016 recorded an average of \logrphk\, = -4.71, while analysis of HARPS observations during 2005-2012 by \citet{GomesdaSilva2021} estimated a median activity level of \logrphk\, = -4.56.}
{From analysis of the Mt. Wilson HK survey data, 
\citet{Donahue1996} reports an average rotation period over 9 seasons of \Prot\, = 21.07\,d, with individual seasonal rotational periods ranging between 20.6 and 22.9 d. }
{ \citet{BoroSaikia2018} report two Ca HK activity cycles with periods \Pcyc\,=\,$15.3\pm0.4$\,yr and \Pcyc\,=\,$7.7\pm0.12$\,yr.}

{Analysis of both RV and H$\alpha$ data returns no detections for this target, but the S-index search yields two significant signals: $P_{I}=1649\pm55$d, $e_{I}=0.42\pm0.12$; $P_{II}=3874\pm1200$d, $e_{II}=0.73\pm0.21$. The first of these signals is likely to be a magnetic activity cycle, based on its long period and signal strength. The second signal is poorly constrained-- the periodogram peak being fit is at 4062.0 days, which is approximately half of the observation baseline. \RVSearch\, struggles to fit a Keplerian orbit to the signal, as there is insufficient data to constrain the orbit very well. This signal may be evidence of a longer period magnetic activity cycle, but additional data is needed to constrain the cycle well.}

\subsubsection{HD 156026 (36 Oph C)}

{36 Oph C (HD 156026, GJ 664, HIP 84478, V2215 Oph, WDS J17153-2636C) is a nearby \citep[5.88 pc; $\varpi = 169.9617\pm0.0311$ mas;][]{GaiaDR3} 
K5V(k) \citep{Gray2006} star which is a very wide separation (731\arcsec.54) companion to the bright K0V+K1V pair 36 Oph A \& B \citep[][]{CayreldeStrobel1989}.}
{The orbital motion of C around AB appears to be detectable astrometrically,
as \citet{Kervella2022} show that C shows a tangential velocity anomaly between the {\it Hipparcos} and {\it Gaia DR3} data of $5.98\,\pm\,1.19$ m\,s$^{-1}$ with a vector of PA = $87^{\circ}.22\pm7^{\circ}.25$ (compare
to the PA between AB and C of PA = $73^{\circ}.83$).}
{The difference in tangential velocities between AB and C is 0.63\,\kms, which is similar to the predicted escape velocity of C from AB (0.61 \kms) \citep{Kervella2022}. }

{Photometric variability at the $\sim$0.02 mag in $V$-band for 36 Oph C was reported by \citet{LloydEvans1987} who estimated a period of 21.0\,d.}
{Independently, \citet{Baliunas1996} report an identical rotation period of 21\,d based on analysis of Mt. Wilson Ca II H \& K observations, and an average activity level
of \logrphk\, = -4.662.}
{\citet{BoroSaikia2018} reports a Ca II H \& K activity cycle period of \Pcyc\,=\,$21.3\pm0.83$\,yr.} 
{36 Oph C appears to be erroneously classified as a RS CVn variable in the General Catalog of Variable Stars \citep{Samus2017} and SIMBAD\footnote{https://simbad.u-strasbg.fr/simbad/}, and while the star is clearly spotted and active, there is no evidence of the star being a chromospherically active binary (i.e. no sign of short-period stellar binary).}
{The radial velocity trend is flat, with scatter at the $\sim$2\,m\,s$^{-1}$ level,
consistent with the 1.57\,m\,s$^{-1}$ jitter previously estimated }by \citet{Isaacson2010}.

{The S-index data shows one significant peak at $P=378.9\pm2.2$\,d, which is likely caused by systematics, as the period is extremely close to one year. Additionally, there are weak peaks in the residual periodogram around 4.9\,d,
$\sim$22\,d and $\sim$25\,d, with the latter two suspiciously near the previously reported 2\,day rotation period. }

\subsubsection{HD 216803 (TW PsA)}

{TW PsA (Fomalhaut B) is a nearby \citep[7.60 pc; $\varpi = 131.5525\pm0.0275$ mas;][]{GaiaDR3} K4Ve \citep{Keenan1989} spectral standard star which is within a very wide, young ($\sim$440 Myr-old) triple system with Fomalhaut A and C (LP 876-10) \citep{Mamajek2013}.}
{The star has essentially solar metallicity and dwarf surface gravity
(\teff\,$=\,4601\pm29$\,K, \logg\,$=\,4.68\pm0.10$, [Fe/H]\,$=\,0.04\pm0.03$)\citep[][]{Soubiran2022}.}
{The star is relatively fast-rotating \citep[\Prot\,=\,10.3\,d, 9.87\,d;][]{Busko1978,Wright2011} and chromospherically active \citep[\logrphk\,=\,-4.44;][]{GomesdaSilva2021}.} 
{\citet{DeRosa2019} reported an astrometric acceleration of TW PsA consistent with a $1.2^{+0.7}_{-0.6}$\,\Mjup\, planet on a $P_{orb}\,=\,25^{+52}_{-21}$ yr orbit based on comparison of the {\it Hipparcos} and {\it Gaia} DR2 astrometry. However it is worth noting that an independent comparison of the {\it Hipparcos} and {\it Gaia} DR2 astrometry by \citet{Kervella2019} yielded  a borderline significance tangential velocity anomaly ($18.67\pm6.39$\,\ms; 2.9$\sigma$), a subsequent analysis using improved DR3 data by \citet{Kervella2022} yielded tighter, but less significant constraints ($2.15\pm1.49$\, \ms; 1.4$\sigma$).}

{Analysis of RV and H$\alpha$ data returns no significant signals for this target. The S-index period search yields several detections: $P_{I}=3.8913\pm0.0002$d, $P_{II}=4.08499\pm0.00049$d, and $P_{III}=2.8\pm0.3$d. However, the S-index data for this target is fairly sparse, so \rvsearch\, is able to fit many different short-period signals to the data easily. We do not believe that any of these signals are physically significant, and disregard them as not physically meaningful.}

\subsection{Targets with No Detections \label{sec:NoDetections}}

Several targets included in this work did not return any significant detections in either the RVs or the activity indicators when run through \rvsearch\,. In some cases, this is due to a lack of data on the given target. Otherwise, there are a few cases in which the target is well-studied, and likely is simply a quiet system. Table \ref{tab:nosignals} lists each of the stars that had no detections, and categorizes them as having insufficient data to make a detection (ID) or well-studied but still contains no signal (NS). Additionally, for each target, we report mean RMS. This works as a valid proxy for stellar variability, to compare with our detection results. {For targets designated ``ID" in the table, we recommend further study for improved completeness in the future.}

\begin{table}[!htbp]
\centering
\caption{Targets with No Significant Signals \label{tab:nosignals}}
\begin{tabular}{lcc}
\hline\hline
Identifier &Classification &RMS [\ms]\\
\hline \vspace{2pt}
{HD 693} & {ID} & {2.81} \\
HD 7570	 & NS	& 5.68\\
HD 23356 & NS* 	&5.27\\
HD 76151 & ID	&8.97\\
{HD 102870} & {NS} & {5.41}\\
HD 131977 &ID &6.93\\
HD 196761 &NS* &4.60\\
\hline\hline
\end{tabular}
\tablenotetext{}{Stars from our sample for which \rvsearch\ did not detect any significant signals. (*: Targets marked with an asterisk have strong signals in their periodogram which almost cross the detection threshold; these are discussed more in depth in Section \ref{sec:NoDetections}). }
\end{table}

Stars marked with an asterisk in Table \ref{tab:nosignals} have signals that are close to but do not quite cross the detection threshold in their RV periodograms. HD 196761 shows strong periodicity around 26-28 days which falls just short of the False Alarm Probability mark. We believe this signal to be evidence of a rotation period for this target. HD 23356 has a strong peak at 2911.6 days. The observation baseline for this target is only about twice this period, so further observation of this target could constrain this signal better. 

\startlongtable
\begin{table*}[!htbp]
\centering
\caption{RV Sensitivity \label{tab:IR_Sensitivity}}
\begin{tabular}{llllll|cc}
\hline\hline
HD & GJ & Mass & Ref. & Lumin. & Ref. & EEID & Doppler Sens.\\
 & & (\Msun) & & (\logl) & & (au) & (\Mearth)\\
\hline \vspace{2pt}
693	 & 10 &	$1.08\pm0.025$    	 &	1	&	 0.477	&	11	&	1.731	&	403.8\\
1581 & 17 &	$1.00\pm0.025$    	 &	1	&	 0.101	&	11	&	1.123	&	9.7\\
2151 & 19 &	$1.141\pm0.0125$  	 &	2	&	 0.541	&	TW	&	1.864	&	44.8\\
4628 & 33 &	$0.75\pm0.02$     	 &	3	&	-0.523	&	11	&	0.548	&	13.4\\
7570 & 55 &	$1.17\pm0.0155$   	 &	2	&	 0.302	&	11	&	1.415	&	88\\
13445 & 86A & $0.797\pm0.024$    &	2	&	-0.389	&	11	&	0.639	&	45.3\\
14412 &	95  & $0.811\pm0.027$    &	2	&	-0.351	&	11	&	0.668	&	24.5\\
16160 &	105A & $0.74\pm0.02$     &	1	&	-0.549	&	11	&	0.532	&	{12.8}\\
20766 &	136	& $0.916\pm0.0275$   &	2	&	-0.100  &	11	&	0.891	&	81\\
20794 &	139	& $0.813\pm0.015$    &	2	&	-0.184	&	11	&	0.809	&	7.4\\
20807 &	138	& $0.955\pm0.0265$   &	2	&	 0.007	&	11	&	1.008	&	21.4\\
22049 &	144	& $0.804\pm0.025$    &	2	&	-0.471	&	11	&	0.582	&	62.7\\
23249 &	150	& $1.17\pm0.003$     &	3	&	 0.500	&	3	&	1.778	&	27.6\\
23356 &	... & $0.80\pm0.02$      &	4	&	-0.515	&	11	&	0.553	&	44.7\\
26965 &	166A & $0.79\pm0.02$     &	3	&	-0.364	&	11	&	0.658	&	11.8\\
30495 &	177	& $1.016\pm0.0225$   &	2	&	-0.015	&	11	&	0.983	&	393.6\\
32147 &	183	& $0.79\pm0.02$      &	1	&	-0.537	&	11	&	0.539	&	9.1\\
38858 &	1085 & $0.95\pm0.05$     &	3	&	-0.083	&	11	&	0.909	&	{17.2}\\
39091 &	9189 & $1.11\pm0.03$     &	1	&	 0.186	&	11	&	1.238	&	51.1\\
50281 &	250A & $0.75\pm0.02$     &	4	&	-0.658	&	11	&	0.469	&	55.9\\
69830 &	302	& $0.89\pm0.03$      &	3	&	-0.216	&	11	&	0.779	&	7.4\\
72673 &	309	& $0.788\pm0.027$    &	2	&	-0.394	&	11	&	0.635	&	9.6\\
75732 &	324A & $0.92\pm0.025$    &	3	&	-0.197	&	11	&	0.797	&	35\\
76151 &	327	& $1.02\pm0.025$     &	1	&	-0.013	&	11	&	0.985	&	17393.5\\
85512 &	370	& $0.704\pm0.019$    &	5	&	-0.778	&	11	&	0.408	&	3.7\\
100623 & 432A &	$0.774\pm0.026$  &	2	&	-0.432	&	11	&	0.608	&	19.2\\
102365 & 442A &	$0.89\pm0.035$   &	3	&	-0.074	&	11	&	0.919	&	13.6\\
102870 & 449 & $1.298\pm0.0415$  &	2	&	 0.576	&	11	&	1.941	&	201.6\\
114613 & 9432 &	$1.27\pm0.02$    &	3	&	 0.626	&	11	&	2.055	&	53.6\\
115617 & 506 & $0.94\pm0.03$     &	3	&	-0.078	&	11	&	0.914	&	15.2\\
125072 & 542 & $0.828\pm0.025$   &	2	&	-0.466	&	11	&	0.585	&	24.9\\
131977 & 570A & $0.77\pm0.03$    &	6	&	-0.653	&	11	&	0.472	&	171.2\\
136352 & 582 & $0.92\pm0.03$     &	3	&	 0.012	&	11	&	1.014	&	9.7\\
140901 & 599A & $0.954\pm0.019$  &	2	&	-0.088	&	11	&	0.904	&	95.3\\
146233 & 616 & $1.003\pm0.025$   &	2	&	 0.039	&	11	&	1.046	&	{17.8}\\
149661 & 631 & $0.859\pm0.0265$  &	2	&	-0.335	&	11	&	0.680	&	110\\
156026 & 664 & $0.6972\pm0.0253$ &	7	&	-0.803	&	11	&	0.397	&	13.5\\
160346 & 688 & $0.785\pm0.025$   &	8	&	-0.480	&	6	&	0.575	&	87.5\\
160691 & 691 & $1.15\pm0.035$    &	1	&	 0.278	&	11	&	1.378	&	27.7\\
188512 & 771A & $1.27\pm0.065$   &	3	& 	 0.780	&	12	&	2.455	&	{255.6}\\
190248 & 780 & $1.001\pm0.033$   &	2	&	 0.097	&	11	&	1.118	&	10.9\\
192310 & 785 & $0.82\pm0.025$    &	3	&	-0.394	&	11	&	0.636	&   7.8\\
196761 & 796 & $0.86\pm0.035$    &	3	&	-0.252	&	11	&	0.748	&	27.1\\
207129 & 838 & $1.07\pm0.035$    &	1	&	 0.082	&	11	&	1.099	&	35.8\\
209100 & 845 & $0.715\pm0.019$   &	10	&	-0.654	&	11	&	0.471	&	{18.2}\\
216803 & 879 & $0.73\pm0.015$    &	8	&	-0.707	&	11	&	0.443	&	176.5\\
\hline\hline
\end{tabular}
\tablenotetext{}{
References:
(1) \citet{Ramirez2012},
(2) \citet{Ramirez2013},
(3) \citet{Brewer2016},
(4) \citet{Maldonado2016},
(5) \citet{Soto2018},
(6) \citet{Luck2017},
(7) \citet{Anders2019},
(8) \citet{Casagrande2011},
(9) \citet{Fekel1983},
(10) \citet{DelgadoMena2019},
(11) \citet{Stassun2019},
(12) \citet{Schofield2019}.}
\end{table*}

\section{Discussion}

\begin{figure}[htb]
\includegraphics[width=.46 \textwidth]{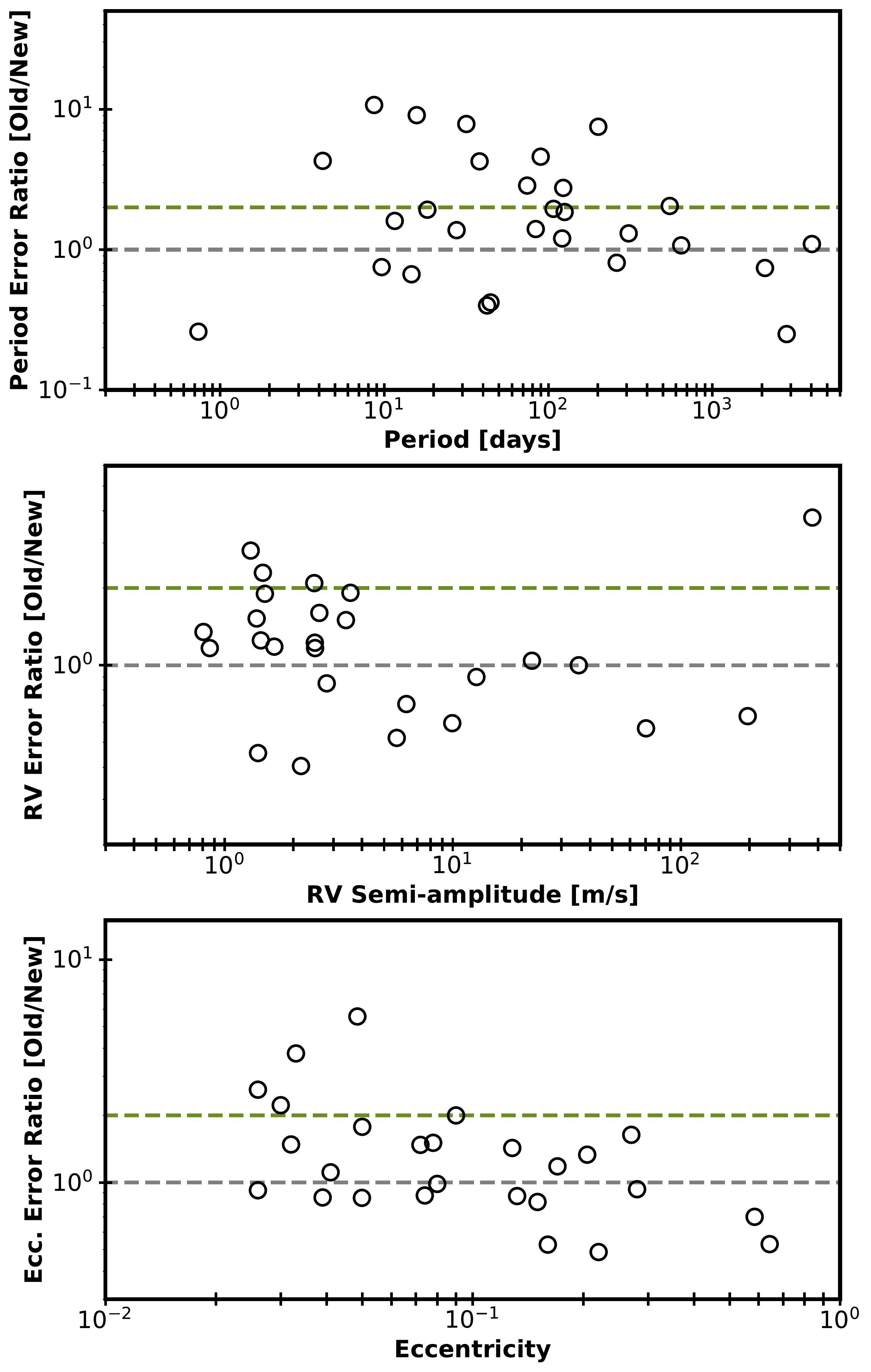}
\caption{Comparison of the uncertainties in previously published works and our updated RV analyses for the planets listed in Table 5 for orbital period (top plot), RV semi-amplitude (middle plot) and orbital eccentricity (bottom plot). The grey dashed lines depict a 1:1 ratio, so planets above the line have more precise results in our analysis while planets below the line have less precise results here than previously published. The green lines denote a factor of two improvement, so planets above the green lines have uncertainties that decreased by 50\%. This happens most commonly in the orbital period comparisons, as the additional months/years of data added here include many more orbits of each planet.
\label{fig:ParameterComparison}}
\end{figure}

In carrying out this study, we sought to characterize the planet detection completeness of nearby, Sun-like stars which have been identified as candidates for future direct imaging observations based upon existing RV observations. 
We compiled archival RV data sets from the \emph{HARPS}, \emph{HIRES}, \emph{UCLES}, \emph{PFS}, and \emph{APF} spectrographs to produce a reasonably complete picture of the existing {precise} RV sensitivity for each star.
Many of the targets in this work are hosts of previously published planetary systems. Yet despite the accumulation of many additional RV data points since their first publication, the majority of these systems' orbital parameters have not been previously updated. 
By utilizing the full range of archival RV data up through present day, we are able to report updated orbital parameters for many of these previously confirmed planetary systems (Table \ref{tab:updates}) and find in many cases that the uncertainties on the planets' periods, RV semi-amplitudes and eccentricities improve when compared to previous publications (Fig. \ref{fig:ParameterComparison}). 

Some select highlights of our updated analyses are summarized below:
\begin{itemize}

    \item We provide the most precise set of orbital parameters yet published for the three Neptune-mass planets orbiting HD 69830.
    
    \item We assert that the 40 day planet HD 20794 c published in \citet{Pepe2011} is due to stellar activity and not a Keplerian signal as its statistical significance has not increased despite the addition of hundreds of new {precise} RV data points.
    
    \item We {show conclusively} that the 58 day planet HD 85512 b published in \citet{Pepe2011} is due to stellar activity and not a Keplerian signal, because the signal changes in period by 10+$\sigma$ over the decade of data collected here.
    
    \item We {present strong evidence} that the 3827 day planet HD 114613 b reported by \citet{Wittenmyer2014} is not Keplerian in nature as its statistical significance decreases despite the addition of hundreds of {precise} RV measurements.
    
    \item We improve the best fit error bars for the period, semi-amplitude, and eccentricity of the SB1 companion to HD 160346 by over an order of magnitude.
    
    \item {We present strong evidence that the planet HD 26965 b 
    ($o^2$ Eri b, 40 Eri b) reported by \citet{Ma2018} is not a planet, and is rather caused by stellar activity. The $42.303$\,d RV signal is nearly identical to a periodicity detected in H$\alpha$ of $P=43.504\pm0.066$\,d, which overlaps previous estimates of the star's rotation \citep[42-43\,d;][]{Baliunas1996,Frick2004}.}
    
    \item {We report two new planet candidates to be further studied and confirmed by future works: HD 192310 RV Signal IV ($P_{IV}=24.559\pm0.016$\,d) and HD 146233 RV Signal III ($P_{III}=19.8777\pm0.0062$\,d).}
\end{itemize}

Our analysis and results thus serve as encouragement for updated analysis on other previously-confirmed planetary systems in which significant amounts of new data have been acquired since publication.

In addition, we report a number of new magnetic activity cycles and signals which are not yet complete enough to be classified, all of which invite further study.

{In this work, o}ur goal was to analyze each star's RV sensitivity completeness, so that we might make recommendations with respect to future work in preparation of a {Direct Imaging (DI)} mission that aims to search for Earth analog planets around these stars. As time on future DI missions is likely to be highly oversubscribed, it is imperative that their target lists be as thoroughly vetted as possible in order to increase these future missions' efficiency and science output. 

\begin{figure}[!htb]
\includegraphics[width=.46 \textwidth]{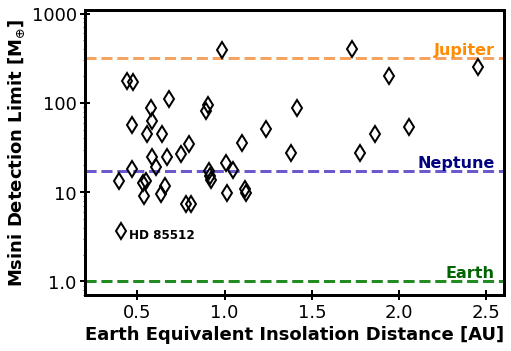}
\caption{50\% detection sensitivity threshold for each target star at its Earth Equivalent Irradiation Distance (EEID) {-- the distance from the host star at which the planet will receive the same amount of energy as the Earth receives from the Sun.} For the majority of targets the existing Doppler sets are not yet sensitive to Neptune-mass planets at their respective EEIDs, which would preclude the formation of stable Earth-analogs, let alone Earth-mass planets themselves.
\label{fig:IR_Summary}}
\end{figure}

\begin{figure}[!htb]
\includegraphics[width=.46 \textwidth]{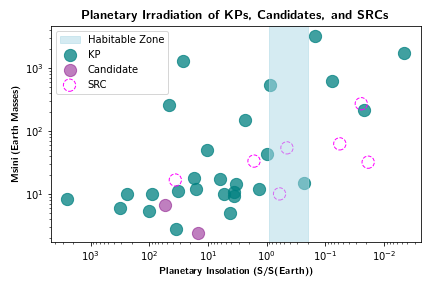}
\caption{{Planet, Candidate, or SRC planetary irradiation relative to Earth ($S/S_\earth$) versus Msini ($M_\earth$). The habitable zone for an Earth analog in a Solar analog system \citep{Kopparapu2014} is marked in the blue shaded region.}
\label{fig:Msini_Insol}}
\end{figure}

One key component of this characterization is to identify the presence of any additional planets in the system and determine whether their orbital parameters preclude the existence of the temperate, terrestrial, planets that the future DI missions seek. If such planets are detected, then these stars should be down weighted in the mission's observing priority list. Figure \ref{fig:IR_Summary}, Figure \ref{fig:Msini_Insol}, and Table \ref{tab:IR_Sensitivity} summarize our findings in this area. While it is clear from Figure \ref{fig:IR_Summary} that even our most well-studied targets do not come close to the 1 \mearth\ limit for a 1 AU orbit, we are at least able to rule out the presence of Neptune to Jupiter mass planets at $\sim$1 AU; such bodies would eliminate the possibility of a dynamically stable Earth analog. {Figure \ref{fig:Msini_Insol} shows the range of \msini\, and planetary insolation of the known planets, candidates, and SRCs in this work relative to the habitable zone for an Earth-like planet around a Sun-like star as defined in \citep{Kopparapu2014}; very few of our detections fall within this region.}

{For the majority of our stars,} the minimum detectable mass planet at 1 AU is well above the mass of Neptune or even Saturn. And in some cases, where the stars have only a handful of existing RV observations, even a stellar companion could remain hidden in the data. We therefore recommend further study of all targets on this list. Future surveys could focus most strongly on those that have the least RV sensitivity in the 1 AU region. 

The stars with the least RV sensitivity, for this study, are those with the smallest number of RV observations. {Our list contains 9 stars with under 50 RV epochs: HD 693 (16 epochs), HD 30495 (50 epochs), HD 76151 (7 epochs), HD 131977 (22 epochs), HD 147584 (1 epoch), HD 160346 (34 epochs), HD 165341 (7 epochs), HD 203608 (1 epoch), and HD 216803 (42 epochs)}. We recommend that future RV surveys focus strongly on these 16 targets, in order to build up their RV baseline and thus increase RV sensitivity. For those targets which are closer to the 1\mearth\ line, we suggest more in-depth analysis of the archival data in order to push this limit.

The final uninformed search and injection/recovery figures created by \RVSearch\, are presented in the accompanying Figure Sets so that targets' results may be examined on an individual basis. The radial velocity data used to perform these fits and analyses will be published in an accompanying machine readable table.

\section{Conclusion}

We expect the detection and characterization of Earth-analog planets to be an exceptionally difficult undertaking due to the challenges presented by observational constraints, instrument systematics, and, most importantly, the variability of the stars themselves. The list of stars that are well suited to future direct imaging searches for such planets is limited due to stringent requirements on the stars' distance from Earth that in turn determines whether a temperate planet orbiting that star falls outside inner working angle of the DI instrument. There are $\sim$100 stars identified by the EPRV WG to meet the criteria both for being a suitable DI target and for being amenable to precision radial velocity observations. We have compiled archival radial velocity time series data from the majority of precision RV spectrographs that have operated in the southern hemisphere over the past two decades for these 50 nearby, Sun-like stars that are likely to be targets of future, space-based, direct imaging missions. 

Our primary objective was to quantify each star's RV completeness via the use of an injection/recovery analysis applied to archival RV data. Our results show that the minimum detectable planet mass at 1 AU ranges from 6.5-818.5 \Mearth\ depending on the star, showcasing the heterogeneous state of the archival RV data collected from these targets. While additional data from the spectrographs included in this study are unlikely to reveal the presence of a 10 c\ms\ signal due to a true Earth analog, there is still room for significant improvements to the stars' RV completeness using these current generation instruments. Future surveys prioritizing those stars for which we are already sensitive to super-Earth/sub-Neptune type planets (M$_{\rm{p}}$ $\sim$ 10-20\Mearth) at 1 AU could increase our sensitivity closer to the 1 \Mearth\, limit. Alternatively, focusing on the stars for which we have the least RV data -- those where giant planets at 1 AU could remain hidden -- could identify currently unknown planetary companions that would preclude the existence of a temperate, terrestrial planet.

In the course of preparing each star's RV time series for the injection/recovery analyses we also performed {an} uninformed search of the RV data to identify and remove any significant signals. In doing so, we recovered 28 previously published planets. The orbital parameters of many of these planets have not been revisited since their original publication date which is often 5-10 years ago. Our updated analysis, which generally includes both additional data from different instruments and a longer observing baseline than previous fits, is able to increase the precision on the planets' {periods, eccentricities and RV semi-amplitudes}. Looking at the ratio of the previously published uncertainties to our updated orbital parameter uncertainties, we find mean uncertainty improvements of {2.7$\times$ in period, 1.3$\times$ in RV semi-amplitude, and 1.4$\times$ in eccentricity}.

The third key component to this work is the identification and characterization of many stars' variability timescales and amplitudes using the same uninformed search methodology applied to each stars' S-index {and, for targets observed by the \UCLES\, instrument, $EW_{H\alpha}$} time series. Understanding a star's rotationally modulated activity signals along with its long term magnetic activity cycles, both of which can mask the presence of low amplitude Keplerian signals, will inform the sampling baseline and cadence necessary in future EPRV surveys to model and mitigate these star-based signals. Our work is not an exhaustive analysis of the stars' activity, but in many cases it does provide an initial or refined characterization of the stars' rotation and magnetic cycles. Future work to better quantify these signals and their development over time is encouraged.

If we will someday require extreme precision RV {follow-up} of planets detected by DI missions around these stars, then it {would well serve} the exoplanet community to begin new observing campaigns of these targets in the near term. Dedicated, high cadence, high precision ($\sigma_{\rm{RV}} \leq 1$\,\ms) RV monitoring will enable the characterization and potentially mitigation of stellar variability signals on time scales of hours to years alongside the the detection of additional, currently unknown planetary companions. Knowledge of how to correctly model and remove signals of both natures will be crucial for any future efforts to measure precise masses for Earth-analog planets.\\

\acknowledgments

KL, JB, and EM were supported by the NASA Exoplanet Exploration Program (ExEP). KL acknowledges support from the Whitman College Independent Study and Senior Thesis programs. The research was carried out in part at the Jet Propulsion Laboratory, California Institute of Technology, under a contract with the National Aeronautics and Space Administration (80NM0018D0004).

The work herein is based on observations obtained at the W. M. Keck Observatory, which is operated jointly by the University of California and the California Institute of Technology, and we thank the UC-Keck and NASA-Keck Time Assignment Committees for their support. We also wish to extend our special thanks to those of Hawaiian ancestry on whose sacred mountain of Mauna Kea we are privileged to be guests. Without their generous hospitality, the Keck observations presented herein would not have been possible. The work herein is also based on observations obtained with the Automated Planet Finder (APF) telescope and its Levy Spectrometer at Lick Observatory, along with data gathered with the 6.5 meter Magellan Telescopes located at Las Campanas Observatory, Chile. We acknowledge the traditional owners of the land on which the Anglo-Australian Telescope (AAT) stands, the Gamilaraay people, and pay our respects to elders past and present. 

Some observations in the paper made use of the High-Resolution Imaging instrument(s) `Alopeke (and/or Zorro). `Alopeke (and/or Zorro) was funded by the NASA Exoplanet Exploration Program and built at the NASA Ames Research Center by Steve B. Howell, Nic Scott, Elliott P. Horch, and Emmett Quigley. `Alopeke (and/or Zorro) was mounted on the Gemini North (and/or South) telescope of the international Gemini Observatory, a program of NSF’s NOIRLab, which is managed by the Association of Universities for Research in Astronomy (AURA) under a cooperative agreement with the National Science Foundation. On behalf of the Gemini partnership: the National Science Foundation (United States), National Research Council (Canada), Agencia Nacional de Investigación y Desarrollo (Chile), Ministerio de Ciencia, Tecnología e Innovación (Argentina), Ministério da Ciência, Tecnologia, Inovações e Comunicações (Brazil), and Korea Astronomy and Space Science Institute (Republic of Korea).

This research has made use of the SIMBAD database, operated at CDS, Strasbourg, France.

\vspace{5mm}
\facilities{UCO/Lick: The APF (Levy spectrograph), 
Magellan: Clay (Planet Finder Spectrograph), 
Keck:I (HIRES),
Gemini North: `Alopeke,
Gemini South: Zorro}


\begin{thebibliography}{}
\expandafter\ifx\csname natexlab\endcsname\relax\def\natexlab#1{#1}\fi
\providecommand{\url}[1]{\href{#1}{#1}}
\providecommand{\dodoi}[1]{doi:~\href{http://doi.org/#1}{\nolinkurl{#1}}}
\providecommand{\doeprint}[1]{\href{http://ascl.net/#1}{\nolinkurl{http://ascl.net/#1}}}
\providecommand{\doarXiv}[1]{\href{https://arxiv.org/abs/#1}{\nolinkurl{https://arxiv.org/abs/#1}}}

\bibitem[{{Adibekyan} {et~al.}(2016){Adibekyan}, {Delgado-Mena}, {Figueira},
  {Sousa}, {Santos}, {Faria}, {Gonz{\'a}lez Hern{\'a}ndez}, {Israelian},
  {Harutyunyan}, {Su{\'a}rez-Andr{\'e}s}, \& {Hakobyan}}]{Adibekyan2016}
{Adibekyan}, V., {Delgado-Mena}, E., {Figueira}, P., {et~al.} 2016, \aap, 591,
  A34, \dodoi{10.1051/0004-6361/201628453}

\bibitem[{{Adibekyan} {et~al.}(2012){Adibekyan}, {Sousa}, {Santos}, {Delgado
  Mena}, {Gonz{\'a}lez Hern{\'a}ndez}, {Israelian}, {Mayor}, \&
  {Khachatryan}}]{Adibekyan2012}
{Adibekyan}, V.~Z., {Sousa}, S.~G., {Santos}, N.~C., {et~al.} 2012, \aap, 545,
  A32, \dodoi{10.1051/0004-6361/201219401}

\bibitem[{{Aguilera-G{\'o}mez} {et~al.}(2018){Aguilera-G{\'o}mez},
  {Ram{\'\i}rez}, \& {Chanam{\'e}}}]{AguileraGomez2018}
{Aguilera-G{\'o}mez}, C., {Ram{\'\i}rez}, I., \& {Chanam{\'e}}, J. 2018, \aap,
  614, A55, \dodoi{10.1051/0004-6361/201732209}

\bibitem[{{Ammler-von Eiff} \& {Reiners}(2012)}]{AmmlervonEiff2012}
{Ammler-von Eiff}, M., \& {Reiners}, A. 2012, \aap, 542, A116,
  \dodoi{10.1051/0004-6361/201118724}

\bibitem[{{Anders} {et~al.}(2019){Anders}, {Khalatyan}, {Chiappini}, {Queiroz},
  {Santiago}, {Jordi}, {Girardi}, {Brown}, {Matijevi{\v{c}}}, {Monari},
  {Cantat-Gaudin}, {Weiler}, {Khan}, {Miglio}, {Carrillo}, {Romero-G{\'o}mez},
  {Minchev}, {de Jong}, {Antoja}, {Ramos}, {Steinmetz}, \& {Enke}}]{Anders2019}
{Anders}, F., {Khalatyan}, A., {Chiappini}, C., {et~al.} 2019, \aap, 628, A94,
  \dodoi{10.1051/0004-6361/201935765}

\bibitem[{{Baliunas} {et~al.}(1996){Baliunas}, {Sokoloff}, \&
  {Soon}}]{Baliunas1996}
{Baliunas}, S., {Sokoloff}, D., \& {Soon}, W. 1996, \apjl, 457, L99,
  \dodoi{10.1086/309891}

\bibitem[{{Baliunas} {et~al.}(1995){Baliunas}, {Donahue}, {Soon}, {Horne},
  {Frazer}, {Woodard-Eklund}, {Bradford}, {Rao}, {Wilson}, {Zhang}, {Bennett},
  {Briggs}, {Carroll}, {Duncan}, {Figueroa}, {Lanning}, {Misch}, {Mueller},
  {Noyes}, {Poppe}, {Porter}, {Robinson}, {Russell}, {Shelton}, {Soyumer},
  {Vaughan}, \& {Whitney}}]{Baliunas1995}
{Baliunas}, S.~L., {Donahue}, R.~A., {Soon}, W.~H., {et~al.} 1995, \apj, 438,
  269, \dodoi{10.1086/175072}

\bibitem[{{Baluev}(2015)}]{Baluev2015}
{Baluev}, R.~V. 2015, \mnras, 446, 1493, \dodoi{10.1093/mnras/stu2150}

\bibitem[{{Batalha} {et~al.}(2019){Batalha}, {Lewis}, {Fortney}, {Batalha},
  {Kempton}, {Lewis}, \& {Line}}]{Batalha2019}
{Batalha}, N.~E., {Lewis}, T., {Fortney}, J.~J., {et~al.} 2019, \apjl, 885,
  L25, \dodoi{10.3847/2041-8213/ab4909}

\bibitem[{{Bayo} {et~al.}(2008){Bayo}, {Rodrigo}, {Barrado Y Navascu{\'e}s},
  {Solano}, {Guti{\'e}rrez}, {Morales-Calder{\'o}n}, \& {Allard}}]{Bayo2008}
{Bayo}, A., {Rodrigo}, C., {Barrado Y Navascu{\'e}s}, D., {et~al.} 2008, \aap,
  492, 277, \dodoi{10.1051/0004-6361:200810395}

\bibitem[{{Beichman} {et~al.}(2005){Beichman}, {Bryden}, {Gautier},
  {Stapelfeldt}, {Werner}, {Misselt}, {Rieke}, {Stansberry}, \&
  {Trilling}}]{Beichman2005}
{Beichman}, C.~A., {Bryden}, G., {Gautier}, T.~N., {et~al.} 2005, \apj, 626,
  1061, \dodoi{10.1086/430059}

\bibitem[{{Benedict} {et~al.}(2022){Benedict}, {McArthur}, {Nelan},
  {Wittenmyer}, {Barnes}, {Smotherman}, \& {Horner}}]{Benedict2022}
{Benedict}, G.~F., {McArthur}, B.~E., {Nelan}, E.~P., {et~al.} 2022, \aj, 163,
  295, \dodoi{10.3847/1538-3881/ac6ac8}

\bibitem[{{Boro Saikia} {et~al.}(2018){Boro Saikia}, {Marvin}, {Jeffers},
  {Reiners}, {Cameron}, {Marsden}, {Petit}, {Warnecke}, \&
  {Yadav}}]{BoroSaikia2018}
{Boro Saikia}, S., {Marvin}, C.~J., {Jeffers}, S.~V., {et~al.} 2018, \aap, 616,
  A108, \dodoi{10.1051/0004-6361/201629518}

\bibitem[{{Bouchy} \& {Carrier}(2001)}]{BouchyCarrier2001}
{Bouchy}, F., \& {Carrier}, F. 2001, \aap, 374, L5,
  \dodoi{10.1051/0004-6361:20010792}

\bibitem[{{Bourrier} {et~al.}(2018){Bourrier}, {Dumusque}, {Dorn}, {Henry},
  {Astudillo-Defru}, {Rey}, {Benneke}, {H{\'e}brard}, {Lovis}, {Demory},
  {Moutou}, \& {Ehrenreich}}]{Bourrier2018}
{Bourrier}, V., {Dumusque}, X., {Dorn}, C., {et~al.} 2018, \aap, 619, A1,
  \dodoi{10.1051/0004-6361/201833154}

\bibitem[{{Brand{\~a}o} {et~al.}(2011){Brand{\~a}o}, {Do{\u{g}}an},
  {Christensen-Dalsgaard}, {Cunha}, {Bedding}, {Metcalfe}, {Kjeldsen},
  {Bruntt}, \& {Arentoft}}]{Brandao2011}
{Brand{\~a}o}, I.~M., {Do{\u{g}}an}, G., {Christensen-Dalsgaard}, J., {et~al.}
  2011, \aap, 527, A37, \dodoi{10.1051/0004-6361/201015370}

\bibitem[{{Brewer} {et~al.}(2016){Brewer}, {Fischer}, {Valenti}, \&
  {Piskunov}}]{Brewer2016}
{Brewer}, J.~M., {Fischer}, D.~A., {Valenti}, J.~A., \& {Piskunov}, N. 2016,
  \apjs, 225, 32, \dodoi{10.3847/0067-0049/225/2/32}

\bibitem[{{Brewer} {et~al.}(2020){Brewer}, {Fischer}, {Blackman}, {Cabot},
  {Davis}, {Laughlin}, {Leet}, {Ong}, {Petersburg}, {Szymkowiak}, {Zhao},
  {Henry}, \& {Llama}}]{Brewer2020}
{Brewer}, J.~M., {Fischer}, D.~A., {Blackman}, R.~T., {et~al.} 2020, \aj, 160,
  67, \dodoi{10.3847/1538-3881/ab99c9}

\bibitem[{{Burgasser} {et~al.}(2000){Burgasser}, {Kirkpatrick}, {Cutri},
  {McCallon}, {Kopan}, {Gizis}, {Liebert}, {Reid}, {Brown}, {Monet}, {Dahn},
  {Beichman}, \& {Skrutskie}}]{Burgasser2000}
{Burgasser}, A.~J., {Kirkpatrick}, J.~D., {Cutri}, R.~M., {et~al.} 2000, \apjl,
  531, L57, \dodoi{10.1086/312522}

\bibitem[{{Burt} {et~al.}(2021){Burt}, {Feng}, {Holden}, {Mamajek}, {Huang},
  {Rosenthal}, {Wang}, {Butler}, {Vogt}, {Laughlin}, {Henry}, {Teske}, {Wang},
  {Crane}, \& {Shectman}}]{Burt2021}
{Burt}, J., {Feng}, F., {Holden}, B., {et~al.} 2021, \aj, 161, 10,
  \dodoi{10.3847/1538-3881/abc2d0}

\bibitem[{{Busko} \& {Torres}(1978)}]{Busko1978}
{Busko}, I.~C., \& {Torres}, C.~A.~O. 1978, \aap, 64, 153

\bibitem[{{Butkovskaya} {et~al.}(2017){Butkovskaya}, {Plachinda}, {Bondar'}, \&
  {Baklanova}}]{Butkovskaya2017}
{Butkovskaya}, V.~V., {Plachinda}, S.~I., {Bondar'}, N.~I., \& {Baklanova},
  D.~N. 2017, Astronomische Nachrichten, 338, 896,
  \dodoi{10.1002/asna.201713396}

\bibitem[{{Butler} {et~al.}(1996){Butler}, {Marcy}, {Williams}, {McCarthy},
  {Dosanjh}, \& {Vogt}}]{Butler1996}
{Butler}, R.~P., {Marcy}, G.~W., {Williams}, E., {et~al.} 1996, \pasp, 108,
  500, \dodoi{10.1086/133755}

\bibitem[{{Butler} {et~al.}(2001){Butler}, {Tinney}, {Marcy}, {Jones}, {Penny},
  \& {Apps}}]{Butler2001}
{Butler}, R.~P., {Tinney}, C.~G., {Marcy}, G.~W., {et~al.} 2001, \apj, 555,
  410, \dodoi{10.1086/321467}

\bibitem[{{Butler} {et~al.}(2006){Butler}, {Wright}, {Marcy}, {Fischer},
  {Vogt}, {Tinney}, {Jones}, {Carter}, {Johnson}, {McCarthy}, \&
  {Penny}}]{Butler2006}
{Butler}, R.~P., {Wright}, J.~T., {Marcy}, G.~W., {et~al.} 2006, \apj, 646,
  505, \dodoi{10.1086/504701}

\bibitem[{{Butler} {et~al.}(2017){Butler}, {Vogt}, {Laughlin}, {Burt},
  {Rivera}, {Tuomi}, {Teske}, {Arriagada}, {Diaz}, {Holden}, \&
  {Keiser}}]{Butler2017}
{Butler}, R.~P., {Vogt}, S.~S., {Laughlin}, G., {et~al.} 2017, \aj, 153, 208,
  \dodoi{10.3847/1538-3881/aa66ca}

\bibitem[{{Casagrande} {et~al.}(2011){Casagrande}, {Sch{\"o}nrich}, {Asplund},
  {Cassisi}, {Ram{\'\i}rez}, {Mel{\'e}ndez}, {Bensby}, \&
  {Feltzing}}]{Casagrande2011}
{Casagrande}, L., {Sch{\"o}nrich}, R., {Asplund}, M., {et~al.} 2011, \aap, 530,
  A138, \dodoi{10.1051/0004-6361/201016276}

\bibitem[{{Casali} {et~al.}(2020){Casali}, {Spina}, {Magrini}, {Karakas},
  {Kobayashi}, {Casey}, {Feltzing}, {Van der Swaelmen}, {Tsantaki},
  {Jofr{\'e}}, {Bragaglia}, {Feuillet}, {Bensby}, {Biazzo}, {Gonneau},
  {Tautvai{\v{s}}ien{\.{e}}}, {Baratella}, {Roccatagliata}, {Pancino}, {Sousa},
  {Adibekyan}, {Martell}, {Bayo}, {Jackson}, {Jeffries}, {Gilmore}, {Randich},
  {Alfaro}, {Koposov}, {Korn}, {Recio-Blanco}, {Smiljanic}, {Franciosini},
  {Hourihane}, {Monaco}, {Morbidelli}, {Sacco}, {Worley}, \&
  {Zaggia}}]{Casali2020}
{Casali}, G., {Spina}, L., {Magrini}, L., {et~al.} 2020, \aap, 639, A127,
  \dodoi{10.1051/0004-6361/202038055}

\bibitem[{{Cayrel de Strobel} {et~al.}(1989){Cayrel de Strobel}, {Perrin},
  {Cayrel}, \& {Lebreton}}]{CayreldeStrobel1989}
{Cayrel de Strobel}, G., {Perrin}, M.~N., {Cayrel}, R., \& {Lebreton}, Y. 1989,
  \aap, 225, 369

\bibitem[{{Cegla} {et~al.}(2019){Cegla}, {Watson}, {Shelyag}, {Mathioudakis},
  \& {Moutari}}]{Cegla2019}
{Cegla}, H.~M., {Watson}, C.~A., {Shelyag}, S., {Mathioudakis}, M., \&
  {Moutari}, S. 2019, \apj, 879, 55, \dodoi{10.3847/1538-4357/ab16d3}

\bibitem[{{Choi} {et~al.}(2016){Choi}, {Dotter}, {Conroy}, {Cantiello},
  {Paxton}, \& {Johnson}}]{Choi2016}
{Choi}, J., {Dotter}, A., {Conroy}, C., {et~al.} 2016, \apj, 823, 102,
  \dodoi{10.3847/0004-637X/823/2/102}

\bibitem[{{Cincunegui} {et~al.}(2007){Cincunegui}, {D{\'\i}az}, \&
  {Mauas}}]{Cincunegui2007}
{Cincunegui}, C., {D{\'\i}az}, R.~F., \& {Mauas}, P.~J.~D. 2007, \aap, 469,
  309, \dodoi{10.1051/0004-6361:20066503}

\bibitem[{{Corsaro} {et~al.}(2012){Corsaro}, {Grundahl}, {Leccia}, {Bonanno},
  {Kjeldsen}, \& {Patern{\`o}}}]{Corsaro2012}
{Corsaro}, E., {Grundahl}, F., {Leccia}, S., {et~al.} 2012, \aap, 537, A9,
  \dodoi{10.1051/0004-6361/201117158}

\bibitem[{{Cosentino} {et~al.}(2012){Cosentino}, {Lovis}, {Pepe}, {Collier
  Cameron}, {Latham}, {Molinari}, {Udry}, {Bezawada}, {Black}, {Born},
  {Buchschacher}, {Charbonneau}, {Figueira}, {Fleury}, {Galli}, {Gallie},
  {Gao}, {Ghedina}, {Gonzalez}, {Gonzalez}, {Guerra}, {Henry}, {Horne},
  {Hughes}, {Kelly}, {Lodi}, {Lunney}, {Maire}, {Mayor}, {Micela}, {Ordway},
  {Peacock}, {Phillips}, {Piotto}, {Pollacco}, {Queloz}, {Rice}, {Riverol},
  {Riverol}, {San Juan}, {Sasselov}, {Segransan}, {Sozzetti}, {Sosnowska},
  {Stobie}, {Szentgyorgyi}, {Vick}, \& {Weber}}]{Cosentino2012}
{Cosentino}, R., {Lovis}, C., {Pepe}, F., {et~al.} 2012, in Society of
  Photo-Optical Instrumentation Engineers (SPIE) Conference Series, Vol. 8446,
  Ground-based and Airborne Instrumentation for Astronomy IV, ed. I.~S.
  {McLean}, S.~K. {Ramsay}, \& H.~{Takami}, 84461V

\bibitem[{{Costes} {et~al.}(2021){Costes}, {Watson}, {de Mooij}, {Saar},
  {Dumusque}, {Cameron}, {Phillips}, {G{\"u}nther}, {Jenkins}, {Mortier}, \&
  {Thompson}}]{Costes2021}
{Costes}, J.~C., {Watson}, C.~A., {de Mooij}, E., {et~al.} 2021, \mnras, 505,
  830, \dodoi{10.1093/mnras/stab1183}

\bibitem[{{Crane} {et~al.}(2006){Crane}, {Shectman}, \& {Butler}}]{Crane2006}
{Crane}, J.~D., {Shectman}, S.~A., \& {Butler}, R.~P. 2006, in Society of
  Photo-Optical Instrumentation Engineers (SPIE) Conference Series, Vol. 6269,
  Society of Photo-Optical Instrumentation Engineers (SPIE) Conference Series,
  ed. I.~S. {McLean} \& M.~{Iye}, 626931

\bibitem[{{Crane} {et~al.}(2010){Crane}, {Shectman}, {Butler}, {Thompson},
  {Birk}, {Jones}, \& {Burley}}]{Crane2010}
{Crane}, J.~D., {Shectman}, S.~A., {Butler}, R.~P., {et~al.} 2010, in Society
  of Photo-Optical Instrumentation Engineers (SPIE) Conference Series, Vol.
  7735, Ground-based and Airborne Instrumentation for Astronomy III, ed. I.~S.
  {McLean}, S.~K. {Ramsay}, \& H.~{Takami}, 773553

\bibitem[{{Crane} {et~al.}(2008){Crane}, {Shectman}, {Butler}, {Thompson}, \&
  {Burley}}]{Crane2008}
{Crane}, J.~D., {Shectman}, S.~A., {Butler}, R.~P., {Thompson}, I.~B., \&
  {Burley}, G.~S. 2008, in Society of Photo-Optical Instrumentation Engineers
  (SPIE) Conference Series, Vol. 7014, Ground-based and Airborne
  Instrumentation for Astronomy II, ed. I.~S. {McLean} \& M.~M. {Casali},
  701479

\bibitem[{{Crass} {et~al.}(2021){Crass}, {Gaudi}, {Leifer}, {Beichman},
  {Bender}, {Blackwood}, {Burt}, {Callas}, {Cegla}, {Diddams}, {Dumusque},
  {Eastman}, {Ford}, {Fulton}, {Gibson}, {Halverson}, {Haywood}, {Hearty},
  {Howard}, {Latham}, {L{\"o}hner-B{\"o}ttcher}, {Mamajek}, {Mortier},
  {Newman}, {Plavchan}, {Quirrenbach}, {Reiners}, {Robertson}, {Roy}, {Schwab},
  {Seifahrt}, {Szentgyorgyi}, {Terrien}, {Teske}, {Thompson}, \&
  {Vasisht}}]{Crass2021}
{Crass}, J., {Gaudi}, B.~S., {Leifer}, S., {et~al.} 2021, arXiv e-prints,
  arXiv:2107.14291.
\newblock \doarXiv{2107.14291}

\bibitem[{{Croll} {et~al.}(2006){Croll}, {Walker}, {Kuschnig}, {Matthews},
  {Rowe}, {Walker}, {Rucinski}, {Hatzes}, {Cochran}, {Robb}, {Guenther},
  {Moffat}, {Sasselov}, \& {Weiss}}]{Croll2006}
{Croll}, B., {Walker}, G. A.~H., {Kuschnig}, R., {et~al.} 2006, \apj, 648, 607,
  \dodoi{10.1086/505792}

\bibitem[{{Cumming} {et~al.}(1999){Cumming}, {Marcy}, \&
  {Butler}}]{Cumming1999}
{Cumming}, A., {Marcy}, G.~W., \& {Butler}, R.~P. 1999, \apj, 526, 890,
  \dodoi{10.1086/308020}

\bibitem[{{da Silva} {et~al.}(2015){da Silva}, {Milone}, \&
  {Rocha-Pinto}}]{daSilva2015}
{da Silva}, R., {Milone}, A. d.~C., \& {Rocha-Pinto}, H.~J. 2015, \aap, 580,
  A24, \dodoi{10.1051/0004-6361/201525770}

\bibitem[{{Dannert} {et~al.}(2022){Dannert}, {Ottiger}, {Quanz}, {Laugier},
  {Fontanet}, {Gheorghe}, {Absil}, {Dandumont}, {Defr{\`e}re}, {Gasc{\'o}n},
  {Glauser}, {Kammerer}, {Lichtenberg}, {Linz}, {Loicq}, \& {LIFE
  Collaboration}}]{Dannert2022}
{Dannert}, F.~A., {Ottiger}, M., {Quanz}, S.~P., {et~al.} 2022, \aap, 664, A22,
  \dodoi{10.1051/0004-6361/202141958}

\bibitem[{{Dawson} \& {Fabrycky}(2010)}]{DawsonFabrycky2010}
{Dawson}, R.~I., \& {Fabrycky}, D.~C. 2010, \apj, 722, 937,
  \dodoi{10.1088/0004-637X/722/1/937}

\bibitem[{{De Rosa} {et~al.}(2019){De Rosa}, {Esposito}, {Hirsch}, {Nielsen},
  {Marley}, {Kalas}, {Wang}, \& {Macintosh}}]{DeRosa2019}
{De Rosa}, R.~J., {Esposito}, T.~M., {Hirsch}, L.~A., {et~al.} 2019, \aj, 158,
  225, \dodoi{10.3847/1538-3881/ab4c9b}

\bibitem[{{Dekker} {et~al.}(2000){Dekker}, {D'Odorico}, {Kaufer}, {Delabre}, \&
  {Kotzlowski}}]{Dekker2000}
{Dekker}, H., {D'Odorico}, S., {Kaufer}, A., {Delabre}, B., \& {Kotzlowski}, H.
  2000, in Society of Photo-Optical Instrumentation Engineers (SPIE) Conference
  Series, Vol. 4008, Optical and IR Telescope Instrumentation and Detectors,
  ed. M.~{Iye} \& A.~F. {Moorwood}, 534--545

\bibitem[{{Delgado Mena} {et~al.}(2019){Delgado Mena}, {Moya}, {Adibekyan},
  {Tsantaki}, {Gonz{\'a}lez Hern{\'a}ndez}, {Israelian}, {Davies}, {Chaplin},
  {Sousa}, {Ferreira}, \& {Santos}}]{DelgadoMena2019}
{Delgado Mena}, E., {Moya}, A., {Adibekyan}, V., {et~al.} 2019, \aap, 624, A78,
  \dodoi{10.1051/0004-6361/201834783}

\bibitem[{{Delrez} {et~al.}(2021){Delrez}, {Ehrenreich}, {Alibert}, {Bonfanti},
  {Borsato}, {Fossati}, {Hooton}, {Hoyer}, {Pozuelos}, {Salmon}, {Sulis},
  {Wilson}, {Adibekyan}, {Bourrier}, {Brandeker}, {Charnoz}, {Deline},
  {Guterman}, {Haldemann}, {Hara}, {Oshagh}, {Sousa}, {Van Grootel}, {Alonso},
  {Anglada-Escud{\'e}}, {B{\'a}rczy}, {Barrado}, {Barros}, {Baumjohann},
  {Beck}, {Bekkelien}, {Benz}, {Billot}, {Bonfils}, {Broeg}, {Cabrera},
  {Collier Cameron}, {Davies}, {Deleuil}, {Delisle}, {Demangeon}, {Demory},
  {Erikson}, {Fortier}, {Fridlund}, {Futyan}, {Gandolfi}, {Garcia Mu{\~n}oz},
  {Gillon}, {Guedel}, {Heng}, {Kiss}, {Laskar}, {Lecavelier des Etangs},
  {Lendl}, {Lovis}, {Maxted}, {Nascimbeni}, {Olofsson}, {Osborn}, {Pagano},
  {Pall{\'e}}, {Piotto}, {Pollacco}, {Queloz}, {Rauer}, {Ragazzoni}, {Ribas},
  {Santos}, {Scandariato}, {S{\'e}gransan}, {Simon}, {Smith}, {Steller},
  {Szab{\'o}}, {Thomas}, {Udry}, \& {Walton}}]{Delrez2021}
{Delrez}, L., {Ehrenreich}, D., {Alibert}, Y., {et~al.} 2021, Nature Astronomy,
  5, 775, \dodoi{10.1038/s41550-021-01381-5}

\bibitem[{{D{\'\i}az} {et~al.}(2018){D{\'\i}az}, {Jenkins}, {Tuomi}, {Butler},
  {Soto}, {Teske}, {Feng}, {Shectman}, {Arriagada}, {Crane}, {Thompson}, \&
  {Vogt}}]{Diaz2018}
{D{\'\i}az}, M.~R., {Jenkins}, J.~S., {Tuomi}, M., {et~al.} 2018, \aj, 155,
  126, \dodoi{10.3847/1538-3881/aaa896}

\bibitem[{{Diego} {et~al.}(1990){Diego}, {Charalambous}, {Fish}, \&
  {Walker}}]{Diego1990}
{Diego}, F., {Charalambous}, A., {Fish}, A.~C., \& {Walker}, D.~D. 1990, in
  Society of Photo-Optical Instrumentation Engineers (SPIE) Conference Series,
  Vol. 1235, Instrumentation in Astronomy VII, ed. D.~L. {Crawford}, 562--576

\bibitem[{{Donahue}(1996)}]{Donahue1996IAU176}
{Donahue}, R.~A. 1996, in Stellar Surface Structure, ed. K.~G. {Strassmeier} \&
  J.~L. {Linsky}, Vol. 176, 261

\bibitem[{{Donahue} {et~al.}(1996){Donahue}, {Saar}, \&
  {Baliunas}}]{Donahue1996}
{Donahue}, R.~A., {Saar}, S.~H., \& {Baliunas}, S.~L. 1996, \apj, 466, 384,
  \dodoi{10.1086/177517}

\bibitem[{{Dumusque} {et~al.}(2011){Dumusque}, {Lovis}, {S{\'e}gransan},
  {Mayor}, {Udry}, {Benz}, {Bouchy}, {Lo Curto}, {Mordasini}, {Pepe}, {Queloz},
  {Santos}, \& {Naef}}]{Dumusque2011}
{Dumusque}, X., {Lovis}, C., {S{\'e}gransan}, D., {et~al.} 2011, \aap, 535,
  A55, \dodoi{10.1051/0004-6361/201117148}

\bibitem[{{Duncan} {et~al.}(1991){Duncan}, {Vaughan}, {Wilson}, {Preston},
  {Frazer}, {Lanning}, {Misch}, {Mueller}, {Soyumer}, {Woodard}, {Baliunas},
  {Noyes}, {Hartmann}, {Porter}, {Zwaan}, {Middelkoop}, {Rutten}, \&
  {Mihalas}}]{Duncan1991}
{Duncan}, D.~K., {Vaughan}, A.~H., {Wilson}, O.~C., {et~al.} 1991, \apjs, 76,
  383, \dodoi{10.1086/191572}

\bibitem[{{Eaton} \& {Poe}(1985)}]{Eaton1985}
{Eaton}, J.~A., \& {Poe}, C.~H. 1985, Information Bulletin on Variable Stars,
  2712, 1

\bibitem[{{Egeland} {et~al.}(2015){Egeland}, {Metcalfe}, {Hall}, \&
  {Henry}}]{Egeland2015}
{Egeland}, R., {Metcalfe}, T.~S., {Hall}, J.~C., \& {Henry}, G.~W. 2015, \apj,
  812, 12, \dodoi{10.1088/0004-637X/812/1/12}

\bibitem[{{Eggenberger} {et~al.}(2007){Eggenberger}, {Udry}, {Chauvin},
  {Beuzit}, {Lagrange}, {S{\'e}gransan}, \& {Mayor}}]{Eggenberger2007}
{Eggenberger}, A., {Udry}, S., {Chauvin}, G., {et~al.} 2007, \aap, 474, 273,
  \dodoi{10.1051/0004-6361:20077447}

\bibitem[{{Eiroa} {et~al.}(2013){Eiroa}, {Marshall}, {Mora}, {Montesinos},
  {Absil}, {Augereau}, {Bayo}, {Bryden}, {Danchi}, {del Burgo}, {Ertel},
  {Fridlund}, {Heras}, {Krivov}, {Launhardt}, {Liseau}, {L{\"o}hne},
  {Maldonado}, {Pilbratt}, {Roberge}, {Rodmann}, {Sanz-Forcada}, {Solano},
  {Stapelfeldt}, {Th{\'e}bault}, {Wolf}, {Ardila}, {Ar{\'e}valo}, {Beichmann},
  {Faramaz}, {Gonz{\'a}lez-Garc{\'\i}a}, {Guti{\'e}rrez}, {Lebreton},
  {Mart{\'\i}nez-Arn{\'a}iz}, {Meeus}, {Montes}, {Olofsson}, {Su}, {White},
  {Barrado}, {Fukagawa}, {Gr{\"u}n}, {Kamp}, {Lorente}, {Morbidelli},
  {M{\"u}ller}, {Mutschke}, {Nakagawa}, {Ribas}, \& {Walker}}]{Eiroa2013}
{Eiroa}, C., {Marshall}, J.~P., {Mora}, A., {et~al.} 2013, \aap, 555, A11,
  \dodoi{10.1051/0004-6361/201321050}

\bibitem[{{Endl} {et~al.}(2012){Endl}, {Robertson}, {Cochran}, {MacQueen},
  {Brugamyer}, {Caldwell}, {Wittenmyer}, {Barnes}, \& {Gullikson}}]{Endl2012}
{Endl}, M., {Robertson}, P., {Cochran}, W.~D., {et~al.} 2012, \apj, 759, 19,
  \dodoi{10.1088/0004-637X/759/1/19}

\bibitem[{ESA(1997)}]{ESA1997}
ESA, ed. 1997, ESA Special Publication, Vol. 1200, {The HIPPARCOS and TYCHO
  catalogues. Astrometric and photometric star catalogues derived from the ESA
  HIPPARCOS Space Astrometry Mission}

\bibitem[{{Faramaz} {et~al.}(2018){Faramaz}, {Bryden}, {Stapelfeldt}, {Booth},
  {Bayo}, {Beust}, {Casassus}, {Cuadra}, {Hales}, {Hughes}, {Olofsson}, {Su},
  \& {Wilner}}]{Faramaz2018}
{Faramaz}, V., {Bryden}, G., {Stapelfeldt}, K.~R., {et~al.} 2018, \mnras, 481,
  44, \dodoi{10.1093/mnras/sty2304}

\bibitem[{{Farihi} {et~al.}(2013){Farihi}, {Bond}, {Dufour}, {Haghighipour},
  {Schaefer}, {Holberg}, {Barstow}, \& {Burleigh}}]{Farihi2013}
{Farihi}, J., {Bond}, H.~E., {Dufour}, P., {et~al.} 2013, \mnras, 430, 652,
  \dodoi{10.1093/mnras/sts677}

\bibitem[{{Fekel} \& {Beavers}(1983)}]{Fekel1983}
{Fekel}, F.~C., J., \& {Beavers}, W.~I. 1983, \apj, 267, 682,
  \dodoi{10.1086/160905}

\bibitem[{{Feng} {et~al.}(2019{\natexlab{a}}){Feng}, {Anglada-Escud{\'e}},
  {Tuomi}, {Jones}, {Chanam{\'e}}, {Butler}, \& {Janson}}]{Feng2019b}
{Feng}, F., {Anglada-Escud{\'e}}, G., {Tuomi}, M., {et~al.} 2019{\natexlab{a}},
  \mnras, 490, 5002, \dodoi{10.1093/mnras/stz2912}

\bibitem[{{Feng} {et~al.}(2019{\natexlab{b}}){Feng}, {Lisogorskyi}, {Jones},
  {Kopeikin}, {Butler}, {Anglada-Escud{\'e}}, \& {Boss}}]{Feng2019a}
{Feng}, F., {Lisogorskyi}, M., {Jones}, H. R.~A., {et~al.} 2019{\natexlab{b}},
  \apjs, 244, 39, \dodoi{10.3847/1538-4365/ab40b6}

\bibitem[{{Feng} {et~al.}(2017{\natexlab{a}}){Feng}, {Tuomi}, \&
  {Jones}}]{Feng2017b}
{Feng}, F., {Tuomi}, M., \& {Jones}, H.~R.~A. 2017{\natexlab{a}}, \mnras, 470,
  4794, \dodoi{10.1093/mnras/stx1126}

\bibitem[{{Feng} {et~al.}(2017{\natexlab{b}}){Feng}, {Tuomi}, {Jones},
  {Barnes}, {Anglada-Escud{\'e}}, {Vogt}, \& {Butler}}]{Feng2017}
{Feng}, F., {Tuomi}, M., {Jones}, H.~R.~A., {et~al.} 2017{\natexlab{b}}, \aj,
  154, 135, \dodoi{10.3847/1538-3881/aa83b4}

\bibitem[{{Fischer}(2018)}]{Fischer2018}
{Fischer}, D.~A. 2018, arXiv e-prints, arXiv:1807.11925.
\newblock \doarXiv{1807.11925}

\bibitem[{{Fischer} {et~al.}(2008){Fischer}, {Marcy}, {Butler}, {Vogt},
  {Laughlin}, {Henry}, {Abouav}, {Peek}, {Wright}, {Johnson}, {McCarthy}, \&
  {Isaacson}}]{Fischer2008}
{Fischer}, D.~A., {Marcy}, G.~W., {Butler}, R.~P., {et~al.} 2008, \apj, 675,
  790, \dodoi{10.1086/525512}

\bibitem[{{Fisher} {et~al.}(1983){Fisher}, {Hall}, {Henry}, {Landis}, {Renner},
  \& {Shore}}]{Fisher1983}
{Fisher}, G.~F., {Hall}, D.~S., {Henry}, G.~W., {et~al.} 1983, Information
  Bulletin on Variable Stars, 2259, 1

\bibitem[{{Flores} {et~al.}(2021){Flores}, {Jaque Arancibia}, {Iba{\~n}ez
  Bustos}, {Buccino}, {Yana Galarza}, {Nu{\~n}ez}, {Miquelarena}, {Alacoria},
  {Saffe}, \& {Mauas}}]{Flores2021}
{Flores}, M., {Jaque Arancibia}, M., {Iba{\~n}ez Bustos}, R.~V., {et~al.} 2021,
  \aap, 645, L6, \dodoi{10.1051/0004-6361/202039902}

\bibitem[{{Foreman-Mackey} {et~al.}(2013){Foreman-Mackey}, {Hogg}, {Lang}, \&
  {Goodman}}]{ForemanMackey2013}
{Foreman-Mackey}, D., {Hogg}, D.~W., {Lang}, D., \& {Goodman}, J. 2013, \pasp,
  125, 306, \dodoi{10.1086/670067}

\bibitem[{{Forveille} {et~al.}(1999){Forveille}, {Beuzit}, {Delfosse},
  {Segransan}, {Beck}, {Mayor}, {Perrier}, {Tokovinin}, \&
  {Udry}}]{Forveille1999}
{Forveille}, T., {Beuzit}, J.-L., {Delfosse}, X., {et~al.} 1999, \aap, 351,
  619.
\newblock \doarXiv{astro-ph/9909342}

\bibitem[{{Frey} {et~al.}(1991){Frey}, {Grim}, {Hall}, {Mattingly}, {Robb},
  {Wood}, \& {Zeigler}}]{Frey1991}
{Frey}, G.~J., {Grim}, B., {Hall}, D.~S., {et~al.} 1991, \aj, 102, 1813,
  \dodoi{10.1086/116005}

\bibitem[{{Frick} {et~al.}(2004){Frick}, {Soon}, {Popova}, \&
  {Baliunas}}]{Frick2004}
{Frick}, P., {Soon}, W., {Popova}, E., \& {Baliunas}, S. 2004, \na, 9, 599,
  \dodoi{10.1016/j.newast.2004.03.005}

\bibitem[{{Fuhrmeister} {et~al.}(2022){Fuhrmeister}, {Czesla}, {Nagel},
  {Reiners}, {Schmitt}, {Jeffers}, {Caballero}, {Shulyak}, {Johnson},
  {Zechmeister}, {Montes}, {L{\'o}pez-Gallifa}, {Ribas}, {Quirrenbach},
  {Amado}, {Galad{\'\i}-Enr{\'\i}quez}, {Hatzes}, {K{\"u}rster}, {Danielski},
  {B{\'e}jar}, {Kaminski}, {Morales}, \& {Zapatero Osorio}}]{Fuhrmeister2022}
{Fuhrmeister}, B., {Czesla}, S., {Nagel}, E., {et~al.} 2022, \aap, 657, A125,
  \dodoi{10.1051/0004-6361/202141733}

\bibitem[{{Fulton} {et~al.}(2018){Fulton}, {Petigura}, {Blunt}, \&
  {Sinukoff}}]{Fulton2018}
{Fulton}, B.~J., {Petigura}, E.~A., {Blunt}, S., \& {Sinukoff}, E. 2018, \pasp,
  130, 044504, \dodoi{10.1088/1538-3873/aaaaa8}

\bibitem[{{Gaia Collaboration} {et~al.}(2020){Gaia Collaboration}, {Brown},
  {Vallenari}, {Prusti}, {de Bruijne}, {Babusiaux}, \& {Biermann}}]{GaiaEDR3}
{Gaia Collaboration}, {Brown}, A.~G.~A., {Vallenari}, A., {et~al.} 2020, arXiv
  e-prints, arXiv:2012.01533.
\newblock \doarXiv{2012.01533}

\bibitem[{{Gaia Collaboration} {et~al.}(2018){Gaia Collaboration}, {Brown},
  {Vallenari}, {Prusti}, {de Bruijne}, {Babusiaux}, {Bailer-Jones}, {Biermann},
  {Evans}, {Eyer}, {Jansen}, {Jordi}, {Klioner}, {Lammers}, {Lindegren},
  {Luri}, {Mignard}, {Panem}, {Pourbaix}, {Randich}, {Sartoretti}, {Siddiqui},
  {Soubiran}, {van Leeuwen}, {Walton}, {Arenou}, {Bastian}, {Cropper},
  {Drimmel}, {Katz}, {Lattanzi}, {Bakker}, {Cacciari}, {Casta{\~n}eda},
  {Chaoul}, {Cheek}, {De Angeli}, {Fabricius}, {Guerra}, {Holl}, {Masana},
  {Messineo}, {Mowlavi}, {Nienartowicz}, {Panuzzo}, {Portell}, {Riello},
  {Seabroke}, {Tanga}, {Th{\'e}venin}, {Gracia-Abril}, {Comoretto},
  {Garcia-Reinaldos}, {Teyssier}, {Altmann}, {Andrae}, {Audard},
  {Bellas-Velidis}, {Benson}, {Berthier}, {Blomme}, {Burgess}, {Busso},
  {Carry}, {Cellino}, {Clementini}, {Clotet}, {Creevey}, {Davidson}, {De
  Ridder}, {Delchambre}, {Dell'Oro}, {Ducourant},
  {Fern{\'a}ndez-Hern{\'a}ndez}, {Fouesneau}, {Fr{\'e}mat}, {Galluccio},
  {Garc{\'\i}a-Torres}, {Gonz{\'a}lez-N{\'u}{\~n}ez}, {Gonz{\'a}lez-Vidal},
  {Gosset}, {Guy}, {Halbwachs}, {Hambly}, {Harrison}, {Hern{\'a}ndez},
  {Hestroffer}, {Hodgkin}, {Hutton}, {Jasniewicz}, {Jean-Antoine-Piccolo},
  {Jordan}, {Korn}, {Krone-Martins}, {Lanzafame}, {Lebzelter}, {L{\"o}ffler},
  {Manteiga}, {Marrese}, {Mart{\'\i}n-Fleitas}, {Moitinho}, {Mora}, {Muinonen},
  {Osinde}, {Pancino}, {Pauwels}, {Petit}, {Recio-Blanco}, {Richards},
  {Rimoldini}, {Robin}, {Sarro}, {Siopis}, {Smith}, {Sozzetti}, {S{\"u}veges},
  {Torra}, {van Reeven}, {Abbas}, {Abreu Aramburu}, {Accart}, {Aerts},
  {Altavilla}, {{\'A}lvarez}, {Alvarez}, {Alves}, {Anderson}, {Andrei},
  {Anglada Varela}, {Antiche}, {Antoja}, {Arcay}, {Astraatmadja}, {Bach},
  {Baker}, {Balaguer-N{\'u}{\~n}ez}, {Balm}, {Barache}, {Barata}, {Barbato},
  {Barblan}, {Barklem}, {Barrado}, {Barros}, {Barstow}, {Bartholom{\'e}
  Mu{\~n}oz}, {Bassilana}, {Becciani}, {Bellazzini}, {Berihuete}, {Bertone},
  {Bianchi}, {Bienaym{\'e}}, {Blanco-Cuaresma}, {Boch}, {Boeche}, {Bombrun},
  {Borrachero}, {Bossini}, {Bouquillon}, {Bourda}, {Bragaglia}, {Bramante},
  {Breddels}, {Bressan}, {Brouillet}, {Br{\"u}semeister}, {Brugaletta},
  {Bucciarelli}, {Burlacu}, {Busonero}, {Butkevich}, {Buzzi}, {Caffau},
  {Cancelliere}, {Cannizzaro}, {Cantat-Gaudin}, {Carballo}, {Carlucci},
  {Carrasco}, {Casamiquela}, {Castellani}, {Castro-Ginard}, {Charlot},
  {Chemin}, {Chiavassa}, {Cocozza}, {Costigan}, {Cowell}, {Crifo}, {Crosta},
  {Crowley}, {Cuypers}, {Dafonte}, {Damerdji}, {Dapergolas}, {David}, {David},
  {de Laverny}, {De Luise}, {De March}, {de Martino}, {de Souza}, {de Torres},
  {Debosscher}, {del Pozo}, {Delbo}, {Delgado}, {Delgado}, {Di Matteo},
  {Diakite}, {Diener}, {Distefano}, {Dolding}, {Drazinos}, {Dur{\'a}n},
  {Edvardsson}, {Enke}, {Eriksson}, {Esquej}, {Eynard Bontemps}, {Fabre},
  {Fabrizio}, {Faigler}, {Falc{\~a}o}, {Farr{\`a}s Casas}, {Federici},
  {Fedorets}, {Fernique}, {Figueras}, {Filippi}, {Findeisen}, {Fonti},
  {Fraile}, {Fraser}, {Fr{\'e}zouls}, {Gai}, {Galleti}, {Garabato},
  {Garc{\'\i}a-Sedano}, {Garofalo}, {Garralda}, {Gavel}, {Gavras}, {Gerssen},
  {Geyer}, {Giacobbe}, {Gilmore}, {Girona}, {Giuffrida}, {Glass}, {Gomes},
  {Granvik}, {Gueguen}, {Guerrier}, {Guiraud}, {Guti{\'e}rrez-S{\'a}nchez},
  {Haigron}, {Hatzidimitriou}, {Hauser}, {Haywood}, {Heiter}, {Helmi}, {Heu},
  {Hilger}, {Hobbs}, {Hofmann}, {Holland}, {Huckle}, {Hypki}, {Icardi},
  {Jan{\ss}en}, {Jevardat de Fombelle}, {Jonker}, {Juh{\'a}sz}, {Julbe},
  {Karampelas}, {Kewley}, {Klar}, {Kochoska}, {Kohley}, {Kolenberg},
  {Kontizas}, {Kontizas}, {Koposov}, {Kordopatis}, {Kostrzewa-Rutkowska},
  {Koubsky}, {Lambert}, {Lanza}, {Lasne}, {Lavigne}, {Le Fustec}, {Le
  Poncin-Lafitte}, {Lebreton}, {Leccia}, {Leclerc}, {Lecoeur-Taibi},
  {Lenhardt}, {Leroux}, {Liao}, {Licata}, {Lindstr{\o}m}, {Lister}, {Livanou},
  {Lobel}, {L{\'o}pez}, {Managau}, {Mann}, {Mantelet}, {Marchal}, {Marchant},
  {Marconi}, {Marinoni}, {Marschalk{\'o}}, {Marshall}, {Martino}, {Marton},
  {Mary}, {Massari}, {Matijevi{\v{c}}}, {Mazeh}, {McMillan}, {Messina},
  {Michalik}, {Millar}, {Molina}, {Molinaro}, {Moln{\'a}r}, {Montegriffo},
  {Mor}, {Morbidelli}, {Morel}, {Morris}, {Mulone}, {Muraveva}, {Musella},
  {Nelemans}, {Nicastro}, {Noval}, {O'Mullane}, {Ord{\'e}novic},
  {Ord{\'o}{\~n}ez-Blanco}, {Osborne}, {Pagani}, {Pagano}, {Pailler},
  {Palacin}, {Palaversa}, {Panahi}, {Pawlak}, {Piersimoni}, {Pineau}, {Plachy},
  {Plum}, {Poggio}, {Poujoulet}, {Pr{\v{s}}a}, {Pulone}, {Racero}, {Ragaini},
  {Rambaux}, {Ramos-Lerate}, {Regibo}, {Reyl{\'e}}, {Riclet}, {Ripepi}, {Riva},
  {Rivard}, {Rixon}, {Roegiers}, {Roelens}, {Romero-G{\'o}mez}, {Rowell},
  {Royer}, {Ruiz-Dern}, {Sadowski}, {Sagrist{\`a} Sell{\'e}s}, {Sahlmann},
  {Salgado}, {Salguero}, {Sanna}, {Santana-Ros}, {Sarasso}, {Savietto},
  {Schultheis}, {Sciacca}, {Segol}, {Segovia}, {S{\'e}gransan}, {Shih},
  {Siltala}, {Silva}, {Smart}, {Smith}, {Solano}, {Solitro}, {Sordo}, {Soria
  Nieto}, {Souchay}, {Spagna}, {Spoto}, {Stampa}, {Steele},
  {Steidelm{\"u}ller}, {Stephenson}, {Stoev}, {Suess}, {Surdej}, {Szabados},
  {Szegedi-Elek}, {Tapiador}, {Taris}, {Tauran}, {Taylor}, {Teixeira},
  {Terrett}, {Teyssandier}, {Thuillot}, {Titarenko}, {Torra Clotet}, {Turon},
  {Ulla}, {Utrilla}, {Uzzi}, {Vaillant}, {Valentini}, {Valette}, {van Elteren},
  {Van Hemelryck}, {van Leeuwen}, {Vaschetto}, {Vecchiato}, {Veljanoski},
  {Viala}, {Vicente}, {Vogt}, {von Essen}, {Voss}, {Votruba}, {Voutsinas},
  {Walmsley}, {Weiler}, {Wertz}, {Wevers}, {Wyrzykowski}, {Yoldas},
  {{\v{Z}}erjal}, {Ziaeepour}, {Zorec}, {Zschocke}, {Zucker}, {Zurbach}, \&
  {Zwitter}}]{GaiaDR2}
---. 2018, \aap, 616, A1, \dodoi{10.1051/0004-6361/201833051}

\bibitem[{{Gaia Collaboration} {et~al.}(2021){Gaia Collaboration}, {Brown},
  {Vallenari}, {Prusti}, {de Bruijne}, {Babusiaux}, {Biermann}, {Creevey},
  {Evans}, {Eyer}, {Hutton}, {Jansen}, {Jordi}, {Klioner}, {Lammers},
  {Lindegren}, {Luri}, {Mignard}, {Panem}, {Pourbaix}, {Randich}, {Sartoretti},
  {Soubiran}, {Walton}, {Arenou}, {Bailer-Jones}, {Bastian}, {Cropper},
  {Drimmel}, {Katz}, {Lattanzi}, {van Leeuwen}, {Bakker}, {Cacciari},
  {Casta{\~n}eda}, {De Angeli}, {Ducourant}, {Fabricius}, {Fouesneau},
  {Fr{\'e}mat}, {Guerra}, {Guerrier}, {Guiraud}, {Jean-Antoine Piccolo},
  {Masana}, {Messineo}, {Mowlavi}, {Nicolas}, {Nienartowicz}, {Pailler},
  {Panuzzo}, {Riclet}, {Roux}, {Seabroke}, {Sordo}, {Tanga}, {Th{\'e}venin},
  {Gracia-Abril}, {Portell}, {Teyssier}, {Altmann}, {Andrae}, {Bellas-Velidis},
  {Benson}, {Berthier}, {Blomme}, {Brugaletta}, {Burgess}, {Busso}, {Carry},
  {Cellino}, {Cheek}, {Clementini}, {Damerdji}, {Davidson}, {Delchambre},
  {Dell'Oro}, {Fern{\'a}ndez-Hern{\'a}ndez}, {Galluccio}, {Garc{\'\i}a-Lario},
  {Garcia-Reinaldos}, {Gonz{\'a}lez-N{\'u}{\~n}ez}, {Gosset}, {Haigron},
  {Halbwachs}, {Hambly}, {Harrison}, {Hatzidimitriou}, {Heiter},
  {Hern{\'a}ndez}, {Hestroffer}, {Hodgkin}, {Holl}, {Jan{\ss}en}, {Jevardat de
  Fombelle}, {Jordan}, {Krone-Martins}, {Lanzafame}, {L{\"o}ffler}, {Lorca},
  {Manteiga}, {Marchal}, {Marrese}, {Moitinho}, {Mora}, {Muinonen}, {Osborne},
  {Pancino}, {Pauwels}, {Petit}, {Recio-Blanco}, {Richards}, {Riello},
  {Rimoldini}, {Robin}, {Roegiers}, {Rybizki}, {Sarro}, {Siopis}, {Smith},
  {Sozzetti}, {Ulla}, {Utrilla}, {van Leeuwen}, {van Reeven}, {Abbas}, {Abreu
  Aramburu}, {Accart}, {Aerts}, {Aguado}, {Ajaj}, {Altavilla}, {{\'A}lvarez},
  {{\'A}lvarez Cid-Fuentes}, {Alves}, {Anderson}, {Anglada Varela}, {Antoja},
  {Audard}, {Baines}, {Baker}, {Balaguer-N{\'u}{\~n}ez}, {Balbinot}, {Balog},
  {Barache}, {Barbato}, {Barros}, {Barstow}, {Bartolom{\'e}}, {Bassilana},
  {Bauchet}, {Baudesson-Stella}, {Becciani}, {Bellazzini}, {Bernet}, {Bertone},
  {Bianchi}, {Blanco-Cuaresma}, {Boch}, {Bombrun}, {Bossini}, {Bouquillon},
  {Bragaglia}, {Bramante}, {Breedt}, {Bressan}, {Brouillet}, {Bucciarelli},
  {Burlacu}, {Busonero}, {Butkevich}, {Buzzi}, {Caffau}, {Cancelliere},
  {C{\'a}novas}, {Cantat-Gaudin}, {Carballo}, {Carlucci}, {Carnerero},
  {Carrasco}, {Casamiquela}, {Castellani}, {Castro-Ginard}, {Castro Sampol},
  {Chaoul}, {Charlot}, {Chemin}, {Chiavassa}, {Cioni}, {Comoretto}, {Cooper},
  {Cornez}, {Cowell}, {Crifo}, {Crosta}, {Crowley}, {Dafonte}, {Dapergolas},
  {David}, {David}, {de Laverny}, {De Luise}, {De March}, {De Ridder}, {de
  Souza}, {de Teodoro}, {de Torres}, {del Peloso}, {del Pozo}, {Delbo},
  {Delgado}, {Delgado}, {Delisle}, {Di Matteo}, {Diakite}, {Diener},
  {Distefano}, {Dolding}, {Eappachen}, {Edvardsson}, {Enke}, {Esquej}, {Fabre},
  {Fabrizio}, {Faigler}, {Fedorets}, {Fernique}, {Fienga}, {Figueras},
  {Fouron}, {Fragkoudi}, {Fraile}, {Franke}, {Gai}, {Garabato},
  {Garcia-Gutierrez}, {Garc{\'\i}a-Torres}, {Garofalo}, {Gavras}, {Gerlach},
  {Geyer}, {Giacobbe}, {Gilmore}, {Girona}, {Giuffrida}, {Gomel}, {Gomez},
  {Gonzalez-Santamaria}, {Gonz{\'a}lez-Vidal}, {Granvik},
  {Guti{\'e}rrez-S{\'a}nchez}, {Guy}, {Hauser}, {Haywood}, {Helmi}, {Hidalgo},
  {Hilger}, {H{\l}adczuk}, {Hobbs}, {Holland}, {Huckle}, {Jasniewicz},
  {Jonker}, {Juaristi Campillo}, {Julbe}, {Karbevska}, {Kervella}, {Khanna},
  {Kochoska}, {Kontizas}, {Kordopatis}, {Korn}, {Kostrzewa-Rutkowska},
  {Kruszy{\'n}ska}, {Lambert}, {Lanza}, {Lasne}, {Le Campion}, {Le Fustec},
  {Lebreton}, {Lebzelter}, {Leccia}, {Leclerc}, {Lecoeur-Taibi}, {Liao},
  {Licata}, {Lindstr{\o}m}, {Lister}, {Livanou}, {Lobel}, {Madrero Pardo},
  {Managau}, {Mann}, {Marchant}, {Marconi}, {Marcos Santos}, {Marinoni},
  {Marocco}, {Marshall}, {Martin Polo}, {Mart{\'\i}n-Fleitas}, {Masip},
  {Massari}, {Mastrobuono-Battisti}, {Mazeh}, {McMillan}, {Messina},
  {Michalik}, {Millar}, {Mints}, {Molina}, {Molinaro}, {Moln{\'a}r},
  {Montegriffo}, {Mor}, {Morbidelli}, {Morel}, {Morris}, {Mulone}, {Munoz},
  {Muraveva}, {Murphy}, {Musella}, {Noval}, {Ord{\'e}novic}, {Orr{\`u}},
  {Osinde}, {Pagani}, {Pagano}, {Palaversa}, {Palicio}, {Panahi}, {Pawlak},
  {Pe{\~n}alosa Esteller}, {Penttil{\"a}}, {Piersimoni}, {Pineau}, {Plachy},
  {Plum}, {Poggio}, {Poretti}, {Poujoulet}, {Pr{\v{s}}a}, {Pulone}, {Racero},
  {Ragaini}, {Rainer}, {Raiteri}, {Rambaux}, {Ramos}, {Ramos-Lerate}, {Re
  Fiorentin}, {Regibo}, {Reyl{\'e}}, {Ripepi}, {Riva}, {Rixon}, {Robichon},
  {Robin}, {Roelens}, {Rohrbasser}, {Romero-G{\'o}mez}, {Rowell}, {Royer},
  {Rybicki}, {Sadowski}, {Sagrist{\`a} Sell{\'e}s}, {Sahlmann}, {Salgado},
  {Salguero}, {Samaras}, {Sanchez Gimenez}, {Sanna}, {Santove{\~n}a},
  {Sarasso}, {Schultheis}, {Sciacca}, {Segol}, {Segovia}, {S{\'e}gransan},
  {Semeux}, {Shahaf}, {Siddiqui}, {Siebert}, {Siltala}, {Slezak}, {Smart},
  {Solano}, {Solitro}, {Souami}, {Souchay}, {Spagna}, {Spoto}, {Steele},
  {Steidelm{\"u}ller}, {Stephenson}, {S{\"u}veges}, {Szabados}, {Szegedi-Elek},
  {Taris}, {Tauran}, {Taylor}, {Teixeira}, {Thuillot}, {Tonello}, {Torra},
  {Torra}, {Turon}, {Unger}, {Vaillant}, {van Dillen}, {Vanel}, {Vecchiato},
  {Viala}, {Vicente}, {Voutsinas}, {Weiler}, {Wevers}, {Wyrzykowski}, {Yoldas},
  {Yvard}, {Zhao}, {Zorec}, {Zucker}, {Zurbach}, \& {Zwitter}}]{GaiaDR3}
---. 2021, \aap, 649, A1, \dodoi{10.1051/0004-6361/202039657}

\bibitem[{{Gaidos} {et~al.}(2000){Gaidos}, {Henry}, \& {Henry}}]{Gaidos2000}
{Gaidos}, E.~J., {Henry}, G.~W., \& {Henry}, S.~M. 2000, \aj, 120, 1006,
  \dodoi{10.1086/301488}

\bibitem[{{Gandolfi} {et~al.}(2018){Gandolfi}, {Barrag{\'a}n}, {Livingston},
  {Fridlund}, {Justesen}, {Redfield}, {Fossati}, {Mathur}, {Grziwa}, {Cabrera},
  {Garc{\'\i}a}, {Persson}, {Van Eylen}, {Hatzes}, {Hidalgo}, {Albrecht},
  {Bugnet}, {Cochran}, {Csizmadia}, {Deeg}, {Eigm{\"u}ller}, {Endl}, {Erikson},
  {Esposito}, {Guenther}, {Korth}, {Luque}, {Monta{\~n}es Rodr{\'\i}guez},
  {Nespral}, {Nowak}, {P{\"a}tzold}, \& {Prieto-Arranz}}]{Gandolfi2018}
{Gandolfi}, D., {Barrag{\'a}n}, O., {Livingston}, J.~H., {et~al.} 2018, \aap,
  619, L10, \dodoi{10.1051/0004-6361/201834289}

\bibitem[{{Garg} {et~al.}(2019){Garg}, {Karak}, {Egeland}, {Soon}, \&
  {Baliunas}}]{Garg2019}
{Garg}, S., {Karak}, B.~B., {Egeland}, R., {Soon}, W., \& {Baliunas}, S. 2019,
  \apj, 886, 132, \dodoi{10.3847/1538-4357/ab4a17}

\bibitem[{{G{\'a}sp{\'a}r} {et~al.}(2013){G{\'a}sp{\'a}r}, {Rieke}, \&
  {Balog}}]{Gaspar2013}
{G{\'a}sp{\'a}r}, A., {Rieke}, G.~H., \& {Balog}, Z. 2013, \apj, 768, 25,
  \dodoi{10.1088/0004-637X/768/1/25}

\bibitem[{{Gaudi} {et~al.}(2020){Gaudi}, {Seager}, {Mennesson}, {Kiessling},
  {Warfield}, {Cahoy}, {Clarke}, {Domagal-Goldman}, {Feinberg}, {Guyon},
  {Kasdin}, {Mawet}, {Plavchan}, {Robinson}, {Rogers}, {Scowen}, {Somerville},
  {Stapelfeldt}, {Stark}, {Stern}, {Turnbull}, {Amini}, {Kuan}, {Martin},
  {Morgan}, {Redding}, {Stahl}, {Webb}, {Alvarez-Salazar}, {Arnold}, {Arya},
  {Balasubramanian}, {Baysinger}, {Bell}, {Below}, {Benson}, {Blais}, {Booth},
  {Bourgeois}, {Bradford}, {Brewer}, {Brooks}, {Cady}, {Caldwell}, {Calvet},
  {Carr}, {Chan}, {Cormarkovic}, {Coste}, {Cox}, {Danner}, {Davis}, {Dewell},
  {Dorsett}, {Dunn}, {East}, {Effinger}, {Eng}, {Freebury}, {Garcia}, {Gaskin},
  {Greene}, {Hennessy}, {Hilgemann}, {Hood}, {Holota}, {Howe}, {Huang}, {Hull},
  {Hunt}, {Hurd}, {Johnson}, {Kissil}, {Knight}, {Kolenz}, {Kraus}, {Krist},
  {Li}, {Lisman}, {Mandic}, {Mann}, {Marchen}, {Marrese-Reading}, {McCready},
  {McGown}, {Missun}, {Miyaguchi}, {Moore}, {Nemati}, {Nikzad}, {Nissen},
  {Novicki}, {Perrine}, {Pineda}, {Polanco}, {Putnam}, {Qureshi}, {Richards},
  {Eldorado Riggs}, {Rodgers}, {Rud}, {Saini}, {Scalisi}, {Scharf}, {Schulz},
  {Serabyn}, {Sigrist}, {Sikkia}, {Singleton}, {Shaklan}, {Smith}, {Southerd},
  {Stahl}, {Steeves}, {Sturges}, {Sullivan}, {Tang}, {Taras}, {Tesch},
  {Therrell}, {Tseng}, {Valente}, {Van Buren}, {Villalvazo}, {Warwick}, {Webb},
  {Westerhoff}, {Wofford}, {Wu}, {Woo}, {Wood}, {Ziemer}, {Arney}, {Anderson},
  {Ma{\'\i}z-Apell{\'a}niz}, {Bartlett}, {Belikov}, {Bendek}, {Cenko},
  {Douglas}, {Dulz}, {Evans}, {Faramaz}, {Feng}, {Ferguson}, {Follette},
  {Ford}, {Garc{\'\i}a}, {Geha}, {Gelino}, {G{\"o}tberg}, {Hildebrandt}, {Hu},
  {Jahnke}, {Kennedy}, {Kreidberg}, {Isella}, {Lopez}, {Marchis}, {Macri},
  {Marley}, {Matzko}, {Mazoyer}, {McCandliss}, {Meshkat}, {Mordasini},
  {Morris}, {Nielsen}, {Newman}, {Petigura}, {Postman}, {Reines}, {Roberge},
  {Roederer}, {Ruane}, {Schwieterman}, {Sirbu}, {Spalding}, {Teplitz},
  {Tumlinson}, {Turner}, {Werk}, {Wofford}, {Wyatt}, {Young}, \&
  {Zellem}}]{Gaudi2020}
{Gaudi}, B.~S., {Seager}, S., {Mennesson}, B., {et~al.} 2020, arXiv e-prints,
  arXiv:2001.06683.
\newblock \doarXiv{2001.06683}

\bibitem[{{Gentile Fusillo} {et~al.}(2019){Gentile Fusillo}, {Tremblay},
  {G{\"a}nsicke}, {Manser}, {Cunningham}, {Cukanovaite}, {Hollands}, {Marsh},
  {Raddi}, {Jordan}, {Toonen}, {Geier}, {Barstow}, \&
  {Cummings}}]{GentileFusillo2019}
{Gentile Fusillo}, N.~P., {Tremblay}, P.-E., {G{\"a}nsicke}, B.~T., {et~al.}
  2019, \mnras, 482, 4570, \dodoi{10.1093/mnras/sty3016}

\bibitem[{{Golimowski} {et~al.}(2000){Golimowski}, {Henry}, {Krist},
  {Schroeder}, {Marcy}, {Fischer}, \& {Butler}}]{Golimowski2000}
{Golimowski}, D.~A., {Henry}, T.~J., {Krist}, J.~E., {et~al.} 2000, \aj, 120,
  2082, \dodoi{10.1086/301567}

\bibitem[{{Golimowski} {et~al.}(1995){Golimowski}, {Nakajima}, {Kulkarni}, \&
  {Oppenheimer}}]{Golimowski1995}
{Golimowski}, D.~A., {Nakajima}, T., {Kulkarni}, S.~R., \& {Oppenheimer}, B.~R.
  1995, \apjl, 444, L101, \dodoi{10.1086/187870}

\bibitem[{{Gomes da Silva} {et~al.}(2014){Gomes da Silva}, {Santos}, {Boisse},
  {Dumusque}, \& {Lovis}}]{GomesdaSilva2014}
{Gomes da Silva}, J., {Santos}, N.~C., {Boisse}, I., {Dumusque}, X., \&
  {Lovis}, C. 2014, \aap, 566, A66, \dodoi{10.1051/0004-6361/201322697}

\bibitem[{{Gomes da Silva} {et~al.}(2021){Gomes da Silva}, {Santos},
  {Adibekyan}, {Sousa}, {Campante}, {Figueira}, {Bossini}, {Delgado-Mena},
  {Monteiro}, {de Laverny}, {Recio-Blanco}, \& {Lovis}}]{GomesdaSilva2021}
{Gomes da Silva}, J., {Santos}, N.~C., {Adibekyan}, V., {et~al.} 2021, \aap,
  646, A77, \dodoi{10.1051/0004-6361/202039765}

\bibitem[{{Gondoin}(2020)}]{Gondoin2020}
{Gondoin}, P. 2020, \aap, 641, A110, \dodoi{10.1051/0004-6361/202038291}

\bibitem[{{Gonzalez} {et~al.}(2010){Gonzalez}, {Carlson}, \&
  {Tobin}}]{Gonzalez2010}
{Gonzalez}, G., {Carlson}, M.~K., \& {Tobin}, R.~W. 2010, \mnras, 403, 1368,
  \dodoi{10.1111/j.1365-2966.2009.16195.x}

\bibitem[{{Gray} {et~al.}(2006){Gray}, {Corbally}, {Garrison}, {McFadden},
  {Bubar}, {McGahee}, {O'Donoghue}, \& {Knox}}]{Gray2006}
{Gray}, R.~O., {Corbally}, C.~J., {Garrison}, R.~F., {et~al.} 2006, \aj, 132,
  161, \dodoi{10.1086/504637}

\bibitem[{{Gray} {et~al.}(2003){Gray}, {Corbally}, {Garrison}, {McFadden}, \&
  {Robinson}}]{Gray2003}
{Gray}, R.~O., {Corbally}, C.~J., {Garrison}, R.~F., {McFadden}, M.~T., \&
  {Robinson}, P.~E. 2003, \aj, 126, 2048, \dodoi{10.1086/378365}

\bibitem[{{Gray} {et~al.}(2001){Gray}, {Napier}, \& {Winkler}}]{Gray2001}
{Gray}, R.~O., {Napier}, M.~G., \& {Winkler}, L.~I. 2001, \aj, 121, 2148,
  \dodoi{10.1086/319956}

\bibitem[{{Halbwachs} {et~al.}(2018){Halbwachs}, {Mayor}, \&
  {Udry}}]{Halbwachs2018}
{Halbwachs}, J.~L., {Mayor}, M., \& {Udry}, S. 2018, \aap, 619, A81,
  \dodoi{10.1051/0004-6361/201833377}

\bibitem[{{Hall}(1976)}]{Hall1976}
{Hall}, D.~S. 1976, in Astrophysics and Space Science Library, Vol.~60, IAU
  Colloq. 29: Multiple Periodic Variable Stars, ed. W.~S. {Fitch}, 287

\bibitem[{{Hall} {et~al.}(2007){Hall}, {Lockwood}, \& {Skiff}}]{Hall2007}
{Hall}, J.~C., {Lockwood}, G.~W., \& {Skiff}, B.~A. 2007, \aj, 133, 862,
  \dodoi{10.1086/510356}

\bibitem[{{Hansen} {et~al.}(2022){Hansen}, {Ireland}, \& {LIFE
  Collaboration}}]{Hansen2022}
{Hansen}, J.~T., {Ireland}, M.~J., \& {LIFE Collaboration}. 2022, \aap, 664,
  A52, \dodoi{10.1051/0004-6361/202243107}

\bibitem[{{Haslebacher} {et~al.}(2022){Haslebacher}, {Demory}, {Demory},
  {Sarazin}, \& {Vidale}}]{Haslebacher2022}
{Haslebacher}, C., {Demory}, M.~E., {Demory}, B.~O., {Sarazin}, M., \&
  {Vidale}, P.~L. 2022, \aap, 665, A149, \dodoi{10.1051/0004-6361/202142493}

\bibitem[{{Hathaway}(2015)}]{Hathaway2015}
{Hathaway}, D.~H. 2015, Living Reviews in Solar Physics, 12, 4,
  \dodoi{10.1007/lrsp-2015-4}

\bibitem[{{Hatzes} {et~al.}(2022){Hatzes}, {Gandolfi}, {Korth}, {Rodler},
  {Sabotta}, {Esposito}, {Barrag{\'a}n}, {Van Eylen}, {Livingston}, {Serrano},
  {Luque}, {Smith}, {Redfield}, {Persson}, {P{\"a}tzold}, {Palle}, {Nowak},
  {Osborne}, {Narita}, {Mathur}, {Lam}, {Kab{\'a}th}, {Johnson}, {Guenther},
  {Grziwa}, {Goffo}, {Fridlund}, {Endl}, {Deeg}, {Csizmadia}, {Cochran},
  {Cuesta}, {Chaturvedi}, {Carleo}, {Cabrera}, {Beck}, \&
  {Albrecht}}]{Hatzes2022}
{Hatzes}, A.~P., {Gandolfi}, D., {Korth}, J., {et~al.} 2022, \aj, 163, 223,
  \dodoi{10.3847/1538-3881/ac5dcb}

\bibitem[{{Haywood} {et~al.}(2016){Haywood}, {Collier Cameron}, {Unruh},
  {Lovis}, {Lanza}, {Llama}, {Deleuil}, {Fares}, {Gillon}, {Moutou}, {Pepe},
  {Pollacco}, {Queloz}, \& {S{\'e}gransan}}]{Haywood2016}
{Haywood}, R.~D., {Collier Cameron}, A., {Unruh}, Y.~C., {et~al.} 2016, \mnras,
  457, 3637, \dodoi{10.1093/mnras/stw187}

\bibitem[{{Heintz} \& {Cantor}(1994)}]{Heintz1994}
{Heintz}, W.~D., \& {Cantor}, B.~A. 1994, \pasp, 106, 363,
  \dodoi{10.1086/133386}

\bibitem[{{Henry} {et~al.}(2002){Henry}, {Walkowicz}, {Barto}, \&
  {Golimowski}}]{Henry2002}
{Henry}, T.~J., {Walkowicz}, L.~M., {Barto}, T.~C., \& {Golimowski}, D.~A.
  2002, \aj, 123, 2002, \dodoi{10.1086/339315}

\bibitem[{{Hinkel} {et~al.}(2017){Hinkel}, {Mamajek}, {Turnbull}, {Osby},
  {Shkolnik}, {Smith}, {Klimasewski}, {Somers}, \& {Desch}}]{Hinkel2017}
{Hinkel}, N.~R., {Mamajek}, E.~E., {Turnbull}, M.~C., {et~al.} 2017, \apj, 848,
  34, \dodoi{10.3847/1538-4357/aa8b0f}

\bibitem[{{Hojjatpanah} {et~al.}(2020){Hojjatpanah}, {Oshagh}, {Figueira},
  {Santos}, {Amazo-G{\'o}mez}, {Sousa}, {Adibekyan}, {Akinsanmi}, {Demangeon},
  {Faria}, {Gomes da Silva}, \& {Meunier}}]{Hojjatpanah2020}
{Hojjatpanah}, S., {Oshagh}, M., {Figueira}, P., {et~al.} 2020, \aap, 639, A35,
  \dodoi{10.1051/0004-6361/202038035}

\bibitem[{{Holberg} {et~al.}(2016){Holberg}, {Oswalt}, {Sion}, \&
  {McCook}}]{Holberg2016}
{Holberg}, J.~B., {Oswalt}, T.~D., {Sion}, E.~M., \& {McCook}, G.~P. 2016,
  \mnras, 462, 2295, \dodoi{10.1093/mnras/stw1357}

\bibitem[{{Horch} {et~al.}(2011){Horch}, {Gomez}, {Sherry}, {Howell}, {Ciardi},
  {Anderson}, \& {van Altena}}]{Horch2011}
{Horch}, E.~P., {Gomez}, S.~C., {Sherry}, W.~H., {et~al.} 2011, \aj, 141, 45,
  \dodoi{10.1088/0004-6256/141/2/45}

\bibitem[{{Horch} {et~al.}(2021){Horch}, {Broderick}, {Casetti-Dinescu},
  {Henry}, {Fekel}, {Muterspaugh}, {Willmarth}, {Winters}, {van Belle},
  {Clark}, \& {Everett}}]{Horch2021}
{Horch}, E.~P., {Broderick}, K.~G., {Casetti-Dinescu}, D.~I., {et~al.} 2021,
  \aj, 161, 295, \dodoi{10.3847/1538-3881/abf9a8}

\bibitem[{{Houk} \& {Cowley}(1975)}]{Houk1975}
{Houk}, N., \& {Cowley}, A.~P. 1975, {University of Michigan Catalogue of
  two-dimensional spectral types for the HD stars. Volume I. Declinations -90
  to -53} ({Dept. of Astronomy, University of Michigan})

\bibitem[{{Howard} \& {Fulton}(2016)}]{Howard2016}
{Howard}, A.~W., \& {Fulton}, B.~J. 2016, \pasp, 128, 114401,
  \dodoi{10.1088/1538-3873/128/969/114401}

\bibitem[{{Howell} {et~al.}(2011){Howell}, {Everett}, {Sherry}, {Horch}, \&
  {Ciardi}}]{Howell2011}
{Howell}, S.~B., {Everett}, M.~E., {Sherry}, W., {Horch}, E., \& {Ciardi},
  D.~R. 2011, \aj, 142, 19, \dodoi{10.1088/0004-6256/142/1/19}

\bibitem[{{Huang} {et~al.}(2018){Huang}, {Burt}, {Vanderburg}, {G{\"u}nther},
  {Shporer}, {Dittmann}, {Winn}, {Wittenmyer}, {Sha}, {Kane}, {Ricker},
  {Vanderspek}, {Latham}, {Seager}, {Jenkins}, {Caldwell}, {Collins},
  {Guerrero}, {Smith}, {Quinn}, {Udry}, {Pepe}, {Bouchy}, {S{\'e}gransan},
  {Lovis}, {Ehrenreich}, {Marmier}, {Mayor}, {Wohler}, {Haworth}, {Morgan},
  {Fausnaugh}, {Ciardi}, {Christiansen}, {Charbonneau}, {Dragomir}, {Deming},
  {Glidden}, {Levine}, {McCullough}, {Yu}, {Narita}, {Nguyen}, {Morton},
  {Pepper}, {P{\'a}l}, {Rodriguez}, {Stassun}, {Torres}, {Sozzetti}, {Doty},
  {Christensen-Dalsgaard}, {Laughlin}, {Clampin}, {Bean}, {Buchhave}, {Bakos},
  {Sato}, {Ida}, {Kaltenegger}, {Palle}, {Sasselov}, {Butler}, {Lissauer},
  {Ge}, \& {Rinehart}}]{Huang2018}
{Huang}, C.~X., {Burt}, J., {Vanderburg}, A., {et~al.} 2018, \apjl, 868, L39,
  \dodoi{10.3847/2041-8213/aaef91}

\bibitem[{{Husser} {et~al.}(2013){Husser}, {Wende-von Berg}, {Dreizler},
  {Homeier}, {Reiners}, {Barman}, \& {Hauschildt}}]{Husser2013}
{Husser}, T.~O., {Wende-von Berg}, S., {Dreizler}, S., {et~al.} 2013, \aap,
  553, A6, \dodoi{10.1051/0004-6361/201219058}

\bibitem[{{Ianna}(1992)}]{Ianna1992}
{Ianna}, P.~A. 1992, in Astronomical Society of the Pacific Conference Series,
  Vol.~32, IAU Colloq. 135: Complementary Approaches to Double and Multiple
  Star Research, ed. H.~A. {McAlister} \& W.~I. {Hartkopf}, 323

\bibitem[{{Ibukiyama} \& {Arimoto}(2002)}]{Ibukiyama2002}
{Ibukiyama}, A., \& {Arimoto}, N. 2002, \aap, 394, 927,
  \dodoi{10.1051/0004-6361:20021157}

\bibitem[{{Isaacson} \& {Fischer}(2010)}]{Isaacson2010}
{Isaacson}, H., \& {Fischer}, D. 2010, \apj, 725, 875,
  \dodoi{10.1088/0004-637X/725/1/875}

\bibitem[{{Jofr{\'e}} {et~al.}(2014){Jofr{\'e}}, {Heiter}, {Soubiran},
  {Blanco-Cuaresma}, {Worley}, {Pancino}, {Cantat-Gaudin}, {Magrini},
  {Bergemann}, {Gonz{\'a}lez Hern{\'a}ndez}, {Hill}, {Lardo}, {de Laverny},
  {Lind}, {Masseron}, {Montes}, {Mucciarelli}, {Nordlander}, {Recio Blanco},
  {Sobeck}, {Sordo}, {Sousa}, {Tabernero}, {Vallenari}, \& {Van
  Eck}}]{Jofre2014}
{Jofr{\'e}}, P., {Heiter}, U., {Soubiran}, C., {et~al.} 2014, \aap, 564, A133,
  \dodoi{10.1051/0004-6361/201322440}

\bibitem[{{Jofr{\'e}} {et~al.}(2015){Jofr{\'e}}, {Heiter}, {Soubiran},
  {Blanco-Cuaresma}, {Masseron}, {Nordlander}, {Chemin}, {Worley}, {Van Eck},
  {Hourihane}, {Gilmore}, {Adibekyan}, {Bergemann}, {Cantat-Gaudin},
  {Delgado-Mena}, {Gonz{\'a}lez Hern{\'a}ndez}, {Guiglion}, {Lardo}, {de
  Laverny}, {Lind}, {Magrini}, {Mikolaitis}, {Montes}, {Pancino},
  {Recio-Blanco}, {Sordo}, {Sousa}, {Tabernero}, \& {Vallenari}}]{Jofre2015}
---. 2015, \aap, 582, A81, \dodoi{10.1051/0004-6361/201526604}

\bibitem[{{Johnson} \& {Morgan}(1953)}]{Johnson1953}
{Johnson}, H.~L., \& {Morgan}, W.~W. 1953, \apj, 117, 313,
  \dodoi{10.1086/145697}

\bibitem[{{Jones} {et~al.}(2002){Jones}, {Paul Butler}, {Tinney}, {Marcy},
  {Penny}, {McCarthy}, {Carter}, \& {Pourbaix}}]{Jones2002}
{Jones}, H. R.~A., {Paul Butler}, R., {Tinney}, C.~G., {et~al.} 2002, \mnras,
  333, 871, \dodoi{10.1046/j.1365-8711.2002.05459.x}

\bibitem[{{Jourdain de Muizon} {et~al.}(1999){Jourdain de Muizon}, {Laureijs},
  {Dominik}, {Habing}, {Metcalfe}, {Siebenmorgen}, {Kessler}, {Bouchet},
  {Salama}, {Leech}, {Trams}, \& {Heske}}]{JourdaindeMuizon1999}
{Jourdain de Muizon}, M., {Laureijs}, R.~J., {Dominik}, C., {et~al.} 1999,
  \aap, 350, 875

\bibitem[{{Kane} {et~al.}(2020){Kane}, {Yal{\c{c}}{\i}nkaya}, {Osborn},
  {Dalba}, {Nielsen}, {Vanderburg}, {Mo{\v{c}}nik}, {Hinkel}, {Ostberg},
  {Esmer}, {Udry}, {Fetherolf}, {Ba{\c{s}}t{\"u}rk}, {Ricker}, {Vanderspek},
  {Latham}, {Seager}, {Winn}, {Jenkins}, {Allart}, {Bailey}, {Bean}, {Bouchy},
  {Butler}, {Campante}, {Carter}, {Daylan}, {Deleuil}, {Diaz}, {Dumusque},
  {Ehrenreich}, {Horner}, {Howard}, {Isaacson}, {Jones}, {Kristiansen},
  {Lovis}, {Marcy}, {Marmier}, {O'Toole}, {Pepe}, {Ragozzine}, {S{\'e}gransan},
  {Tinney}, {Turnbull}, {Wittenmyer}, {Wright}, \& {Wright}}]{Kane2020}
{Kane}, S.~R., {Yal{\c{c}}{\i}nkaya}, S., {Osborn}, H.~P., {et~al.} 2020, \aj,
  160, 129, \dodoi{10.3847/1538-3881/aba835}

\bibitem[{{Kass} \& {Raftery}(1995)}]{KassRaftery1995}
{Kass}, R., \& {Raftery}, A. 1995, Journal of the American Statistical
  Association, 90, 773

\bibitem[{{Katoh} {et~al.}(2013){Katoh}, {Itoh}, {Toyota}, \&
  {Sato}}]{Katoh2013}
{Katoh}, N., {Itoh}, Y., {Toyota}, E., \& {Sato}, B. 2013, \aj, 145, 41,
  \dodoi{10.1088/0004-6256/145/2/41}

\bibitem[{{Keenan} \& {McNeil}(1989)}]{Keenan1989}
{Keenan}, P.~C., \& {McNeil}, R.~C. 1989, \apjs, 71, 245,
  \dodoi{10.1086/191373}

\bibitem[{{Kervella} {et~al.}(2019){Kervella}, {Arenou}, {Mignard}, \&
  {Th{\'e}venin}}]{Kervella2019}
{Kervella}, P., {Arenou}, F., {Mignard}, F., \& {Th{\'e}venin}, F. 2019, \aap,
  623, A72, \dodoi{10.1051/0004-6361/201834371}

\bibitem[{{Kervella} {et~al.}(2022){Kervella}, {Arenou}, \&
  {Th{\'e}venin}}]{Kervella2022}
{Kervella}, P., {Arenou}, F., \& {Th{\'e}venin}, F. 2022, \aap, 657, A7,
  \dodoi{10.1051/0004-6361/202142146}

\bibitem[{{Kipping}(2013)}]{Kipping2013}
{Kipping}, D.~M. 2013, \mnras, 434, L51, \dodoi{10.1093/mnrasl/slt075}

\bibitem[{{Kiyaeva} {et~al.}(2008){Kiyaeva}, {Kiselev}, \&
  {Izmailov}}]{Kiyaeva2008}
{Kiyaeva}, O.~V., {Kiselev}, A.~A., \& {Izmailov}, I.~S. 2008, Astronomy
  Letters, 34, 405, \dodoi{10.1134/S1063773708060054}

\bibitem[{{Konrad} {et~al.}(2022){Konrad}, {Alei}, {Quanz}, {Angerhausen},
  {Carri{\'o}n-Gonz{\'a}lez}, {Fortney}, {Grenfell}, {Kitzmann},
  {Molli{\`e}re}, {Rugheimer}, {Wunderlich}, \& {LIFE
  Collaboration}}]{Konrad2022}
{Konrad}, B.~S., {Alei}, E., {Quanz}, S.~P., {et~al.} 2022, \aap, 664, A23,
  \dodoi{10.1051/0004-6361/202141964}

\bibitem[{Kopparapu {et~al.}(2014)Kopparapu, Ramirez, SchottelKotte, Kasting,
  Domagal-Goldman, \& Eymet}]{Kopparapu2014}
Kopparapu, R.~K., Ramirez, R.~M., SchottelKotte, J., {et~al.} 2014, The
  Astrophysical Journal, 787, L29, \dodoi{10.1088/2041-8205/787/2/l29}

\bibitem[{{Krist} {et~al.}(2010){Krist}, {Stapelfeldt}, {Bryden}, {Rieke},
  {Su}, {Chen}, {Beichman}, {Hines}, {Rebull}, {Tanner}, {Trilling}, {Clampin},
  \& {G{\'a}sp{\'a}r}}]{Krist2010}
{Krist}, J.~E., {Stapelfeldt}, K.~R., {Bryden}, G., {et~al.} 2010, \aj, 140,
  1051, \dodoi{10.1088/0004-6256/140/4/1051}

\bibitem[{{Lippincott}(1973)}]{Lippincott1973}
{Lippincott}, S.~L. 1973, \aj, 78, 303, \dodoi{10.1086/111418}

\bibitem[{{Lloyd Evans} \& {Koen}(1987)}]{LloydEvans1987}
{Lloyd Evans}, T., \& {Koen}, M.~C.~J. 1987, South African Astronomical
  Observatory Circular, 11, 21

\bibitem[{{Lockwood} {et~al.}(2007){Lockwood}, {Skiff}, {Henry}, {Henry},
  {Radick}, {Baliunas}, {Donahue}, \& {Soon}}]{Lockwood2007}
{Lockwood}, G.~W., {Skiff}, B.~A., {Henry}, G.~W., {et~al.} 2007, \apjs, 171,
  260, \dodoi{10.1086/516752}

\bibitem[{{Lovis} \& {Fischer}(2010)}]{Lovis2010}
{Lovis}, C., \& {Fischer}, D. 2010, in Exoplanets, ed. S.~{Seager} (University
  of Arizona Press), 27--53

\bibitem[{{Lovis} {et~al.}(2006){Lovis}, {Mayor}, {Pepe}, {Alibert}, {Benz},
  {Bouchy}, {Correia}, {Laskar}, {Mordasini}, {Queloz}, {Santos}, {Udry},
  {Bertaux}, \& {Sivan}}]{Lovis2006}
{Lovis}, C., {Mayor}, M., {Pepe}, F., {et~al.} 2006, \nat, 441, 305,
  \dodoi{10.1038/nature04828}

\bibitem[{{Lovis} {et~al.}(2011){Lovis}, {Dumusque}, {Santos}, {Bouchy},
  {Mayor}, {Pepe}, {Queloz}, {S{\'e}gransan}, \& {Udry}}]{Lovis2011}
{Lovis}, C., {Dumusque}, X., {Santos}, N.~C., {et~al.} 2011, arXiv e-prints,
  arXiv:1107.5325.
\newblock \doarXiv{1107.5325}

\bibitem[{{Luck}(2017)}]{Luck2017}
{Luck}, R.~E. 2017, \aj, 153, 21, \dodoi{10.3847/1538-3881/153/1/21}

\bibitem[{{Luhn} {et~al.}(2020){Luhn}, {Wright}, {Howard}, \&
  {Isaacson}}]{Luhn2020}
{Luhn}, J.~K., {Wright}, J.~T., {Howard}, A.~W., \& {Isaacson}, H. 2020, \aj,
  159, 235, \dodoi{10.3847/1538-3881/ab855a}

\bibitem[{{Ma} {et~al.}(2018){Ma}, {Ge}, {Muterspaugh}, {Singer}, {Henry},
  {Gonz{\'a}lez Hern{\'a}ndez}, {Sithajan}, {Jeram}, {Williamson}, {Stassun},
  {Kimock}, {Varosi}, {Schofield}, {Liu}, {Powell}, {Cassette}, {Jakeman},
  {Avner}, {Grieves}, {Barnes}, {Zhao}, {Gilda}, {Grantham}, {Stafford},
  {Savage}, {Bland}, \& {Ealey}}]{Ma2018}
{Ma}, B., {Ge}, J., {Muterspaugh}, M., {et~al.} 2018, \mnras, 480, 2411,
  \dodoi{10.1093/mnras/sty1933}

\bibitem[{{Mahdi} {et~al.}(2016){Mahdi}, {Soubiran}, {Blanco-Cuaresma}, \&
  {Chemin}}]{Mahdi2016}
{Mahdi}, D., {Soubiran}, C., {Blanco-Cuaresma}, S., \& {Chemin}, L. 2016, \aap,
  587, A131, \dodoi{10.1051/0004-6361/201527472}

\bibitem[{{Makarov}(2010)}]{Makarov2010}
{Makarov}, V.~V. 2010, \apj, 715, 500, \dodoi{10.1088/0004-637X/715/1/500}

\bibitem[{Makarov {et~al.}(2021)Makarov, Zacharias, \& Finch}]{Makarov2021}
Makarov, V.~V., Zacharias, N., \& Finch, C.~T. 2021, Looking for astrometric
  signals below 20 m/s: A candidate exo-Jupiter in $\delta$ Pav.
\newblock \doarXiv{2105.03244}

\bibitem[{{Maldonado} {et~al.}(2012){Maldonado}, {Eiroa}, {Villaver},
  {Montesinos}, \& {Mora}}]{Maldonado2012}
{Maldonado}, J., {Eiroa}, C., {Villaver}, E., {Montesinos}, B., \& {Mora}, A.
  2012, \aap, 541, A40, \dodoi{10.1051/0004-6361/201218800}

\bibitem[{{Maldonado} \& {Villaver}(2016)}]{Maldonado2016}
{Maldonado}, J., \& {Villaver}, E. 2016, \aap, 588, A98,
  \dodoi{10.1051/0004-6361/201527883}

\bibitem[{{Maldonado} {et~al.}(2013){Maldonado}, {Villaver}, \&
  {Eiroa}}]{Maldonado2013}
{Maldonado}, J., {Villaver}, E., \& {Eiroa}, C. 2013, \aap, 554, A84,
  \dodoi{10.1051/0004-6361/201321082}

\bibitem[{{Mamajek} \& {Hillenbrand}(2008)}]{Mamajek2008}
{Mamajek}, E.~E., \& {Hillenbrand}, L.~A. 2008, \apj, 687, 1264,
  \dodoi{10.1086/591785}

\bibitem[{{Mamajek} {et~al.}(2013){Mamajek}, {Bartlett}, {Seifahrt}, {Henry},
  {Dieterich}, {Lurie}, {Kenworthy}, {Jao}, {Riedel}, {Subasavage}, {Winters},
  {Finch}, {Ianna}, \& {Bean}}]{Mamajek2013}
{Mamajek}, E.~E., {Bartlett}, J.~L., {Seifahrt}, A., {et~al.} 2013, \aj, 146,
  154, \dodoi{10.1088/0004-6256/146/6/154}

\bibitem[{{Mann} {et~al.}(2015){Mann}, {Feiden}, {Gaidos}, {Boyajian}, \& {von
  Braun}}]{Mann2015}
{Mann}, A.~W., {Feiden}, G.~A., {Gaidos}, E., {Boyajian}, T., \& {von Braun},
  K. 2015, \apj, 804, 64, \dodoi{10.1088/0004-637X/804/1/64}

\bibitem[{{Marcy}(2007)}]{Marcy2007}
{Marcy}, G. 2007, in In the Spirit of Bernard Lyot: The Direct Detection of
  Planets and Circumstellar Disks in the 21st Century, 21

\bibitem[{{Marshall} {et~al.}(2011){Marshall}, {L{\"o}hne}, {Montesinos},
  {Krivov}, {Eiroa}, {Absil}, {Bryden}, {Maldonado}, {Mora}, {Sanz-Forcada},
  {Ardila}, {Augereau}, {Bayo}, {Del Burgo}, {Danchi}, {Ertel}, {Fedele},
  {Fridlund}, {Lebreton}, {Gonz{\'a}lez-Garc{\'\i}a}, {Liseau}, {Meeus},
  {M{\"u}ller}, {Pilbratt}, {Roberge}, {Stapelfeldt}, {Th{\'e}bault}, {White},
  \& {Wolf}}]{Marshall2011}
{Marshall}, J.~P., {L{\"o}hne}, T., {Montesinos}, B., {et~al.} 2011, \aap, 529,
  A117, \dodoi{10.1051/0004-6361/201116673}

\bibitem[{{Mason} {et~al.}(2001){Mason}, {Wycoff}, {Hartkopf}, {Douglass}, \&
  {Worley}}]{Mason2001}
{Mason}, B.~D., {Wycoff}, G.~L., {Hartkopf}, W.~I., {Douglass}, G.~G., \&
  {Worley}, C.~E. 2001, \aj, 122, 3466, \dodoi{10.1086/323920}

\bibitem[{{Mawet} {et~al.}(2019){Mawet}, {Hirsch}, {Lee}, {Ruffio}, {Bottom},
  {Fulton}, {Absil}, {Beichman}, {Bowler}, {Bryan}, {Choquet}, {Ciardi},
  {Christiaens}, {Defr{\`e}re}, {Gomez Gonzalez}, {Howard}, {Huby}, {Isaacson},
  {Jensen-Clem}, {Kosiarek}, {Marcy}, {Meshkat}, {Petigura}, {Reggiani},
  {Ruane}, {Serabyn}, {Sinukoff}, {Wang}, {Weiss}, \& {Ygouf}}]{Mawet2019}
{Mawet}, D., {Hirsch}, L., {Lee}, E.~J., {et~al.} 2019, \aj, 157, 33,
  \dodoi{10.3847/1538-3881/aaef8a}

\bibitem[{{Mayor} {et~al.}(2003){Mayor}, {Pepe}, {Queloz}, {Bouchy},
  {Rupprecht}, {Lo Curto}, {Avila}, {Benz}, {Bertaux}, {Bonfils}, {Dall},
  {Dekker}, {Delabre}, {Eckert}, {Fleury}, {Gilliotte}, {Gojak}, {Guzman},
  {Kohler}, {Lizon}, {Longinotti}, {Lovis}, {Megevand}, {Pasquini}, {Reyes},
  {Sivan}, {Sosnowska}, {Soto}, {Udry}, {van Kesteren}, {Weber}, \&
  {Weilenmann}}]{Mayor2003}
{Mayor}, M., {Pepe}, F., {Queloz}, D., {et~al.} 2003, The Messenger, 114, 20

\bibitem[{{Mermilliod}(2006)}]{Mermilliod1991}
{Mermilliod}, J.~C. 2006, VizieR Online Data Catalog, II/168

\bibitem[{{Metcalfe} {et~al.}(2013){Metcalfe}, {Buccino}, {Brown}, {Mathur},
  {Soderblom}, {Henry}, {Mauas}, {Petrucci}, {Hall}, \& {Basu}}]{Metcalfe2013}
{Metcalfe}, T.~S., {Buccino}, A.~P., {Brown}, B.~P., {et~al.} 2013, \apjl, 763,
  L26, \dodoi{10.1088/2041-8205/763/2/L26}

\bibitem[{{Meunier} {et~al.}(2010){Meunier}, {Desort}, \&
  {Lagrange}}]{Meunier2010}
{Meunier}, N., {Desort}, M., \& {Lagrange}, A.~M. 2010, \aap, 512, A39,
  \dodoi{10.1051/0004-6361/200913551}

\bibitem[{{Meunier} {et~al.}(2022){Meunier}, {Kretzschmar}, {Gravet}, {Mignon},
  \& {Delfosse}}]{Meunier2022}
{Meunier}, N., {Kretzschmar}, M., {Gravet}, R., {Mignon}, L., \& {Delfosse}, X.
  2022, \aap, 658, A57, \dodoi{10.1051/0004-6361/202142120}

\bibitem[{{Meunier} \& {Lagrange}(2019)}]{Meunier2019}
{Meunier}, N., \& {Lagrange}, A.~M. 2019, \aap, 625, L6,
  \dodoi{10.1051/0004-6361/201935099}

\bibitem[{{Meunier} \& {Lagrange}(2020)}]{MeunierLagrange2020}
---. 2020, \aap, 638, A54, \dodoi{10.1051/0004-6361/201937354}

\bibitem[{{Meunier} {et~al.}(2015){Meunier}, {Lagrange}, {Borgniet}, \&
  {Rieutord}}]{Meunier2015}
{Meunier}, N., {Lagrange}, A.~M., {Borgniet}, S., \& {Rieutord}, M. 2015, \aap,
  583, A118, \dodoi{10.1051/0004-6361/201525721}

\bibitem[{{Meunier} {et~al.}(2017){Meunier}, {Lagrange}, {Mbemba Kabuiku},
  {Alex}, {Mignon}, \& {Borgniet}}]{Meunier2017}
{Meunier}, N., {Lagrange}, A.~M., {Mbemba Kabuiku}, L., {et~al.} 2017, \aap,
  597, A52, \dodoi{10.1051/0004-6361/201629052}

\bibitem[{{Mishenina} {et~al.}(2012){Mishenina}, {Soubiran}, {Kovtyukh},
  {Katsova}, \& {Livshits}}]{Mishenina2012}
{Mishenina}, T.~V., {Soubiran}, C., {Kovtyukh}, V.~V., {Katsova}, M.~M., \&
  {Livshits}, M.~A. 2012, \aap, 547, A106, \dodoi{10.1051/0004-6361/201118412}

\bibitem[{{Mishenina} {et~al.}(2004){Mishenina}, {Soubiran}, {Kovtyukh}, \&
  {Korotin}}]{Mishenina2004}
{Mishenina}, T.~V., {Soubiran}, C., {Kovtyukh}, V.~V., \& {Korotin}, S.~A.
  2004, \aap, 418, 551, \dodoi{10.1051/0004-6361:20034454}

\bibitem[{{Mittag} {et~al.}(2013){Mittag}, {Schmitt}, \&
  {Schr{\"o}der}}]{Mittag2013}
{Mittag}, M., {Schmitt}, J.~H.~M.~M., \& {Schr{\"o}der}, K.~P. 2013, \aap, 549,
  A117, \dodoi{10.1051/0004-6361/201219868}

\bibitem[{{Montes} {et~al.}(2018){Montes}, {Gonz{\'a}lez-Peinado}, {Tabernero},
  {Caballero}, {Marfil}, {Alonso-Floriano}, {Cort{\'e}s-Contreras},
  {Gonz{\'a}lez Hern{\'a}ndez}, {Klutsch}, \& {Moreno-J{\'o}dar}}]{Montes2018}
{Montes}, D., {Gonz{\'a}lez-Peinado}, R., {Tabernero}, H.~M., {et~al.} 2018,
  \mnras, 479, 1332, \dodoi{10.1093/mnras/sty1295}

\bibitem[{{Morel} {et~al.}(2004){Morel}, {Micela}, {Favata}, \&
  {Katz}}]{Morel2004}
{Morel}, T., {Micela}, G., {Favata}, F., \& {Katz}, D. 2004, \aap, 426, 1007,
  \dodoi{10.1051/0004-6361:20041047}

\bibitem[{{Mortier} \& {Collier Cameron}(2017)}]{Mortier2017}
{Mortier}, A., \& {Collier Cameron}, A. 2017, \aap, 601, A110,
  \dodoi{10.1051/0004-6361/201630201}

\bibitem[{{Mugrauer}(2019)}]{Mugrauer2019}
{Mugrauer}, M. 2019, \mnras, 490, 5088, \dodoi{10.1093/mnras/stz2673}

\bibitem[{{National Academies of Sciences, Engineering, and
  Medicine}(2018)}]{ESS}
{National Academies of Sciences, Engineering, and Medicine}. 2018, {Exoplanet
  Science Strategy} ({Washington, DC}: {The National Academies Press}).
\newblock \url{https://doi.org/10.17226/25187}

\bibitem[{{National Academies of Sciences, Engineering, and
  Medicine}(2021)}]{Astro2020}
---. 2021, {Pathways to Discovery in Astronomy and Astrophysics for the 2020s}
  (The National Academies Press)

\bibitem[{{Nissen} {et~al.}(2020){Nissen}, {Christensen-Dalsgaard},
  {Mosumgaard}, {Silva Aguirre}, {Spitoni}, \& {Verma}}]{Nissen2020}
{Nissen}, P.~E., {Christensen-Dalsgaard}, J., {Mosumgaard}, J.~R., {et~al.}
  2020, \aap, 640, A81, \dodoi{10.1051/0004-6361/202038300}

\bibitem[{{Nordlund} {et~al.}(2009){Nordlund}, {Stein}, \&
  {Asplund}}]{Nordlund2009}
{Nordlund}, {\r{A}}., {Stein}, R.~F., \& {Asplund}, M. 2009, Living Reviews in
  Solar Physics, 6, 2, \dodoi{10.12942/lrsp-2009-2}

\bibitem[{{Noyes} {et~al.}(1984){Noyes}, {Hartmann}, {Baliunas}, {Duncan}, \&
  {Vaughan}}]{Noyes1984}
{Noyes}, R.~W., {Hartmann}, L.~W., {Baliunas}, S.~L., {Duncan}, D.~K., \&
  {Vaughan}, A.~H. 1984, \apj, 279, 763, \dodoi{10.1086/161945}

\bibitem[{{Oelkers} {et~al.}(2018){Oelkers}, {Rodriguez}, {Stassun}, {Pepper},
  {Somers}, {Kafka}, {Stevens}, {Beatty}, {Siverd}, {Lund}, {Kuhn}, {James}, \&
  {Gaudi}}]{Oelkers2018}
{Oelkers}, R.~J., {Rodriguez}, J.~E., {Stassun}, K.~G., {et~al.} 2018, \aj,
  155, 39, \dodoi{10.3847/1538-3881/aa9bf4}

\bibitem[{{Ol{\'a}h} {et~al.}(2016){Ol{\'a}h}, {K{\H{o}}v{\'a}ri}, {Petrovay},
  {Soon}, {Baliunas}, {Koll{\'a}th}, \& {Vida}}]{Olah2016}
{Ol{\'a}h}, K., {K{\H{o}}v{\'a}ri}, Z., {Petrovay}, K., {et~al.} 2016, \aap,
  590, A133, \dodoi{10.1051/0004-6361/201628479}

\bibitem[{{Pepe} {et~al.}(2002){Pepe}, {Mayor}, {Rupprecht}, {Avila},
  {Ballester}, {Beckers}, {Benz}, {Bertaux}, {Bouchy}, {Buzzoni}, {Cavadore},
  {Deiries}, {Dekker}, {Delabre}, {D'Odorico}, {Eckert}, {Fischer}, {Fleury},
  {George}, {Gilliotte}, {Gojak}, {Guzman}, {Koch}, {Kohler}, {Kotzlowski},
  {Lacroix}, {Le Merrer}, {Lizon}, {Lo Curto}, {Longinotti}, {Megevand},
  {Pasquini}, {Petitpas}, {Pichard}, {Queloz}, {Reyes}, {Richaud}, {Sivan},
  {Sosnowska}, {Soto}, {Udry}, {Ureta}, {van Kesteren}, {Weber}, {Weilenmann},
  {Wicenec}, {Wieland}, {Christensen-Dalsgaard}, {Dravins}, {Hatzes},
  {K{\"u}rster}, {Paresce}, \& {Penny}}]{Pepe2002}
{Pepe}, F., {Mayor}, M., {Rupprecht}, G., {et~al.} 2002, The Messenger, 110, 9

\bibitem[{{Pepe} {et~al.}(2007){Pepe}, {Correia}, {Mayor}, {Tamuz}, {Couetdic},
  {Benz}, {Bertaux}, {Bouchy}, {Laskar}, {Lovis}, {Naef}, {Queloz}, {Santos},
  {Sivan}, {Sosnowska}, \& {Udry}}]{Pepe2007}
{Pepe}, F., {Correia}, A.~C.~M., {Mayor}, M., {et~al.} 2007, \aap, 462, 769,
  \dodoi{10.1051/0004-6361:20066194}

\bibitem[{{Pepe} {et~al.}(2011){Pepe}, {Lovis}, {S{\'e}gransan}, {Benz},
  {Bouchy}, {Dumusque}, {Mayor}, {Queloz}, {Santos}, \& {Udry}}]{Pepe2011}
{Pepe}, F., {Lovis}, C., {S{\'e}gransan}, D., {et~al.} 2011, \aap, 534, A58,
  \dodoi{10.1051/0004-6361/201117055}

\bibitem[{{Pepe} {et~al.}(2021){Pepe}, {Cristiani}, {Rebolo}, {Santos},
  {Dekker}, {Cabral}, {Di Marcantonio}, {Figueira}, {Lo Curto}, {Lovis},
  {Mayor}, {M{\'e}gevand}, {Molaro}, {Riva}, {Zapatero Osorio}, {Amate},
  {Manescau}, {Pasquini}, {Zerbi}, {Adibekyan}, {Abreu}, {Affolter}, {Alibert},
  {Aliverti}, {Allart}, {Allende Prieto}, {{\'A}lvarez}, {Alves}, {Avila},
  {Baldini}, {Bandy}, {Barros}, {Benz}, {Bianco}, {Borsa}, {Bourrier},
  {Bouchy}, {Broeg}, {Calderone}, {Cirami}, {Coelho}, {Conconi}, {Coretti},
  {Cumani}, {Cupani}, {D'Odorico}, {Damasso}, {Deiries}, {Delabre},
  {Demangeon}, {Dumusque}, {Ehrenreich}, {Faria}, {Fragoso}, {Genolet},
  {Genoni}, {G{\'e}nova Santos}, {Gonz{\'a}lez Hern{\'a}ndez}, {Hughes},
  {Iwert}, {Kerber}, {Knudstrup}, {Landoni}, {Lavie}, {Lillo-Box}, {Lizon},
  {Maire}, {Martins}, {Mehner}, {Micela}, {Modigliani}, {Monteiro}, {Monteiro},
  {Moschetti}, {Murphy}, {Nunes}, {Oggioni}, {Oliveira}, {Oshagh}, {Pall{\'e}},
  {Pariani}, {Poretti}, {Rasilla}, {Rebord{\~a}o}, {Redaelli}, {Santana
  Tschudi}, {Santin}, {Santos}, {S{\'e}gransan}, {Schmidt}, {Segovia},
  {Sosnowska}, {Sozzetti}, {Sousa}, {Span{\`o}}, {Su{\'a}rez Mascare{\~n}o},
  {Tabernero}, {Tenegi}, {Udry}, \& {Zanutta}}]{Pepe2021}
{Pepe}, F., {Cristiani}, S., {Rebolo}, R., {et~al.} 2021, \aap, 645, A96,
  \dodoi{10.1051/0004-6361/202038306}

\bibitem[{{Perdelwitz} {et~al.}(2021){Perdelwitz}, {Mittag}, {Tal-Or},
  {Schmitt}, {Caballero}, {Jeffers}, {Reiners}, {Schweitzer}, {Trifonov},
  {Ribas}, {Quirrenbach}, {Amado}, {Seifert}, {Cifuentes},
  {Cort{\'e}s-Contreras}, {Montes}, {Revilla}, \&
  {Skrzypinski}}]{Perdelwitz2021}
{Perdelwitz}, V., {Mittag}, M., {Tal-Or}, L., {et~al.} 2021, \aap, 652, A116,
  \dodoi{10.1051/0004-6361/202140889}

\bibitem[{{Pourbaix} {et~al.}(2004){Pourbaix}, {Tokovinin}, {Batten}, {Fekel},
  {Hartkopf}, {Levato}, {Morrell}, {Torres}, \& {Udry}}]{Pourbaix2004}
{Pourbaix}, D., {Tokovinin}, A.~A., {Batten}, A.~H., {et~al.} 2004, \aap, 424,
  727, \dodoi{10.1051/0004-6361:20041213}

\bibitem[{{Quanz} {et~al.}(2022){Quanz}, {Ottiger}, {Fontanet}, {Kammerer},
  {Menti}, {Dannert}, {Gheorghe}, {Absil}, {Airapetian}, {Alei}, {Allart},
  {Angerhausen}, {Blumenthal}, {Buchhave}, {Cabrera},
  {Carri{\'o}n-Gonz{\'a}lez}, {Chauvin}, {Danchi}, {Dandumont}, {Defr{\'e}re},
  {Dorn}, {Ehrenreich}, {Ertel}, {Fridlund}, {Garc{\'\i}a Mu{\~n}oz},
  {Gasc{\'o}n}, {Girard}, {Glauser}, {Grenfell}, {Guidi}, {Hagelberg},
  {Helled}, {Ireland}, {Janson}, {Kopparapu}, {Korth}, {Kozakis}, {Kraus},
  {L{\'e}ger}, {Leedj{\"a}rv}, {Lichtenberg}, {Lillo-Box}, {Linz}, {Liseau},
  {Loicq}, {Mahendra}, {Malbet}, {Mathew}, {Mennesson}, {Meyer}, {Mishra},
  {Molaverdikhani}, {Noack}, {Oza}, {Pall{\'e}}, {Parviainen}, {Quirrenbach},
  {Rauer}, {Ribas}, {Rice}, {Romagnolo}, {Rugheimer}, {Schwieterman},
  {Serabyn}, {Sharma}, {Stassun}, {Szul{\'a}gyi}, {Wang}, {Wunderlich},
  {Wyatt}, \& {LIFE Collaboration}}]{Quanz2022}
{Quanz}, S.~P., {Ottiger}, M., {Fontanet}, E., {et~al.} 2022, \aap, 664, A21,
  \dodoi{10.1051/0004-6361/202140366}

\bibitem[{{Quirrenbach}(2010)}]{Quirrenbach2010}
{Quirrenbach}, A. 2010, in Exoplanets, ed. S.~{Seager} (University of Arizona
  Press), 157--174

\bibitem[{{Radick} {et~al.}(2018){Radick}, {Lockwood}, {Henry}, {Hall}, \&
  {Pevtsov}}]{Radick2018}
{Radick}, R.~R., {Lockwood}, G.~W., {Henry}, G.~W., {Hall}, J.~C., \&
  {Pevtsov}, A.~A. 2018, \apj, 855, 75, \dodoi{10.3847/1538-4357/aaaae3}

\bibitem[{{Ram{\'\i}rez} {et~al.}(2013){Ram{\'\i}rez}, {Allende Prieto}, \&
  {Lambert}}]{Ramirez2013}
{Ram{\'\i}rez}, I., {Allende Prieto}, C., \& {Lambert}, D.~L. 2013, \apj, 764,
  78, \dodoi{10.1088/0004-637X/764/1/78}

\bibitem[{{Ram{\'\i}rez} {et~al.}(2012){Ram{\'\i}rez}, {Fish}, {Lambert}, \&
  {Allende Prieto}}]{Ramirez2012}
{Ram{\'\i}rez}, I., {Fish}, J.~R., {Lambert}, D.~L., \& {Allende Prieto}, C.
  2012, \apj, 756, 46, \dodoi{10.1088/0004-637X/756/1/46}

\bibitem[{{Ram{\'\i}rez} {et~al.}(2014){Ram{\'\i}rez}, {Mel{\'e}ndez}, {Bean},
  {Asplund}, {Bedell}, {Monroe}, {Casagrande}, {Schirbel}, {Dreizler}, {Teske},
  {Tucci Maia}, {Alves-Brito}, \& {Baumann}}]{Ramirez2014}
{Ram{\'\i}rez}, I., {Mel{\'e}ndez}, J., {Bean}, J., {et~al.} 2014, \aap, 572,
  A48, \dodoi{10.1051/0004-6361/201424244}

\bibitem[{{Reinhold} {et~al.}(2013){Reinhold}, {Reiners}, \&
  {Basri}}]{Reinhold2013}
{Reinhold}, T., {Reiners}, A., \& {Basri}, G. 2013, \aap, 560, A4,
  \dodoi{10.1051/0004-6361/201321970}

\bibitem[{{Robertson} {et~al.}(2013){Robertson}, {Endl}, {Cochran}, \&
  {Dodson-Robinson}}]{Robertson2013}
{Robertson}, P., {Endl}, M., {Cochran}, W.~D., \& {Dodson-Robinson}, S.~E.
  2013, \apj, 764, 3, \dodoi{10.1088/0004-637X/764/1/3}

\bibitem[{{Robertson} {et~al.}(2014){Robertson}, {Mahadevan}, {Endl}, \&
  {Roy}}]{Robertson2014}
{Robertson}, P., {Mahadevan}, S., {Endl}, M., \& {Roy}, A. 2014, Science, 345,
  440, \dodoi{10.1126/science.1253253}

\bibitem[{{Rosenthal} {et~al.}(2021){Rosenthal}, {Fulton}, {Hirsch},
  {Isaacson}, {Howard}, {Dedrick}, {Sherstyuk}, {Blunt}, {Petigura}, {Knutson},
  {Behmard}, {Chontos}, {Crepp}, {Crossfield}, {Dalba}, {Fischer}, {Henry},
  {Kane}, {Kosiarek}, {Marcy}, {Rubenzahl}, {Weiss}, \&
  {Wright}}]{Rosenthal2021}
{Rosenthal}, L.~J., {Fulton}, B.~J., {Hirsch}, L.~A., {et~al.} 2021, arXiv
  e-prints, arXiv:2105.11583.
\newblock \doarXiv{2105.11583}

\bibitem[{{Saar} \& {Brandenburg}(1999)}]{Saar1999}
{Saar}, S.~H., \& {Brandenburg}, A. 1999, \apj, 524, 295,
  \dodoi{10.1086/307794}

\bibitem[{{Saar} \& {Donahue}(1997)}]{SaarDonahue1997}
{Saar}, S.~H., \& {Donahue}, R.~A. 1997, \apj, 485, 319, \dodoi{10.1086/304392}

\bibitem[{{Saar} \& {Osten}(1997)}]{Saar1997}
{Saar}, S.~H., \& {Osten}, R.~A. 1997, \mnras, 284, 803,
  \dodoi{10.1093/mnras/284.4.803}

\bibitem[{{Samus'} {et~al.}(2017){Samus'}, {Kazarovets}, {Durlevich},
  {Kireeva}, \& {Pastukhova}}]{Samus2017}
{Samus'}, N.~N., {Kazarovets}, E.~V., {Durlevich}, O.~V., {Kireeva}, N.~N., \&
  {Pastukhova}, E.~N. 2017, Astronomy Reports, 61, 80,
  \dodoi{10.1134/S1063772917010085}

\bibitem[{{Santos} {et~al.}(2001){Santos}, {Israelian}, \&
  {Mayor}}]{Santos2001}
{Santos}, N.~C., {Israelian}, G., \& {Mayor}, M. 2001, \aap, 373, 1019,
  \dodoi{10.1051/0004-6361:20010648}

\bibitem[{{Santos} {et~al.}(2004){Santos}, {Israelian}, \&
  {Mayor}}]{Santos2004}
---. 2004, \aap, 415, 1153, \dodoi{10.1051/0004-6361:20034469}

\bibitem[{{Schnupp} {et~al.}(2010){Schnupp}, {Bergfors}, {Brandner}, {Daemgen},
  {Fischer}, {Marcy}, {Henning}, {Hippler}, \& {Janson}}]{Schnupp2010}
{Schnupp}, C., {Bergfors}, C., {Brandner}, W., {et~al.} 2010, \aap, 516, A21,
  \dodoi{10.1051/0004-6361/201014740}

\bibitem[{{Schofield} {et~al.}(2019){Schofield}, {Chaplin}, {Huber},
  {Campante}, {Davies}, {Miglio}, {Ball}, {Appourchaux}, {Basu}, {Bedding},
  {Christensen-Dalsgaard}, {Creevey}, {Garc{\'\i}a}, {Handberg}, {Kawaler},
  {Kjeldsen}, {Latham}, {Lund}, {Metcalfe}, {Ricker}, {Serenelli}, {Silva
  Aguirre}, {Stello}, \& {Vanderspek}}]{Schofield2019}
{Schofield}, M., {Chaplin}, W.~J., {Huber}, D., {et~al.} 2019, \apjs, 241, 12,
  \dodoi{10.3847/1538-4365/ab04f5}

\bibitem[{{Schrijver}(1987)}]{Schrijver1987}
{Schrijver}, C.~J. 1987, \aap, 172, 111

\bibitem[{{Schwabe}(1843)}]{Schwabe1843}
{Schwabe}, M. 1843, Astronomische Nachrichten, 20, 283,
  \dodoi{10.1002/asna.18430201706}

\bibitem[{{Scott} {et~al.}(2021){Scott}, {Howell}, {Gnilka}, {Stephens},
  {Salinas}, {Matson}, {Furlan}, {Horch}, {Everett}, {Ciardi}, {Mills}, \&
  {Quigley}}]{Scott2021}
{Scott}, N.~J., {Howell}, S.~B., {Gnilka}, C.~L., {et~al.} 2021, Frontiers in
  Astronomy and Space Sciences, 8, 138, \dodoi{10.3389/fspas.2021.716560}

\bibitem[{{Seager} {et~al.}(2018){Seager}, {Kasdin}, \& {Starshade Rendezvous
  Probe Team}}]{Seager2018}
{Seager}, S., {Kasdin}, J., \& {Starshade Rendezvous Probe Team}. 2018, in
  American Astronomical Society Meeting Abstracts, Vol. 231, American
  Astronomical Society Meeting Abstracts \#231, 121.09

\bibitem[{Seifahrt {et~al.}(2018)Seifahrt, Stürmer, Bean, \&
  Schwab}]{Seifahrt2018}
Seifahrt, A., Stürmer, J., Bean, J.~L., \& Schwab, C. 2018, MAROON-X: A Radial
  Velocity Spectrograph for the Gemini Observatory,  arXiv,
  \dodoi{10.48550/ARXIV.1805.09276}.
\newblock \url{https://arxiv.org/abs/1805.09276}

\bibitem[{{Sierchio} {et~al.}(2014){Sierchio}, {Rieke}, {Su}, \&
  {G{\'a}sp{\'a}r}}]{Sierchio2014}
{Sierchio}, J.~M., {Rieke}, G.~H., {Su}, K.~Y.~L., \& {G{\'a}sp{\'a}r}, A.
  2014, \apj, 785, 33, \dodoi{10.1088/0004-637X/785/1/33}

\bibitem[{Simpson {et~al.}(2010)Simpson, Baliunas, Henry, \&
  Watson}]{Simpson2010}
Simpson, E.~K., Baliunas, S.~L., Henry, G.~W., \& Watson, C.~A. 2010, Monthly
  Notices of the Royal Astronomical Society, 408, 1666–1679,
  \dodoi{10.1111/j.1365-2966.2010.17230.x}

\bibitem[{{Soriano} \& {Vauclair}(2010)}]{Soriano2010}
{Soriano}, M., \& {Vauclair}, S. 2010, \aap, 513, A49,
  \dodoi{10.1051/0004-6361/200911862}

\bibitem[{{Soto} \& {Jenkins}(2018)}]{Soto2018}
{Soto}, M.~G., \& {Jenkins}, J.~S. 2018, \aap, 615, A76,
  \dodoi{10.1051/0004-6361/201731533}

\bibitem[{{Soubiran} {et~al.}(2022){Soubiran}, {Brouillet}, \&
  {Casamiquela}}]{Soubiran2022}
{Soubiran}, C., {Brouillet}, N., \& {Casamiquela}, L. 2022, \aap, 663, A4,
  \dodoi{10.1051/0004-6361/202142409}

\bibitem[{{Soubiran} \& {Girard}(2005)}]{Soubiran2005}
{Soubiran}, C., \& {Girard}, P. 2005, \aap, 438, 139,
  \dodoi{10.1051/0004-6361:20042390}

\bibitem[{{Soubiran} {et~al.}(2018){Soubiran}, {Jasniewicz}, {Chemin},
  {Zurbach}, {Brouillet}, {Panuzzo}, {Sartoretti}, {Katz}, {Le Campion},
  {Marchal}, {Hestroffer}, {Th{\'e}venin}, {Crifo}, {Udry}, {Cropper},
  {Seabroke}, {Viala}, {Benson}, {Blomme}, {Jean-Antoine}, {Huckle}, {Smith},
  {Baker}, {Damerdji}, {Dolding}, {Fr{\'e}mat}, {Gosset}, {Guerrier}, {Guy},
  {Haigron}, {Jan{\ss}en}, {Plum}, {Fabre}, {Lasne}, {Pailler}, {Panem},
  {Riclet}, {Royer}, {Tauran}, {Zwitter}, {Gueguen}, \& {Turon}}]{Soubiran2018}
{Soubiran}, C., {Jasniewicz}, G., {Chemin}, L., {et~al.} 2018, \aap, 616, A7,
  \dodoi{10.1051/0004-6361/201832795}

\bibitem[{{Sousa} {et~al.}(2008){Sousa}, {Santos}, {Mayor}, {Udry},
  {Casagrande}, {Israelian}, {Pepe}, {Queloz}, \& {Monteiro}}]{Sousa2008}
{Sousa}, S.~G., {Santos}, N.~C., {Mayor}, M., {et~al.} 2008, \aap, 487, 373,
  \dodoi{10.1051/0004-6361:200809698}

\bibitem[{{Sousa} {et~al.}(2018){Sousa}, {Adibekyan}, {Delgado-Mena}, {Santos},
  {Andreasen}, {Ferreira}, {Tsantaki}, {Barros}, {Demangeon}, {Israelian},
  {Faria}, {Figueira}, {Mortier}, {Brand{\~a}o}, {Montalto}, {Rojas-Ayala}, \&
  {Santerne}}]{Sousa2018}
{Sousa}, S.~G., {Adibekyan}, V., {Delgado-Mena}, E., {et~al.} 2018, \aap, 620,
  A58, \dodoi{10.1051/0004-6361/201833350}

\bibitem[{{Spina} {et~al.}(2016){Spina}, {Mel{\'e}ndez}, \&
  {Ram{\'\i}rez}}]{Spina2016}
{Spina}, L., {Mel{\'e}ndez}, J., \& {Ram{\'\i}rez}, I. 2016, \aap, 585, A152,
  \dodoi{10.1051/0004-6361/201527429}

\bibitem[{{Spina} {et~al.}(2018){Spina}, {Mel{\'e}ndez}, {Karakas}, {dos
  Santos}, {Bedell}, {Asplund}, {Ram{\'\i}rez}, {Yong}, {Alves-Brito}, {Bean},
  \& {Dreizler}}]{Spina2018}
{Spina}, L., {Mel{\'e}ndez}, J., {Karakas}, A.~I., {et~al.} 2018, \mnras, 474,
  2580, \dodoi{10.1093/mnras/stx2938}

\bibitem[{{Stassun} {et~al.}(2019){Stassun}, {Oelkers}, {Paegert}, {Torres},
  {Pepper}, {De Lee}, {Collins}, {Latham}, {Muirhead}, {Chittidi},
  {Rojas-Ayala}, {Fleming}, {Rose}, {Tenenbaum}, {Ting}, {Kane}, {Barclay},
  {Bean}, {Brassuer}, {Charbonneau}, {Ge}, {Lissauer}, {Mann}, {McLean},
  {Mullally}, {Narita}, {Plavchan}, {Ricker}, {Sasselov}, {Seager}, {Sharma},
  {Shiao}, {Sozzetti}, {Stello}, {Vanderspek}, {Wallace}, \&
  {Winn}}]{Stassun2019}
{Stassun}, K.~G., {Oelkers}, R.~J., {Paegert}, M., {et~al.} 2019, \aj, 158,
  138, \dodoi{10.3847/1538-3881/ab3467}

\bibitem[{{Takeda} {et~al.}(2005){Takeda}, {Ohkubo}, {Sato}, {Kambe}, \&
  {Sadakane}}]{Takeda2005}
{Takeda}, Y., {Ohkubo}, M., {Sato}, B., {Kambe}, E., \& {Sadakane}, K. 2005,
  \pasj, 57, 27, \dodoi{10.1093/pasj/57.1.27}

\bibitem[{{The LUVOIR Team}(2019)}]{LUVOIRFinalReport}
{The LUVOIR Team}. 2019, arXiv e-prints, arXiv:1912.06219.
\newblock \doarXiv{1912.06219}

\bibitem[{{Th{\'e}venin} {et~al.}(2005){Th{\'e}venin}, {Kervella}, {Pichon},
  {Morel}, {di Folco}, \& {Lebreton}}]{Thevenin2005}
{Th{\'e}venin}, F., {Kervella}, P., {Pichon}, B., {et~al.} 2005, \aap, 436,
  253, \dodoi{10.1051/0004-6361:20042075}

\bibitem[{{Tian} {et~al.}(2020){Tian}, {El-Badry}, {Rix}, \&
  {Gould}}]{Tian2020}
{Tian}, H.-J., {El-Badry}, K., {Rix}, H.-W., \& {Gould}, A. 2020, \apjs, 246,
  4, \dodoi{10.3847/1538-4365/ab54c4}

\bibitem[{{Tinney} {et~al.}(2011){Tinney}, {Butler}, {Jones}, {Wittenmyer},
  {O'Toole}, {Bailey}, \& {Carter}}]{Tinney2011}
{Tinney}, C.~G., {Butler}, R.~P., {Jones}, H. R.~A., {et~al.} 2011, \apj, 727,
  103, \dodoi{10.1088/0004-637X/727/2/103}

\bibitem[{{Tokovinin}(2014)}]{Tokovinin2014}
{Tokovinin}, A. 2014, \aj, 147, 86, \dodoi{10.1088/0004-6256/147/4/86}

\bibitem[{{Tokovinin}(1991)}]{Tokovinin1991}
{Tokovinin}, A.~A. 1991, \aaps, 91, 497

\bibitem[{{Trifonov} {et~al.}(2020){Trifonov}, {Tal-Or}, {Zechmeister},
  {Kaminski}, {Zucker}, \& {Mazeh}}]{Trifonov2020}
{Trifonov}, T., {Tal-Or}, L., {Zechmeister}, M., {et~al.} 2020, \aap, 636, A74,
  \dodoi{10.1051/0004-6361/201936686}

\bibitem[{{Trifonov} {et~al.}(2021){Trifonov}, {Caballero}, {Morales},
  {Seifahrt}, {Ribas}, {Reiners}, {Bean}, {Luque}, {Parviainen}, {Pall{\'e}},
  {Stock}, {Zechmeister}, {Amado}, {Anglada-Escud{\'e}}, {Azzaro}, {Barclay},
  {B{\'e}jar}, {Bluhm}, {Casasayas-Barris}, {Cifuentes}, {Collins}, {Collins},
  {Cort{\'e}s-Contreras}, {de Leon}, {Dreizler}, {Dressing}, {Esparza-Borges},
  {Espinoza}, {Fausnaugh}, {Fukui}, {Hatzes}, {Hellier}, {Henning}, {Henze},
  {Herrero}, {Jeffers}, {Jenkins}, {Jensen}, {Kaminski}, {Kasper},
  {Kossakowski}, {K{\"u}rster}, {Lafarga}, {Latham}, {Mann}, {Molaverdikhani},
  {Montes}, {Montet}, {Murgas}, {Narita}, {Oshagh}, {Passegger}, {Pollacco},
  {Quinn}, {Quirrenbach}, {Ricker}, {Rodr{\'\i}guez L{\'o}pez}, {Sanz-Forcada},
  {Schwarz}, {Schweitzer}, {Seager}, {Shporer}, {Stangret}, {St{\"u}rmer},
  {Tan}, {Tenenbaum}, {Twicken}, {Vanderspek}, \& {Winn}}]{Trifonov2021}
{Trifonov}, T., {Caballero}, J.~A., {Morales}, J.~C., {et~al.} 2021, Science,
  371, 1038, \dodoi{10.1126/science.abd7645}

\bibitem[{{Trilling} {et~al.}(2008){Trilling}, {Bryden}, {Beichman}, {Rieke},
  {Su}, {Stansberry}, {Blaylock}, {Stapelfeldt}, {Beeman}, \&
  {Haller}}]{Trilling2008}
{Trilling}, D.~E., {Bryden}, G., {Beichman}, C.~A., {et~al.} 2008, \apj, 674,
  1086, \dodoi{10.1086/525514}

\bibitem[{{Tsantaki} {et~al.}(2013){Tsantaki}, {Sousa}, {Adibekyan}, {Santos},
  {Mortier}, \& {Israelian}}]{Tsantaki2013}
{Tsantaki}, M., {Sousa}, S.~G., {Adibekyan}, V.~Z., {et~al.} 2013, \aap, 555,
  A150, \dodoi{10.1051/0004-6361/201321103}

\bibitem[{{Udry} {et~al.}(2019){Udry}, {Dumusque}, {Lovis}, {S{\'e}gransan},
  {Diaz}, {Benz}, {Bouchy}, {Coffinet}, {Lo Curto}, {Mayor}, {Mordasini},
  {Motalebi}, {Pepe}, {Queloz}, {Santos}, {Wyttenbach}, {Alonso}, {Collier
  Cameron}, {Deleuil}, {Figueira}, {Gillon}, {Moutou}, {Pollacco}, \&
  {Pompei}}]{Udry2019}
{Udry}, S., {Dumusque}, X., {Lovis}, C., {et~al.} 2019, \aap, 622, A37,
  \dodoi{10.1051/0004-6361/201731173}

\bibitem[{{Valenti} \& {Fischer}(2005)}]{Valenti2005}
{Valenti}, J.~A., \& {Fischer}, D.~A. 2005, \apjs, 159, 141,
  \dodoi{10.1086/430500}

\bibitem[{{van Leeuwen}(2007)}]{vanLeeuwen2007}
{van Leeuwen}, F. 2007, \aap, 474, 653, \dodoi{10.1051/0004-6361:20078357}

\bibitem[{{van Maanen}(1938)}]{vanMaanen1938}
{van Maanen}, A. 1938, \apj, 88, 27, \dodoi{10.1086/143957}

\bibitem[{{Vidotto} {et~al.}(2014){Vidotto}, {Gregory}, {Jardine}, {Donati},
  {Petit}, {Morin}, {Folsom}, {Bouvier}, {Cameron}, {Hussain}, {Marsden},
  {Waite}, {Fares}, {Jeffers}, \& {do Nascimento}}]{Vidotto2014}
{Vidotto}, A.~A., {Gregory}, S.~G., {Jardine}, M., {et~al.} 2014, \mnras, 441,
  2361, \dodoi{10.1093/mnras/stu728}

\bibitem[{{Vogt} {et~al.}(1994){Vogt}, {Allen}, {Bigelow}, {Bresee}, {Brown},
  {Cantrall}, {Conrad}, {Couture}, {Delaney}, {Epps}, {Hilyard}, {Hilyard},
  {Horn}, {Jern}, {Kanto}, {Keane}, {Kibrick}, {Lewis}, {Osborne},
  {Pardeilhan}, {Pfister}, {Ricketts}, {Robinson}, {Stover}, {Tucker}, {Ward},
  \& {Wei}}]{Vogt1994}
{Vogt}, S.~S., {Allen}, S.~L., {Bigelow}, B.~C., {et~al.} 1994, in Society of
  Photo-Optical Instrumentation Engineers (SPIE) Conference Series, Vol. 2198,
  Instrumentation in Astronomy VIII, ed. D.~L. {Crawford} \& E.~R. {Craine},
  362

\bibitem[{{Vogt} {et~al.}(2010){Vogt}, {Wittenmyer}, {Butler}, {O'Toole},
  {Henry}, {Rivera}, {Meschiari}, {Laughlin}, {Tinney}, {Jones}, {Bailey},
  {Carter}, \& {Batygin}}]{Vogt2010}
{Vogt}, S.~S., {Wittenmyer}, R.~A., {Butler}, R.~P., {et~al.} 2010, \apj, 708,
  1366, \dodoi{10.1088/0004-637X/708/2/1366}

\bibitem[{{Vogt} {et~al.}(2014){Vogt}, {Radovan}, {Kibrick}, {Butler},
  {Alcott}, {Allen}, {Arriagada}, {Bolte}, {Burt}, {Cabak}, {Chloros},
  {Cowley}, {Deich}, {Dupraw}, {Earthman}, {Epps}, {Faber}, {Fischer}, {Gates},
  {Hilyard}, {Holden}, {Johnston}, {Keiser}, {Kanto}, {Katsuki}, {Laiterman},
  {Lanclos}, {Laughlin}, {Lewis}, {Lockwood}, {Lynam}, {Marcy}, {McLean},
  {Miller}, {Misch}, {Peck}, {Pfister}, {Phillips}, {Rivera}, {Sandford},
  {Saylor}, {Stover}, {Thompson}, {Walp}, {Ward}, {Wareham}, {Wei}, \&
  {Wright}}]{Vogt2014}
{Vogt}, S.~S., {Radovan}, M., {Kibrick}, R., {et~al.} 2014, \pasp, 126, 359,
  \dodoi{10.1086/676120}

\bibitem[{{Watson} {et~al.}(2011){Watson}, {Littlefair}, {Diamond}, {Collier
  Cameron}, {Fitzsimmons}, {Simpson}, {Moulds}, \& {Pollacco}}]{Watson2011}
{Watson}, C.~A., {Littlefair}, S.~P., {Diamond}, C., {et~al.} 2011, \mnras,
  413, L71, \dodoi{10.1111/j.1745-3933.2011.01036.x}

\bibitem[{{Willamo} {et~al.}(2020){Willamo}, {Hackman}, {Lehtinen},
  {K{\"a}pyl{\"a}}, {Olspert}, {Viviani}, \& {Warnecke}}]{Willamo2020}
{Willamo}, T., {Hackman}, T., {Lehtinen}, J.~J., {et~al.} 2020, \aap, 638, A69,
  \dodoi{10.1051/0004-6361/202037666}

\bibitem[{{Wilson}(1968)}]{Wilson1968}
{Wilson}, O.~C. 1968, \apj, 153, 221, \dodoi{10.1086/149652}

\bibitem[{{Wittenmyer} {et~al.}(2006){Wittenmyer}, {Endl}, {Cochran}, {Hatzes},
  {Walker}, {Yang}, \& {Paulson}}]{Wittenmyer2006}
{Wittenmyer}, R.~A., {Endl}, M., {Cochran}, W.~D., {et~al.} 2006, \aj, 132,
  177, \dodoi{10.1086/504942}

\bibitem[{{Wittenmyer} {et~al.}(2017){Wittenmyer}, {Jones}, {Zhao}, {Marshall},
  {Butler}, {Tinney}, {Wang}, \& {Johnson}}]{Wittenmyer2017}
{Wittenmyer}, R.~A., {Jones}, M.~I., {Zhao}, J., {et~al.} 2017, \aj, 153, 51,
  \dodoi{10.3847/1538-3881/153/2/51}

\bibitem[{{Wittenmyer} {et~al.}(2014){Wittenmyer}, {Horner}, {Tinney},
  {Butler}, {Jones}, {Tuomi}, {Salter}, {Carter}, {Koch}, {O'Toole}, {Bailey},
  \& {Wright}}]{Wittenmyer2014}
{Wittenmyer}, R.~A., {Horner}, J., {Tinney}, C.~G., {et~al.} 2014, \apj, 783,
  103, \dodoi{10.1088/0004-637X/783/2/103}

\bibitem[{{Wright} {et~al.}(2011){Wright}, {Drake}, {Mamajek}, \&
  {Henry}}]{Wright2011}
{Wright}, N.~J., {Drake}, J.~J., {Mamajek}, E.~E., \& {Henry}, G.~W. 2011,
  \apj, 743, 48, \dodoi{10.1088/0004-637X/743/1/48}

\bibitem[{{Xuan} \& {Wyatt}(2020)}]{Xuan2020}
{Xuan}, J.~W., \& {Wyatt}, M.~C. 2020, \mnras, 497, 2096,
  \dodoi{10.1093/mnras/staa2033}

\bibitem[{{Zacharias} {et~al.}(2015){Zacharias}, {Finch}, {Subasavage},
  {Bredthauer}, {Crockett}, {Divittorio}, {Ferguson}, {Harris}, {Harris},
  {Henden}, {Kilian}, {Munn}, {Rafferty}, {Rhodes}, {Schultheiss}, {Tilleman},
  \& {Wieder}}]{Zacharias2015}
{Zacharias}, N., {Finch}, C., {Subasavage}, J., {et~al.} 2015, \aj, 150, 101,
  \dodoi{10.1088/0004-6256/150/4/101}

\bibitem[{{Zechmeister} {et~al.}(2013){Zechmeister}, {K{\"u}rster}, {Endl}, {Lo
  Curto}, {Hartman}, {Nilsson}, {Henning}, {Hatzes}, \&
  {Cochran}}]{Zechmeister2013}
{Zechmeister}, M., {K{\"u}rster}, M., {Endl}, M., {et~al.} 2013, \aap, 552,
  A78, \dodoi{10.1051/0004-6361/201116551}

\bibitem[{{Zechmeister} {et~al.}(2018){Zechmeister}, {Reiners}, {Amado},
  {Azzaro}, {Bauer}, {B{\'e}jar}, {Caballero}, {Guenther}, {Hagen}, {Jeffers},
  {Kaminski}, {K{\"u}rster}, {Launhardt}, {Montes}, {Morales}, {Quirrenbach},
  {Reffert}, {Ribas}, {Seifert}, {Tal-Or}, \& {Wolthoff}}]{Zechmeister2018}
{Zechmeister}, M., {Reiners}, A., {Amado}, P.~J., {et~al.} 2018, \aap, 609,
  A12, \dodoi{10.1051/0004-6361/201731483}

\bibitem[{{Zechmeister} {et~al.}(2020){Zechmeister}, {Reiners}, {Amado},
  {Azzaro}, {Bauer}, {B{\'e}jar}, {Caballero}, {Guenther}, {Hagen}, {Jeffers},
  {Kaminski}, {K{\"u}rster}, {Launhardt}, {Montes}, {Morales}, {Quirrenbach},
  {Reffert}, {Ribas}, {Seifert}, \& {Tal-Or}}]{Zechmeister2020}
---. 2020, {SERVAL: SpEctrum Radial Velocity AnaLyser}.
\newblock \doeprint{2006.011}

\end{thebibliography}
\end{document}